\input harvmac
\newcount\tocnew\tocnew=1\newcount\tocopage
\newbox\tocbox\newdimen\tocsize
\def\tocstrut{{\vrule height8.5pt depth3.5pt width0pt}} 
\def\tocref#1#2{\tocbanner#1               
    \global\setbox\tocbox=\vbox{
    \box\tocbox\vbox{
    \line{\tocstrut{#2#1}~\dotfill~\folio} 
    }}\tocsuffix}
\def\tocline#1{\tocbanner
    \global\setbox\tocbox=\vbox{
    \box\tocbox\vbox{
    \line{\tocstrut{#1}}
    }}\tocsuffix}
\def\tocbanner{\ifnum\tocnew=1\tocstart\fi
    \ifnum\tocnew=3\tocont\fi}
\def\tocsuffix{\ifdim\tocsize<\ht\tocbox\tocgen
    \global\tocnew=3\fi}
\def\tocstart{\global\tocsize=.95\vsize   
    \global\setbox\tocbox=\vbox{
    \centerline{\tocstrut\bf Table Of Contents} 
    \line{\tocstrut\hfil}                 
    }\global\tocnew=2}
\def\tocont{\global\setbox\tocbox=\vbox{
    \centerline{\tocstrut Table Of Contents (Continued)} 
    }\global\tocnew=2}
\def\tocgen{\ifnum\tocnew=1
    \message{No TOC entries found.}\else
    \tocopage=\pageno\pageno=0\message{(TOC}
    \shipout\box\tocbox\message{)}
    \pageno=\tocopage\global\tocnew=1\fi}

%
\message{S-Tables Macro v1.0, ACS, TAMU (RANHELP@VENUS.TAMU.EDU)}
%
%
\newhelp\stablestylehelp{You must choose a style between 0 and 3.}%
\newhelp\stablelinehelp{You should not use special hrules when
stretching
a table.}%
\newhelp\stablesmultiplehelp{You have tried to place an S-Table
inside another S-Table.  I would recommend not going on.}%
%
%
\newdimen\stablesthinline
\stablesthinline=0.4pt
\newdimen\stablesthickline
\stablesthickline=1pt
%
%
\newif\ifstablesborderthin
\stablesborderthinfalse
\newif\ifstablesinternalthin
\stablesinternalthintrue
\newif\ifstablesomit
\newif\ifstablemode
\newif\ifstablesright
\stablesrightfalse
%
%
\newdimen\stablesbaselineskip
\newdimen\stableslineskip
\newdimen\stableslineskiplimit
%
%
\newcount\stablesmode
\newcount\stableslines
\newcount\stablestemp
\stablestemp=3
\newcount\stablescount
\stablescount=0
\newcount\stableslinet
\stableslinet=0
%
%
%
\newcount\stablestyle
\stablestyle=0
%
%
\def\stablesleft{\quad\hfil}%
\def\stablesright{\hfil\quad}%
%
%
\catcode`\|=\active%
%
%
\newcount\stablestrutsize
\newbox\stablestrutbox
\setbox\stablestrutbox=\hbox{\vrule height10pt depth5pt width0pt}
\def\stablestrut{\relax\ifmmode%
                         \copy\stablestrutbox%
                       \else%
                         \unhcopy\stablestrutbox%
                       \fi}%
%
%
\newdimen\stablesborderwidth
\newdimen\stablesinternalwidth
\newdimen\stablesdummy
\newcount\stablesdummyc
\newif\ifstablesin
\stablesinfalse
%
%
%
%
%
\def\stablesadj{%
  \ifcase\stablestyle%
    \hbox to \hsize\bgroup\hss\vbox\bgroup%
  \or%
    \hbox to \hsize\bgroup\vbox\bgroup%
  \or%
    \hbox to \hsize\bgroup\hss\vbox\bgroup%
  \or%
    \hbox\bgroup\vbox\bgroup%
  \else%
    \errhelp=\stablestylehelp%
    \errmessage{Invalid style selected, using default}%
    \hbox to \hsize\bgroup\hss\vbox\bgroup%
  \fi}%
\def\stablesend{\egroup%
  \ifcase\stablestyle%
    \hss\egroup%
  \or%
    \hss\egroup%
  \or%
    \egroup%
  \or%
    \egroup%
  \else%
    \hss\egroup%
  \fi}%
\def\stablestart{%
  \ifstablesin%
    \errhelp=\stablesmultiplehelp%
    \errmessage{An S-Table cannot be placed within an S-Table!}%
  \fi
  \global\stablesintrue%
  \global\advance\stablescount by 1%
  \message{<S-Tables Generating Table \number\stablescount}%
  \begingroup%
  \stablestrutsize=\ht\stablestrutbox%
  \advance\stablestrutsize by \dp\stablestrutbox%
  \ifstablesborderthin%
    \stablesborderwidth=\stablesthinline%
  \else%
    \stablesborderwidth=\stablesthickline%
  \fi%
  \ifstablesinternalthin%
    \stablesinternalwidth=\stablesthinline%
  \else%
    \stablesinternalwidth=\stablesthickline%
  \fi%
  \tabskip=0pt%
  \stablesbaselineskip=\baselineskip%
  \stableslineskip=\lineskip%
  \stableslineskiplimit=\lineskiplimit%
  \offinterlineskip%
  \def\borderrule{\vrule width \stablesborderwidth}%
  \def\internalrule{\vrule width \stablesinternalwidth}%
  \def\thinline{\noalign{\hrule height \stablesthinline}}%
  \def\thickline{\noalign{\hrule height \stablesthickline}}%
  \def\trule{\omit\leaders\hrule height \stablesthinline\hfill}%
  \def\ttrule{\omit\leaders\hrule height \stablesthickline\hfill}%
  \def\tttrule##1{\omit\leaders\hrule height ##1\hfill}%
  \def\stablesel{&\omit\global\stablesmode=0%
    \global\advance\stableslines by 1\borderrule\hfil\cr}%
  \def\el{\stablesel&}%
  \def\elt{\stablesel\thinline&}%
  \def\eltt{\stablesel\thickline&}%
  \def\elttt##1{\stablesel\noalign{\hrule height ##1}&}%
  \def\elspec{&\omit\hfil\borderrule\cr\omit\borderrule&%
              \ifstablemode%
              \else%
                \errhelp=\stablelinehelp%
                \errmessage{Special ruling will not display properly}%
              \fi}%
  \def\stmultispan##1{\mscount=##1 \loop\ifnum\mscount>3
\stspan\repeat}%
  \def\stspan{\span\omit \advance\mscount by -1}%
  \def\multicolumn##1{\omit\multiply\stablestemp by ##1%
     \stmultispan{\stablestemp}%
     \advance\stablesmode by ##1%
     \advance\stablesmode by -1%
     \stablestemp=3}%
  \def\multirow##1{\stablesdummyc=##1\parindent=0pt\setbox0\hbox\bgroup%
    \aftergroup\emultirow\let\temp=}
  \def\emultirow{\setbox1\vbox to\stablesdummyc\stablestrutsize%
    {\hsize\wd0\vfil\box0\vfil}%
    \ht1=\ht\stablestrutbox%
    \dp1=\dp\stablestrutbox%
    \box1}%
%
  \def\stpar##1{\vtop\bgroup\hsize ##1%
     \baselineskip=\stablesbaselineskip%
     \lineskip=\stableslineskip%

\lineskiplimit=\stableslineskiplimit\bgroup\aftergroup\estpar\let\temp=}%
  \def\estpar{\vskip 6pt\egroup}%
  \def\stparrow##1##2{\stablesdummy=##2%
     \setbox0=\vtop to ##1\stablestrutsize\bgroup%
     \hsize\stablesdummy%
     \baselineskip=\stablesbaselineskip%
     \lineskip=\stableslineskip%
     \lineskiplimit=\stableslineskiplimit%
     \bgroup\vfil\aftergroup\estparrow%
     \let\temp=}%
  \def\estparrow{\vfil\egroup%
     \ht0=\ht\stablestrutbox%
     \dp0=\dp\stablestrutbox%
     \wd0=\stablesdummy%
     \box0}%
  \def|{\global\advance\stablesmode by 1&&&}%
  \def\|{\global\advance\stablesmode by 1&\omit\vrule width 0pt%
         \hfil&&}%
\def\vt{\global\advance\stablesmode
by 1&\omit\vrule width \stablesthinline%
          \hfil&&}%
  \def\vtt{\global\advance\stablesmode by 1&\omit\vrule width
\stablesthickline%
          \hfil&&}%
  \def\vttt##1{\global\advance\stablesmode by 1&\omit\vrule width ##1%
          \hfil&&}%
  \def\vtr{\global\advance\stablesmode by 1&\omit\hfil\vrule width%
           \stablesthinline&&}%
  \def\vttr{\global\advance\stablesmode by 1&\omit\hfil\vrule width%
            \stablesthickline&&}%
\def\vtttr##1{\global\advance\stablesmode
 by 1&\omit\hfil\vrule width ##1&&}%
  \stableslines=0%
  \stablesomitfalse}
\def\stablesdef{\bgroup\stablestrut\borderrule##\tabskip=0pt plus 1fil%
  &\stablesleft##\stablesright%
  &##\ifstablesright\hfill\fi\internalrule\ifstablesright\else\hfill\fi%
  \tabskip 0pt&&##\hfil\tabskip=0pt plus 1fil%
  &\stablesleft##\stablesright%
  &##\ifstablesright\hfill\fi\internalrule\ifstablesright\else\hfill\fi%
  \tabskip=0pt\cr%
  \ifstablesborderthin%
    \thinline%
  \else%
    \thickline%
  \fi&%
}%
\def\endtable{\advance\stableslines by 1\advance\stablesmode by 1%
   \message{- Rows: \number\stableslines, Columns:
\number\stablesmode>}%
   \stablesel%
   \ifstablesborderthin%
     \thinline%
   \else%
     \thickline%
   \fi%
   \egroup\stablesend%
\endgroup%
\global\stablesinfalse}
%

\overfullrule=0pt \abovedisplayskip=12pt plus 3pt minus 3pt
\belowdisplayskip=12pt plus 3pt minus 3pt

\noblackbox
\input epsf
\newcount\figno
\figno=0
\def\fig#1#2#3{
\par\begingroup\parindent=0pt\leftskip=1cm\rightskip=1cm\parindent=0pt
\baselineskip=11pt \global\advance\figno by 1 \midinsert
\epsfxsize=#3 \centerline{\epsfbox{#2}} \vskip 12pt
\centerline{{\bf Figure \the\figno:} #1}\par
\endinsert\endgroup\par}
\def\figlabel#1{\xdef#1{\the\figno}}

\def\IR{\relax{\rm I\kern-.18em R}}


\font\cmss=cmss10 \font\cmsss=cmss10 at 7pt
\def\rlx{\relax\leavevmode}
\def\inbar{\vrule height1.5ex width.4pt depth0pt}
\def\IC{\relax\,\hbox{$\inbar\kern-.3em{\rm C}$}}
\def\IN{\relax{\rm I\kern-.18em N}}
\def\IP{\relax{\rm I\kern-.18em P}}
\def\IR{\relax{\rm I\kern-.18em R}}
\def\ZZ{\rlx\leavevmode\ifmmode\mathchoice{\hbox{\cmss Z\kern-.4em Z}}
 {\hbox{\cmss Z\kern-.4em Z}}{\lower.9pt\hbox{\cmsss Z\kern-.36em Z}}
 {\lower1.2pt\hbox{\cmsss Z\kern-.36em Z}}\else{\cmss Z\kern-.4em
 Z}\fi}
\def\IZ{\relax\ifmmode\mathchoice
{\hbox{\cmss Z\kern-.4em Z}}{\hbox{\cmss Z\kern-.4em Z}}
{\lower.9pt\hbox{\cmsss Z\kern-.4em Z}} {\lower1.2pt\hbox{\cmsss
Z\kern-.4em Z}}\else{\cmss Z\kern-.4em Z}\fi}

\def\narrowplus{\kern -.04truein + \kern -.03truein}
\def\narrowminus{- \kern -.04truein}
\def\narrowminussub{\kern -.02truein - \kern -.01truein}

\def\O{{\cal O}}

\def\frac#1#2{{#1\over #2}}

\def\IZ{\relax\ifmmode\mathchoice
{\hbox{\cmss Z\kern-.4em Z}}{\hbox{\cmss Z\kern-.4em Z}}
{\lower.9pt\hbox{\cmsss Z\kern-.4em Z}} {\lower1.2pt\hbox{\cmsss
Z\kern-.4em Z}}\else{\cmss Z\kern-.4em Z}\fi}
\def\IB{\relax{\rm I\kern-.18em B}}
\def\IC{{\relax\hbox{$\inbar\kern-.3em{\rm C}$}}}
\def\ID{\relax{\rm I\kern-.18em D}}
\def\IE{\relax{\rm I\kern-.18em E}}
\def\IF{\relax{\rm I\kern-.18em F}}
\def\IG{\relax\hbox{$\inbar\kern-.3em{\rm G}$}}
\def\IGa{\relax\hbox{${\rm I}\kern-.18em\Gamma$}}
\def\IH{\relax{\rm I\kern-.18em H}}
\def\II{\relax{\rm I\kern-.18em I}}
\def\IK{\relax{\rm I\kern-.18em K}}
\def\IP{\relax{\rm I\kern-.18em P}}

\font\cmss=cmss10 \font\cmsss=cmss10 at 7pt
\def\IR{\relax{\rm I\kern-.18em R}}

\def\1{{\bf 1}}
\def\3{{\bf 3}}
\def\7{{\bf 7}}
\def\2{{\bf 2}}
\def\8{{\bf 8}}

\def\hat{\widehat}
\def\quabla{{\sqcap}\!\!\!\!{\sqcup}}

\def\o{\over}
\def\IP{\relax{\rm I\kern-.18em P}}

\def\cE{{\cal E}}
\def\cF{{\cal F}}
\def\cI{{\cal I}}
\def\O{{\cal O}}

\def\det{{\rm det}}
\def\Ext{{\rm Ext}}
\def\uExt{\underline{\rm Ext}}
\def\Hom{{\rm Hom}}
\def\uHom{\underline{{\rm Hom}}}

%

%
%
\def\eqnn#1{\xdef #1{(\secsym\the\meqno)}\writedef{#1\leftbracket#1}%
\global\advance\meqno by1\wrlabeL#1}
\def\eqna#1{\xdef #1##1{\hbox{$(\secsym\the\meqno##1)$}}
\writedef{#1\numbersign1\leftbracket#1{\numbersign1}}%
\global\advance\meqno by1\wrlabeL{#1$\{\}$}}
\def\eqn#1#2{\xdef #1{(\secsym\the\meqno)}\writedef{#1\leftbracket#1}%
\global\advance\meqno by1$$#2\eqno#1\eqlabeL#1$$}



\lref\vagudi{
 R.~Dijkgraaf, S.~Gukov, A.~Neitzke and C.~Vafa,
  ``Topological M-theory as unification of form theories of gravity,'' hep-th/0411073.
}

\lref\vafai{C.~Vafa,
``Superstrings and topological strings at large N,''
J.\ Math.\ Phys.\  {\bf 42}, 2798 (2001), hep-th/0008142.}

\lref\civ{F.~Cachazo, K.~A.~Intriligator and C.~Vafa,
``A large N duality via a geometric transition,''
Nucl.\ Phys.\ B {\bf 603}, 3 (2001), hep-th/0103067.}

\lref\nekrasovtop{L.~Baulieu, A.~S.~Losev and N.~A.~Nekrasov,
  ``Target space symmetries in topological theories. I,''
  JHEP {\bf 0202}, 021 (2002), hep-th/0106042.}

\lref\swncg{N.~Seiberg and E.~Witten,
  ``String theory and noncommutative geometry,''
  JHEP {\bf 9909}, 032 (1999), hep-th/9908142.}

\lref\cveticone{M.~Cvetic, G.~W.~Gibbons, H.~Lu and C.~N.~Pope,
 ``Cohomogeneity one manifolds of Spin(7) and G(2) holonomy,''
 Phys.\ Rev.\ D {\bf 65}, 106004 (2002), hep-th/0108245;
``M-theory conifolds,''
 Phys.\ Rev.\ Lett.\  {\bf 88}, 121602 (2002), hep-th/0112098;
``A G(2) unification of the deformed and resolved conifolds,''
Phys.\ Lett.\ B {\bf 534}, 172 (2002), hep-th/0112138.}

\lref\cvetictwo{M.~Cvetic, G.~W.~Gibbons, H.~Lu and C.~N.~Pope,
``New complete non-compact Spin(7) manifolds,''
 Nucl.\ Phys.\ B {\bf 620}, 29 (2002), hep-th/0103155.}

\lref\brand{A.~Brandhuber, J.~Gomis, S.~S.~Gubser and S.~Gukov,
``Gauge theory at large N and new G(2) holonomy metrics,''
Nucl.\ Phys.\ B {\bf 611}, 179 (2001), hep-th/0106034.}

\lref\gkp{
  S.~B.~Giddings, S.~Kachru and J.~Polchinski,
  ``Hierarchies from fluxes in string compactifications,''
  Phys.\ Rev.\ D {\bf 66}, 106006 (2002), hep-th/0105097.}

\lref\Kgukov{S.~Gukov, S.~Kachru, X.~Liu and L.~McAllister,
  ``Heterotic moduli stabilization with fractional Chern-Simons invariants,''
  Phys.\ Rev.\ D {\bf 69}, 086008 (2004)
  hep-th/0310159.}

\lref\fawad{S.~F.~Hassan,
``T-duality, space-time spinors and R-R fields in curved backgrounds,''
Nucl.\ Phys.\ B {\bf 568}, 145 (2000), hep-th/9907152.}

\lref\brandtwo{A.~Brandhuber,
``G(2) holonomy spaces from invariant three-forms,''
Nucl.\ Phys.\ B {\bf 629}, 393 (2002), hep-th/0112113.}

\lref\salamontwo{R. ~Bryant, S.~Salamon,
``On the construction of some complete metrics with exceptional
holonomy,'' Duke Math. J. {\bf 58} (1989) 829;
G.~W.~Gibbons, D.~N.~Page and C.~N.~Pope,
``Einstein Metrics On S**3 R**3 And R**4 Bundles,''
Commun.\ Math.\ Phys.\  {\bf 127}, 529 (1990).}

\lref\kovalev{A. Kovalev,
``Twisted connected sums and special Riemannian holonomy,''
math-DG/0012189.}

\lref\joyce{D. ~Joyce,
``Compact Riemannian 7 manifolds with holonomy $G_2$ I, J. Diff. Geom.
{\bf 43} (1996) 291;
II: J. Diff. Geom.{\bf 43} (1996) 329.}

\lref\grayone{A. Gray, L. Hervella,
``The sixteen classes of almost Hermitian manifolds and their linear
invariants,'' Ann. Mat. Pura Appl.(4) {\bf 123} (1980) 35.}

\lref\salamon{ S. Chiossi, S. Salamon,
``The intrinsic torsion of $SU(3)$ and $G_2$ structures,''
 Proc. conf. Differential Geometry Valencia 2001 [math.DG/0202282].}

\lref\gauntlett{J.~P.~Gauntlett, D.~Martelli and D.~Waldram,
``Superstrings with intrinsic torsion,'' hep-th/0302158.}

\lref\gukov{S.~Gukov,
``Solitons, superpotentials and calibrations,''
Nucl.\ Phys.\ B {\bf 574}, 169 (2000), hep-th/9911011.}

\lref\ach{B.~S.~Acharya and B.~Spence,
``Flux, supersymmetry and M theory on 7-manifolds,''
arXiv:hep-th/0007213.}

\lref\bw{C.~Beasley and E.~Witten,
 ``A note on fluxes and superpotentials in M-theory compactifications on
manifolds of G(2) holonomy,'' JHEP {\bf 0207}, 046 (2002),
hep-th/0203061.}

\lref\behr{K.~Behrndt and C.~Jeschek,
``Fluxes in M-theory on 7-manifolds and G structures,''
JHEP {\bf 0304}, 002 (2003), hep-th/0302047;
``Fluxes in M-theory on 7-manifolds: G-structures and
superpotential,'' hep-th/0311119.}

\lref\bertwo{K.~Behrndt and C.~Jeschek,
``Superpotentials from flux compactifications of M-theory,'', hep-th/0401019.}

\lref\gray{M. Fernandez and A. Gray, ``Riemannian manifolds with
structure group $G_2$,'' Ann. Mat. Pura. Appl. {\bf 32} (1982),
19-45.}

\lref\graytwo{M. Fernandez and L. Ugarte,
``Dolbeault cohomology for $G_2$ manifolds,''
Geom. Dedicata, {\bf 70} (1998) 57.}

\lref\ivan{T.~Friedrich and S.~Ivanov,
 ``Parallel spinors and connections with skew-symmetric torsion
  in string theory,'' math.dg/0102142;
T.~Friedrich and S.~Ivanov,
``Killing spinor equations in dimension 7 and geometry of integrable
 $G_2$-manifolds,'' math.dg/0112201.
P.~Ivanov and S.~Ivanov,
``SU(3)-instantons and $G_2$, Spin(7)-heterotic string solitons,'' math.dg/0312094.}

\lref\tp{G.~Papadopoulos and A.~A.~Tseytlin, ``Complex geometry of conifolds
and 5-brane wrapped on 2-sphere,''
Class.\ Quant.\ Grav.\  {\bf 18}, 1333 (2001).hep-th/0012034.}

\lref\papad{S.~Ivanov and G.~Papadopoulos,
  ``A no-go theorem for string warped compactifications,''
  Phys.\ Lett.\ B {\bf 497}, 309 (2001).}

\lref\smit{B.~de Wit, D.~J.~Smit and N.~D.~Hari Dass,
  ``Residual Supersymmetry Of Compactified D = 10 Supergravity,''
  Nucl.\ Phys.\ B {\bf 283}, 165 (1987).}

\lref\lust{G.~L.~Cardoso, G.~Curio, G.~Dall'Agata, D.~Lust, P.~Manousselis and G.~Zoupanos,
``Non-Kaehler string backgrounds and their five torsion classes,''
Nucl.\ Phys.\ B {\bf 652}, 5 (2003), hep-th/0211118.}

\lref\louis{S.~Gurrieri, J.~Louis, A.~Micu and D.~Waldram,
``Mirror symmetry in generalized Calabi-Yau compactifications,''
Nucl.\ Phys.\ B {\bf 654}, 61 (2003), hep-th/0211102.}

\lref\intril{
  K.~A.~Intriligator,
  ``'Integrating in' and exact superpotentials in 4-d,''
  Phys.\ Lett.\ B {\bf 336}, 409 (1994), hep-th/9407106.}

\lref\rstrom{A.~Strominger, ``Superstrings with torsion,'' Nucl.\
Phys.\ B {\bf 274}, 253 (1986).}

\lref\mal{J.~M.~Maldacena,
``The large N limit of superconformal field theories and supergravity,''
Adv.\ Theor.\ Math.\ Phys.\  {\bf 2}, 231 (1998)
[Int.\ J.\ Theor.\ Phys.\  {\bf 38}, 1113 (1999), hep-th/9711200.}

\lref\ks{I.~R.~Klebanov and M.~J.~Strassler,
``Supergravity and a confining gauge theory: Duality cascades and
chiSB-resolution of naked singularities,'' JHEP {\bf 0008}, 052 (2000), hep-th/0007191.}

\lref\klebts{I.~R.~Klebanov and A.~A.~Tseytlin,
  ``Gravity duals of supersymmetric SU(N) x SU(N+M) gauge theories,''
  Nucl.\ Phys.\ B {\bf 578}, 123 (2000), hep-th/0002159.}

\lref\mn{J.~M.~Maldacena and C.~Nunez,
``Towards the large N limit of pure N = 1 super Yang Mills,''
Phys.\ Rev.\ Lett.\  {\bf 86}, 588 (2001), hep-th/0008001.}

\lref\gcon{S.~Ferrara, L.~Girardello and H.~P.~Nilles,
  ``Breakdown Of Local Supersymmetry Through Gauge Fermion Condensates,''
  Phys.\ Lett.\ B {\bf 125}, 457 (1983);
M.~Dine, R.~Rohm, N.~Seiberg and E.~Witten,
  ``Gluino Condensation In Superstring Models,''
  Phys.\ Lett.\ B {\bf 156}, 55 (1985);
J.~P.~Derendinger, L.~E.~Ibanez and H.~P.~Nilles,
  ``On The Low-Energy D = 4, ${\cal N}=1$ Supergravity Theory
Extracted From The D = 10,
  ${\cal N}=1$ Superstring,''
  Phys.\ Lett.\ B {\bf 155}, 65 (1985).}

\lref\dabbie{A.~Sen,
  ``Orbifolds of M-Theory and String Theory,''
  Mod.\ Phys.\ Lett.\ A {\bf 11}, 1339 (1996), hep-th/9603113;
``Duality and Orbifolds,''

  Nucl.\ Phys.\ B {\bf 474}, 361 (1996), hep-th/9604070;
A.~Dabholkar,
  ``Lectures on orientifolds and duality,''
{\it Trieste 1997, High energy physics and cosmology 128-191},
 hep-th/9804208.}

\lref\dvo{
  R.~Dijkgraaf and C.~Vafa,
  ``A perturbative window into non-perturbative physics,''
  hep-th/0208048.}

\lref\dvt{
  R.~Dijkgraaf and C.~Vafa,
  ``Matrix models, topological strings, and supersymmetric gauge theories,''
  Nucl.\ Phys.\ B {\bf 644}, 3 (2002),hep-th/0206255.}

\lref\vafai{C.~Vafa,
``Superstrings and topological strings at large N,''
J.\ Math.\ Phys.\  {\bf 42}, 2798 (2001), hep-th/0008142.}

\lref\civ{F.~Cachazo, K.~A.~Intriligator and C.~Vafa,
``A large N duality via a geometric transition,''
Nucl.\ Phys.\ B {\bf 603}, 3 (2001), hep-th/0103067.}

\lref\syz{A.~Strominger, S.~T.~Yau and E.~Zaslow,
``Mirror symmetry is T-duality,''
Nucl.\ Phys.\ B {\bf 479}, 243 (1996), hep-th/9606040.}

\lref\tduality{E.~Bergshoeff, C.~M.~Hull and T.~Ortin,
``Duality in the type II superstring effective action,''
Nucl.\ Phys.\ B {\bf 451}, 547 (1995), hep-th/9504081;
P.~Meessen and T.~Ortin,
``An Sl(2,Z) multiplet of nine-dimensional type II supergravity theories,''
Nucl.\ Phys.\ B {\bf 541}, 195 (1999), hep-th/9806120.}

\lref\eot{J.~D.~Edelstein, K.~Oh and R.~Tatar,
``Orientifold,
 geometric transition and large N duality for SO/Sp gauge  theories,''
JHEP {\bf 0105}, 009 (2001), hep-th/0104037.}

\lref\dotu{K.~Dasgupta, K.~Oh and R.~Tatar, {``Geometric
transition, large N dualities and MQCD dynamics,''} Nucl.\ Phys.\
B {\bf 610}, 331 (2001), hep-th/0105066; {``Open/closed string
dualities and Seiberg duality from geometric transitions in
M-theory,''} JHEP {\bf 0208}, 026 (2002), hep-th/0106040.}

\lref\dotd{K.~Dasgupta, K.~h.~Oh, J.~Park and R.~Tatar, ``Geometric
transition versus cascading solution,'' JHEP {\bf 0201}, 031
(2002), hep-th/0110050.}

\lref\ohta{K.~Ohta and T.~Yokono,
``Deformation of conifold and intersecting branes,''
JHEP {\bf 0002}, 023 (2000), hep-th/9912266.}

\lref\dott{K.~h.~Oh and R.~Tatar,
``Duality and confinement
in N = 1 supersymmetric theories from geometric  transitions,''
Adv.\ Theor.\ Math.\ Phys.\  {\bf 6}, 141 (2003), hep-th/0112040.}

\lref\edelstein{J.~D.~Edelstein and C.~Nunez,
``D6 branes and M-theory geometrical transitions from gauged  supergravity,''
JHEP {\bf 0104}, 028 (2001), hep-th/0103167.}

\lref\chsw{P.~Candelas, G.~T.~Horowitz, A.~Strominger and E.~Witten,
  ``Vacuum Configurations For Superstrings,''
  Nucl.\ Phys.\ B {\bf 258}, 46 (1985).}

\lref\candelas{P.~Candelas and X.~C.~de la Ossa, ``Comments on
conifolds,'' Nucl.\ Phys.\ B {\bf 342}, 246 (1990).}

\lref\minakas{
   P. Kaste, Ruben Minasian, M. Petrini, A. Tomasiello
  ``Nontrivial RR two-form field strength and SU(3)-structure''
   Fortsch.Phys. {\bf 51} 764 (2003), hep-th/0412187.
}

\lref\minasianone{R.~Minasian and D.~Tsimpis,
``Hopf reductions, fluxes and branes,''
Nucl.\ Phys.\ B {\bf 613}, 127 (2001), hep-th/0106266.}

\lref\imamura{Y.~Imamura,
``Born-Infeld action and Chern-Simons term from Kaluza-Klein monopole in
M-theory,''
Phys.\ Lett.\ B {\bf 414}, 242 (1997), hep-th/9706144;
A.~Sen,
``Dynamics of multiple Kaluza-Klein monopoles in M and string theory,''
Adv.\ Theor.\ Math.\ Phys.\  {\bf 1}, 115 (1998), hep-th/9707042;
``A note on enhanced gauge symmetries in M- and string theory,''
JHEP {\bf 9709}, 001 (1997), hep-th/9707123.}

\lref\robbins{K.~Dasgupta, G.~Rajesh, D.~Robbins and S.~Sethi,
``Time-dependent warping, fluxes, and NCYM,''
JHEP {\bf 0303}, 041 (2003), hep-th/0302049;
K.~Dasgupta and M.~Shmakova,
``On branes and oriented B-fields,''
Nucl.\ Phys.\ B {\bf 675}, 205 (2003), hep-th/0306030.}

\lref\svw{S.~Sethi, C.~Vafa and E.~Witten,
``Constraints on low-dimensional string compactifications,''
Nucl.\ Phys.\ B {\bf 480}, 213 (1996), hep-th/9606122;
K.~Dasgupta and S.~Mukhi,
``A note on low-dimensional string compactifications,''
Phys.\ Lett.\ B {\bf 398}, 285 (1997), hep-th/9612188.}

\lref\kachruone{S.~Kachru, M.~B.~Schulz, P.~K.~Tripathy and S.~P.~Trivedi,
``New supersymmetric string compactifications,''
JHEP {\bf 0303}, 061 (2003), hep-th/0211182;
S.~Kachru, M.~B.~Schulz and S.~Trivedi,
``Moduli stabilization from fluxes in a simple IIB orientifold,''
JHEP {\bf 0310}, 007 (2003), hep-th/0201028.}

\lref\hitchin{N. ~Hitchin,
``Stable forms and special metrics'',
Contemp. Math., {\bf 288}, Amer. Math. Soc. (2000).}

\lref\giveon{S.~S.~Gubser,
 ``Supersymmetry and F-theory realization of the deformed conifold with
three-form flux,'' hep-th/0010010;
A.~Giveon, A.~Kehagias and H.~Partouche,
``Geometric transitions, brane dynamics and gauge theories,''
JHEP {\bf 0112}, 021 (2001), hep-th/0110115.}

\lref\bonan{E. Bonan,
``Sur le varietes remanniennes a groupe d'holonomie $G_2$ ou Spin(7),''
C. R. Acad. Sci. paris {\bf 262} (1966) 127.}



\lref\amv{M.~Atiyah, J.~M.~Maldacena and C.~Vafa,
``An M-theory flop as a large N duality,''
J.\ Math.\ Phys.\  {\bf 42}, 3209 (2001), hep-th/0011256.}

\lref\realm{S.~Alexander, K.~Becker, M.~Becker, K.~Dasgupta,
A.~Knauf and R.~Tatar,
  ``In the realm of the geometric transitions,''
  Nucl.\ Phys.\ B {\bf 704}, 231 (2005), hep-th/0408192.}

\lref\cvetic{M.~Cvetic, G.~W.~Gibbons, H.~Lu and C.~N.~Pope,
  ``Ricci-flat metrics, harmonic forms and brane resolutions,''
  Commun.\ Math.\ Phys.\  {\bf 232}, 457 (2003), hep-th/0012011.}

\lref\hulltown{C.~M.~Hull and P.~K.~Townsend,
  ``Finiteness And Conformal Invariance In Nonlinear Sigma Models,''
  Nucl.\ Phys.\ B {\bf 274}, 349 (1986);
``The Two Loop Beta Function For Sigma Models With Torsion,''
  Phys.\ Lett.\ B {\bf 191}, 115 (1987).}

\lref\hullwit{C.~M.~Hull and E.~Witten,
  ``Supersymmetric Sigma Models And The Heterotic String,''
  Phys.\ Lett.\ B {\bf 160}, 398 (1985).}

\lref\gross{D.~J.~Gross, J.~A.~Harvey, E.~J.~Martinec and R.~Rohm,
  ``The Heterotic String,''
  Phys.\ Rev.\ Lett.\  {\bf 54}, 502 (1985);
``Heterotic String Theory. 1. The Free Heterotic String,''
  Nucl.\ Phys.\ B {\bf 256}, 253 (1985);
``Heterotic String Theory. 2. The Interacting Heterotic String,''
  Nucl.\ Phys.\ B {\bf 267}, 75 (1986).}

\lref\syz{A.~Strominger, S.~T.~Yau and E.~Zaslow,
``Mirror symmetry is T-duality,''
Nucl.\ Phys.\ B {\bf 479}, 243 (1996), hep-th/9606040.}

\lref\tduality{E.~Bergshoeff, C.~M.~Hull and T.~Ortin,
``Duality in the type II superstring effective action,''
Nucl.\ Phys.\ B {\bf 451}, 547 (1995), hep-th/9504081;
P.~Meessen and T.~Ortin,
``An Sl(2,Z) multiplet of nine-dimensional type II supergravity theories,''
Nucl.\ Phys.\ B {\bf 541}, 195 (1999), hep-th/9806120.}

\lref\adoptone{J.~D.~Edelstein, K.~Oh and R.~Tatar,
``Orientifold, geometric transition and large N duality for SO/Sp
gauge theories,'' JHEP {\bf 0105}, 009 (2001), hep-th/0104037.}

\lref\grifhar{P.~Griffiths and J.~Harris, \underbar{Principles
of Algebraic Geometry}, Wiley, New York 1978.}

\lref\hart{R.~Hartshorne, \underbar{Algebraic Geometry}, Springer-Verlag,
Berlin, 1977.}

\lref\cortismith{A.~Corti and I.~Smith, ``Conifold transitions and Mori
theory'', math.SG/0501043.}

\lref\ohta{K.~Ohta and T.~Yokono,
``Deformation of conifold and intersecting branes,''
JHEP {\bf 0002}, 023 (2000), hep-th/9912266.}

\lref\adoptfo{K.~h.~Oh and R.~Tatar, ``Duality and confinement in
N = 1 supersymmetric theories from geometric transitions,'' Adv.\
Theor.\ Math.\ Phys.\  {\bf 6}, 141 (2003), hep-th/0112040.}

\lref\edelstein{J.~D.~Edelstein and C.~Nunez, ``D6 branes and
M-theory geometrical transitions from gauged supergravity,'' JHEP
{\bf 0104}, 028 (2001), hep-th/0103167.}

\lref\candelas{P.~Candelas and X.~C.~de la Ossa,
``Comments On Conifolds,''
Nucl.\ Phys.\ B {\bf 342}, 246 (1990).}

\lref\tsimpis{R.~Minasian and D.~Tsimpis,
  ``On the geometry of non-trivially embedded branes,''
  Nucl.\ Phys.\ B {\bf 572}, 499 (2000), hep-th/9911042.}

\lref\minasianone{R.~Minasian and D.~Tsimpis,
``Hopf reductions, fluxes and branes,''
Nucl.\ Phys.\ B {\bf 613}, 127 (2001), hep-th/0106266.}

\lref\pandoz{L.~A.~Pando Zayas and A.~A.~Tseytlin,
``3-branes on resolved conifold,''
JHEP {\bf 0011}, 028 (2000), hep-th/0010088.}

\lref\rBB{K.~Becker and M.~Becker, ``M-Theory on
eight-manifolds,'' Nucl.\ Phys.\ B {\bf 477}, 155 (1996),
hep-th/9605053.}

\lref\bbdgs{K.~Becker, M.~Becker, P.~S.~Green, K.~Dasgupta and
E.~Sharpe, ``Compactifications of heterotic strings on
non-K\"ahler complex manifolds. II,'' Nucl.\ Phys.\ B {\bf 678},
19 (2004), hep-th/0310058.}

\lref\bbdg{K.~Becker, M.~Becker, K.~Dasgupta and P.~S.~Green,
``Compactifications of heterotic theory on non-K\"ahler complex
manifolds. I,'' JHEP {\bf 0304}, 007 (2003), hep-th/0301161.}

\lref\hull{C.~M.~Hull,
  ``Compactifications Of The Heterotic Superstring,''
  Phys.\ Lett.\ B {\bf 178}, 357 (1986); C.~M.~Hull and E.~Witten,
  ``Supersymmetric Sigma Models And The Heterotic String,''
  Phys.\ Lett.\ B {\bf 160}, 398 (1985); C.~M.~Hull,
  ``Superstring Compactifications With Torsion And Space-Time Supersymmetry,''
Print-86-0251 (CAMBRIDGE), Published in Turin Superunif.1985:347.}

\lref\gates{J.~M.~Gates, C.~M.~Hull and M.~Rocek,
  ``Twisted Multiplets And New Supersymmetric Nonlinear Sigma Models,''
  Nucl.\ Phys.\ B {\bf 248}, 157 (1984).}

\lref\bbdp{K.~Becker, M.~Becker, K.~Dasgupta and S.~Prokushkin,
  ``Properties of heterotic vacua from superpotentials,''
  Nucl.\ Phys.\ B {\bf 666}, 144 (2003), hep-th/0304001.}

\lref\GP{E.~Goldstein and S.~Prokushkin,

 ``Geometric model for complex non-K\"ahler manifolds with SU(3) structure,''
  Commun.\ Math.\ Phys.\  {\bf 251}, 65 (2004), hep-th/0212307.}

\lref\townsend{P.~K.~Townsend,
``D-branes from M-branes,''
Phys.\ Lett.\ B {\bf 373}, 68 (1996), hep-th/9512062.}

\lref\sav{K.~Dasgupta, G.~Rajesh and S.~Sethi,
``M theory, orientifolds and G-flux,''
JHEP {\bf 9908}, 023 (1999), hep-th/9908088.}

\lref\beckerD{K.~Becker and K.~Dasgupta,
``Heterotic strings with torsion,''
JHEP {\bf 0211}, 006 (2002), hep-th/0209077.}

\lref\lisheng{K.~Becker and L.~S.~Tseng,
  ``Heterotic flux compactifications and their moduli,'' hep-th/0509131.}

\lref\ks{I.~R.~Klebanov and M.~J.~Strassler,
 ``Supergravity and a confining gauge theory: Duality cascades
  and $\chi_{SB}$-resolution of naked singularities,''
JHEP {\bf 0008}, 052 (2000), hep-th/0007191.}

\lref\radu{I.~Bena, R.~Roiban and R.~Tatar,
  ``Baryons, boundaries and matrix models,''
  Nucl.\ Phys.\ B {\bf 679}, 168 (2004), hep-th/0211271;~R.~Roiban, R.~Tatar and J.~Walcher,
``Massless flavor in geometry and matrix models,''
Nucl.\ Phys.\ B {\bf 665}, 211 (2003), hep-th/0301217;~K.~Landsteiner, C.~I.~Lazaroiu and R.~Tatar,
``(Anti)symmetric matter and superpotentials from IIB orientifolds,''
JHEP {\bf 0311}, 044 (2003), hep-th/0306236;~``Chiral field Theories, Konishi Anomalies and Matrix Models'',
JHEP {\bf 0402}, 044 (2004), hep-th/0307182,~``Chiral field theories from conifolds,''
JHEP {\bf 0311}, 057 (2003), hep-th/0310052;~K.~Landsteiner, C.~I.~Lazaroiu and R.~Tatar,
``Puzzles for matrix models of chiral field theories,'' hep-th/0311103.}

\lref\gv{R.~Gopakumar and C.~Vafa,
``On the gauge theory/geometry correspondence,''
Adv.\ Theor.\ Math.\ Phys.\  {\bf 3}, 1415 (1999), hep-th/9811131.}

\lref\dvu{R.~Dijkgraaf and C.~Vafa,
``Matrix models, topological strings, and supersymmetric gauge theories,''
Nucl.\ Phys.\ B {\bf 644}, 3 (2002), hep-th/0206255.}

\lref\ps{J.~Polchinski and M.~J.~Strassler, ``The String Dual of a
Confining Four-Dimensional Gauge Theory ,'' hep-th/0003136.}

\lref\susskind{L.~Susskind,
``The anthropic landscape of string theory,'' hep-th/0302219.}

\lref\banks{T.~Banks, M.~Dine and E.~Gorbatov,
``Is there a string theory landscape?,'' hep-th/0309170.}

\lref\hori{K.~Hori and C.~Vafa,
``Mirror symmetry,'' hep-th/0002222;
M.~Aganagic, A.~Klemm and C.~Vafa,
``Disk instantons, mirror symmetry and the duality web,''
Z.\ Naturforsch.\ A {\bf 57}, 1 (2002), hep-th/0105045;
M.~Aganagic, A.~Klemm, M.~Marino and C.~Vafa,
``Matrix model as a mirror of Chern-Simons theory,''
JHEP {\bf 0402}, 010 (2004), hep-th/0211098.}

\lref\agavafa{M.~Aganagic and C.~Vafa,
  ``G(2) manifolds, mirror symmetry and geometric engineering,''
 hep-th/0110171.}

\lref\karch{M.~Aganagic, A.~Karch, D.~Lust and A.~Miemiec,
``Mirror symmetries for brane configurations and branes at singularities,''
Nucl.\ Phys.\ B {\bf 569}, 277 (2000), hep-th/9903093.}

\lref\ksschub{S.~Katz and S.A.\ Stromme, ``Schubert: a maple package
for intersection theory'', http://www.mi.uib.no/schubert/}

\lref\dmconi{K.~Dasgupta and S.~Mukhi,
``Brane constructions, conifolds and M-theory,''
Nucl.\ Phys.\ B {\bf 551}, 204 (1999), hep-th/9811139.}

\lref\gtone{M.~Becker, K.~Dasgupta, A.~Knauf and R.~Tatar,
  ``Geometric transitions, flops and non-K\"ahler manifolds. I,''
  Nucl.\ Phys.\ B {\bf 702}, 207 (2004), hep-th/0403288.}

\lref\toappear{S.~Alexander, K.~Becker, M.~Becker, K.~Dasgupta,
A.~Knauf, R.~Tatar,
``In the realm of the geometric transitions,'' hep-th/0408192.}

\lref\dvd{R.~Dijkgraaf and C.~Vafa,
``A perturbative window into non-perturbative physics,'' hep-th/0208048.}

\lref\civu{F.~Cachazo, S.~Katz and C.~Vafa,
``Geometric transitions and N = 1 quiver theories,'' hep-th/0108120.}

\lref\civd{F.~Cachazo, B.~Fiol, K.~A.~Intriligator, S.~Katz and C.~Vafa,
``A geometric unification of dualities,'' Nucl.\ Phys.\ B {\bf 628}, 3 (2002),
hep-th/0110028.}

\lref\radu{R.~Roiban, R.~Tatar and J.~Walcher,
``Massless flavor in geometry and matrix models,''
Nucl.\ Phys.\ B {\bf 665}, 211 (2003), hep-th/0301217.}

\lref\radd{K.~Landsteiner, C.~I.~Lazaroiu and R.~Tatar,
``(Anti)symmetric matter and superpotentials from IIB orientifolds,''
JHEP {\bf 0311}, 044 (2003), hep-th/0306236.}

\lref\radt{K.~Landsteiner, C.~I.~Lazaroiu and R.~Tatar,
``Chiral field theories from conifolds,''
JHEP {\bf 0311}, 057 (2003), hep-th/0310052.}

\lref\radp{K.~Landsteiner, C.~I.~Lazaroiu and R.~Tatar,
``Puzzles for matrix models of chiral field theories,'' hep-th/0311103.}

\lref\ber{M.~Bershadsky, S.~Cecotti, H.~Ooguri and C.~Vafa,
 ``Kodaira-Spencer theory of gravity and exact results for quantum string amplitudes,''
Commun.\ Math.\ Phys.\  {\bf 165}, 311 (1994), hep-th/9309140.}

\lref\bismut{J. M. Bismut,
``A local index theorem for non-K\"ahler manifolds,''
Math. Ann. {\bf 284} (1989) 681.}

\lref\monar{F. Cabrera, M. Monar and A. Swann,
``Classification of $G_2$ structures,''
J. London Math. Soc. {\bf 53} (1996) 98;
F. Cabrera,
``On Riemannian manifolds with $G_2$ structure,''
Bolletino UMI A {\bf 10} (1996) 98.}

\lref\kath{Th. Friedrich and I. Kath,
``7-dimensional compact Riemannian manifolds with killing spinors,''
Comm. Math. Phys. {\bf 133} (1990) 543;
Th. Friedrich, I. Kath, A. Moroianu and U. Semmelmann,
``On nearly parallel $G_2$ structures,'' J. geom. Phys. {\bf 23} (1997) 259;
S. Salamon,
``Riemannian geometry and holonomy groups,''
Pitman Res. Notes Math. Ser., {\bf 201} (1989);
V. Apostolov and S. Salamon,
``K\"ahler reduction of metrics with holonomy $G_2$,''
math-DG/0303197.}

\lref\ooguri{H.~Ooguri and C.~Vafa,
``The C-deformation of gluino and non-planar diagrams,''
Adv.\ Theor.\ Math.\ Phys.\  {\bf 7}, 53 (2003), hep-th/0302109;
``Gravity induced C-deformation,''
Adv.\ Theor.\ Math.\ Phys.\  {\bf 7}, 405 (2004), hep-th/0303063.}

\lref\tv{T.~R.~Taylor and C.~Vafa,
``RR flux on Calabi-Yau and partial supersymmetry breaking,''
Phys.\ Lett.\ B {\bf 474}, 130 (2000), hep-th/9912152.}

\lref\wittencoho{E.~Witten,
  ``Topological Sigma Models,''
  Commun.\ Math.\ Phys.\  {\bf 118}, 411 (1988);
``Mirror manifolds and topological field theory,''
 hep-th/9112056; ``Topological Quantum Field Theory,''
  Commun.\ Math.\ Phys.\  {\bf 117}, 353 (1988).}

\lref\toprev{M.~Marino,
  ``Les Houches lectures on matrix models and topological strings,''
  hep-th/0410165; A.~Neitzke and C.~Vafa,
  ``Topological strings and their physical applications,''
  hep-th/0410178; M.~Vonk,
  ``A mini-course on topological strings,'' hep-th/0504147.}

\lref\hitchin{N.~Hitchin,
  ``Generalized Calabi-Yau manifolds,''
  Quart.\ J.\ Math.\ Oxford Ser.\  {\bf 54}, 281 (2003) math.dg/0209099.}

\lref\gualtieri{M.~Gualtieri,
``Generalised Complex Geometry,''
math.DG/0401221.}

\lref\yaufu{J.~X.~Fu and S.~T.~Yau,
  ``Existence of supersymmetric Hermitian metrics with torsion on non-Kaehler
  manifolds,'' hep-th/0509028.}

\lref\yauli{J.~Li and S.~T.~Yau,
  ``The existence of supersymmetric string theory with torsion,''
  hep-th/0411136; ``Hermitian Yang-Mills Connection On Nonkahler Manifolds,''
in {\it San Diego 1986, Proceedings, Mathematical Aspects of String Theory} 560-573.}

\lref\kapuli{A.~Kapustin and Y.~Li,
  ``Topological sigma-models with H-flux and twisted generalized complex
  manifolds,''
  hep-th/0407249.}

\lref\sensethi{A.~Sen and S.~Sethi,
  ``The mirror transform of type I vacua in six dimensions,''
  Nucl.\ Phys.\ B {\bf 499}, 45 (1997)
  hep-th/9703157; J.~de Boer, R.~Dijkgraaf, K.~Hori,
A.~Keurentjes, J.~Morgan, D.~R.~Morrison and S.~Sethi,
  ``Triples, fluxes, and strings,''
  Adv.\ Theor.\ Math.\ Phys.\  {\bf 4}, 995 (2002)
 hep-th/0103170; D.~R.~Morrison and S.~Sethi,
  ``Novel type I compactifications,''
  JHEP {\bf 0201}, 032 (2002)
  hep-th/0109197.}

\lref\wittentop{E.~Witten,
  ``Mirror manifolds and topological field theory,''
  hep-th/9112056;  In Yau, S.T. (ed.): {\it Mirror symmetry I} 121.}

\lref\malikov{F. ~Malikov, V. ~Schechtman, A. ~Vaintrob,
``Chiral de Rham complex'', math.AG/9803041.}

\lref\schect{ F. ~ Malikov, V. ~Schechtman,
``Chiral de Rham complex. II,''
math.AG/9901065; ``Chiral Poincar\'e duality,'' math.AG/9905008;
V.~ Gorbounov, F. ~Malikov, V. ~Schechtman,
``Gerbes of chiral differential operators,'' math.AG/9906117.}

\lref\zerotwo{J.~Distler,
  ``Resurrecting (2,0) Compactifications,''
  Phys.\ Lett.\ B {\bf 188}, 431 (1987);
J.~Distler and B.~R.~Greene,
  ``Aspects Of (2,0) String Compactifications,''
  Nucl.\ Phys.\ B {\bf 304}, 1 (1988);
J.~Distler and S.~Kachru,
  ``(0,2) Landau-Ginzburg theory,''
  Nucl.\ Phys.\ B {\bf 413}, 213 (1994), hep-th/9309110;
J.~Distler and S.~Kachru,
  ``Duality of (0,2) string vacua,''
  Nucl.\ Phys.\ B {\bf 442}, 64 (1995), hep-th/9501111;
T.~M.~Chiang, J.~Distler and B.~R.~Greene,
  ``Some features of (0,2) moduli space,''
  Nucl.\ Phys.\ B {\bf 496}, 590 (1997), hep-th/9702030.}

\lref\reczt{A.~Adams, A.~Basu and S.~Sethi,
  ``(0,2) duality,''
  Adv.\ Theor.\ Math.\ Phys.\  {\bf 7}, 865 (2004), hep-th/0309226;
S.~Katz and E.~Sharpe,
  ``Notes on certain (0,2) correlation functions,'' hep-th/0406226.}

\lref\wittwo{E.~Witten,
  ``Two-dimensional models with (0,2) supersymmetry: Perturbative aspects,''
  hep-th/0504078.}

\lref\zumino{B.~Zumino,
  ``Supersymmetry And Kahler Manifolds,''
  Phys.\ Lett.\ B {\bf 87}, 203 (1979).}

\lref\horwit{P.~Horava and E.~Witten,
  ``Heterotic and type I string dynamics from eleven dimensions,''
  Nucl.\ Phys.\ B {\bf 460}, 506 (1996), hep-th/9510209;
``Eleven-Dimensional Supergravity on a Manifold with Boundary,''
  Nucl.\ Phys.\ B {\bf 475}, 94 (1996), hep-th/9603142.}

\lref\dmone{K.~Dasgupta and S.~Mukhi,
  ``Orbifolds of M-theory,''
  Nucl.\ Phys.\ B {\bf 465}, 399 (1996), hep-th/9512196;
E.~Witten,
  ``Five-branes and M-theory on an orbifold,''
  Nucl.\ Phys.\ B {\bf 463}, 383 (1996), hep-th/9512219.}

\lref\kimn{J.~P.~Gauntlett, N.~Kim, D.~Martelli and D.~Waldram,
  ``Wrapped fivebranes and ${\cal N} = 2$ super Yang-Mills theory,''
  Phys.\ Rev.\ D {\bf 64}, 106008 (2001), hep-th/0106117;
F.~Bigazzi, A.~L.~Cotrone and A.~Zaffaroni,
  ``${\cal N} = 2$ gauge theories from wrapped five-branes,''
  Phys.\ Lett.\ B {\bf 519}, 269 (2001), hep-th/0106160;
R.~Apreda, F.~Bigazzi, A.~L.~Cotrone, M.~Petrini and A.~Zaffaroni,
  ``Some comments on ${\cal N} = 1$ gauge theories from wrapped branes,''
  Phys.\ Lett.\ B {\bf 536}, 161 (2002), hep-th/0112236;
P.~Di Vecchia, A.~Lerda and P.~Merlatti,
  ``${\cal N} = 1$ and ${\cal N} = 2$ super Yang-Mills theories from wrapped branes,''
  Nucl.\ Phys.\ B {\bf 646}, 43 (2002), hep-th/0205204;
F.~Bigazzi, A.~L.~Cotrone, M.~Petrini and A.~Zaffaroni,
  ``Supergravity duals of supersymmetric four dimensional gauge theories,''
  Riv.\ Nuovo Cim.\  {\bf 25N12}, 1 (2002), hep-th/0303191.}

\lref\axell{G.~Curio and A.~Krause,
  ``Four-flux and warped heterotic M-theory compactifications,''
  Nucl.\ Phys.\ B {\bf 602}, 172 (2001), hep-th/0012152.}

\lref\axelk{G.~Curio and A.~Krause,
  ``G-fluxes and non-perturbative stabilisation of heterotic M-theory,''
  Nucl.\ Phys.\ B {\bf 643}, 131 (2002), hep-th/0108220;
M.~Becker, G.~Curio and A.~Krause,
  ``De Sitter vacua from heterotic M-theory,''
  Nucl.\ Phys.\ B {\bf 693}, 223 (2004), hep-th/0403027;
G.~Curio, A.~Krause and D.~Lust,
  ``Moduli stabilization in the heterotic / IIB discretuum,'' hep-th/0502168.}

\lref\sezgin{A.~Salam and E.~Sezgin,
  ``$SO(4)$ Gauging Of N=2 Supergravity In Seven-Dimensions,''
  Phys.\ Lett.\ B {\bf 126}, 295 (1983).}

\lref\buchkov{E.~I.~Buchbinder and B.~A.~Ovrut,
  ``Vacuum stability in heterotic M-theory,''
  Phys.\ Rev.\ D {\bf 69}, 086010 (2004), hep-th/0310112.}

\lref\jefgreg{R.~Gregory, J.~A.~Harvey and G.~W.~Moore,
  ``Unwinding strings and T-duality of Kaluza-Klein and H-monopoles,''
  Adv.\ Theor.\ Math.\ Phys.\  {\bf 1}, 283 (1997), hep-th/9708086.}

\lref\senbanks{A.~Sen,
  ``F-theory and Orientifolds,''
  Nucl.\ Phys.\ B {\bf 475}, 562 (1996), hep-th/9605150;
T.~Banks, M.~R.~Douglas and N.~Seiberg,
  ``Probing F-theory with branes,''
  Phys.\ Lett.\ B {\bf 387}, 278 (1996), hep-th/9605199.}

\lref\senF{A.~Sen,
  ``Orientifold limit of F-theory vacua,''
  Phys.\ Rev.\ D {\bf 55}, 7345 (1997), hep-th/9702165;
  ``Orientifold limit of F-theory vacua,''
  Nucl.\ Phys.\ Proc.\ Suppl.\  {\bf 68}, 92 (1998), hep-th/9709159.}

\lref\ouyang{P.~Ouyang,
  ``Holomorphic D7-branes and flavored N = 1 gauge theories,''
  Nucl.\ Phys.\ B {\bf 699}, 207 (2004), hep-th/0311084.}

\lref\orione{S.~Chakravarty, K.~Dasgupta, O.~J.~Ganor and G.~Rajesh,
  ``Pinned branes and new non Lorentz invariant theories,''
  Nucl.\ Phys.\ B {\bf 587}, 228 (2000), hep-th/0002175.}

\lref\difuli{A.~Bergman and O.~J.~Ganor,
  ``Dipoles, twists and noncommutative gauge theory,''
  JHEP {\bf 0010}, 018 (2000), hep-th/0008030;
K.~Dasgupta, O.~J.~Ganor and G.~Rajesh,
  ``Vector deformations of N = 4 super-Yang-Mills theory, pinned branes,  and
  arched strings,''
  JHEP {\bf 0104}, 034 (2000), hep-th/0010072;
A.~Bergman, K.~Dasgupta, O.~J.~Ganor, J.~L.~Karczmarek and G.~Rajesh,
  ``Nonlocal field theories and their gravity duals,''
  Phys.\ Rev.\ D {\bf 65}, 066005 (2002), hep-th/0103090;
K.~Dasgupta and M.~M.~Sheikh-Jabbari,
  ``Noncommutative dipole field theories,''
  JHEP {\bf 0202}, 002 (2002), hep-th/0112064.}

\lref\fawad{S.~F.~Hassan,
  ``T-duality, space-time spinors and R-R fields in curved backgrounds,''
  Nucl.\ Phys.\ B {\bf 568}, 145 (2000), hep-th/9907152.}

\lref\ganorha{O.~J.~Ganor and A.~Hanany,
  ``Small $E_8$ Instantons and Tensionless Non-critical Strings,''
  Nucl.\ Phys.\ B {\bf 474}, 122 (1996), hep-th/9602120.}

\lref\dall{G.~Dall'Agata and N.~Prezas,
  ``${\cal N} = 1$ geometries for M-theory and type IIA strings with fluxes,''
  Phys.\ Rev.\ D {\bf 69}, 066004 (2004), hep-th/0311146;
  ``Scherk-Schwarz reduction of M-theory on G2-manifolds with fluxes,''
  JHEP {\bf 0510}, 103 (2005), hep-th/0509052.}

\lref\lilia{L.~Anguelova and D.~Vaman,
  ``${\cal R}^4$ corrections to heterotic M-theory,'' hep-th/0506191;
P.~Manousselis, N.~Prezas and G.~Zoupanos,
  ``Supersymmetric compactifications of heterotic strings with fluxes and
  condensates,'' hep-th/0511122.}

\lref\nekra{N.~Nekrasov,~H.~Ooguri and C.~Vafa, `S-duality and
Topological Strings,'' hep-th/0403167.}

\lref\lustu{G.~L.~Cardoso, G.~Curio, G.~Dall'Agata and D.~Lust,
``BPS action and superpotential for heterotic string compactifications  with
fluxes,'' JHEP {\bf 0310}, 004 (2003),hep-th/0306088.}

\lref\lustd{G.~L.~Cardoso, G.~Curio, G.~Dall'Agata and D.~Lust,
``Heterotic string theory on non-K\"ahler manifolds with H-flux
and gaugino condensate,'' hep-th/0310021.}

\lref\bd{M.~Becker and K.~Dasgupta, ``K\"ahler versus non-K\"ahler
compactifications,'' hep-th/0312221.}

\lref\micu{S.~Gurrieri and A.~Micu,
``Type IIB theory on half-flat manifolds,''
Class.\ Quant.\ Grav.\  {\bf 20}, 2181 (2003), hep-th/0212278.}

\lref\dal{G.~Dall'Agata and N.~Prezas,
``N = 1 geometries for M-theory and type IIA strings with fluxes,''
hep-th/0311146.}

\lref\douo{M.~R.~Douglas,``The statistics of string / M theory vacua,''
JHEP {\bf 0305}, 046 (2003), hep-th/0303194.}

\lref\wittenchern{E.~Witten,``Chern-Simons gauge theory as a
string theory,'' hep-th/9207094.}

\lref\ms{D.~Martelli and J.~Sparks,``G Structures, fluxes and
calibrations in M Theory,'' hep-th/0306225.}

\lref\grana{A.~Butti, M.~Grana, R.~Minasian, M.~Petrini and A.~Zaffaroni,
    ``The baryonic branch of Klebanov-Strassler solution: A supersymmetric
family
    of SU(3) structure backgrounds,''
    JHEP {\bf 0503} (2005) 069, hep-th/0412187.}

\lref\fg{A.~F.~Frey and A.~Grana,``Type IIB solutions with
interpolating supersymetries ,'' hep-th/0307142.}

\lref\bbs{K.~Becker, M.~Becker and R.~Sriharsha,``PP-waves,
M-theory and fluxes,'' hep-th/0308014.}

\lref\minu{P. Kaste,~R. Minasian,~A. Tomasiello,''Supersymmetric M theory
Compactifications with
Fluxes on Seven-Manifolds and G Structures'', JHEP {\bf 0307} 004, 2003,
hep-th/0303127.}

\lref\minasianone{
  A.~Butti, M.~Grana, R.~Minasian, M.~Petrini and A.~Zaffaroni,
  ``The baryonic branch of Klebanov-Strassler solution: A supersymmetric family
  of SU(3) structure backgrounds,''
  JHEP {\bf 0503}, 069 (2005), hep-th/0412187.
}

\lref\minasian{S.~Fidanza, R.~Minasian and A.~Tomasiello,
  ``Mirror symmetric SU(3)-structure manifolds with NS fluxes,''
  Commun.\ Math.\ Phys.\  {\bf 254}, 401 (2005), hep-th/0311122;
M.~Grana, R.~Minasian, M.~Petrini and A.~Tomasiello,
  ``Supersymmetric backgrounds from generalized Calabi-Yau manifolds,''
  JHEP {\bf 0408}, 046 (2004), hep-th/0406137;
``Type II strings and generalized Calabi-Yau manifolds,''
  Comptes Rendus Physique {\bf 5}, 979 (2004), hep-th/0409176;
`Generalized structures of N = 1 vacua,'' hep-th/0505212.}

\lref\lindstrom{U.~Lindstrom,
  ``Generalized N = (2,2) supersymmetric non-linear sigma models,''
  Phys.\ Lett.\ B {\bf 587}, 216 (2004), hep-th/0401100;
U.~Lindstrom, R.~Minasian, A.~Tomasiello and M.~Zabzine,
  ``Generalized complex manifolds and supersymmetry,''
  Commun.\ Math.\ Phys.\  {\bf 257}, 235 (2005), hep-th/0405085;
U.~Lindstrom,
  ``Generalized complex geometry and supersymmetric non-linear sigma models,''
  hep-th/0409250;
U.~Lindstrom, M.~Rocek, R.~von Unge and M.~Zabzine,
  ``Generalized Kaehler geometry and manifest N = (2,2) supersymmetric
  nonlinear sigma-models,'' hep-th/0411186.}

\lref\urangapark{J.~Park, R.~Rabadan and A.~M.~Uranga,
  ``Orientifolding the conifold,''
  Nucl.\ Phys.\ B {\bf 570}, 38 (2000), hep-th/9907086.}

\lref\uranga{A.~M.~Uranga,
  ``Brane configurations for branes at conifolds,''
  JHEP {\bf 9901}, 022 (1999), hep-th/9811004.}

\lref\dm{K.~Dasgupta and S.~Mukhi,
  ``Brane constructions, conifolds and M-theory,''
  Nucl.\ Phys.\ B {\bf 551}, 204 (1999), hep-th/9811139;
``Brane constructions, fractional branes and anti-de Sitter domain walls,''
  JHEP {\bf 9907}, 008 (1999), hep-th/9904131.}

\lref\beru{K.~Behrndt and M.~Cvetic, ``General N=1 Supersymmetric Flux
Vacua of
(Massive) Type IIA String Theory'', hep-th/0403049.}

\lref\dmftheory{K.~Dasgupta and S.~Mukhi,
  ``F-theory at constant coupling,''
  Phys.\ Lett.\ B {\bf 385}, 125 (1996), hep-th/9606044.}

\lref\dalu{G.~Dall'Agata, ``On Supersymmetric Solutions of Type IIB
Supergravity with General Fluxes'', hep-th/0403220.}

\Title{\vbox{\hbox{hep-th/0511099}}}
{\vbox{ \vskip-10in
\hbox{\centerline{Geometric Transitions,}}\hbox{\centerline{
Flops and Non-K\"ahler
Manifolds: II}}}}

\vskip-.4in \centerline{\bf Melanie Becker$^{1,2}$,~ Keshav
Dasgupta$^3$, ~Sheldon Katz$^4$}
\centerline{\bf Anke Knauf$^{5,6}$, ~Radu Tatar$^7$} \vskip.1in
\centerline{\it${}^1$~Jefferson Physical Laboratory, Harvard University,
Cambridge, MA 02138, USA}
\centerline{\it ${}^2$ Physics Department, Texas A\& M University,
College Station,
TX 77843, USA}
\centerline{\tt melanie@schwinger.harvard.edu}
\centerline{\it ${}^3$ Rutherford Physics Building, McGill University,
Montreal, QC H3A 2T8, Canada}
\centerline{\tt keshav@hep.physics.mcgill.ca}
\centerline{\it ${}^4$ Depts. of Math \& Phys, University of Illinois at
Urbana-Champaign, IL 61801, USA}
\centerline{\tt katz@math.uiuc.edu}
\centerline{\it ${}^5$~Department of Physics, University of
Maryland, College Park, MD 20742}
\centerline{\it ${}^6$II Institut f\"ur Theoretische Physik,
Universit\"at Hamburg 22761 Hamburg, Germany} \vskip.02in
\centerline{\tt anke@umd.edu}
\centerline{\it ${}^7$ Dept. of Maths Sciences, Liverpool
University, Liverpool, L69 3BX, England, U.K.}
\centerline{\tt Radu.Tatar@liverpool.ac.uk}



\centerline{\bf Abstract}

\noindent We continue our study of geometric transitions in type II and
heterotic theories. In type IIB theory we
discuss an F-theory setup which clarifies many of our earlier assumptions
and allows us to study gravity duals of ${\cal N} = 1$ gauge theories with
arbitrary global symmetry group $G$.
We also point out the subtle differences between global and local
metrics, and show that in many cases the global descriptions are
far more complicated than discussed earlier. We determine the full
global description in type I/heterotic theory.

\noindent In type IIA, our analysis gives rise to a local
non-K\"ahler metric whose global description involves a particular
orientifold action with gauge fluxes localised on branes.
We are also able to identify the three form
fields that allow for a smooth flop in the M-theory lift.
We briefly discuss the issues of generalised complex structures
in type IIB theory and possible half-twisted models in the heterotic
duals of our type II models. In a companion paper we will present details
on the topological aspects of these models.

\Date{}


\centerline{\bf Contents}\nobreak\medskip{\baselineskip=12pt
\parskip=0pt\catcode`\@=11

\noindent {1.} {Introduction and Summary} \leaderfill{1} \par
\noindent {2.} {Resolved Conifold Revisited} \leaderfill{8} \par
\noindent \quad{2.1.} {The Full Global Picture} \leaderfill{10} \par
\noindent \quad{2.2.} {Field Theory Interpretations} \leaderfill{16} \par
\noindent \quad{2.3.} {Topological Aspects and Excursions into Generalised Geometries} \leaderfill{17} \par
\noindent {3.} {Global Heterotic Models} \leaderfill{23} \par
\noindent \quad{3.1.} {Heterotic (0,2) Models: Brief Discussions} \leaderfill{23} \par
\noindent \quad{3.2.} {Heterotic (0,2) Models: Detailed Discussions} \leaderfill{26} \par
\noindent {4.} {Pulling Rank 2 Vector Bundles through Conifold Transitions} \leaderfill{41} \par
\noindent \quad{4.1.} {The Serre Construction} \leaderfill{42} \par
\noindent \quad{4.2.} {Conifold Transitions} \leaderfill{45} \par
\noindent {5.} {M-Theory and Non-K\"ahler Manifolds} \leaderfill{46} \par
\noindent {6.} {Global Type IIA Backgrounds} \leaderfill{49} \par
\noindent \quad{6.1.} {Type IIA Background Revisited} \leaderfill{50} \par
\noindent \quad{6.2.} {More Subtleties} \leaderfill{53} \par
\noindent {7.} {Torsion Classes} \leaderfill{66} \par
\noindent \quad{7.1.} {$SU(3)$ and $G_2$-Structure Manifolds} \leaderfill{67} \par
\noindent \quad{7.2.} {Torsion Classes Before Geometric Transitions} \leaderfill{70} \par
\noindent \quad{7.3.} {Torsion Classes After Geometric Transitions} \leaderfill{72} \par
\noindent {8.} {Discussion and Future Directions} \leaderfill{74} \par
\catcode`\@=12 \bigbreak\bigskip}

\newsec{Introduction and Summary}

The study of geometric transitions has been an extensive field of research in the past five years.
It started with the seminal papers \ks\mn\vafai\ where the idea of
connecting open string theory with branes to
closed string theory with fluxes was spelled out. The upshot was a powerful recipe to
derive effective field theory
results from string theory compactification with branes and fluxes.  Spectacular results
have been obtained subsequently
in many directions. For example, earlier development of the string flux compactifications
\sav\gkp\ now applied to this scenario,
showed a new ``democratic'' way  to think about effective theories
coming from string compactifications.

The direction opened in \vafai, based on the previous work on
topological strings \gv, had a powerful
impact on stretching the skills to use string theory to describe explicitly
large classes of effective field
theories. By matching effective field theory quantities such as gluino condensate
and meson vacuum expectation values with
geometrical objects, one can use compactification details to get new insights
into the strongly coupled field theories.
There were several avenues to accomplish that. One was to use matrix models to
obtain effective potentials and their
extrema \dvo\dvt. Another one was to translate geometrical results regarding period
matrices into field theory language
\civ\adoptone\civu\civd. A dynamical way to undertand the above match is to
use an M theory description where the
branes and geometries have a description in terms of unique M5 branes
\dotu\dotd\adoptfo\radu.

More recently it has become clear that some aspects of geometric transitions in \vafai\
have to be reconsidered
to accomodate some ``exotic'' NS forms that are needed to have a precise match between
field theory and geometric quantities.  This was pointed out in \louis\ for
compact manifolds and has been
succesfully implemented for geometric transitions in \gtone\realm. This
implementation requires a
departure from the compactification with $SU(3)$ {\it holonomy}, i.e. standard
Calabi-Yau manifolds, to a
larger class of manifolds with
$SU(3)$ {\it structure}, i.e. non-K\"ahler manifolds.
The same happens in M theory where the $G_2$ holonomy condition should be dropped
in favor of a
more general $G_2$ structure condition.

However the above development of geometric transition \vafai, although spectacular in many
respects, lacked the preciseness of the equivalent Klebanov-Strassler \ks\ or Maldacena-Nunez \mn\
models because of the absence of a complete supergravity description of the system. An attempt to
get the full supergravity description was started in a series of papers \gtone, \realm\ by following a
duality chain that used some aspects of T-duality and mirror symmetry advocated in \syz. The program
of following this chain by a step-by-step application of known dualities turned out to be highly non-trivial
because the duality rules were suited to some regime of parameter space that were opposite to the
regime that we would be interested in for studying geometric transitions. Nevertheless, with some subtle
manipulations we obtained the right metric in type II and M-theories. However, the
metric that we got in \gtone\ and \realm\ turned out to be only a {\it local} description of
a much more involved {\it global} framework. The global framework involves non-trivial orientifold actions
giving rise to other D-branes and orientifold planes in addition to the already present
wrapped five branes. Our ideas can be summarized as follows.

The starting point of our program is a type IIB solution which is {\it locally} the one of \pandoz.
Recall that the solution of \pandoz\ describes D5 branes wrapped on the resolution two-cycle of a resolved
conifold, exactly what one wants for an IR description of a ${\cal N} = 1$ gauge theory.
The full {\it global}
solution of \pandoz\ is not supersymmetric as discussed in \cvetic, \gtone. 
Therefore, to get
a supersymmetric background
we embed this solution into an F-theory setup.
In terms of type IIB language this is done by
considering extra $D7/O7$ branes on a particular six dimensional manifold whose local
metric (i.e. in the region where
the effects of background $D7/O7$ branes are negligible) looks similar to the metric
of \pandoz, but now written
in terms of {\it local coordinates}. From the F-theory point of view this is simply a
compactification on a fourfold
with $G$-fluxes \sav \beckerD. We will present the details of the construction of
this fourfold in sec.\ 2.1. In
short, the fourfold is a $T^2$ fibration over a six-dimensional
base that locally resembles a resolved conifold with fractional
$D3$ branes\foot{In \gtone\ we constructed a fourfold with non--degenerating $T^2$ fiber. From our discussion, this
would capture only the local behavior of the system as the degeneration points of the fiber (alternatively the
positions of the seven branes) are not in the local neighborhood.}.
The base is not a Calabi-Yau manifold because $c_1 \ne 0$ but will be K\"ahler\foot{The base could be
Calabi-Yau only at one special point $-$ the orientifold point \senF. Clearly we will not be in this region
if we want to move the seven branes far away.}.
This is similar to the embedding of the Klebanov-Strassler
background \ks\ in F-theory \gkp. In this case, the F-theory base is again
approximately a conifold, and the $G$-fluxes
appear in IIB as $H_{NS}$ and $H_{RR}$ fluxes.

In the light of these considerations, the field theory results should also be
reinterpreted to accomodate the
new branes and orientifolds.
It is inherent to geometric transitions that the gluino condensate
in the brane side is identified with geometrical
structures in the flux side. For the ${\cal N} =1$ theory on the D5 branes
this means identification of
$\Lambda^3$ with the size of $S^3$ cycles.
In the setup presented here, additional fundamental matter is
introduced. In type IIB
this corresponds to the D7 branes situated on top of the O7-planes. Their distance from
the resolution cycles is
the mass of the corresponding fundamental quarks\foot{There are {\it two} different matter multiplets (or flavor
degrees of freedom) here in the UV. One of them are the matter multiplets coming from the strings that stretch
between full and fractional D3 branes. These form the {\it bi-fundamental} matters, and are in general
charged under both the gauge groups in the ultaviolet. The second one come from the strings stretching between
full and fractional D3 branes, and the seven branes. They are the {\it fundamental} matters. At the far IR, when the
D3 branes cascade away leaving only the fractional D3 branes, the theory will only have fundamental matter multiplets
and no bi-fundamental matters.
In this paper we will concentrate only in this region, leaving a more detailed discussion of the UV behavior
for the sequel.}.

Consider now a scenario where the D5 branes wrapped
on the resolution two-cycle are close to {\it one} orientifold plane.
Clearly, then the other
three orientifold O7 planes will be too far away from the wrapped D5 branes to be considered a part of the
effective theory, so effectively there are only 4 relevant D7 branes on top of one O7 plane.
This implies for the field theory that the global picture represents an
${\cal N} =1~ SU(N)$ gauge theory with matter in the fundamental representation
and flavor group $SU(2)^4$, the quark masses of the flavors being $m$ and the scale
of the theory being $\Lambda_{\rm global}$. The global picture is obtained by starting with
${\cal N} =1~ Sp(2N)$ gauge theory and giving an expectation value to the adjoint Higgs which breaks
$Sp(2N)$ to $SU(N)$ and the global group from $SO(8)$ to $SU(2)^4$. The local picture therefore represents a
pure ${\cal N} =1~ SU(N)$
theory which is the effective theory after integrating out the massive quarks with a scale
\eqn\scale{\Lambda_{\rm local}^{3 N} = m^{4}  \Lambda_{\rm global}^{3 N - 4}.}
Thus, to go from a local picture to a global picture one has to integrate in
the massive quarks as in \intril.
We will soon see that the local picture is forced upon us by the identifications of geometric transitions.

The local geometry versus global geometry issue is also discussed in \gkp\
where the local Klebanov-Strassler solution \ks\ gets the extra flavor group
$G$ globally.
This makes the Klebanov-Strassler solution an effective theory.
The only difference is that in their case the solution of \ks\ is supersymmetric,
whereas our starting point, the solution of \pandoz\ is not.
This can be understood from the fact that \ks\ uses integer and fractional D3 branes,
the latter being wrapped D5 on
vanishing cycles. This is a susy configuration.
When the vanishing cycle becomes finite (the conical singularity is resolved),
the fractional D3 branes become
genuine D5 wrapped on 2-cycles which\foot{We will also refer to these branes as fractional D3 branes.},
together with the integer D3 branes,
are in general a non-susy system\foot{As we will discuss in sec. 2.1, it doesn't really matter if
there might exist
such a solution that preserves
supersymmetry (i.e allows primitive three form fluxes). Of course, to discuss
the geometric transition
of ${\cal N} = 1$ $SU(N)$ theory with fundamental flavor transforming under a group
$G$, we {\it have} to introduce $D7/O7$ branes along with
the fractional and whole
$D3$ branes \dmftheory\ (see also \ouyang\ where somewhat equivalent construction, but only with $D7$ branes,
are made to study Klebanov-Strassler model \ks\ with fundamental flavors). 
Our local
metric studied earlier in \gtone, \realm\ is much more robust and it {\it only} depends on the topology of the
resolved conifold.}.

From the type IIB solution one can infer a mirror IIA solution \gtone. The result is
a family of non-K\"ahler manifolds,
the non-K\"ahlerity coming from the presence of type IIB NS flux. This
means that the geometric transitions of  \vafai\ can actually be generalized to non--K\"ahler manifolds,
as presented in \gtone.

The non-K\"ahlerity in type IIA theory is proportional to the NS flux so the
non-K\"ahler geometry can be considered as a function of the type IIB
NS flux. However,
in type IIB we know from \civ\ that the NS flux is proportional to the coupling
constant of the dual gauge theory,
and therefore the non-K\"ahlerity of type IIA solutions are inversely
proportional to the coupling constant of the field theory. Thus we see two possible
limits:

\noindent (a) The NS flux is (almost) zero.
Then $1/g^2$ in field theory is also near zero which means we are in the
extreme IR. So the K\"ahler limit of our solution is dual to the extreme IR in the
field theory.

\noindent (b) The NS flux is large so the deviation from K\"ahlerity is big.
Then $1/g^2$ in the field theory is large which means that
we are in the far UV of the field theory.

Our solution maps naturally to the one discussed in \grana,
where an interpolating solution between the Klebanov-Strassler (KS)
solution \ks\ and the Maldacena-Nunez (MN) solution \mn\ was considered, with the MN solution
obtained in the extreme IR. Both these backgrounds are in type IIB.
In the limit (a), our solution for wrapped D5 branes will only have RR flux. This is
similar to the  MN solution.
In the limit (b) and any intermediate case between (a) and (b), our solution will
have both NS and RR flux. This is a
Klebanov-Strassler type solution\foot{A word of caution here. In \realm\ we showed that
the gravity dual of
${\cal N} = 1$ $SU(N)$ gauge theory looks very similar to the Klebanov-Strassler \ks\
background with
three form fluxes. However there were some small differences which we will elaborate on
later.}.
The difference between our analysis and \grana\ is the fact that we have an
interpolating IIA solution,
which appears very naturally in the form of the non-K\"ahler deformation of the
deformed conifold.

The distinction global/local picture translates onto the type IIA case. This will be
discussed in detail in sec.\ 6.
The wrapped $D5$ branes become  wrapped $D6$ branes and
the stacks of $D7/O7$ become stacks of $D6/O6$ which are now located at fixed points
of a specific orientifold
operation. The global group is
$SO(4)^8$.
In terms of field theory, the global picture represents an  ${\cal N} =1$ $SU(N)$ theory
with matter
in the fundamental representation and a {\it maximal} flavor group $SU(2)^{16}$. If in
the global picture, the field theory has the same scale
$\Lambda_{\rm global}$ and all the flavors have the same mass $m$, then the local scale
is again given by the relation
\scale. This means that the mirror symmetry passes a field theory consistency condition,
namely, the scales of the
effective theories are the same in type IIA and type IIB.

We will discuss in sec.\ 7 how our IIA solutions fit into the class of known susy backgrounds with
torsion. For example, one type of solutions are discussed in \minakas\
with nontrivial RR two-form and nontrivial dilaton.
Here the nonzero components of the torsion classes are $W_2^{+}, W_4, W_5$. The second
type of solutions are the
half-flat solutions which were first discussed in \louis\
and are obtained as mirrors of type IIB solutions with NS flux. For the
half-flat solutions $W_1^{+}, W_2^{+}, W_3$ are nonzero. The half-flat solutions were
lifted to $G_2$ manifolds by using
Hitchin's Hamiltonian flow  in \vagudi.

Our solution will be neither of the type discussed in \minakas\ nor
half-flat. Due to the elaborate F--theory setup in IIB we will have fluxes turned on
the lift to M--theory that destroy the $G_2$ holonomy. We therefore expect only a $G_2$
structure manifold\foot{Thus the holonomy could even be $SU(2)$ in some of these models. However, susy
will be determined {\it not} by the holonomy but by the structure. So $G_2$ structure but $SU(2)$
 holonomy will preserve
${\cal N} = 1$ susy in 4 dimensions.}.
Furthermore, we only have the local metrics available, so our torsion class analysis should
not be taken too seriously. We cannot make statements about the global topology of our
IIA backgrounds. The local solutions are not susy and will therefore not obey any
classification for susy torsional backgrounds such as \minakas\ or \louis.

In contrast, in the heterotic theory we will propose, in sec.\ 3,  a global background that reduces in the
local limit to the one proposed in \realm.
The global background will look very similar to the MN background \mn, although
it can in general differ from the MN background \mn\ by warp
factors in the metric, but satisfies the torsional relation at least in the limit where it agrees with MN.
The method to construct heterotic solutions is to start with a slightly different F--theory
setup.
We can trade the $D7/O7$ stacks
for $D9/O9$. This way we reach a
type I picture related to a heterotic picture by an S-duality. This mapping implies a
possible geometric transition for
heterotic strings, as observed in \realm. The global
heterotic solution is not neccesarily dual to a global type IIB solution as this would
imply going away from
the orientifold limit.
 Since the type II  transitions can be understood by
studying the topological sector,
we can use a half-twisted version to understand the transition for the (0,2) theories
\zerotwo\reczt\wittwo.
This will be addressed in sec.\ 3.1 and sec.\ 3.2.

We will also study heterotic vector bundles in sec.\ 4. They are associated with the global group $G$
that we encounter in IIB. The
global symmetries appear as
gauge symmetries of the $SO(32)$ bundle in the heterotic side. In sec.\ 4.2 we will
discuss possible ways to pull
bundles across a conifold transition.

As we discussed above,
the global type IIB picture leads us to the global type IIA model. This is basically the mirror of the
type IIB background (once we take into account all the subtleties). In type IIA, however, we encounter some
interesting scenarios. First, we find that the global picture is generated from a specific orientifold action.
Due to the inherent orientifold action we find some gauge fluxes that are secretly localised on the
D6 branes. Some part of these fluxes may even give rise to three-from fluxes in the internal space, although their
effect in the local region of interest may be negligible. Globally they might have some effects.  Secondly, the
same orientifold action produces three-form potentials that are not pure gauge. We find this a novel aspect of
our model. The three form potentials are responsible for a smooth flop operation in the M-theory lift of
our model. This was anticipated in \amv\ and \edelstein, and here we provide the first concrete realisation.
These details will be presented in sec.\ 6; and in sec.\ 7 we will argue how these fluxes determine the
torsion classes both before and after geometric transitions in type IIA theory.

Before we go into the detail discussions of the above aspects, we should point out a simple T-dual version
in type IIA of the full type IIB global picture. First, one shouldn't confuse this T-dual model with the
{\it mirror} type IIA model. The mirror is a different duality that captures the {\it full} content of our type IIB
model. The brane dual that we are talking of are in the same spirit as \uranga, \dm\ and it captures only some
aspects of our full type IIB global picture. Nevertheless, many of the UV expectations will be clear from the T-dual.
First, the resolved conifold will T-dualise into two intersecting NS5 branes, and the fractional D3 branes will 
become
half-D4 branes. This is same as our model in \dotu\ and \dotd. The D7/O7 system will T-dualise into a D6/O6
system that will be parallel to one of the NS5 branes and in fact will coincide with the NS5 branes. This is somewhat
similar to the model advocated in \urangapark\ and in \ouyang. We differ from \urangapark\ in two respects: one,
our brane
construction is the T-dual of a resolved conifold, and two, we have both full D4 and half-D4 branes. The
model of \urangapark, that is T-dual to a conifold,  
takes only full D4 branes, and therefore could study gauge theories with vanishing Beta-function
only. We also differ from \ouyang\ because \ouyang\ studies the T-dual of a Klebanov-Tseytlin model \klebts\
with D7 branes as flavors. As we know \klebts\ behaves badly at the IR because of naked-singularity there, and
Klebanov-Strassler \ks\ will be
the right model to study IR of a confining gauge theory. So the model advocated by \ouyang\ can say
nothing about the IR behavior that we would be interested in. Plus the flavors are given by only D7 branes in
\ouyang, that in general should be viewed from an F-theory perspective which is lacking in \ouyang\foot{These
issues were in fact already pointed out in \ouyang, although no solutions were presented there.}.

In the rest of the paper we will solve all the issues that could in principle plague the models of
\urangapark\ and \ouyang. As discussed above, we can show much more than merely giving the global story. We can
discuss the full dynamics of geometric transitions in {\it all} string theories, including M-theory.

We start by revisiting the root of all of these issues: the resolved conifold.

\newsec{Resolved conifold revisited}

The original study of open-closed string duality in type II theory
starts with D6 branes wrapping a three cycle of a non-compact
deformed conifold. Naively one might expect the deformed conifold
to be a complex K\"ahler manifold with a non-zero three cycle.
However as discussed earlier in \gtone\ this is not quite correct,
and the manifold that actually would solve the string equations of
motion is a non-K\"ahler deformation of the deformed conifold. It
also turns out that the manifold has no integrable complex
structure, but only has an almost complex structure. This is
consistent with the prediction of \vafai. The non-K\"ahlerity of
the underlying metric can be easily seen from its explicit
form\foot{In deriving the metric, we took a simpler model where all
the spheres were replaced by tori with periodic coordinates ($x,
\theta_1$) and ($y,\theta_2$). The coordinate $z$ formed a
non-trivial $U(1)$ fibration over the $T^2$ base. The replacement
of spheres by two tori is directly motivated from the
corresponding brane constructions of \uranga,\dm, where
non-compact NS5 branes required the existence of tori instead of
spheres in the T-dual picture. In fact there is a deeper reason behind this
choice. The
metrics that we studied earlier in \gtone\ and \realm\ are actually
{\it local} descriptions of a much more elaborate global story.
As we will show, the local
description {\it only} allows tori.
We will soon provide
a derivation that will justify all the assumptions taken. In fact
we will see that most of the assumptions are imposed on us by some
stringent requirements on the type IIB metric.}
\eqn\fiitaamet{\eqalign{& ds_{IIA}^2 = g_1~\left[(dz -
{b}_{z\mu}~dx^\mu) + \Delta_1~{\rm cot}~\hat\theta_1~(dx -
b_{x\theta_i}~d\theta_i) + \Delta_2~{\rm cot}~ \hat\theta_2~(dy -
b_{y\theta_j}~d\theta_j)+ ..\right]^2 \cr & + g_2~ {[} d\theta_1^2
+ (dx - b_{x\theta_i}~d\theta_i)^2] + g_3~[ d\theta_2^2 + (dy -
b_{y\theta_j}~d\theta_j)^2{]} +  g_4~{\rm sin}~\psi~{[}(dx -
b_{x\theta_i}~d\theta_i)~d \theta_2 \cr & ~~~~~~~~~ + (dy -
b_{y\theta_j}~d\theta_j)~d\theta_1 {]} + ~g_4~{\rm
cos}~\psi~{[}d\theta_1 ~d\theta_2 - (dx - b_{x\theta_i}~d\theta_i)
(dy - b_{y\theta_j}~d\theta_j)].}} where the coefficients $g_i$
and the coordinates $\theta_i, \hat\theta_i$ etc. are defined in
\gtone. The background has
non-trivial gauge fields (that form the sources of the wrapped D6
branes) and a non-zero string coupling (which could in principle
be small).

Existence of such an exact supergravity background helps us to obtain
the corresponding mirror type IIB background. One would expect that
this can be easily achieved
using the mirror rules of \syz. It turns out however that
the mirror rules of \syz, as discussed in \realm\ and \gtone, do not
quite suffice\foot{The mirror transformations used in \gtone\ was to take a
specific type IIB background and get the metric \fiitaamet.}.
A detailed analysis of this is
presented in \realm.
As discussed in \gtone\ and \realm, we have to be careful
about various subtle issues while doing the mirror transformations:

\noindent (a) The mirror rules of \syz\ tells us that {\it any} Calabi-Yau
manifold with a mirror admits, at least {\sl locally}, a $T^3$ fibration over a
three dimensional base. This seems to fail for the deformed
conifold as it does not possess enough isometries to represent it as a $T^3$
fibration\foot{There is a subtlety about the local and global pictures that we will clarify
soon.}.
On the other hand, a resolved conifold does have a well defined $T^3$
torus over a 3-d base, which can be exploited to get the mirror (see also
\agavafa). It also turns out that the $T^3$ torus is a lagrangian
submanifold \karch, so a mirror transformations will not break any
supersymmetry.

\noindent (b) Viewing the mirror transformation naively as three T-dualities
along the $T^3$ torus {\it does~not} give the right mirror metric. There are
various issues here. The rules of \syz\ tell us that the mirror transformation
would only work when the three dimensional base is very large. The configuration that
we have is exactly opposite of the case \syz. Recall that
our configuration lies at the end of a much larger cascading theory.
By UV/IR correspondences, this means that the
base manifold is very small. Furthermore we are at the {\it tip} of the
geometric transition and therefore we have to be in a situation with very
small base (in fact very small fiber too). In \realm\ and in \gtone\ we showed
that we could still apply the rules of \syz\ if we impose a non-trivial large
complex structure on the underlying $T^3$ torus. The complex structure
can be integrable or non-integrable. Using an integrable complex structure, we
showed in \realm\ and in \gtone\ that we can come remarkably close to getting
the right mirror metric. Our conjecture there was that if we use a
non-integrable complex structure we can get the right mirror manifold.

It seems therefore natural to start with the manifold that exhibits three isometry
directions --- the resolved conifold. We can, however, not use the metric for D5
branes wrapping the $S^2$ of a resolved conifold as derived in \pandoz, because
it breaks all supersymmetry \cvetic.  The metric that we
proposed in \gtone\ (where we kept the harmonic functions undetermined) is very
close to the metric of \pandoz\ but differs in some subtle way:

\noindent (a) The type IIB resolved conifold metric that we proposed in
\gtone\ is a D5 wrapping a two cycle that {\it preserves} supersymmetry.
We will discuss this issue in more detail below.

\noindent (b) As explained in \gtone, \realm, our IIB manifold
also has seven branes (and possibly orientifold planes) along with
the type IIB three-form fluxes. The metric constructed in \pandoz\
doesn't have seven branes but allows three-form fluxes.

\noindent The {\it local} behavior of the type IIB metric is expressed in terms
of non-trivial complex structures $\tau_1$ and $\tau_2$ as $dz_1 = dx - \tau_1 d\theta_1$ and
$dz_2 = dy - \tau_2 d\theta_2$. The local metric then reads
\eqn\iibmet{ds^2 = (dz + \Delta_1 ~ {\rm cot}~\theta_1 ~ dx  +
\Delta_2 ~{\rm cot}~\theta_2 ~dy)^2 + \vert dz_1 \vert^2 + \vert dz_2\vert^2}
where all the warp factors can locally be absorbed in to the coordinate differentials.
In this formalism the metric may naively look similar to the one studied in
\pandoz\ but the global picture is completely different from the
one proposed by \pandoz.

\subsec{The full global picture}

To consider the full global picture, let us consider an F-theory
compactification on a fourfold which is a non-trivial $T^2$
fibration over a resolved conifold base. For simplicity we will
consider a compact base, although the non-compact base would be
easy to generalize. To preserve charge conservation, therefore we
will remove the D5 branes for the time being. Two results are
immediately obvious:

\noindent (a) Since the fourfold is a Calabi-Yau, the base will no longer
remain a Calabi-Yau resolved conifold, and will have non zero first Chern
class. The non zero first Chern class comes precisely from the seven branes in
the picture. Although not Calabi-Yau, the
base is still K\"ahler\foot{Thus when the seven branes are kept at large distances, the torus
fibration will look effectively trivial in local region away from the seven branes.
Here the manifold will effectively simulate vanishing first Chern condition. This
will be useful later when we will try to give a topological theory description
for our background.}.

\noindent (b) In the presence of fluxes the metric of the fourfold
only changes by an overall conformal factor and therefore modulo this
subtlety, we can consider a Calabi-Yau fourfold for our case.

In the following we will also be able to shed light simultaneously on the
fourfold {\it after} the geometric transition in type IIB theory.

We start by considering F theory on a base $B$ (whose local metric is
given in \iibmet),
so that $X\to B$ is an elliptically
fibered Calabi-Yau fourfold.  Suppose that $B$ contains a smooth curve
$E\simeq P^1$
with normal bundle ${\cal O}(-1)\oplus {\cal O}(-1)$,
and that there is a conifold
transition from $B$ to $B'$ obtained by contracting the $P^1$ to a
conifold and then smoothing.  This gives another elliptically fibered
Calabi-Yau fourfold $X'\to B'$.

We compute the Euler characteristic of $X$ by topology.
Recall that the Weierstrass model for $X$
is obtained as a Calabi-Yau hypersurface in the projective bundle
\eqn\proj{
P={\cal P}({\cal O}_B\oplus{\cal O}_B(2K_B)\oplus{\cal O}_B(3K_B)).}
with equation
\eqn\weier{
y^2=4x^3-g_2x-g_3}
where $y$ is a coordinate on the bundle
${\cal O}_B(3K_B)$
and $x$ is a coordinate on the bundle ${\cal O}_B(2K_B)$ (hence they
live in the dual bundles
${\cal O}_B(-3K_B)$ and ${\cal O}_B(-2K_B)$ respectively).  The
third homogeneous coordinate $z$ in the $P^2$ fiber corresponding to the
factor ${\cal O}_B$ has been suppressed.
We also have that $g_k$ is a section of
${\cal O}_B(-2kK_B)$ for $k=2,3$.

The elliptic fiber is generically smooth, but is a cuspidal cubic over
points of $B$ where $g_2=g_3=0$, and is a nodal cubic over points
of $B$ where the discriminant\foot{Not to be confused with $\Delta_i$, the
warp factors appearing in the metrics.}
$\Delta=g_2^3-27g_3^2=0$ but $g_2$ and $g_3$ are not
simultaneously zero.  The zero locus
of $\Delta$ is a complex surface $S$ containing the curve $D$ defined
by $g_2=g_3=0$.

Since the Euler characteristic of a smooth cubic curve, nodal cubic curve,
resp. cuspidal cubic curve are respectively $0,1,2$ we obtain
\eqn\eulerf{
\chi(X)=0\cdot\chi(B-S)+1\cdot\chi(S-D)+2\cdot\chi(D)=\chi(S)+\chi(D).}
We now make the following claim: $\chi(X')=\chi(X)$, i.e the Euler
characteristics do not change under geometric transition.

To see this, note first that by the adjunction formula, $K_B$ is
trivial after restriction to $E$.  That means that $g_2$ and $g_3$
restricted to $E$ are sections of the trivial bundle, i.e. constants.
Choosing these constants to be nonzero and generic, we see that
$S\cap E$ is empty.

For $X'\to B'$ we adopt corresponding notation $S',\ C',\ g_2',\ g_3'$.
We can also contract $E$ to a conifold $B_0$ and adopt similar
notation $S_0,\ D_0,\ (g_2)_0,\ (g_3)_0$.  Note that $g_2$ and $g_3$
uniquely fix $(g_2)_0$ and $(g_3)_0$ by assigning to the conifold point
the (constant) value of $g_2$ resp.\ $g_3$ on $E$.  Since $S\cap E$ is
empty, we learn that $S_0$ does not contain the conifold, hence
$S_0\simeq S$ and $D_0\simeq D$.  Thus $\chi(X)=\chi(X_0)$.  Now we see that
$S'$ and $D'$ are deformations of $S_0$ and $D_0$ which avoid the conifold,
so have the same Euler characteristic.  It follows that
$\chi(X')=\chi(X_0)=\chi(X)$.

To illustrate this a bit more, let us consider an
{example}. Consider a singular quadric hypersurface in $P^4$
with coordinates $(x_0\ldots,x_4)$ defined
by the equation
\eqn\quapf{
x_1^2+x_2^2+x_3^2+x_4^2=0.}
This is $B_0$. The point $(1,0,0,0,0)$ is the conifold.  It has
two K\"ahler small resolutions, related by flops.  One of them,
call it $B$, can be described by blowing up the plane defined by
$x_1=x_2=0$.  This also leads to a toric description if desired.
The conifold transition is completed by taking the equation of
$B'$ to be: \eqn\conift{ x_1^2+x_2^2+x_3^2+x_4^2=tx_0^2} for any
nonzero value of $t$. The Euler characteristics can be computed by
{\it schubert} \ksschub, obtaining 19728 for $\chi(X)$ and $\chi(X')$. This
is perfectly consistent with the result derived in \dotd\ (see
also \dotu) where the F-theory background for a fourfold after
geometric transition was derived. The resulting Euler
characteristics evaluated there were precisely 19728, confirming
the generic derivation presented here.

The above picture is not complete unless we devise a way to add
five branes in the setup. This would mean that we take a
non-compact base. The analysis of adding five branes to the
picture has been given in \gtone. The novelty of this derivation
is that we are no longer
required to consider a local picture where the fourfold is viewed
as a product manifold (as in \gtone). The full global picture
implies the existence of seven branes on a six dimensional manifold
that has a blown up $P^1$ with wrapped D5 branes on it. This
therefore raises the question: can we show now that the
correct local metric for our case is \iibmet\ as we have the full
global picture at hand?

Knowing the local type IIB metric before the geometric transition
will clarify many of the assumptions that we used in \gtone.
The metric that has been proposed in \pandoz\ is not the full global
picture because it does not contain information about seven branes etc.
in the metric (not to mention the issue of supersymmetry). It could at best
be valid locally, but {\it not} in the global coordinates
the metric has been formulated in. We will therefore
rewrite the metric of \pandoz\ using only local coordinates. The
proposed metric of D5 wrapped on a ${\bf P}^1$ of a resolved
conifold is given by \pandoz: \eqn\metresconi{\eqalign{ds^2 = &
~h^{-1/2} ds^2_{0123} + h^{1/2} \Big[\gamma' d\tilde r^2 + {1\o 4}
\gamma' \tilde r^2 (d\tilde\psi + {\rm cos}~\tilde\theta_1
d\tilde\phi_1 + {\rm cos}~\tilde\theta_2 d\tilde\phi_2)^2 + \cr &
~~~+ {1\o 4}\gamma(d\tilde\theta_1^2 + {\rm sin}^2~\tilde\theta_1
d\tilde\phi_1^2) + {1\o 4}(\gamma + 4a^2) (d\tilde\theta_2^2 +
{\rm sin}^2~\tilde\theta_2 d\tilde\phi_2^2)\Big],}} where the
coefficients etc. are defined in \pandoz, \gtone. Globally this
metric breaks supersymmetry, and the harmonic function $h$ doesn't
contain the information about the seven branes (and orientifold
seven planes). To rewrite the metric, let us define the
coordinates in the following way: \eqn\defofcoord{\eqalign{&
\tilde\psi = \langle\psi\rangle + {2 z\o
\sqrt{\gamma'_0\sqrt{h_0}}}, ~~~~~~~~~~~~ \tilde\phi_2 =
\langle\phi_2\rangle +{2y \o \sqrt{(\gamma_0+4a^2)\sqrt{h_0}}~{\rm
sin}~ \langle\theta_2\rangle} \cr
 & \tilde\phi_1 =
\langle\phi_1\rangle +{2x \o \sqrt{\gamma_0\sqrt{h_0}}~{\rm
sin}~\langle\theta_1\rangle}, ~~~~~~~~~~~ \tilde r = r_0 + {r\o
\sqrt{\gamma'_0\sqrt{h_0}}} \cr & \tilde\theta_1 =
\langle\theta_1\rangle + {2\theta_1\o \sqrt{\gamma_0\sqrt{h_0}}},
~~~~~~~~ ~~~~ \tilde\theta_2 = \langle\theta_2\rangle + {2\theta_2
\o \sqrt{(\gamma_0+4a^2)\sqrt{h_0}}}}} where $h_0 = h(r_0),
\gamma_0 = \gamma(r_0)$ are constants measured at $r = r_0$. The
coordinates ($x, y, \theta_1, \theta_2, z, r$) are small
deviations from their respective expectation values. Therefore
writing the metric using these coordinates will only give
information about a small local region. The metric after
coordinate redefinitions \defofcoord\ will take the following
form: \eqn\metformj{ds^2 = dr^2 + (dz + \Delta_1^0 {\rm
cot}~\langle\theta_1\rangle~ dx  + \Delta_2^0 {\rm
cot}~\langle\theta_2\rangle~dy)^2 + (d\theta_1^2 + dx^2) +
(d\theta_2^2 + dy^2) + ....} where we have absorbed $\gamma'_0\sqrt{h_0}$ 
in the defination of $dr$, and the dotted terms are higher
orders in ($x, y, \theta_1, \theta_2, z, r$). Furthermore the coefficients
$\Delta_1^0, \Delta_2^0$ appearing in \metformj\ are defined as:
\eqn\deltas{\Delta_1^0 \equiv \Delta_1(r_0) = \sqrt{\gamma_0' \o
\gamma_0}~r_0~,~~~~~~\Delta_2^0 \equiv \Delta_2(r_0) = \sqrt{\gamma_0' \o (\gamma_0+4
a^2)~}~r_0}
Comparing the metric \metformj\ with the one from \gtone, i.e \iibmet, we see that
\metformj\ is the limit of \iibmet\ when $\theta_1 \to \langle\theta_1\rangle$ and
$\theta_2 \to \langle\theta_2\rangle$ (along with $r = r_0$) in \iibmet!
Thus \iibmet\ is the correct local
description of D5 branes wrapped on ${\bf P}^1$ of a resolved conifold that would
preserve supersymmetry\foot{Thus the replacement $\Delta_i^0(r_0)~{\rm cot}~\langle\theta_i\rangle ~\to ~
\Delta_i(r_0)~{\rm cot}~\theta_i$ in \metformj\ is the local back-reactions of the added seven branes to 
finally convert \metformj\ to \iibmet\ and preserve supersymmetry. When the 
F-theory torus do not degenerate in local neighborhood, like what we took in
\gtone, then the metric \iibmet\ has very small $\theta_i$ dependences 
which means we are in the limit \metformj. Thus $\langle\theta_i\rangle \to \theta_i$
keeping $r = r_0$, is exactly what we meant by
{\it delocalization} in \gtone.}. 
Not only that, now we see why we expect two tori instead of
two spheres in our metric \iibmet. Locally a sphere is similar to a degenerating torus, and
therefore expanding the sphere metric with coordinates ($\phi_i, \theta_i$) ($i = 1,2$)
about a point ($\langle\phi_i\rangle, \langle\theta_i\rangle$) we get the metric of a
torus. In fact
the tori in \metformj\ are with complex structures
$\tau_{1,2} = i$, which we boosted to obtain the appropriate large complex structure limit
\gtone.
Finally, as we have argued earlier, \iibmet\ is also consistent with the T-dual brane
picture \dm, \uranga\ where
it is natural to obtain flat tori from brane configurations. Thus all of these fit
perfectly now.

Before moving ahead, let us entertain another possible scenario. The metric \metresconi\ is 
non-supersymmetric, and therefore susceptible to quantum corrections. A generic quantum correction
may change all the warp factors, and could even make the two spheres asymmetric (or even squashed). 
However if we remove the wrapped D5 branes, then the metric \metresconi\ is the metric of a resolved 
conifold that could become supersymmetric in the absence of fluxes. Question now is how robust is
our local description \iibmet? Our local description along with the seven branes should preserve 
susy even in the presence of fluxes, as we argued above. To check this, let us assume that instead of
\metresconi\ we start with another more generic metric whose explicit form can be written as
\eqn\newmetnow{\eqalign{ds^2 = &
~F_0(\tilde r)~ ds^2_{0123} + F_1(\tilde r) ~d\tilde r^2 + 
F_2(\tilde r)~ (d\tilde\psi + {\rm cos}~\tilde\theta_1
d\tilde\phi_1 + {\rm cos}~\tilde\theta_2 d\tilde\phi_2)^2 + \cr &
~~~+ \Big[F_3(\tilde r)~d\tilde\theta_1^2 + F_4(\tilde r)~{\rm sin}^2~\tilde\theta_1
d\tilde\phi_1^2\Big] + \Big[F_5(\tilde r) ~d\tilde\theta_2^2 + F_6(\tilde r)~
{\rm sin}^2~\tilde\theta_2 d\tilde\phi_2^2\Big]}}
where $F_i(\tilde r)$ are the warp factors, and the two spheres are now asymmetric and squashed. The 
above metric \newmetnow\ could either be interpreted as the deformation of \metresconi\ after quantum 
corrections, or as a generic type IIB metric on a resolved conifold with wrapped D5 branes. Whatever 
be our interpretation, far away from the wrapped D5 branes we should simply see the resolved
conifold part, as the effect of wrapped D5 branes would be negligible. Furthermore primitivity is also restored at 
large distances, and this would imply that the warp factors should approach the values set by \metresconi\ with 
$h(\tilde r) ~\to~ 1$. We are of course not so concerned about the global behavior as we want to study the local
metric so that we can put in D7 branes and O7 planes in the system. The local metric can be easily
extracted from \newmetnow\ using the small expansion trick that we had devised earlier. In the present case, the
small expansion will be
\eqn\expone{\eqalign{&
\tilde\psi = \langle\psi\rangle + {z\o
\sqrt{F_2(r_0)}}, ~~~~~~~~~~~~ \tilde\phi_2 =
\langle\phi_2\rangle +{y \o \sqrt{F_6(r_0)}~{\rm
sin}~ \langle\theta_2\rangle} \cr
 & \tilde\phi_1 =
\langle\phi_1\rangle +{x \o \sqrt{F_4(r_0)}~{\rm
sin}~\langle\theta_1\rangle}, ~~~~~~~~~~~ \tilde r = r_0 + {r\o
\sqrt{F_1(r_0)}} \cr & \tilde\theta_1 =
\langle\theta_1\rangle + {\theta_1\o \sqrt{F_3(r_0)}},
~~~~~~~~ ~~~~ \tilde\theta_2 = \langle\theta_2\rangle + {\theta_2
\o \sqrt{F_5(r_0)}}}}
where it is assumed that none of the warp factors $F_i(\tilde r)$ would vanish near $\tilde r~\to~r_0$. Furthermore 
our small expansion for the warp factor 
\eqn\wurp{F_i(\tilde r)~ = ~ F_i(r_0) ~+~ {r\o \sqrt{F_1(r_0)}}~{\del F_i \o \del \tilde r}\Big\vert_{\tilde r = r_0}
~\approx F_i(r_0) ~~{\rm for}~~r \to 0}
is thus justifiable. Therefore plugging \expone\ into \newmetnow, we can easily show that the local metric is 
exactly of the form \metformj\ with the only difference being that the $\Delta_i^0$ are now defined as
\eqn\deltanow{\Delta_1^0 ~ = ~ {F_2(r_0)\o \sqrt{F_4(r_0)}}, ~~~~~~~~~ 
\Delta_2^0 ~ = ~ {F_2(r_0)\o \sqrt{F_6(r_0)}}}
which reduce to \deltas\ once we remove the quantum corrections. 

Our small calculation discussed above
actually has some profound consequences. It tells us that \iibmet\ is not only the right 
local metric for our case, but also that \iibmet\ is not affected by the details of the quantum corrections envisioned 
on \metresconi. All the possible corrections simply change the warp factors $\Delta_1$ and $\Delta_2$. But there is 
an even more intriguing possibilty related to footnote 5 that we mentioned earlier. Imagine we do manage to find
a supersymmetric wrapped D5 brane system. The new metric should look like \newmetnow\ precisely because \newmetnow\ 
captures all the essentials of a metric of D5 wrapped on two cycles of a resolved conifold. Thus the quantum corrected
metric for \metresconi\ and the expected susy solution will bear strong resemblance. However in both cases the local 
metric is exactly \iibmet\ so our local description with the metric \iibmet\ will not change and we can consider
seven branes in the system to study ${\cal N} = 1$ gauge theory with flavors. Integrating out the flavors 
(i.e seven branes are far away or non-degenerating F-theory torus of \gtone) is equivalent 
to considering the metric \metformj\ locally and this is what was taken in \gtone\foot{To make this 
precise, the ($r_0, \langle\theta_i\rangle$) in eq. 5.2
of \gtone\ is the ($r_0, \langle\theta_i\rangle$) of \defofcoord\ here also.}.

\subsec{Field Theory Interpretation}

In geometric transitions, gluino condensates are identified with sizes of
geometrical cycles. In the global picture discussed in the previous
section, this identification will be between
the gluino condensate describing the IR of the ${\cal N} = 1~ SU(N)$ theory and the size of the $S^3$ cycle in the
deformed geometry.

To understand what happens in the local picture let us start from the discussion of type I
in \senbanks. On one wrapped D5 brane the field theory is ${\cal N} = 1~ Sp(2)$ with some
twisting of the normal bundle in order to account for the wrapping on a 2--cycle. It also contains 32 $Sp(2)$
``half-hypermultiplets'' which will become massive due to Wilson lines. After a T-duality to type IIB,
we get a D5 wrapped on a $P^1$ cycle. If one concentrates on the vicinity of a single fixed point and 
the D7 branes which
are located around this fixed point,
this gives rise to an $Sp(2)$ or $SU(2)$ theory with 4 massive flavors and a
superpotential
\eqn\supsen{W=\sum_{i} m_i Q_i \bar{Q}_i~+~Q_i A \bar{Q}_i,~~~~~~~~~i=1,\cdots,4}
Because of the Wilson lines in type I, the gauge group $Sp(2)$ breaks to $U(1)$ and the masses of the
fundamental flavors will change as $m_i \pm a$, where $a$ is the expectation value of an adjoint Higgs field.

The difference between global and local picture then amounts to whether or not we consider the
fundamental flavors to be part of the gauge theory\foot{The other flavors corresponding to the other
three fixed points are already integrated out from an even richer theory. We will
consider an intermediate scale to accomodate
only one flavor from a specific fixed point.}. If we integrate out the flavors, then
the IR gauge coupling of the effective theory will be constant.

If we do not integrate out the flavors, the
gauge coupling constant will be given by
\eqn\taunc{\tau(z)~=~\tau_0+{1 \o 2 \pi i}\left(\sum_{i=1}^{4} {\rm log}~(z-m_i^2)-4 {\rm log}~ z\right)} where
$z$ is the complex direction orthogonal to the D7 branes. As $\tau$ depends on the size of cycles in the metric \fiitaamet, functions like $g_2$ are part of $\tau$ and will
depend nontrivially on
$z$, too. But the D7 branes are extended in directions both parallel and orthogonal to the $P^1$ cycle in the
resolved conifold, therefore they will also be both parallel and orthogonal to the $S^3$ cycle in the
deformed conifold. The coupling $\tau$ will then
depend on both the integral over A-cycle and B-cycle. The identification
between $\tau$ and $\int~(\chi + i e^{-2\phi})~ H_{NS}$ will change if $\tau$ is given by \taunc.
This is to be compared with the case in \radu\ where the flavors were introduced as infinite D5 branes wrapped on
2-cycles which do not intersect the $P^1$ cycles and which become part of the $S^3$ after the transition.

This modifies the approach of
\vafai\ if the flavors introduced by D7 branes are still part of the
theory, but the original geometric transition (on the level of the effective field theory) is obtained if the flavors are integrated out.

In the case of $N$ D5 branes in type I (and several D5 branes in IIB), we find gauge group $Sp(2N)$, which is
then broken by the adjoint Higgs field to $SU(N)$. Again, the usual
geometric transition arises if all massive flavors
are integrated out.

The local picture constructed in \gtone\ and in \realm\ captures almost all the details of geometric
transitions envisioned by \vafai, as in this limit the effect of D7/O7 branes is invisible.
If the flavors corresponding to D7 branes are
integrated in, then the picture of \vafai\ has to be extended accordingly
to accomodate the flavor degrees of freedom. In
this paper, unless mentioned otherwise,
we will stick with the simplest model where the D7 branes and orientifold seven planes are far from the
resolution two-cycle of our manifold.

\subsec{Topological aspects and excursions into generalized
geometries}

The analysis of supersymmetry in type IIB theory that we gave above therefore
relies heavily on the correct choices of NS and RR fluxes (including the
RR fluxes from seven branes). On the other hand, the topological sector
of this model, which would be --- say after geometric transition --- a
B-model on a deformed conifold has {\it no} dependence on the
RR fields. Does that mean that we are allowed to choose any
arbitrary background $H_{RR}$ fields?
We will at least show that the RR fields are quantized without leaning on the assumption
that they are governed by the number of branes before geometric transition (because that would
already assume correctness of the geometric
transition).

An alternative verification of the quantization of RR fields can
be done by considering which correlation function we
expect to measure. Taking into account the $H_{NS}$ fields,
the fermionic part of a Green-Schwarz superstring can be written
as \eqn\gsstring{S^g_{\rm fermionic} = 4i \psi^p \Delta_+ \psi^p +
4i \psi^{\dot q} \Delta_- \psi^{\dot q} + R_{ijkl}
\sigma^{ij}_{\dot p \dot q} \sigma^{kl}_{rs}\psi^{\dot p}
\psi^{\dot q} \psi^{r}\psi^{s}} where $(\psi^{p}, \psi^{\dot q})$
are the two inequivalent spinor representations of the transverse
$D_4$ and the sigma matrices are defines as $\sigma^{ij}_{\dot p
\dot q} \equiv \Gamma^{[i}_{r[\dot p}\Gamma^{j]}_{\dot q]r}$ with
a similar definition for the other components. The Gamma matrix
has $8 \times  8$ blocks given as $\Gamma^i \equiv  \pmatrix{0&
\Gamma^i_{p\dot q}\cr \Gamma^i_{\dot r s}& 0}$ which are used to
define the $\sigma$'s above. In this notation we have to specify
what we mean by the covariant derivative $\Delta_\pm$ and
$R_{ijkl}$. The most generic definition of the covariant
derivative will be given as \hull,\bbdg: \eqn\lapid{\Delta_\pm
\psi^{q(\dot q)} = \del_\pm \psi^{q(\dot q)} + {1\o 2}
\left(\omega - {1\o 2} H \right)^{ab} \sigma^{pq(\dot p \dot
q)}_{ab} \psi^{p(\dot p)}} where ${1\o 2} H$ forms the torsion and
we have chosen the torsional connection $\omega_+ \equiv \omega -
{1\o 2} H$, and not the other one i.e $\omega_-$ (see \bbdg\ for
details on this). Observe that in the absence of $H_{NS}$ there is
no such ambiguity and the GS superstring can be defined
unambiguously.

The curvature $R_{ijkl}$ that we defined above is actually
measured w.r.t. the connection $\omega_+$. We could also measure
the curvature w.r.t. the other connection $\omega_-$. If we define
$R^+_{ijkl} \equiv R_{ijkl}$ and $R^-_{ijkl}$ as the other
curvature\foot{The definition of these curvatures are
$R^{(\pm)i}_{jkl} = \del_k \Gamma^{(\pm)i}_{lj} +
\Gamma^{(\pm)i}_{km}\Gamma^{(\pm)m}_{lj} - (k \leftrightarrow l)$
where $\Gamma^{\pm}$ is the torsional connection.}, then the
following identity can be easily verified:
\eqn\curviden{R^+_{ijkl} = R^-_{ijkl} - 2H_{[ijk,l]}} which also
serves as the definition of the torsion \hull, \bbdg, \bbdgs. We want to
concentrate on the topological nature of these models.

An efficient way to study the topological theory is to
convert the GS fermions into NS forms, i.e measure the
fermions as worldsheet fermions with ($p,q$) $U(1)$ charges. This
will elucidate the (2,2) nature of the lagrangian. Thus we will
have \eqn\ferm{(\psi^p, \psi^{\dot q})~ \to ~ (\psi_\pm^i,
\bar\psi_{\pm}^j) ~ \in ~ \Bigg( \left(K^{1/2}_\Sigma, \bar
K^{1/2}_\Sigma\right) \otimes \phi^\ast\left(T^{(1,0)}{\cal
M}\right), \left(K^{1/2}_\Sigma, \bar K^{1/2}_\Sigma\right)
\otimes \phi^\ast\left(T^{(0,1)}{\cal M}\right) \Bigg)} where
$\Sigma$ is the two dimensional world-sheet and the background
manifold is given by ${\cal M}$ with a particular choice of almost
complex structure. Therefore the 2d local complex coordinates
would be ($z, \bar z$) and the corresponding 6d ones would be
($T^{(1,0)}{\cal M}, T^{(0,1)}{\cal M}$). The manifold ${\cal M}$
therefore admits an almost complex structure which may or may not be
integrable. In type IIB ${\cal M}$ will be a {\it conformally}
K\"ahler manifold with an integrable complex structure \realm,
whereas in type IIA, ${\cal M}$ will in general be a non-K\"ahler
manifold
with a non-integrable complex structure \gtone. The sigma model
action can now be re-written from \gsstring\ to
\eqn\nsstring{S^n_{\rm fermionic} = 2i (g_{i\bar j}+B_{i\bar j})
\left(\bar\psi^j_-\Delta_+\psi^i_- +
\bar\psi^j_+\Delta_-\psi^i_+\right) + R^+_{i\bar j k \bar l}
\psi^i_+\psi^k_-\bar\psi^j_+\bar\psi^l_-} where the curvature etc.
are measured w.r.t. the right connection. Observe the appearance
of ${\bf g + B}$ in the above action. This takes care of the
torsion.

The topological twist in the absence of torsion
acts on the bundle in the following way: it converts
one of the fermion to a world-sheet one-form and the other fermion to a world-sheet
zero form \wittencoho.
In particular its action is:
\eqn\twist{(\psi^i_+, \psi^i_-) ~~ \to ~~ (\psi^i_z, \chi^i) \in \left(K_\Sigma \otimes
\phi^\ast(T^{(1,0)}{\cal M}), \phi^\ast(T^{(1,0)}{\cal M})\right)}
with a similar kind of action for the other pair. This implies that the correlation
function of operators $<{\cal O}_1{\cal O}_2{\cal O}_3....{\cal O}_n>$ is given by the
integral of the top form on the manifold ${\cal M}$. Generically, operators in
the twisted sigma model are of the form\foot{For a review of topological correlation function
the readers may want to refer to the original work of Witten \wittencoho, or some of the
recent review papers \toprev.}
\eqn\oper{{\cal O}_f = f_{i_1...i_p\bar j_1...\bar j_q} \chi^{i_1}...\chi^{i_p}\chi^{\bar j_1}
...\chi^{\bar j_q} = f_{i_1...i_p\bar j_1...\bar j_q} dZ^{i_1}...dZ^{i_p} d\bar Z^{j_1}...
d\bar Z^{j_q}}
which is basically a ($p,q$) form on the manifold ${\cal M}$ because ($Z^i, \bar Z^j$) are the
complex coordinates on the manifold determined by an almost complex structure.

In the presence of torsion the issue of a topological twist is much more complicated.
(2,2) nonlinear sigma models (before twist) with $H \equiv
H_{NS} = dB \ne 0$ were first discussed in \gates. In fact the earlier mentioned
two different
spin-connections $\omega_\pm$ will become useful now\foot{The
heterotic side of the story is completed in \bbdg,\bbdgs.}. Due to
the existence of $\omega_\pm$, there are two different complex
structures compatible w.r.t. to the choices of torsional
spin-connections \gates. These complex structures can then be used
to decompose the fermionic components of the (2,2) sigma model
\kapuli. The zero mode distribution will be similar to the case without torsion:
\eqn\zerom{k_\mp \equiv {\rm dim~ker}~\Delta_\pm -
{\rm dim~ker}~\Delta^\dagger_\pm} implying vanishing first
Chern-classes\foot{This doesn't mean that the target manifold has
to be K\"ahler. In \bbdg, \bbdgs\ numerous non-K\"ahler manifolds
have been constructed with vanishing first Chern class.}. Once
such constraints are observed, the twisting proceeds in the standard
way (see \kapuli\ for recent discussion where the techniques of
\wittentop\ is used to do the twisting). A generic
operator in such a twisted model looks similar to \oper, and is therefore related to the cohomological states
of the underlying manifold.

The above form of the correlation function and the operators
solve the RR--flux question we raised. Since the
correlators involve ($p,q$) forms, they are related to $H^{p,q}$
of the manifold ${\cal M}$. Its elements are quantized and
normalizable forms of the manifold, and therefore the RR forms
will be given by quantized normalizable three forms. And indeed,
these forms are related to the {\it number} of wrapped D5 branes
in the dual picture\foot{The dual picture that we are talking
about here is only the far IR story {\it a-la} \ks. We can also
build a setup with spatially varying NS and RR fluxes
(or infinite sequences of flop transitions in the dual) like
\civd\ but will not do so here.}.

The analysis presented so far only skims the surface of a much
more rigorous structure. Not only the twisting can be elaborated
for sigma models with ($2,2$) supersymmetry, but they also have a
deeper connection to generalized complex structures developed
by Hitchin \hitchin\ and Gualtieri \gualtieri.
These will be studied in the sequel to this
paper. We will also try to
connect the type II solutions that we presented in \gtone\ and
\realm\ with the Hitchin--Gualtieri system. A more recent paper
analyzing some aspect of this connection is \kapuli.

Before we proceed with the analysis of topological
aspects of ($2,2$) models, let us make a few observations. These
will be useful to relate the analysis to heterotic ($0,2$) models.
First, the ($2,2$) action can be written in a concise way using
superfield notation as \gates: \eqn\superfield{S = \int
d^2\sigma d^2\theta\Big[ g_{ij}D^\alpha {\bf \Phi}^i~ D_\alpha
{\bf \Phi}^j + B_{ij} D^\alpha {\bf \Phi}^i~ (\sigma^3 D)_\alpha
{\bf \Phi}^j\Big] + ...} where $\sigma^3$ is the third Pauli
matrix and the dotted terms involve the four--fermi terms. As
is clear from the above action, the superfields ${\bf \Phi}^i$, $i
= 1, .., d$ realize only the ($1,1$) part of the action.
Additional supersymmetry can be implemented w.r.t. the two allowed
choices of the complex structure $J^i_{\pm j}$ as\foot{We are
using both $D_\pm$ and $\Delta_\pm$ to write the derivatives in a
sigma-model. The distinction between them should be clear from the
context.} \eqn\susie{\delta_\epsilon {\bf \Phi}^i =
-i(\epsilon_+D_-{\bf \Phi}^j)J^i_{+j} + i(\epsilon_-D_+{\bf
\Phi}^j)J^i_{-j}}
\susie\ is a combination of two different
variations: one for chiral superfields in a (2,2) sigma model, and the
other one for a twisted\foot{Not to be confused with the topological
twist that we perform on this model.} chiral superfield (see sec. 8 of \gates).
This means
that the very existence of $B_{ij}$ fields in the background
implies a (2,2) sigma model action {\it coupled} to twisted chiral
fields\foot{As is obvious from the above analysis, in the absence
of $B$-field the target manifold has to be K\"ahler for \susie\ to
exist \zumino. The presence of $B$ fields allows two different
complex structures $J_{\pm j}^i$ that are covariantly constant
w.r.t. to the two connections $\omega_{\mp}$, respectively. The two
complex structures appearing in the susy variation \susie\ are
perfectly consistent with the fact that chiral and twisted chiral
superfield susy transformations require two {\it different}
complex structures. These two choices of complex structure differ
by the Pauli matrix $\sigma^3$, and this is the reason why
$\sigma^3$ appears in the sigma-model action \superfield. All
these details fit nicely with the expected theory of chiral and
twisted-chiral superfields, as shown in \gates. Furthermore, as noted in \kapuli, this bi--Hermitian structure
satisfies the definition for a twisted generalized K\"ahler target \gualtieri.}. This is
basically the content of the action \superfield\ as twisted chiral
fields (say $\eta^p$) couple with chiral fields (say
$\Phi^q$) as \gates: \eqn\twistrojo{S_{\rm coupling} = \int
d^2\sigma d^2\theta \Big( K_{p\bar q} (D^a \Phi^p) (\sigma^3 D)_a
{\bar\eta}^{\bar q} + K_{\bar p q} (D^a \bar\Phi^{\bar p})
(\sigma^3 D)_a {\eta}^{q}\Big)} which is precisely the $B$ field
coupling in \superfield\ if we interpret ${\bf \Phi}^i$ as a
superfield with both chiral and twisted chiral components.

The above observation has in fact interesting consequences for
the connection of this model with generalized complex geometry
developed recently by Hitchin \hitchin\ and Gualtieri \gualtieri.
This connection is explored in a series of papers
\lindstrom,\minasian. We only give a brief account here and leave the detailed discussion for future work.
The starting point
is the observation repeatedly mentioned above,
namely: the action \superfield\ as constructed naively shows (1,1)
world--sheet supersymmetry. The additional supersymmetry is implemented via the
transformation \susie\ which eventually enhances it to (2,2).

Where does the mathematical structure of
generalized complex geometry fit into this? Note
that once we have the (2,2) action in this form we can
easily relate it to the fermionic term in \gsstring\ or
\nsstring. Defining the bosonic part of ${\bf \Phi}^i$ as
$X^i$, the action \superfield\ takes the following form:
\eqn\supn{S = {1\o 8\pi \alpha'} \int d^2\sigma \Big[(g_{ij} +
B_{ij}) \del_+ X^i \del_- X^j + {1\o 4}S^g_{\rm fermionic}\Big].}
Written this way, the action will require no other
corrections\foot{For example Chern-Simons corrections.} and
consequently be anomaly free. Observe that the right moving sector has eight
fermions denoted as $\psi^p$ in \gsstring. On the other
hand, the left moving sector also has eight fermions
denoted as $\psi^{\dot q}$. Together they give rise to the (2,2)
world sheet action. The bosonic part (i.e. the $X^i$ part) of the
above action can be re--written in the following way:
\eqn\poisson{S_{\rm bosonic} = \int p_\mu \wedge dX^\mu + {1\o 2}
\theta^{\mu\nu} \eta_\mu \wedge \eta_\nu + {1\o 2} G^{\mu\nu}
\eta_\mu \wedge \ast\eta_\nu + {1\o 2} B} where $\eta_\mu$ is a
one--form on the world--sheet and $G_{\mu\nu}, \theta_{\mu\nu}$ are
related to the {\it open--string} data of a Seiberg--Witten
non--commutative theory \swncg. The above form of the action first
appeared in \nekrasovtop\ and is subsequently used by \lindstrom.

Supersymmetrizing \poisson\ is not difficult. The original action
\supn\ is supersymmetrized using $S^g_{\rm fermionic}$ terms. Here
we see that susy requires the following substitutions
\eqn\susysubs{\del_\pm X^i~~ \to ~~ D_\pm {\bf \Phi}^i, ~~~~~
\eta_\mu ~~ \to ~~ \Psi_{\pm \mu}, ~~~~~ (G_{\mu\nu},
\theta_{\mu\nu}) ~~\to ~~ {\Big(}E_{(\mu\nu)},
E_{[\mu\nu]}{\Big)}.} We can now write the most generic action
using the ingredients of \susysubs\ and \poisson. This will
typically look like: \eqn\susylook{S_{\rm gen} = \int d^2\sigma
d^2\theta \Big( a_1~D_+{\bf \Phi}^\mu \Psi_{-\mu} + a_2~D_-{\bf
\Phi}^\mu \Psi_{+\mu} + a_3 ~\Psi_{+\mu}\Psi_{-\nu}E^{\mu\nu} +
a_4~D_+{\bf \Phi}^\mu D_-{\bf \Phi}^\nu B_{\mu\nu}\Big)} where
$a_i$ are some unknown coefficients. In writing the above action
we strictly followed \poisson. This would explain the
non--existence of terms like $D_+{\bf \Phi}^\mu D_-{\bf \Phi}^\nu
E_{\mu\nu}$ in the action.
In \lindstrom\ it was pointed out that that to achieve (1,1) supersymmetry,
the coefficients in \susylook\ have to be precisely \eqn\aivalues{a_1 ~=~ 1, ~~~~~~ a_2 ~=~ -1,
~~~~~~a_3 ~=~ -1, ~~~~~~a_4 ~=~ 1.}
Above technique therefore
demonstrates a simple way to obtain a supersymmetric action from the bosonic
action of \nekrasovtop. The next question will is: under what
conditions does \susylook\ have (2,2) supersymmetry? Since we have
already fixed the unknown coefficients $a_i$, the only other way
is to look for target space geometry. This is exactly where the
mathematical construction of generalized complex geometries fits
in! It turns out, as was discussed by Lindstrom in \lindstrom\
(see the first paper in the list), the target manifold has to be a bi--Hermitian or, equivalently, twisted
generalized K\"ahler manifold \gualtieri\ to attain the full (2,2) world--sheet
supersymmetry.

The discussion how the manifolds constructed in \gtone\ and
\realm\ fit into this (twisted) generalized complex framework will be left for the
sequel to this paper. Our aim
in the next section is to study the heterotic side of the story, i.e. (0,2) models. We will
be able to provide a full global description of the particular setup initiated in \realm.
But before doing so, let us conclude this section by pointing out one obvious result: integrating
out $\Psi_{\pm \mu}$ in \susylook\ reproduces back \supn.

\newsec{Global Heterotic models}

In the type II analysis in \gtone, \realm\ and above we showed that the global descriptions for
the geometric transition backgrounds are particularly involved. In type IIB we gave
a local metric, and the global background is a K\"ahler manifold with additional D7 branes and O7 planes. On the
other hand, the type IIA background is locally and globally non--K\"ahler.
We now want to discuss the type I/Heterotic
models whose local descriptions were presented in \realm. These backgrounds turned out to be, not surprisingly,
non--K\"ahler (but complex) manifolds. In the following we present global descriptions of these
models.

\subsec{Heterotic (0,2) models: brief discussion}

To develop (0,2) models in the context of complex
structures and not generalized complex structures, we have to
go back to \supn. In this action we have the freedom
to add non--interacting fields. This ruins the
carefully balanced (2,2) supersymmetry of this model. We can
use this to our advantage by adding non--interacting fields {\it
only} in the left--moving sector. This breaks the
left moving supersymmetry, and one might therefore hope
to obtain an action for (0,2) models from \supn, at least {\it
classically}. On the other hand, a possible (0,2) action is also restricted
because this will be the action for heterotic string. It turns out,
there are few allowed changes one can do to find the
classical (0,2) action from a given (2,2) action:

\noindent $\bullet$ Keep the right moving sector unchanged, i.e. $\psi^p$ remain as before.

\noindent $\bullet$ In the left moving sector, replace $\psi^{\dot q}$ by eight
fermions $\Psi^a$, $a = 1, ... 8$. Also add 24 additional non--interacting
fermions $\Psi^b$, $b = 9, ... 32$.

\noindent $\bullet$ Replace $\omega_+$ by gauge fields $A$, i.e. embed the {\it torsional}
spin connection into the gauge connection.

The above set of transformations will convert the classical (2,2)
action given in \supn\ to a classical (0,2) one. One might,
however, wonder about the Bianchi identity in the heterotic theory.
The type IIB three--form fields are closed, whereas heterotic
three--form fields satisfy the Bianchi identity. One immediate
reconciliation would be that  because of the {\it embedding}
$\omega_+ = A$, the heterotic three--form should be closed. This
may seem like an admissible solution to the problem, but because
of subtleties mentioned in \bbdg, \smit, \papad\ the above
embedding will not allow any compact non--K\"ahler manifolds to appear in
the heterotic theory. Therefore, as a first approximation,
we will assume an embedding of the
form \eqn\bedd{ \omega_+ ~ = ~ A ~ + ~{\cal O}(\alpha').} Using
this the new action with (0,2) supersymmetry becomes:
\eqn\snow{\eqalign{S = {1\o 8\pi \alpha'}& \int d^2\sigma
\Big[(g_{ij} + B_{ij}) \del_+ X^i \del_- X^j + i \psi^p (\Delta_+
\psi)^p + i \Psi^A (\Delta_-\Psi)^A + \cr & ~~~~~~~~~~~~ +~ {1\o
2}F_{ij(AB)} \sigma^{ij}_{pq} ~\psi^p \psi^q \Psi^A \Psi^B + {\cal
O}(\alpha')\Big]}} where due to the embedding \bedd, $F^a_{ij}$
forms the Yang--Mills field strength measured w.r.t. Lie algebra
matrices $T^a_{AB}$. The fermion indices are $A = 1,..., 32$, which means
there are 32 fermions, and hence $T^a$ form tensors of
rank 16. The reader can easily identify the above action as an
action for a heterotic sigma model with torsion \hull, \bbdg, \bbdgs,
\rstrom, \hullwit, \gross, \hulltown. The action of the
Laplacian \lapid\ changes accordingly to
\eqn\lapidnow{\eqalign{& \Delta_-\Psi^A = \del_-\Psi^A + A^{AB}_i
(\del_-X^i)~\Psi^B\cr &  \Delta_+\psi^p = \del_+\psi^p + {1\o
2}(\omega_+)^{ab} \sigma^{pq}_{ab} \psi^q \cr & H_{ijk} = {1\o
2}\left(B_{ij,k} + B_{jk,i} + B_{kj,i}\right)\,.}} As expected, this
set of actions  \lapidnow\ determines the (0,1) supersymmetric heterotic sigma
model \hullwit. This is similar to the (1,1) action for the type
II case. The full (0,2) susy will be determined by additional
actions on the fields (exactly as for the (1,1) case before).

The above set of manipulations
that convert a classical (2,2) action to a classical (0,2)
one will help us to understand various things about the heterotic theory using data of type
II theories. In particular we would like to ask the following questions:

\noindent $\bullet$ Can we use this to find new torsional background in heterotic theory?
In \realm\ we provided a possible background in heterotic theory
which {\it might} show geometric transition. However, this background was duality chased
from the orientifold corner of type IIB theory. This
means, in particular, that quantum corrections would modify both type II as well as
the heterotic background once we shift the type IIB background from the orientifold
point. Although it is possible to infer the quantum corrections to the type IIB background (i.e.
from F-theory), the corresponding correction in the heterotic picture is not easy to
determine.
Therefore, using above manipulations we might be able to infer a possible background
in heterotic theory that would show geometric transition.

\noindent $\bullet$ From above manipulations we also observe
that both metric and $B_{ij}$ fields can be taken to the
heterotic side. However, this is only possible if the original
(2,2) background does not have any $H_{RR}$ fields. In the presence
of $H_{RR}$ the simple manipulations that we performed cannot give
a (0,2) or a (0,1) model. We are therefore particularly interested in
type II models that allow only for an NS three--form, like MN \mn.
When both NS and RR backgrounds are present, it might still be possible to perform above manipulations if one can find an equivalent U--dual background. This U--dual background will in general not be K\"ahler
(not even conformally K\"ahler). Observe that these U-dualities {\it do not}
require the original background to be at the orientifold point.
This is therefore different from the analysis performed in \sav,
\beckerD\foot{Another interesting question would be to allow for both
NS and RR background in the S--dual type I picture. This is a
highly restrictive scenario as the allowed values of NS fluxes in
the S--dual type I picture are only two discrete choices
\sensethi.}.

\noindent $\bullet$ Once a particular heterotic background is found one has to address
the issue of vector bundles\foot{The vector bundles that we are
interested in are a subgroup of the $SO(32)$ bundle of the heterotic theory. In fact, there
are two possible bundles here. If the proposed heterotic background
is dual to an $SU(N)$ gauge theory, where $N$ is the number of wrapped NS5 branes, then the
$SU(N)$ gauge group may also appear in the geometry. This would be a
much more involved case. If a dual theory exists, we have to determine
the full UV behavior of this theory. We can assume (as in the type II cases) that
the geometric dual just describes the far IR of a much more detailed theory, and  $-$ if
we also assume that the theory will eventually confine $-$  then $SU(N)$ will be broken. This means we only have to consider $SO(32)$ (or possibly its
subgroup because of the embedding \bedd). More details will follow.}.
Two sets of questions arise:

\noindent (a) What are the allowed vector bundles on the manifold? How do we study the
stability of these bundles?

\noindent (b) How do we pull bundles through a geometric transition (or conifold transition)?

\noindent The first issue, of the existence of the bundle, can be inferred from the
detailed analysis given (at least for the $U(1)$ case) in \bbdgs, \lust. However, the
situation here may become a little simpler than the one in \bbdgs, because
of the embedding \bedd. This embedding implies that for a given complex structure
$J$
\eqn\comp{ i \del \bar\del J ~ = ~ 0 + {\cal O}(\alpha')\,.}
This means we can study the stability of bundles using the
recent analysis of Li and Yau \yauli\ and Fu and Yau \yaufu.
The second issue, of pulling the bundles through
a conifold transition, is important if we want to present a heterotic model
that does show geometric transition. In \bbdgs\ it was shown how bundles can be
pulled through flops. Here we would like to study a more subtle phenomena: the
analysis of the bundles after geometric transition.

\noindent $\bullet$ Our manipulations can now be
used to understand half--twisted sigma models from type II
twisted models. The half--twist is performed
only to the right moving sector of the sigma model \snow.
This was presented first in \wittentop. Our aim is to
understand in more detail the Chiral de Rham complex CDR
\malikov. Because of the embedding \bedd, we expect a connection
between the chiral differential operators CDO \malikov, \schect\
and the chiral de Rham complex. This issue has also recently been
addressed by Witten \wittwo. CDO's are relevant for the study of conformal
field theories for models that are {\it twisted} or
{\it half-twisted}. Therefore, all the relevant operators in the theory
are written as BRST exact operators w.r.t. the BRST charge $Q$. In this
way the CFT (which is also a TFT) is different from the usual CFT
for string sigma models. As discussed in
\wittwo, under the embedding \bedd, the CDO is then a half--twisted
version of a sigma model with (2,2) supersymmetry called the CDR.


\subsec{Heterotic (0,2) models: detailed discussion}

To use the sigma-model analysis that we presented in the above
section, we have to determine a particular heterotic background
and show that such an equivalent background can also appear in the
type IIB set-up. In our earlier paper \realm\ we provided a
heterotic background that has non-trivial dilaton, metric and
torsion. The explicit vielbeins for this background {\it after}
geometric transition were given in \realm\ as:
\eqn\vielsnow{\eqalign{& e^1 = (h_2 + a_2^2 h_1)^{-{1\o 2}}
(C'D')^{-{1\o 4}} (dy - b_{yj}~ d\zeta^j), ~~~~ e^2 = (h_4 + a_1^2
h_1)^{-{1\o 2}} (C'D')^{-{1\o 4}} (dx - b_{xi} ~d\zeta^i) \cr &
e^3 = {1\o \sqrt 2}(h_3 h_4)^{1\o 2} (C'D')^{-{1\o 4}}(d\theta_1 +
\gamma_4 d\theta_2), ~~~~ e^4 = {1\o \sqrt 2}(h_3 h_4)^{1\o 2}
(C'D')^{-{1\o 4}}(d\theta_1 + \gamma_5 d\theta_2)\cr & e^5 =
\gamma^{'{1\o 2}} H^{1\o 4} (C'D')^{-{1\o 4}} dr, ~~~~~ e^6 =
h_1^{1\o 2} (C'D')^{-{1\o 4}} dz}} where, as one might recall from
\realm, the vielbeins $e^1$ and $e^2$ are simple because ${\rm Re}
~\tau_1 = 0$, whereas the vielbeins $e^3$ and $e^4$ contain mixed
components because ${\rm Re}~\tau_2 \ne 0$, where
\eqn\chichi{d\chi_1 \equiv dx + \tau_1 dy, ~~~~~~~~ d\chi_2 \equiv
d\theta_1 + \tau_2 d\theta_2.} It was pointed out that for the
background after transition the complex structure is specified by
${\rm Re}~\tau_1 = 0, {\rm Re}~\tau_2 \ne 0$, whereas the opposite
holds true for the background before transition. The variables
$\gamma_4$ and $\gamma_5$ appearing in \vielsnow\ are defined as:
\eqn\gamviel{\gamma_4 = {\rm Re}~\tau_2 \pm {\rm Im}~\tau_2, ~~~~
\gamma_5 = {\rm Re}~\tau_2 \mp {\rm Im}~\tau_2} with $\tau_2$
defined above in \chichi. Similarly the other variables were given
in \realm\ as: \eqn\hideterm{\eqalign{& h_1 = {e^{-2\phi}\o
\alpha_0 CD}, ~~~ h_2 = \alpha_0(C + e^{2\phi} E^2), ~~~ h_4 =
\alpha_0(D + e^{2\phi} F^2) \cr
& h_3 = {C - \beta_1^2 E^2 \o \alpha_0(D + e^{2\phi} F^2)}, ~ a_1
= -\alpha_0 e^{2\phi} ED, ~ a_2 = -\alpha_0 e^{2\phi} FC, ~ \alpha
= {1\o 1+E^2 + F^2}\cr &  \alpha_0 = {1\o CD + (CF^2 + D E^2)
e^{2\phi}},~~~~ \beta_1 = {\sqrt{\alpha_0} e^{2\phi}\o \sqrt{
e^{2\phi} CD - {(C + e^{2\phi} E^2) (1-D^2) F^{-2}}}}\cr & C_\pm ~ = ~
{\alpha\o 2}\left(1+F^2 \pm {EF\o \sqrt{(1+E^2)/(1+F^2)}}\right) , ~~~~~ C~=~C_-, ~~~~ D ~=~ C_+
\cr & E = \Delta_1~\cot\theta_1, ~~~~~~ F =
\Delta_2~\cot\theta_2, ~~~~~~ r~ =~ r_0,~~~~~~ 
\theta_i ~ = ~ {\rm arbitrary}}} 
where $h_i, C, D$ and the dilaton $\phi$ are all
evaluated at fixed radial
coordinates $r = r_0$ and $C', D'$ are defined for fixed $\theta_i = \langle\theta_i\rangle$ also. 
$\gamma(r^2)$ satisfies
$3\gamma'\gamma(\gamma+4 a^2)=2 r^2$ where $\gamma'=\partial
\gamma/\partial r^2$ (again both measured at $r = r_0$). 
$H$ is a warp factor due to D5 and D7-branes
as well as O7--planes, but it was not further specified in \realm.
Using the above definitions the metric after geometric transition
is simply $ds^2 = \eta_{ab} e^a e^b$ with $e^a = e^a_\mu dx^\mu$.
This is equivalent to: \eqn\tyhet{ds^2 = {\cal A}_1~dz^2 + {\cal
A}_2~(dy - b_{yj}~ d\zeta^j)^2 + {\cal A}_3 ~(dx - b_{xi}
~d\zeta^i)^2  + {\cal A}_4~\vert d\chi_2 \vert^2 + {\cal A}_5~
dr^2} with the coefficients ${\cal A}_i$ are written in terms of
$h_i, a_i$ etc. as \eqn\defcala{\eqalign{& {\cal A}_2 = {1\o (h_2
+ a_2^2 h_1)\sqrt{C'D'}}, ~~~~~ {\cal A}_3 = {1\o (h_4 + a_1^2
h_1)\sqrt{C'D'}}\cr &
 {\cal A}_1 = {h_1 \o \sqrt{C'D'}}, ~~~ {\cal A}_4 = {h_3 h_4 \o
\sqrt{C'D'}}, ~~~ {\cal A}_5 =  {\gamma' \sqrt{H} \o
\sqrt{C'D'}}}} The metric \tyhet\ is somewhat similar to the type
I picture developed earlier in \realm. The torsion and the
coupling are \eqn\torcnow{\eqalign{& H_{\rm het}~ \equiv \tilde H
~=~ \tilde{\cal H}^b_{xz\theta_1}~ dy ~\wedge ~dz ~\wedge
~d\theta_2 - \tilde{\cal H}^b_{yz\theta_2}~ dx ~\wedge ~dz ~\wedge
~d\theta_1 ~ + \cr & ~~~~~~~~~~ + ~\tilde{\cal H}^b_{xzr}~ dy
~\wedge ~ dz ~\wedge~ dr~ -~ \tilde{\cal H}^b_{yzr}~dx ~\wedge ~dz
~\wedge ~dr \cr & ~~~~~ g^{\rm het} = { 1\o \sqrt{C'D'}}}} The
above background is of course {\it not} the complete background.
We need stable vector bundles satisfying torsional DUY equations.
This will be dealt a little later. Here we make the following
observations:

\noindent $\bullet$ In this background the dilaton is proportional
to the warp factors appearing in the metric. This is consistent
with the expectations for a torsional background
\hull,\rstrom,\sav, \beckerD, \bbdg, \lisheng.

\noindent $\bullet$
As emphasized in \realm, it could be possible
that the dual background before geometric transition is related to
$NS5$ branes wrapped on two cycles of the non-K\"ahler manifold.
These stem from D5--branes wrapped on a two cycle of a resolved
conifold (which survives the orientifold operation) in IIB. After
the geometric transition we get a heterotic background which has
only fluxes but no branes. The question here is whether we can
shrink the two cycle on which we have wrapped $NS5$ branes and get
another background with fluxes. At a more fundamental level, can
the closed string background with fluxes compute anything of the
world volume dynamics on the wrapped $NS5$ branes? To get an
answer to this question, we have to carefully study the heterotic
background.

\noindent $\bullet$ We would also like to give a heterotic
background that is {\it away} from the orientifold point (in IIB).
This is a tricky question, because the U-dualities by which we
obtained our heterotic background are only defined for type IIB
theory at the orientifold point. To find a more generic heterotic
background we could rely on the sigma--model derivation. It states
that a specific torsional background in type IIB theory may be
lifted to heterotic theory as long as there are no RR background
fluxes. One such RR--flux free background is known in type IIB
theory: it is the Maldacena-Nunez torsional background \mn.
However, this cannot be lifted naively because the three--form is
closed and the background has no vector bundles. We will show that
a similar background can be found in heterotic theory for some
specific choices of $b_{ij}$ in \tyhet.

The issue of vector bundles is important. It can be easily shown
that the action \snow\ is invariant under supersymmetry
transformation by a spinor $\sigma$ provided
\eqn\susycol{F^a_{ij}~ \Gamma^{[i} ~ \Gamma^{j]} ~ = ~ 0, ~~~~~~
\del\sigma = {1\o 2} \omega_-^{ab}~\Gamma^{[b} ~ \Gamma^{a]}
\sigma.} This ensures ${\cal N} = 1$ supersymmetry for the
background \tyhet. Observe also the appearance of the other
spin-connection $\omega_-$ in the susy transformations. This stems
from the relation of the (0,2) sigma model with the (2,2) sigma
model. The reader will also recognize that the first condition in
\susycol\ is $J^{ij} F^a_{ij} = 0$ for a fundamental form $J_{ij}$
that is covariantly constant with respect to a connection with
torsion.

At this point one can also entertain the following puzzle: if the
metric \tyhet\ is related to some wrapped NS5 branes in the dual
theory, then two different theories on the NS5 branes seem
possible: a (2,0) theory with self-dual $B^+_{\mu\nu}$
field\foot{This is the world volume $B^+$ field, and shouldn't be
confused with heterotic torsion.} and a (1,1) gauge theory. It
turns out that U-dualities from type IIB background always lead to
the (1,1) theory and not the (2,0) theory\foot{The (2,0) theory on
the other hand can be derived directly from M-theory on an
interval {\it a-la} Horava--Witten \horwit. In fact SO(32)
heterotic theory yields a (1,1) five--brane theory and $E_8\times
E_8$ gives the (2,0) theory \ganorha,\dmone.}.

We now turn to the issue of finding a heterotic metric away from
the type IIB orientifold limit. First, we see that the metric
\tyhet\ can be re--written in the following way:
\eqn\tyhetnow{\eqalign{ds^2 ~ = & {\cal A}_1 ~dz^2 + {\cal A}_2
~dy^2 + {\cal A}_3 ~dx^2 + {\cal A}_4 ~\vert d\chi_2 \vert^2 +
{\cal A}_2 ~ b^2_{yj} (d\zeta^j)^2 + \cr & + {\cal A}_3 ~ b^2_{xi}
(d\zeta^i)^2 - 2 \left[{\cal A}_2~b_{yj}~dy~d\zeta^j + {\cal
A}_3~b_{xi}~dx~d\zeta^i\right]}} where $\zeta^i = \theta_i$ are
the coordinates along angular directions. We allow for a generic
choice of type IIB $B_{NS}$ fields at the orientifold point given
by: \eqn\bfiib{b ~ = ~ b_{x\theta_1}~dx \wedge d\theta_1 +
b_{x\theta_2}~dx \wedge d\theta_2 + b_{y\theta_1}~dy \wedge
d\theta_1 + b_{y\theta_2}~dy \wedge d\theta_2.} Apparently, this
choice of $B_{NS}$ field is more generic than the one for the type
IIB background in \realm. Does the type IIB theory allow such a
choice of $B_{NS}$ field? For an arbitrary choice of type IIB
metric the answer is of course no. However, using the construction
in \realm, we see that the type IIB geometry at the orientifold
point is $T^2 \times T^2 \times S^1 \times R$, with $R$ being the
direction parameterized by $dr^2$. It was shown in
\realm\ that this kind of type IIB geometry and a $B_{NS}$ field
with choice ($b_{x\theta_1}, b_{y\theta_2}$) only, requires the
two $T^2$ tori ($\chi_1, \chi_2$) in \chichi\ to have the
following complex structures (after geometric transition):
\eqn\cotwton{\tau_1 ~ = ~ i \sqrt{h_2+a_2^2h_1 \o h_4+a_1^2h_1},
~~~~ \tau_2 = {1 \o h_3 h_4} \left[a_1 a_2 h_1 + i \sqrt{h_2 h_3
h_4 - a_1^2 a_2^2 h_1^2} \right].} We see that one of the tori is
a square torus, whereas the other is not. Such a compactification
is a non--compact version of $T^6$ flux compactifications studied
in \sav, \beckerD\foot{Any orientifold action will not be visible
in the metric as each of these terms are invariant under any such
actions.}. Therefore it allows for all possible fluxes that are
{\it invariant} under orbifold and orientifold actions. The choice
of $B_{NS}$ field \bfiib\ is clearly invariant under such actions,
and is therefore a valid choice for the IIB background.

We now focus on relating the heterotic background \tyhetnow\ from
\realm\ to a background similar to Maldacena--Nunez with the
metric of a deformed conifold and $B_{NS}$ field, but no RR--flux.
A first step would be to achieve a metric with two square tori,
accompanied by an appropriate change in fluxes.
Here we will assume that we can consistently employ the other
choice of IIB $B_{NS}$ in \bfiib, i.e \eqn\bnow{b ~ = ~
b_{x\theta_2}~dx \wedge d\theta_2 + b_{y\theta_1}~dy \wedge
d\theta_1} along with the $\chi_2$ torus in the heterotic metric
\tyhetnow\ converted to a square one.
The supersymmetry of the background will be restored by a modified
type IIB $B_{RR}$ field, so that both ($\chi_1, \chi_2$) tori are
now square.

Once we change $B_{RR}$ in type IIB, the heterotic torsion will
change completely from the one in \torcnow. Similarly the
coefficients in the metric \tyhetnow\ will have to be
re--evaluated. The question is if we can determine a valid
torsional background now\foot{The dilaton will be determined from
the warp factors in the metric.}.

In the following we will show that the metric and the torsion can be determined
to satisfy the torsional relation from the superpotential \bbdp,
\lustu, \lustd\ \eqn\toreqn{ H ~ = ~ e^{2\phi}~\ast
d\left(e^{-2\phi}~J\right)} with the dilaton $\phi$, the torsion
$H$ and the fundamental 2-form $J$. Alternatively, once we know a
particular metric --- and the dilaton from the warp factors
--- the torsion can be evaluated via \toreqn. We
will also be able to infer possible deviation of the metric that
would lead to ${\cal N} = 2$ spacetime supersymmetry.

Our starting metric is \tyhetnow, where we account for the changes
made to $B_{NS}$ \bnow\ and the complex structure of the $\chi_2$
torus by letting the coefficients ${\cal A}_i$ be arbitrary. Let
$\tau_2 = i\vert \tau \vert$ in \chichi. The choice \bnow\ leads
to $dy~ d\theta_1$ and $dx~ d\theta_2$ cross--terms in the metric.
Requiring those to have the same prefactor can be achieved by
\eqn\relbA{ {\cal A}_3 ~ = ~ {\cal
A}_2~b_{y\theta_1}~b^{-1}_{x\theta_2}.} Furthermore imposing equal
prefactors on $dy^2$ and $d\theta_2^2$ leads to \eqn\relbAA{ {\cal
A}_4 ~ = ~ \vert \tau \vert^{-2}{\cal A}_2 \left(1 -
b_{y\theta_1}~b_{x\theta_2}\right).}
After that the metric becomes: \eqn\menow{\eqalign{ds^2 ~ = ~&
{\cal A}_1~dz^2 + {\cal A}_2~\Big[(dy^2 + d\theta_2^2) +
b_{y\theta_1}\cdot b^{-1}_{x\theta_2}~ dx^2 + \left(\vert \tau
\vert^{-2} - \vert \tau \vert^{-2} b_{y\theta_1}\cdot
b_{x\theta_2} + b^2_{y\theta_1}\right)~d\theta_1^2  \cr &
~~~~~~~~~~ - 2~b_{y\theta_1}\left( dy~d\theta_1 + dx~d\theta_2
\right)\Big] + {\cal A}_5~dr^2 + ds^2_{0123}.}}
If we furthermore restrict also $dx^2$ and $d\theta_1^2$ to have
equal coefficients, the $b$ field becomes constrained, too:
\eqn\consab{{\cal A}_3 ~ = ~ {{\cal A}_2 \o\vert \tau \vert^2},
~~~~ {\cal A}_4 ~ = ~{\cal A}_2 \left({1\o \vert \tau \vert^2} -
b^2_{y\theta_1}\right), ~~~~ b_{x\theta_2} ~ = ~\vert \tau \vert^2
b_{y \theta_1}} where ${\cal A}_1$ is still arbitrary. The final
metric, after we do all these substitutions, becomes
\eqn\methe{ds^2 = {\cal A}_5~dr^2 + {\cal A}_1~dz^2 + {\cal
A}_2~\Big[(dy^2 + d\theta_2^2) + {1\o \vert\tau\vert^2}(dx^2 +
d\theta_1^2) -2 ~b_{y\theta_1}\left( dy~d\theta_1 + dx~d\theta_2
\right)\Big]} along with flat Minkowski spacetime. To bring
\consab\ into some familiar form, we perform the following set of
transformations: \eqn\setoftrans{\eqalign{& y ~~ \to ~~ {\rm
sin}~\langle\psi\rangle~y + {\rm cos}~\langle\psi\rangle~\theta_2
\cr & \theta_2 ~~ \to ~~ -{\rm cos}~\langle\psi\rangle~ y + {\rm
sin}~\langle\psi\rangle~\theta_2 \cr & z ~~ \to ~~ z + a_1~{\rm
cot}~\langle\theta_1\rangle~x + b_1~{\rm
cot}~\langle\theta_2\rangle~y}} with ($\theta_1, x, r$) remaining
unchanged. We have also denoted the expectation values of the
angles as $\langle \Theta\rangle$, with $\Theta =$ ($\theta_2,
\psi, \theta_1$) and $a_1,b_1$ are constants. Furthermore, in our
notation $d\psi$ is related to $dz$ (see sec. 5 of \gtone\ for
notations).

The metric \methe, after inserting in the set of transformations
\setoftrans, takes the following form: \eqn\methenow{\eqalign{
ds^2 ~ = ~&{\cal A}_1~\Big(dz + a_1~{\rm
cot}~\langle\theta_1\rangle~dx + b_1~{\rm
cot}~\langle\theta_2\rangle~dy\Big)^2 + {\cal A}_2~\Big[(dy^2 +
d\theta_2^2) + {1\o \vert\tau\vert^2}(dx^2 + d\theta_1^2)\Big] ~ +
\cr & -2~{\cal A}_2 ~b_{y\theta_1}\Big[{\rm
sin}~\langle\psi\rangle (dy~d\theta_1 +dx~d\theta_2) + {\rm
cos}~\langle\psi\rangle (d\theta_1~d\theta_2 - dx~dy)\Big] +{\cal
A}_5~dr^2.}} One observes that for non--constant $\Theta$, the
metric will be exactly a {\it warped} deformed conifold\foot{This
doesn't mean that the metric will be conformally Ricci-flat. By
{\it warped} deformed conifold we mean generic warping and not
just conformally warped. We will soon distinguish between the
various possible warpings allowed in these set-ups.}. In the
following we will therefore try to answer the following questions:

\vskip.1in

\noindent $\bullet$ When can we have non-constant $\Theta$?

\noindent $\bullet$ What is the background dilaton?

\noindent $\bullet$ What will be the torsion for this background?

\noindent $\bullet$ What are the stable vector bundles allowed for
such a torsional background?

\vskip.1in

\noindent Another related question would be to ask for the allowed
deformations of the above metric \methenow\ that would preserve
${\cal N} =2$ supersymmetry. Of course once we make $\Theta$
non-constant, both dilaton and torsion will have to change so that
the background still preserves ${\cal N} =1$ supersymmetry.
Therefore deforming {\it away} from ${\cal N} =1$ supersymmetry
should be non--trivial.

Coming back to the metric \methenow, we see that \methenow\ has a
close relation with Maldacena--Nunez (MN) ${\cal N} =1$
supergravity solution. The MN solution \mn\ is a warped deformed
conifold, with the metric taking the following form\foot{The
metric written in \mn\ uses left-invariant one-forms. One can
easily show the following result by writing the metric in
components.}: \eqn\mnmet{\eqalign{ds^2_{\rm MN} ~ & = ~ N~dr^2 +
{N \o 4} \left(d\psi + {\rm cos}~\tilde\theta_1~d\phi_1 + {\rm
cos}~\tilde\theta_2~d\phi_2\right)^2  + \cr & +  {N \o 4}
\left(e^{2g} + a^2\right) \left(d\tilde\theta_2^2 + {\rm
sin}^2\tilde\theta_2~d\phi_2^2\right) +
 {N \o 4}\left(d\tilde\theta_1^2 + {\rm sin}^2\tilde\theta_1~d\phi_1^2\right) + \cr
& -  {Na\o 2}\Big[{\rm cos}~\psi (d\tilde\theta_1 d\tilde\theta_2
- {\rm sin}~\tilde\theta_1  {\rm sin}~\tilde\theta_2~d\phi_1
d\phi_2) + {\rm sin}~\psi ({\rm
sin}~\tilde\theta_1~d\phi_1~d\tilde\theta_2 + {\rm
sin}~\tilde\theta_2~d\phi_2~d\tilde\theta_1)\Big]}} where we have
used ($\tilde\theta_1, \tilde\theta_2$) in \mnmet\ to distinguish
it from ($\theta_1, \theta_2$) used in \methenow. In fact to see
the relation between \mnmet\ and \methenow\ we have to use {\it
local} coordinates. Let us therefore use the following
substitutions: \eqn\foldef{\eqalign{& \psi = \langle\psi\rangle +
z, ~~ \tilde\theta_1 = \langle\theta_1\rangle + \theta_1,
~~\tilde\theta_2 = \langle\theta_2\rangle + \theta_2 \cr & \phi_1
= \langle\phi_1\rangle + {x \o {\rm sin}~\langle\theta_1\rangle},
~~~~~~~~ \phi_2 = \langle\phi_2\rangle + {y \o {\rm
sin}~\langle\theta_2\rangle}}} along with the following
definitions for $a$ and $g$ (we follow the notation of \grana):
\eqn\aandg{a(r) ~ = ~ -{2r \o {\rm sinh}~2r}, ~~~~~~ e^{2g} ~ = ~4
r~{\rm coth}~2r - {4 r^2 \o {\rm sinh}^2~2r} - 1} which are in
fact necessary to make \mnmet\ a solution to the string equation
of motion. Recall however that the MN solution \mnmet\ is a metric
in type IIB theory with {\rm closed} three--form field $H_{NS}$
and non--trivial dilaton. This solution has vanishing $H_{RR}$,
axion and five--form. To establish the connection between \mnmet\
and \methenow, let us substitute \foldef\ in \mnmet\ letting
$(\theta_1, \theta_2, x, y) \to 0$. Under these assumptions
\mnmet\ takes the following form (ignoring numerical factors of
${N \o 4}$ for simplicity): \eqn\mnmetnow{\eqalign{ds^2_{\rm MN}
~& = ~ \Big(dz + {\rm cot}~\langle\theta_1\rangle~dx + {\rm
cot}~\langle\theta_2\rangle~dy\Big)^2 + \left(e^{2g}+a^2\right)
(dy^2 + d\theta_2^2) + (dx^2 + d\theta_1^2) ~ + \cr & ~~~~~~~
-2~a\Big[{\rm sin}~\langle\psi\rangle (dy~d\theta_1 +dx~d\theta_2)
+ {\rm cos}~\langle\psi\rangle (d\theta_1~d\theta_2 - dx~dy)\Big]
+ 4~dr^2 + ......}} where the dotted terms are corrections that
are higher orders in ($\theta_1, \theta_2, x, y$).

Comparing \mnmetnow\ with our metric \methenow, we see that they
are of the same form up to higher order corrections on \mnmetnow.
Therefore one might be tempted to conjecture that deforming away
from the point ($\langle\theta_{1,2}\rangle,
\langle\phi_{1,2}\rangle$) should give the global description of
the heterotic metric \methenow. One possible solution for the
coefficients ${\cal A}_i$ in \methenow\ (or in \tyhetnow) that
allows for non-constant $\Theta$ would then be: \eqn\avalve{{\cal
A}_1 = {\cal A}_3 = {{\cal A}_5 \o 4} = {N \o 4}, ~~~~ {\cal A}_2
= {N(e^{2g}+ a^2)\o 4}, ~~~~ {\cal A}_4 = {N e^{2g}\o e^{2g}+a^2}}
where we have re--introduced the numerical factor of ${N\o 4}$.
Finally, the type IIB $B_{NS}$ field and the ($\chi_1, \chi_2$)
tori will have the following form: \eqn\bandtori{\eqalign{& b =
a~dx \wedge d\theta_2 + {a\o e^{2g} + a^2}~ dy \wedge d\theta_1
\cr & d\chi_1 = dx + i~dy, ~~~ d\chi_2 = d\theta_1 +
i\sqrt{e^{2g}+a^2}~d\theta_2.}} Under the above choices of
coefficients, our metric \methenow\ agrees exactly with the MN
metric \mnmet. Of course we should remind the readers that we have
not yet {\it derived} the coefficients of \methenow\ from our
duality chain. We haven't identified a dual background before
geometric transition that would allow us to do so. So \avalve\ is
simply a possibility at this point (albeit a strong one).

It is also worth mentioning that the MN background was derived for
the IR regime (for small $r$). The same is true for our local IIB
metric. However, once we infer the local heterotic metric and then
lift it to a global background, there is no a--priori restriction
to be in the IR. Our global solution should be valid for UV as
well, but since MN fails in this regime, we will not specify the
coefficients in the global metric. The UV limit for the MN
background has been found in \grana\ and we will come back to this
issue later.

In the case where the coefficients ${\cal A}_i$ in \methenow\ are
{\it different} from the one considered in \avalve, our ansatz
would be the global form of the metric \methenow\ with
non--constant $\Theta$. The heterotic metric is finally given by:
\eqn\hetfinal{\eqalign{ds^2_{\rm het}  & = ~ ds^2_{0123} + {\cal
A}_1~ \left(d\psi + a_1~{\rm cos}~\theta_1~d\phi_1 + b_1~{\rm
cos}~\theta_2~d\phi_2\right)^2 + {\cal A}_3~ \left(d\theta_1^2 +
{\rm sin}^2\theta_1~d\phi_1^2\right) + \cr & + {\cal
A}_2~\left(d\theta_2^2 + {\rm sin}^2\theta_2~d\phi_2^2\right) - 2
~{\cal A}_2~b_{y\theta_1}\Big[{\rm cos}~\psi~ (d\theta_1
~d\theta_2 - {\rm sin}~\theta_1 ~ {\rm sin}~\theta_2~d\phi_1~
d\phi_2) + \cr & ~~~~~~~~~~~~~ + {\rm sin}~\psi~ ({\rm
sin}~\theta_1~d\phi_1~d\theta_2 + {\rm
sin}~\theta_2~d\phi_2~d\theta_1)\Big] + {\cal A}_5~dr^2}} where
($\psi, \theta_{1,2}, \phi_{1,2}$) are the global coordinates; and
we have kept the warp factors ${\cal A}_i$ arbitrary. Notice that
the above metric written in terms of global coordinates has a
lesser number of isometries than \methenow\ written in terms of
local coordinates. In fact \methenow\ is a $T^3$ fibration over a
three--dimensional base, i.e in SYZ \syz\ form, whereas \hetfinal\
cannot be brought into SYZ--form at all. This observation may also
resolve one puzzle regarding mirror transformation of a resolved
or deformed conifold. The resolved conifold has a natural $T^3$
torus built inside it irrespective of whether we use global or
local coordinates. On the other hand a deformed conifold is {\it
not} a $T^3$ fibration over a three dimensional base. But when we
use local coordinates to write the metric of a deformed conifold,
we regain some of the isometries, and then a SYZ--form for the
deformed conifold can be written. This is precisely the reason for
the success of the duality chain proposed in \gtone\ and \realm.
Since this duality chain involves T--dualities, it seems
impossible that one could {\it loose} isometries along the way.
Thus, the metrics of \gtone, \realm\ all have three isometries in
terms of local coordinates used therein. This is perfectly
consistent with the arguments in this article.

The above metric \hetfinal\ is another example of a warped
deformed conifold. When the warp factors ${\cal A}_i$ are
different from \avalve\ this metric does {\it not} coincide with
the MN metric \mnmet. We should now compare this with other available
warped--deformed--conifold solutions:

\vskip.1in

$\bullet$ The Klebanov-Strassler (KS) metric of \ks, and

$\bullet$ The Geometric transition dual metric of \realm, \gtone.

\vskip.1in

\noindent Both of these are in type IIB theory with background NS
and RR fluxes. Furthermore whereas the KS solution \ks\ is written
in terms of global coordinates ($\psi, \theta_{1,2}, \phi_{1,2},
r$), the solution presented in \realm, \gtone\ are written only in
terms of {\it local} coordinates ($z, x, y, \theta_{1,2}$). The
full global picture involves seven branes and orientifold planes,
as discussed earlier. To compare all the different pictures, let
us cite the metric of a simple (i.e. Ricci--flat) deformed
conifold \candelas, \ohta, \tsimpis: \eqn\conimet{\eqalign{ds_{\rm
DC}^2~& =~ {2\o 3K^2}~ \left(d\psi + {\rm cos}~\theta_1~d\phi_1 +
{\rm cos}~\theta_2~d\phi_2\right)^2 + K~{\rm cosh}~\rho ~
\left(d\theta_1^2 + {\rm sin}^2\theta_1~d\phi_1^2\right) + \cr &+
K~{\rm cosh}~\rho ~\left(d\theta_2^2 + {\rm
sin}^2\theta_2~d\phi_2^2\right) -2K~\Big[{\rm cos}~\psi~
(d\theta_1 ~d\theta_2 - {\rm sin}~\theta_1 ~ {\rm
sin}~\theta_2~d\phi_1~ d\phi_2) + \cr & ~~~~~~~~~~~~~ + {\rm
sin}~\psi~ ({\rm sin}~\theta_1~d\phi_1~d\theta_2 + {\rm
sin}~\theta_2~d\phi_2~d\theta_1)\Big] + {2\o 3K^2}~d\rho^2}} where
$\rho$ is the ``radial'' coordinate, and $K = K(\rho) =
{\left({\rm sinh}~2\rho - 2\rho\right)^{1/3}\o 2^{1/3}{\rm
sinh}~\rho}$ (see \candelas, \ks, \ohta, \tsimpis\ for more
details). The metric \conimet\ shows the same warp factors for the
two spheres parameterized by ($\theta_1, \phi_1$) and ($\theta_2,
\phi_2$). So does the KS solution \ks. On the other hand, all the
other solutions, Maldacena-Nunez \mnmet, the geometric transition
solutions of \realm, \gtone, and the heterotic solution \hetfinal\
have different warp factors for the two spheres (or the two tori
in case of \realm, \gtone). For example, the IIB solution for
gravity dual presented in \realm\ is given (in local coordinates)
by: \eqn\metnba{\eqalign{ds^2_{IIB}~=~&~h_1(dz + a_1~dx + a_2~dy)^2 +
h_2 (dy^2 + d\theta_2^2) + h_4(dx^2 + h_3~d\theta_1^2) ~+ \cr &~~+
h_5\big[{\rm sin}~\psi~(dx~d\theta_2 + dy~d\theta_1) + {\rm
cos}~\psi~ (d\theta_1~d\theta_2 - dx~dy)\big]
+\gamma'\sqrt{H}~dr^2}} where $h_i$ and $a_i$ were defined in
\hideterm. The global picture for this solution has not been
worked out yet. But comparing locally shows that the solution of
\realm, i.e. \metnba, forms a new class of supergravity dual with
markedly different warping behavior of a deformed conifold metric.
This results from the different, more elaborate set-up of \realm\
and \gtone\ compared to \ks\ and \mn.

We still have to determine the torsion for our background
\hetfinal, where we will use \avalve\ since we know it results in
a valid background in terms of global coordinates\foot{In other
words, we neglect the UV regime for the time being. The derivation
would be equivalent, the appropriate coefficients for the
vielbeins can be found in \grana.}. We need the fundamental
2--form $J$ in terms of vielbeins $e_i$ and the dilaton. Since the
background has changed, the vielbeins are not those in \vielsnow.
Instead we choose: \eqn\gmvielb{\eqalign{&
 e^1 ~=~ \sqrt{N}~dr, ~~~e^5 ~=~ {\sqrt{N}\o 2}~e^g~d\theta_2, ~~
   e^2~=~{\sqrt{N} \o 2}~\left(d\psi+\cos\theta_1\,d\phi_1+\cos\theta_2\,
   d\phi_2\right)\cr
 & e^3 ~=~ {\sqrt{N}\o 2}~\left(\sin\psi \sin\theta_1~d\phi_1
   +\cos\psi~ d\theta_1-a~d\theta_2\right)\cr
 & e^4 ~=~ -{\sqrt{N}\o 2}~\Big[{\cal B}~e^g\sin\theta_2~d\phi_2
   + {\cal A} (\cos\psi\sin\theta_1~d\phi_1-\sin\psi~d\theta_1
   + a\sin\theta_2~d\phi_2)\Big]\cr
 & e^6 ~=~ -{\sqrt{N}\o 2}~\Big[{\cal A}~e^g\sin\theta_2~d\phi_2
   - {\cal B} (\cos\psi\sin\theta_1~d\phi_1-\sin\psi~d\theta_1
   +a\sin\theta_2~d\phi_2)\Big]}}
which gives rise to the metric \hetfinal\ with ${\cal A}_i$
defined as in \avalve. It was pointed out in \grana\ that these
are the correct vielbeins for observing the $SU(3)$ structure of
this background and they were first given in \tp. They are not
quite those of the deformed conifold since our background
\hetfinal, as noted earlier, is neither a Ricci--flat deformed nor
a conformally deformed conifold\foot{Although not a conformally deformed conifold, our model
will have many of the characteristic features of a deformed conifold as can be seen from \hetfinal. For
example there will be a three-cycle that can undergo a geometric transition. The three cycle will also
have a similar Hopf-fibration structure as a deformed conifold. This is clear from the $d\psi$ fibration structure
of \hetfinal. These similarities will be exploited in the next section to study the vector bundles
across conifold transitions.}.
The ${\cal A}, {\cal B}$ used in
\gmvielb\ satisfy ${\cal A}^2+{\cal B}^2=1$, with ${\cal A}, {\cal
B}$ given by \eqn\AandB{
 {\cal A}~=~\coth 2r-2r {\rm csch}^2 2r~,~~~~~
   {\cal B}~=~{\rm csch} 2r~\sqrt{-1+4r\coth
   2r -4r^2{\rm csch}^2 2r}~.}
The fundamental two--form is evaluated with a choice of complex
structure such that $J=(e^1\wedge e^2+e^3\wedge e^4+e^5\wedge
e^6)$. This amounts to \eqn\Jmn{\eqalign{ &
 J ~=~ {N\o 2}~dr\wedge (d\psi+\cos\theta_1~d\phi_1
   +\cos\theta_2~d\phi_2)~  \cr
 & -{N \o 4}~{\cal A}\sin\theta_1~d\theta_1\wedge
   d\phi_1 - {N\o 4}~\left(-{\cal A}^2a+{\cal A} e^{2g}-2{\cal B}
   ae^g\right) \sin\theta_2 ~d\theta_2\wedge
   d\phi_2 ~  \cr
 & +{N \o 4}~\left({\cal A}a+{\cal B}e^g\right)~
   \left[\sin\psi(d\theta_1\wedge d\theta_2-\sin\theta_1\sin\theta_2~
   d\phi_1\wedge d\phi_2)\right.\cr
 & ~~~~~~~~~~~~~~~~~~~~~~\left.+\cos\psi(\sin\theta_1~d\theta_2\wedge
   d\phi_1-\sin\theta_2~d\theta_1\wedge d\phi_2)\right].}}
From the value of $J$ one can easily see that the manifold is non-K\"ahler. This
is of course expected because the local metric is also non-K\"ahler with torsion.
The background dilaton can be extracted from the warped metric
\hetfinal\ or from \mn, \tp, and is given by
 \eqn\dilatmn{
 e^{2\Phi}~=~{e^{g + 2\Phi_0}\o {\rm sinh}~ 2r}}
where $\Phi_0$ is some constant value that could be fixed from the
U--dual type IIB background. With the dilaton \dilatmn\ one
computes \eqn\dephiJ{
 d(e^{-2\Phi}J)~=~e^{-2\Phi}\left(-2{\partial\Phi\o \partial r}~dr\wedge
   J+dJ\right)~.}
The Hodge dual of this expression is most easily found in terms of
vielbeins, since in a non--coordinate basis simply: \eqn\hodgedual{
 *~(e^{\alpha_1}\wedge e^{\alpha_2}\wedge e^{\alpha_3}) ~=~ {1 \o 3!}~\epsilon^{\alpha_1
 \alpha_2\alpha_3}_{~~~~~~~~\mu_1\mu_2\mu_3}~e^{\mu_1}\wedge e^{\mu_2}\wedge
 e^{\mu_3}~.} We choose the orientation so that $\epsilon^{123456}=1$.
Inverting \gmvielb\ and replacing the coordinate differentials by
vielbeins one finds \eqn\hvielb{\eqalign{
 & e^{2\Phi}~*~d(e^{-2\Phi}J)~=~{1\o \sqrt{N}~F_2(r)}~\left[{F_2(r)~(1+8r^2-\cosh
   4r)~(4r-{\rm sinh} 4r)\o F_1(r)~{\rm sinh}^2 2r}~ e^1\wedge e^2\wedge e^6 \right.\cr
 & \left.+{2~(-1+2r\coth 2r) \o \sinh 2r}~
   e^1\wedge e^3\wedge e^5 + {(1+8r^2-\cosh 4r)\o \sinh^3 2r}~e^1\wedge e^4\wedge
   e^6 ~  \right. \cr
 & \left. +{F_2^2(r)\o {\rm sinh} r~\cosh r}~e^2\wedge e^3\wedge
   e^6 + \left(-{r\o {\rm sinh}^2 r}+{1\o {\rm sinh} r~\cosh
   r}-{r\o \cosh^2 r}\right)~e^2\wedge e^4\wedge e^5 ~
   \right.  \cr
 & \left.+ ~{(-4r+{\rm sinh} 4r)\o \sinh^2 2r}~e^3\wedge
 e^4\wedge e^6\right]}}
with $F_1(r)$ and $F_2(r)$ defined by\foot{Again these relations are motivated from the similarity
of \hetfinal\ with the MN background \mn. We may not consider the values taken by \mn, and represent all
our results using the unknown warp-factors ${\cal A}_i$. This way we can allow a non-trivial vector bundle
for our background. More details on this will be presented elsewhere.}
\eqn\FandG{\eqalign{&
 F_1(r)~=~-1+8r^2+\cosh 4r-4r~{\rm sinh} 4r \cr
& F_2(r)~=~\sqrt{-1+4r(\coth 2r-r~{\rm csch}^2~2r)}}}
The 3--form \hvielb\ is the torsion for our background \hetfinal\
with dilaton \dilatmn. In terms of global coordinates ($r,
\theta_i, \phi_i, \psi$) the torsion $H$ is given as
\eqn\hmn{\eqalign{
   H ~ = &~e^{2\Phi}~*~d(e^{-2\Phi}J) \cr
   =& ~-{Na' \o 4}~\cos\psi~dr\wedge(d\theta_1\wedge
      d\theta_2-\sin\theta_1\sin\theta_2~d\phi_1\wedge d\phi_2)~+\cr
    & -~ {Na' \o 4}~\sin\psi~dr\wedge (\sin\theta_2~d\theta_1\wedge
      d\phi_2-\sin\theta_1~d\theta_2\wedge d\phi_1)~+\cr
    & +~{Na \o 4}~\sin\psi~d\theta_1\wedge d\theta_2\wedge
      (d\psi+\cos\theta_1~d\phi_1+\cos\theta_2~d\phi_2)~+\cr
    & -~{N\o 4}~(\sin\theta_1\cos\theta_2 - a\cos\psi\cos\theta_1 \sin\theta_2)
      ~d\theta_1\wedge d\phi_1\wedge d\phi_2 ~+\cr
    & -~{N\o 4}~(\sin\theta_2\cos\theta_1 - a\cos\psi\cos\theta_2 \sin\theta_1)
      ~d\theta_2\wedge d\phi_1\wedge d\phi_2 ~+\cr
    & -~{N\o 4}~\sin\theta_1~d\theta_1\wedge d\phi_1\wedge d\psi
      +{N\o 4}~\sin\theta_2~d\theta_2\wedge d\phi_2\wedge d\psi ~+\cr
    & -~ {Na\o 4}\cos\psi~(\sin\theta_2~d\theta_1\wedge d\phi_2\wedge d\psi
      -\sin\theta_1~d\theta_2\wedge d\phi_1\wedge d\psi)~+ \cr
    & -~ {Na \o 4}~\sin\psi\sin\theta_1\sin\theta_2~d\phi_1\wedge d\phi_2\wedge
      d\psi}}
with $a'=\partial a/\partial r$. At this point the attentive
reader might question the existence of torsion and warped solution
from the point of view of \smit, \papad, \bbdg. We seem to have
used an embedding where $\omega_+ = A$, i.e. the gauge connection
$A$ is embedded in the torsional spin connection, and not the one
of \bedd. We hasten to point out that this is justified as long as
the underlying manifold is {\it non--compact}. For compact
non--K\"ahler manifolds the situation is much more subtle and
delicate as was pointed out in the series of papers \bbdg, \bbdp,
\bbdgs.

\noindent To complete the analysis of the proposed heterotic
background \hetfinal\ with torsion \hmn\ and dilaton \dilatmn\ we
need to consider two more issues:

\vskip.1in

$\bullet$ The existence of vector bundles, and

$\bullet$ The global behavior of $b_{y\theta_1}, b_{x\theta_2}$ and $\vert\tau\vert$.

\vskip.1in

\noindent The vector bundles will be studied in the next section.
Here we concentrate on the behavior of $b_{y\theta_1}$ and
$b_{x\theta_2}$. In the process we will also be able to study the
complex structure $\vert\tau\vert$ in detail.

The global behavior of type IIB $B$--fields $b_{y\theta_1}$ and
$b_{x\theta_2}$ can only be determined after we solve the
background equation of motions. However, the situation is
complicated because of the additional seven branes and orientifold
planes. That was one of the motivations for using local
coordinates in the first place, because we lack the global metric in
IIB.

Nevertheless, the global behavior of the IIB $B$--fields can be
extracted from the relation between type IIB and heterotic
picture. Recall that the global heterotic metric \hetfinal\
contains the global IIB $B$--fields, and was obtained by
connecting the local pictures in both theories and then using the
similarity of the heterotic metric with Maldacena-Nunez \mn\ to
obtain the global picture. In our case of interest, a background
with only NS flux, we know MN to be a valid solution in the IR.
Comparing \hetfinal\ with the MN metric determines $B$\foot{One
might as well solve the heterotic equations of motion (see also
\grana).}. For small $r$: \eqn\bfgpo{\eqalign{& b_{x\theta_2} ~ =
~-1 + {2\o 3}r^2 - {14\o 45} r^4 + {\cal O}(r^{6}) \cr &
b_{y\theta_1}~ =~-1 + {10\o3}r^2-{446\o 45}r^4+ {\cal O}(r^{6})}}
Near $r \to 0$ both $B$--field components are constant as one
might have expected. Having determined the $B$ field we can also
fix the $\chi_2$-torus in \chichi. The complex structure is given
as \eqn\chicom{\vert\tau\vert^2 ~ = ~ 1 + {8\o 3} r^2 - {32 \o 45}
r^4 +{\cal O}(r^{6})} which tells us how the ($\theta_1,
\theta_2$) torus varies as we move along the radial direction. In
fact, near $r\to 0$: $\tau_2 \equiv i\vert\tau\vert = i$ which,
along with $\tau_1 = i$, completely specifies the IR behavior in
IIB.

The discussion in \mn\ does not extend to the UV regime. Here we
can rely on the analysis of \grana\ which embeds the MN background
in a class of interpolating solutions between MN and KS. Using
their results we can obtain the large $r$ behavior of the
$B$--fields: \eqn\buv{\eqalign{& b_{x\theta_2} ~ = ~ -2~e^{-2r} +
  a_{\rm uv}~(2r-1)~e^{-{10r\o 3}} - {1\o 2}~ a_{\rm
  uv}^2~(2r-1)^2~e^{-{14r\o 3}}+ {\cal O}(e^{-6r}) \cr
& b_{y\theta_1}~ =~ -2~e^{-2r} - a_{\rm uv}~(2r-1)~e^{-{10r\o 3}} -
{1\o 2}~ a_{\rm uv}^2~(2r-1)^2~e^{-{14r\o 3}}+ {\cal O}(e^{-6r})}}
where $a_{\rm uv}=-\infty$ corresponds to MN in the interpolating
scenario. The complex structure then results in
\eqn\tauuv{\vert\tau\vert^2~=~~1+2~e^{-4r}- a_{\rm
uv}~(2r-1)~e^{-{4r\o 3}} + {1\o 2}~a_{\rm uv}^2 ~(2r-1)^2~e^{-{8r\o
3}}+ {\cal O}(e^{-{16r\o3}}).} Notice that for $r\to\infty$ the
$B$--fields vanish and the complex structure approaches again
$\tau_2=i$ and $\tau_1=i$.

This finalizes the study of the heterotic background \hetfinal.
Before we move on to study vector bundles we would like to comment
on possible deformations to ${\cal N} = 2$ susy. Let us go back
and consider \menow, written in terms of local coordinates. We
have already observed that in this metric the relative
coefficients of $dy^2$ and $d\theta_2^2$ are the same, whereas the
relative coefficients of $dx^2$ and $d\theta_1^2$ are different.
Instead of requiring the latter also to be equal let us keep them
inhomogeneous. We furthermore assume that the $B$--fields go to
zero in the following way: \eqn\bzero{b_{y\theta_1}~~\to ~~ 
{g}_1\epsilon, ~~~~~~~ b_{x\theta_2}~~\to ~~ {g}_2\epsilon} with
$\epsilon$ approaching zero. Under this requirement the ratio of
$b_{y\theta_1}$ and $b_{x\theta_2}$ approaches ${b_{y\theta_1} \o
b_{x\theta_2}} = {{g}_1 \o {g}_2}$ although individually they are very
small everywhere. Using \bzero\ and the coordinate transformation
\setoftrans\ we obtain the following local form of the metric in
the limit $\epsilon\to 0$: \eqn\ntwomet{\eqalign{ ds^2 ~ = ~&{\cal
A}_1~\Big(dz + a_1~{\rm cot}~\langle\theta_1\rangle~dx + b_1~{\rm
cot}~\langle\theta_2\rangle~dy\Big)^2 + {\cal A}_2~(dy^2 +
d\theta_2^2) ~ + \cr & ~~~~~~~~~ + {\cal A}_2 \left({{g}_1\o {g}_2}~
dx^2 + {1\o \vert\tau\vert^2}~d\theta_1^2\right)
 + {\cal A}_5~dr^2 + ds^2_{0123}}}
where ${\cal A}_i$ can be generic functions of the local
coordinates, and $a_1, b_1$ are constants\foot{For ${\cal N} =1$
case studied earlier, $a_1 ~=~b_1~=~ 1$.}. If we set $a_1 ~= ~0,~
b_1~=~1$, we arrive at a metric that is the local form of the
${\cal N} =2$ metric studied by \kimn. We can therefore conjecture
our metric \ntwomet\ to have the global form (recall \foldef\ for
the relation between local and global coordinates):
\eqn\twoglob{\eqalign{ ds^2 ~ = ~&{\cal A}_1~\Big(d\psi +
~{\rm cos}~\theta_2~d\phi_2\Big)^2 + {\cal A}_2~
(d\theta_2^2 + {\rm sin}^2 \theta_2~d\phi_2^2)~ + \cr & ~~~~~~~~~
+ {\cal A}_2 \left({1\o \vert\tau\vert^2}~d\theta_1^2 +
{{g}_1\o {g}_2}~{\rm sin}^2 \theta_1~d\phi_1^2\right)
 + {\cal A}_5~dr^2 + ds^2_{0123}.}}
To determine the precise values of the various coefficients
we can compare \twoglob\ to the ${\cal N}=2$ metric of \kimn\ (see
also \sezgin): \eqn\kimnmet{\eqalign{ds^2 ~ = ~& ds^2_{0123} +
{e^{-x}\o g_c^2~\Omega}~{\rm cos}^2 \theta_1~\Big(d\psi + ~{\rm
cos}~\theta_2~d\phi_2\Big)^2 + F(r)~ (d\theta_2^2 + {\rm sin}^2
\theta_2~d\phi_2^2)~ + \cr & ~ + F(r)~ \left({1\o g_c^2~
F(r)}~d\theta_1^2 + {e^x\o \Omega~ g_c^2~ F(r)}~ {\rm sin}^2
\theta_1~d\phi_1^2\right)
 + g_c^2~ e^{2x} \left({\del F\o \del r}\right)^2 dr^2}}
where $g_c$ is the {\it coupling} constant and $F(r)$ is a
function of the radial coordinate $r$ \kimn. The other two
variables, $\Omega$ and $e^x$ are defined in terms of the
coordinates, $g_c$, $F(r)$ and an integration constant $\kappa$,
as (see \kimn\ for derivations): \eqn\defOx{\Omega ~ = ~ e^x~{\rm
cos}^2 \theta_1 + e^{-x}~{\rm sin}^2 \theta_1, ~~~ e^{-2x} ~ = ~ 1
- {1+ \kappa~e^{-2g_c^2 F} \o 2 g_c^2 F.}} Comparing \twoglob\ and
\kimnmet, it is obvious that with the correct identifications of the
coefficients ${\cal A}_i$ and ${g}_1, {g}_2, \vert\tau\vert$ we
can obtain an ${\cal N} = 2$ deformation of our background
\hetfinal. The appropriate dilaton $\Phi$ and torsion $H$ for this
case have already been considered in \kimn\ and therefore we will
not re--derive the torsion for this background. The interested
reader may find all the details in \kimn.

One final comment on the metrics \twoglob\ and \kimnmet: Notice
that the relative coefficients of $d\theta_1^2$ and ${\rm sin}^2
\theta_1 d\phi_1^2$ are different. This is reminiscent of a
similar behavior that we encountered for the type IIB ${\cal N}
=1$ metric \metnba\ (see also \realm\ for derivation). In \metnba\
the local metric showed equal coefficients for ($dy, d\theta_2$)
but different relative coefficients for ($dx, d\theta_1$). Since
we encounter the same behavior for the ${\cal N} =2$ case herein,
this might be an indication that the metric \metnba\ actually
shows ${\cal N} =2$ supersymmetry, and only the NS and RR fluxes
break the supersymmetry to ${\cal N} =1$. If the above statement
is true (we have not verified this yet) this would be perfectly
consistent with the predictions for geometric transitions
considered in \tv, \civ, \civd, \civu, \dotd, \dotu, \adoptfo,
\adoptone\ where it was clearly stated that the underlying 
manifolds preserve ${\cal N} =2$ supersymmetry and the fluxes break susy from
${\cal N} = 2$ to ${\cal N} = 1$.


\newsec{Pulling Rank~2 vector bundles through conifold transitions}

In the previous subsection we gave a detailed construction of a heterotic
background on a manifold that resembles a warped deformed conifold. To
complete the story we need to study vector bundles on the manifold, and also
discuss how the bundles are pulled through conifold transitions. In the
following therefore we will address these questions in some detail.

Since the analysis will involve some mathematics that might not be too
well-known to physicists, we will start by discussing
some standard facts about rank~2 vector bundles. For simplicity of exposition, we specialize
to vector bundles on Calabi-Yau threefolds.


\subsec{The~Serre~construction}


\noindent Consider a holomorphic rank~2 vector bundle $E$ on a Calabi-Yau
threefold $X$ satisfying anomaly cancellation.  Pick a holomorphic
section $s\in H^0(X,E)$, and suppose that the set of zeros of $s$ is a
smooth complex curve $C\subset X$.\foot{The existence of such a
section is in a sense not really a restrictive assumption.  If $\O(1)$
as usual denotes the hyperplane bundle on $X$ given by the realization
of $X$ in some $\IP^n$, then $E\otimes\O(N)$ admits such a section for
$N$ sufficiently large.}  Multiplication and wedging by $s$ gives an
exact sequence
$$0\to \O_X\to \O_X(E)\to \O_X(\det E).$$
Here and in the sequel, $\O_X(E)$ as usual denotes the sheaf of
holomorphic sections of the vector bundle $E$.  The notion of a sheaf
gives a framework for simultaneously discussing sections over arbitrary
open sets.

Note that $\det E$ is
the trivial bundle by the condition $c_1(E)=0$, so we can rewrite the
above as
$$0\to \O_X\to \O_X(E)\to \O_X.$$
The sections in the image of the
rightmost map clearly vanish on $C$.  Denoting the sheaf of functions vanishing
on $C$ by $\cI_C$,\foot{{\it Warning:} $\cI_C$ is not a line bundle.}
the above exact sequence induces a short exact sequence
\eqn\easext{0\to \O_X\to \O_X(E)\to \cI_C\to 0.}

The {\it Serre construction} allows us to reverse this process to give
a construction of $E$ from $C$ and some extra data.  The Serre construction
is explained in \grifhar.

The exact sequence \easext\ describes $\O_X(E)$ as an extension of
$\cI_C$ by $\O_X$.  Such extensions are classified by the Ext group
$\Ext^1(\cI_C,\O_X)$.  There are the same groups as have appeared
in the physics literature in describing open string states at large radius.

Ext groups enjoy a number of properties, explained in \grifhar.

\noindent $\bullet$ If $\cE$ and $\cF$ are any sheaves, then
$\Ext^0(\cE,\cF)\simeq\Hom(\cE,\cF)$.

\noindent $\bullet$ If $E$ is a vector bundle and $\cF$ is any sheaf, then
$\Ext^i(\O(E),\cF)\simeq H^i(X,O(E^*)\otimes\cF)$.

\noindent $\bullet$ If $0\to \cE_1\to \cE\to \cE_2\to 0$ is a short exact
sequence of sheaves and $\cF$ is any sheaf, then there are long exact
sequences
$$\cdots\to \Ext^i(\cE_2,\cF)\to \Ext^i(\cE,\cF)\to\Ext^i(\cE_1,\cF)
\to \Ext^{i+1}(\cE_2,\cF)\to\cdots$$
and
$$
\cdots\to\Ext^i(\cF,\cE_1)\to\Ext^i(\cF,\cE)\to\Ext^i(\cF,\cE_2)\to
\Ext^{i+1}(\cF,\cE_1)\to\cdots$$

\smallskip
The ideal sheaf $\cI_C$ fits into the short exact sequence
$$0\to \cI_C\to \O_X{\to} \O_C\to 0,$$
where the last nontrivial map is the restriction map.
It follows that the desired Ext group $\Ext^1(\cI_C,\O_X)$ fits
into the short exact sequence
$$\Ext^1(\O_X,\O_X)\to \Ext^1(\cI_C,\O_X)\to\Ext^2(\O_C,\O_X)\to
\Ext^2(\O_X,\O_X).
$$
By the properties of Ext, we have that $\Ext^i(\O_X,\O_X)\simeq H^i(X,\O_X)$.
But $H^1(X,\O_X)=H^1(X,\O_X)=0$ for Calabi-Yau threefolds.  We conclude that
\eqn\extonetwo{\Ext^1(\cI_C,\O_X)\simeq\Ext^2(\O_C,\O_X).}

\smallskip
We now need some mathematical results which generalize Serre duality and
the adjunction formula.  For this, we first need to introduce local Exts.  As a
preliminary, we discuss the case of $\Ext^0$ first, i.e. $\Hom$.  Given
bundles $E$ and $F$, we have the group $\Hom(E,F)$, but we also have
a local version, namely the bundle $E^*\otimes F$.  In sheaf language,
we would write the sheaf $\uHom(\O(E),\O(F))$, which is nothing but the
sheaf $\O(E^*\otimes F)$.  This is the same thing as $\uExt^0(\O(E),\O(F))$
by definition.
The connection between local and global Hom (or equivalently $\Ext^0$) is
$$H^0(X,\uHom(\O(E),\O(F)))\simeq \Hom(\O(E),\O(F)).$$
An analogous equality is true for arbitrary sheaves which are not necessarily
the sheaves of sections of a holomorphic vector bundle.

Given sheaves $\cE,\cF$, we have the local Ext sheaves
$\uExt^i(\cE,\cF)$.  These enjoy several properties, some analogous to
the properties of $Ext$ groups.  These properties are all explained
in \grifhar.

\noindent $\bullet$ If $E$ and $F$ are vector bundles, then
$\uExt^0(\O(E),\O(F))\simeq\O(E^*\otimes F)$.

\noindent $\bullet$ If $E$ is a vector bundle and $\cF$ is any sheaf, then
$\Ext^i(\O(E),\cF)=0$ for $i>0$.  This is a local version of a corresponding
property for Ext groups, since higher cohomology vanishes on disks.

\noindent $\bullet$ If $0\to \cE_1\to \cE\to \cE_2\to 0$ is a short exact
sequence of sheaves and $\cF$ is any sheaf, then there are long exact
sequences of sheaves
$$\cdots\to \uExt^i(\cE_2,\cF)\to \uExt^i(\cE,\cF)\to\uExt^i(\cE_1,\cF)
\to \uExt^{i+1}(\cE_2,\cF)\to\cdots$$
and
$$
\cdots\to\uExt^i(\cF,\cE_1)\to\uExt^i(\cF,\cE)\to\uExt^i(\cF,\cE_2)\to
\uExt^{i+1}(\cF,\cE_1)\to\cdots$$
But it is not the case that $H^0(X,\uExt^i(\O(E),\O(F)))\simeq
\Ext^i(\O(E),\O(F))$ for $i>0$.  Instead there is a more complicated
relation between local and global Ext groups described by a spectral
sequence.  Rather than beginning a lengthy digression into the formalism of
spectral sequences, we content ourselves with a simple consequence which
suffices for our purposes.

\noindent $\bullet$ If $\cE,\cF$ are sheaves and $\uExt^j(\cE,\cF)=0$ for
$j<i$, then $H^0(X,\uExt^i(\O(E),\O(F)))\simeq
\Ext^i(\O(E),\O(F))$.

We now apply this to duality.  We have, as explained in \hart,
$$\uExt^i(\O_C,\O_X)\simeq\cases{\Omega^1_C\qquad i=2\cr
0\qquad i < 2}$$
So the final bullet above applies, and we conclude that
$$\Ext^1(\cI_C,\O_X)\simeq\Ext^2(\O_C,\O_X)\simeq H^0(C,\Omega^1_C).
$$
Putting this all together, $E$ can be recovered from the curve $C$ and
a holomorphic 1-form on $C$.

Now let $C$ be an arbitrary smooth curve and $\omega$ a holomorphic 1-form
on $C$.  Reversing the above construction, we get an extension
\eqn\sheafasext{0\to \O_X\to \cE\to \cI_C\to 0}
whose extension class in $\Ext^1(\cI_C,\O_X)$ corresponds to $\omega$
under the above isomorphisms.  We cannot conclude that $\cE$ is a
vector bundle, only that it is a sheaf.  Furthermore, a local
calculation shows that $\cE$ fails to be (the sheaf of sections of) a
vector bundle precisely at the points of $C$ where $\omega$ vanishes.
If $C$ is a smooth curve of genus $g$ and $\omega$ is not identically
zero, then there are $2g-2$ such points (including multiplicity).

We conclude that we get a vector bundle only for $g=1$.  But as elliptic
curves are abundant on Fano or Calabi-Yau threefolds $B$, we learn that
rank~2 bundles are straightforward to construct on these threefolds.

\smallskip
We now have to satisfy anomaly cancellation, which we presume to be of
the form $c_i(\cE)=c_i(B)$ for $i=1,2$.
From \sheafasext, we get
$c_1(\cE)=0$ and we compute that $c_2(\cE)$ is represented by the
homology class of the curve $C$.
So we clearly cannot use
the above form of the Serre construction if $c_1(B)\ne0$.  But we can
easily cancel the anomaly by tensoring $\cE$ by an appropriate line bundle
in many cases.  And since most bundles cannot be produced by
the Serre construction, there is no reason to doubt that the anomaly can
be cancelled for most or even all $B$ by using more general bundles.
The above comments notwithstanding,
we will stick with the Serre construction for now so that we can be more
precise.

To cancel the $c_1$ anomaly, we have
to replace $\cE$ by $\cE\otimes L$, where $L$ is a line bundle with
$2c_1(L)=c_1(B)$.  Note that not every element of integral cohomology
is divisible
by~2, so $L$ need not exist.  But in a rough sense, ``about half'' of the $B$'s
have this property.  For example, for a hypersurface $B\subset{\bf P}^4$ of
odd degree $2n+1$, we have $c_1(B)=\O(4-2n)$ so we can take $L=\O(2-n)$.

Now
$$c_2(\cE\otimes L)=c_2(\cE)+c_1(\cE)\cdot c_1(L)+c_1(L)^2
=c_2(\cE)+c_1(L)^2,$$
so we just have to take any elliptic
curve $C$ representing the cohomology class $c_2(B)-c_1(L)^2$,
and anomaly cancellation is
satisfied.  It is very easy to find such curves, at least as long as
$c_2(B)-c_1(B)^2/4$ is non-negative, since elliptic curves abound
on Fano and Calabi-Yau threefolds $B$.

\subsec{Conifold Transitions}

It is now a simple matter to use the results of the previous section to pull
the bundle $\cE\otimes L$ through the conifold transition, at least
as far as the complex structure is concerned.  The metric will be left
for future work.

We first note how the Chern classes of $B$ change under a conifold transition
\cortismith :
\eqn\cherncon{c_1(B)=c_1(B'),\qquad c_2(B)=c_2(B')-[E],}
where we are using the notation introduced in sec.\ 2.1,
i.e.\ $E\subset B$ is the
${\bf P}^1$ that contracts to the conifold $B_0$, which then smooths to $B'$.
If we let $C\subset B$ and $C'\subset B'$ be the elliptic curves used in the
Serre construction to construct bundles $\cE,\ \cE'$, it follows from
\cherncon\ and the calculation of the Chern classes in the previous section
that we must require $C=C'-E$, and then we will be done.

This situation is easy to arrange.  We let $C_0\subset B_0$
be an elliptic curve containing the conifold with the required cohomology
class.  Then $C_0$ deforms to the
desired elliptic curve $C'\subset B'$.  On the other end of the transition,
the preimage of $C_0$ under the contraction map $B\to B_0$ is a reducible
curve of the form $C+E$, where $C$ projects isomorphically to $C_0$, hence is
elliptic.  So $C=C'-E$ in the required sense, and we get bundles
$\cE\otimes L$ and $\cE'\otimes L$ which satisfy the required properties.
Note that we have used \cherncon\ to identify the appropriate bundle $L$
on $B$ with another bundle on $B'$ which has also been denoted by $L$.


\newsec{M-theory and non-K\"ahler manifolds}

\noindent So far we have discussed the issues of vector bundles for the
kind of heterotic compactifications related to geometric transitions.
In this section we want to address a more generic question: Under
what condition does a F-theory (or an equivalent M-theory)
compactification) with fluxes allow K\"ahler or non-K\"ahler
manifolds with vector bundles? In other words, how does the
back--reaction of vector bundles affect the underlying heterotic
geometries?

A brief discussion of this issue appeared in \bd, where
it was discussed that localized and non--localized fluxes from M-theory play an
important role in determing the precise back--reaction effects on the
underlying geometry. The conclusion was that the heterotic manifold
is always non--K\"ahler when
there are only non--localized M--theory fluxes. On the other hand, in the presence of only localized fluxes, the manifold may or may not be K\"ahler. The
generic formula for the non--K\"ahlerity is given by the relation:
\eqn\nonkah{dJ ~ = ~ \alpha'\ast\left[\Omega_3(\omega_+)-\Omega_3(A)\right]
+ e^{-2\phi} d\phi \wedge J}
where $A$ is the one-form gauge bundle and $\phi$ is the heterotic dilaton.
Apparently, even in the absence of a background three--form,
the manifold can become non--K\"ahler due to vector bundles and non--trivial
dilaton. Thus, K\"ahlerity is restored only when
\eqn\kaha{\omega_+ ~ = ~ A, ~~~~~~ \phi ~ = ~ {\rm constant}}
which is precisely the condition studied in \chsw! Here we have derived the
condition by demanding a K\"ahler compactification from the generic
equation for $dJ$.

Let us now imagine that we do not turn on any non--localized gauge
fluxes and at the same time do not allow the standard embedding.
Then naively we would expect to get a
non--K\"ahler manifold with the non--K\"ahlerity coming precisely
from the difference $\Omega_3(\omega_+) - \Omega_3(A)$ (and the dilaton).
In fact, the three--form fluxes will typically look like \bbdp, \bbdgs
\eqn\thrform{H ~ = ~ f + {\alpha'\o 2} ~{\rm Tr}~\Big( \omega_0  \wedge
\tilde f \wedge \tilde f + \tilde f \wedge {\cal R}_{\omega_0} +
{1\o 2} \tilde f \wedge d\tilde f  - {1\o 6} \tilde f \wedge \tilde f
\wedge \tilde f\Big) + {\cal O}(\alpha'^2)}
where $f = \alpha'[\Omega_3(\omega_0) - \Omega_3(A)]$ and
$\omega_0$ is the {\it gravitational}
spin-connection at zeroth order in $\alpha'$ (see \bbdgs\ for
more details). We have also defined $\tilde f$ as a one--form created from
$f$ using the vielbeins, and
\eqn\romega{{\cal R}_{\omega_0} ~ = ~ d\omega_0 + {2\o 3}
\omega_0 \wedge \omega_0.}
The above three--form back-reacts on the geometry to make the space non--K\"ahler.
Thus, vector bundles without the standard embedding seem to be
allowed only on non--K\"ahler manifolds.
This would
seem to contradict the result of \Kgukov\ where it was found that
a fractional gauge Chern--Simons term can appear in an ordinary K\"ahler
compactification.
However, this apparent puzzle can be resolved by
taking the background {\it gaugino condensate} into
account\foot{The gaugino condensate contributes to the (3,0) and the
(0,3) part of the three--form, and can break susy. But we will consider a condensate that 
preserves susy.}. This can be explained as follows:

The existence of torsion in the heterotic theory is a direct consequence
of $G$--fluxes in M--theory \sav, \axell, \beckerD. As mentioned above,
there are two kinds of fluxes: localized and non--localized ones \bbdg, \bbdgs. M--theory
anomaly cancellation (i.e. non--zero Euler number) requires
fluxes for consistency. Considering only non--localised fluxes
gives rise to heterotic torsion. But this picture is in general
not complete. This is because localized fluxes at the orbifold singularities
are inevitable consequences of the existence of {\it normalizable}
harmonic forms near the singularities \robbins. The
situation becomes more involved because of the presence of
non--localized fluxes that back--react on the harmonic forms by
reverse--backreacting on the geometry. In a generic situation, the
harmonic forms may not be easy to evaluate. However for our case this
could be done \robbins. A more
standard analysis of how both kinds of fluxes conspire to give the
right Bianchi identity in the heterotic theory has been presented in
\bbdg. We will not go into details here, and conclude that
flux compactifications in M--theory generically lead to vector
bundles in the heterotic side, and in special cases, also to
torsion\foot{Recall that the
complex three--form that appears in heterotic theory in the
presence of torsion has to be imaginary self--dual (ISD) to
preserve supersymmetry in four dimensions. This implies the
background equation \nonkah.
This equation is more general than the constraint derived in
\rstrom,\hull, \smit\ and it reduces to the known form when the
manifold is complex \bbdp.
This equation makes the non--K\"ahler nature of the manifold
manifest. Observe, that if we scale the metric then the three--form
${H}$ scales linearly, too. On the other hand, from the Bianchi
identity we observe that the three--form does not scale (at least
to the lowest order in $\alpha'$). This implies that the radial
modulus should get stabilized \bbdg, \bbdp.
This argument, although correct, is rather naive at this point. The fact that
the Bianchi identity does not scale is only true for the K\"ahler
case. In the non--K\"ahler case the three form appears on
both sides of the identity (as we saw in the derivation of the
correct background three--form). Therefore
the correct way to study the potential for the radial modulus
would be to evaluate the three--form flux order by order in
$\alpha'$ and use the kinetic term to calculate the potential.
This was done in \bbdp, \bbdgs.}.

To conclude, the manifold can still become
non--K\"ahler in the absence of flux via the relation \nonkah.
In \Kgukov\ the $\Omega_3(\omega)$ term was
cancelled by one of the Chern-Simons terms of the gauge fields.
Therefore, the non--K\"ahlerity in this model arose from
vector bundles of one of the $E_8$ gauge groups. To obtain a Calabi--Yau
space we need $dJ = 0$. This can only be true with
an additional contribution to the superpotential of \bbdp, \lustu.
This additional contribution has been worked out in \lustd\ following the
work of \gcon\ (see also \lilia\ for some recent works). 
The equation for non--K\"ahlerity is now given by
\eqn\nonnow{dJ ~ =~ \alpha'\ast\left[\Omega_3(\omega_+)-\Omega_3(A)\right]
- \alpha'\ast\langle{\bar\chi}^A \Gamma \chi^A\rangle
+ e^{-2\phi} d\phi \wedge J}
where $\chi^A$ are the gaugino fields of heterotic theory.

The above equation can in principle have solutions with
$dJ = 0$ if the dilaton is chosen to be constant. In fact, both the
gaugino condensate term and the anomaly term are of the same order in
$\alpha'$ if we ignore the ${\cal O}(\alpha'^2)$ terms from \thrform.
If we cancel the $\Omega_3$ term with one of the $E_8$ terms $\Omega_3(A_1)$ in \nonnow, then $dJ$
vanishes (for a constant dilaton)
iff we choose a negative sign for the condensate. This could
in principle resolve the
puzzle raised for \Kgukov. Furthermore, the above identification
restricts the possible values for the gaugino condensate.

To summarize,
a dynamical way to study a flux background
is to take our proposed superpotential
and
solve for $dJ$ using the various contributions (tree level,
perturbative and non-perturbative). If $dJ = 0$ with integrable
complex structure we obtain K\"ahler CY compactifications. All other cases, i.e.
when $dJ \ne 0$ with integrable or non--integrable complex structure, will
correspond to non--K\"ahler compactifications. In this case the radius is stabilized at tree level. In the
K\"ahler case, the radius is fixed non--perturbatively
\axelk,\buchkov,\Kgukov. In both cases the $\sigma$--model conformal
invariance is restored at {\it that} particular radius.

Thus, along with the mathematical analysis of vector bundles in sec. 4,
we have a nearly complete picture of the gravity dual in heterotic theory.
Recently, a detailed study of
vector bundles on {\it compact} non--K\"ahler manifolds that
are some $T^2$ bundle over a $K3$ base (the examples studied in \sav, \beckerD, \axell,
\GP, \bbdg, \bbdp, \bbdgs) was presented in \yaufu\
(see also \yauli\ for earlier works on the subject from a mathematical point
of view and \lust, \bbdgs\ for a physical point of view).





\newsec{Global type IIA background}





We have now sufficiently elaborated on the heterotic and the
type IIB setup. In type IIB the
{\it local} metric is the metric of D5 branes wrapped on a two-- cycle of a resolved conifold {\it a-la} \pandoz. The {\it global} picture
is obtained from an F--theory background that we discussed in sec. 2.1,
and has F--theory seven--branes distributed in some particular way
determined by an underlying Weierstrass equation. The whole configuration
preserves ${\cal N} = 1$ supersymmetry.

We will now discuss the global type IIA background, which will turn out to have intersecting D6--branes and O6--planes on a geometry that looks locally like the non--K\"ahler deformed conifold constructed in \gtone. However, the F--theory setup will change the allowed fluxes, which have to be invariant under a certain orientifold action. This will affect the metric only in a minor way, but we will show that we find fluxes which cannot be completely gauged away. This is in contrast to the ${\cal M}$--theory lift advocated in \gtone, where our non--K\"ahler manifold lifted to a purely geometrical background in 11 dimensions. The background constructed here will lift to an 11--dimensional background with $G$--fluxes. We comment on the implications for $G_2$ structure versus $G_2$ holonomy in the next section.

\subsec{Type IIA background revisited}

The full {\it global} type IIA is now determined as a mirror of the global IIB background.
The local metric was already discussed in \gtone\ and \realm. The
mirror D6 branes in the local picture
now wrap the three cycle of a non--K\"ahler deformed
conifold. Globally there are additional D--branes and O--planes.
They are the mirror
of the type IIB D7/O7 system.
To specify the
precise coordinates of the branes in type IIA picture, let us consider
carefully the orientations of the branes in type IIB case. In the local
type IIB background, the resolved conifold has two two--tori oriented
along ($x, \theta_1$) and ($y, \theta_2$). The $U(1)$ fibration is
given by the coordinate $z$, and the radial direction is $r$. In this local
framework, let us assume that the D5 branes wrap the torus ($y, \theta_2$)
and are stretched along spacetime coordinates $x^{0, 1, 2, 3}$. A mirror
transformation results in D6 branes oriented along ($z, x , \theta_2$)
and stretched along the usual spacetime directions $x^{0,1,2,3}$. The local
non-K\"ahler metric of a {\it deformed} conifold is given by \fiitaamet.
The three--cycle is a $U(1)$ fibration over a compact
two--dimensional space specified by ($\alpha, \beta$) such that
\eqn\migr{ds_2^2 = \tilde g_2~d\theta_1(d\theta_1 - 2\alpha) +
g_3 ~dy(dy - 2\beta)}
forms {\it another} two dimensional subspace inside the non--K\"ahler
deformed conifold\foot{In the language of the non--K\"ahler metric,
the $U(1)$ fibration is over the {\it full} four--dimensional subspace
and is not restricted to the 2d base ($\alpha, \beta$).}.
If we denote the type IIB $B_{NS}$ as
($b_{x\theta_1},~ b_{y\theta_2}$) $\equiv$ ($b_1,~ b_2$), then the
differential forms ($\alpha, \beta$) and the function $\tilde g_2$ are
\eqn\abg{\alpha ~ = ~  {\rm sin}(2 ~{\rm tan}^{-1} b_1)~dx, ~~~~~
\beta ~ = ~  b_2 ~dy, ~~~~~ \tilde g_2 ~ = ~ g_2(1+b_1^2)}
with $g_2, g_3$ defined in \gtone, \realm\ and in \fiitaamet.
The metric of the three--cycle can now be written as
$ds_3^2 = G_{mn} ~dy^m dy^n$ with $y^m = (z, x, \theta_2)$, and
\eqn\twoacomp{\eqalign{G & = \pmatrix{G_{zz} & G_{zx} &
G_{z\theta_2}\cr \noalign{\vskip -0.20
cm}  \cr G_{xz} & G_{xx} & G_{x\theta_2}
\cr \noalign{\vskip -0.20 cm}  \cr G_{\theta_2 z} & G_{\theta_2 x}
& G_{\theta_2 \theta_2 }} \cr \noalign{\vskip
-0.25 cm}  \cr & = \pmatrix{g_1 & g_1~\Delta_1 ~{\rm cot}~\hat\theta_1
& - g_1~b_2~\Delta_2 ~{\rm cot}~\hat\theta_2
\cr \noalign{\vskip -0.20 cm}  \cr
 g_1~\Delta_1 ~{\rm cot}~\hat\theta_1 & g_2 +
g_1~\Delta^2_1 ~{\rm cot}^2~\hat\theta_1 &
- g_1~b_2~\Delta_1~\Delta_2 ~{\rm cot}~\hat\theta_1 ~{\rm cot}~\hat\theta_2\cr
\noalign{\vskip -0.20 cm}  \cr
- g_1~b_2~\Delta_2 ~{\rm cot}~\hat\theta_2 &
- g_1~b_2~\Delta_1~\Delta_2 ~{\rm cot}~\hat\theta_1 ~{\rm cot}~\hat\theta_2
 & \tilde g_3 + g_1~b_2^2~\Delta^2_2 ~{\rm cot}^2~\hat\theta_2}}}
with $\tilde g_3$ defined as $\tilde g_3 = g_3 (1 + b_2^2)$. \twoacomp\ is the precise metric
of the three--cycle on which we have wrapped D6 branes in the local
geometry.

To study the full global picture, we have to contemplate a few
possible scenarios
in type IIB theory, all resulting from different torus fibrations of the
F--theory fourfold:

\noindent (1) The F--theory torus is fibered non--trivially over the
two--dimensional torus parametrized by ($x, y$). This would mean
that the seven branes wrap directions ($z, \theta_1, \theta_2$)
and are stretched
along the non--compact directions ($r, x^{0,1,2,3}$).

\noindent (2) The F--theory torus is fibered non--trivially over a
different two dimensional torus parametrised by ($x, \theta_1$). Here the
seven branes would wrap directions ($y, \theta_2, z$), and are
stretched along the same non--compact directions as above.

\noindent (3) The F--theory torus is fibered non--trivially over a compact
two dimensional surface parametrised by ($\theta_1, \theta_2$). The
seven branes in this scenario will wrap directions ($x, y, z$) and are
stretched along the other non--compact directions as above.

Three T--dualities along ($x, y, z$) will act non--trivially in all these
cases. Let us consider case (1) first. Two T--dualities along
($x, y$) will take us to type I $SO(32)$ theory. This is precisely the
background studied in \realm. An S--duality to this
background will give us the required heterotic manifold, whose global
geometry we studied in section 3.

A further T--duality along $z$ direction
to the type I geometry will take us to type IIA. In fact,
the type IIA background is an orientifold background. But since an orientifold
operation is {\it not} a symmetry of type IIA theory, we should also accompany
the orientifold action $\Omega$ with a space reversal along the $z$ direction,
${\cal I}_z: z ~\to ~ -z$. Thus the duality chain for this case is
\eqn\dulumu{{\rm Type~I~~on}~~~{S_z^1\o \Omega}~~~
{}^{~T_z}_{\longrightarrow}~~~
{\rm Type ~IIA~~on}~~~ {S_z^1\o \Omega\cdot {\cal I}_z\cdot(-1)^{F_L}}}
where $T_z$ denote T--duality along the $z$ direction and
$(-1)^{F_L}$ changes the sign of the left--moving Ramond sector.
The two fixed points of ${\cal I}_z$ are the two
$O8$ planes, their charge being cancelled by D8--branes \horwit, \dabbie.

The global picture has therefore wrapped D6--branes on three--cycle of a
non--K\"ahler manifold, along with a $D8-O8$ system at the two
fixed points of $z$ (or $\psi$ in global coordinates). The $D8-O8$ system
intersects the D6 branes along a compact two dimensional torus ($x, \theta_2$)
and the 3+1 dimensional non--compact spacetime $x^{0,1,2,3}$.

In case (2), a mirror transformation gives rise to another type IIA background
modded out by an orientifold action.
The precise duality chain is
\eqn\dulumulu{{\rm Type ~IIB~~on}~~~~{T^5\o \Omega\cdot {\cal I}_{x\theta_1}
\cdot (-1)^{F_L}}~~{}^{T_{xyz}}_{{\longrightarrow}}~~
{\rm Type ~IIA~~on}~~~~
{T^5\o\Omega\cdot {\cal I}_{yz\theta_1}\cdot (-1)^{F_L}}}
where the $T^5$ torus is given by $T^5 = T^2_{(x\theta_1)}~ \times~
T^2_{(y\theta_2)}~\times S^1_z$, with $S^1_z$ forming the $U(1)$
fibration discussed above.
The fixed points of this action are the $O6$ planes whose charges are
cancelled by D6--branes. This $D6-O6$ system intersects with the other
wrapped D6 branes along the same compact two--torus ($x, \theta_2$) and the
3+1 dimensional non--compact spacetime.

Finally, in case (3), we expect the type IIA orientifold operation to be
$\Omega\cdot{\cal I}_{xyz\theta_1\theta_2}\cdot (-1)^{F_L}$. The fixed points
of this action are the
$O4$ planes which need $D4$ branes to cancel the total charge locally. These
four--branes are oriented along the non--compact directions
($r, x^{0,1,2,3}$), and therefore overlap with the wrapped D6 branes only
along the spacetime directions.

Let us now study the metric.
Case (1) with intersecting $D6-D8-O8$ system
cannot be the global completion of our type IIA background. Recall, that the
original type IIB (local) metric is {\it not}
invariant under the orientifold operation of case (1) \realm.
There are cross terms in the metric that are not invariant
under orientation reversal. The invariant part of the
metric can eventually give us a valid heterotic background, but it lacks the cross terms characteristic for the deformed conifold. This will not lead to the non--K\"ahler deformed conifold we require for the geometric transtition in IIA \foot{Note, that even an attempt to obtain a IIA background in this scenario is futile, because the global metric in type I/heterotic lost its isometry along $z$ (or $\psi$). Therefore we cannot perform the last T--duality that would lead to a IIA background. But, as explained above, even if it was possible to T--dualize along non--isometry directions (see e.g. \jefgreg) the resulting background would not be the one we require in IIA.}.

One can also easily rule out case (3),
where the F--theory fibration is along the
compact direction ($\theta_1, \theta_2$) in the local picture. A
coordinate reversal transformation
${\cal I}_{\theta_1 \theta_2}: \theta_i \to -\theta_i$ does not appear to
be a symmetry of the local metric \iibmet. So the metric
cannot provide the type IIA global picture.

This leaves us with case (2) where the F--theory torus is
fibered over the ($x, \theta_1$) torus. Not only is the IIB metric
\iibmet\ invariant under ($x, \theta_1$) ~ $\to$~ ($-x, -\theta_1$), but also the two tori with non--trivial complex
structure (boosted by $f_{1,2}$). This means that
after a mirror transformation we will have a global intersecting $D6-O6-D6$
system in type IIA with a local configuration of wrapped D6 branes
on the three cycle of a non--K\"ahler deformed conifold. Since no part
of the metric is projected out under orientifold operation, this will serve as
the full global picture in type IIA.

Having obtained a complete global picture in terms of
$D7/O7$ in type IIB
or $D6/O6$ in type IIA, we now turn to the question what the global
picture means in terms of the underlying ${\cal N} =1$ gauge theory.
The answer, as anticipated in section 2.2., is simple because in type
IIB the existence of $D7/O7$ implies {\it global} symmetries
in the gauge theory \dmftheory. In our case the global symmetry will be
$SU(2)^{16}$ instead of the expected $SO(8)^{4}$ because of Wilson lines. This means that we
are actually studying ${\cal N} =1$ $SU(N)$ gauge theory with fundamental flavors transforming under
the global symmetry $SU(2)^{16}$.


\subsec{More subtleties}

The above conclusion, although correct, is unfortunately premature.
The underlying orientifolding/orbifolding operation makes the situation
rather tricky.

First, observe that the type IIB orientifold operation has a non--trivial
effect on the $B_{NS}$ and $B_{RR}$ fields. The original choice of
$B_{NS}$ fields ($b_{x\theta_1}, b_{y\theta_2}$) are projected out. This
would mean that the fibration structures in the type IIA local metric would
also have to change. In fact the three cases studied earlier give rise
to the following choices of type IIB $B_{NS}$ fields:

 $\bullet$ Case (1): $\Omega$ and ${\cal I}_{xy}$ allow $b_{x\theta_1},
b_{x\theta_2}, b_{y\theta_1}, b_{y\theta_2}$. We used these choices to
determine the heterotic manifold after geometric transition in earlier
sections.

$\bullet$ Case (2): $\Omega$ and ${\cal I}_{x \theta_1}$ only allow
$b_{x\theta_2}$ and $b_{y\theta_1}$ \foot{The orientifold action also allows for other components of
$B_{NS}$ fields like $b_{xz}, b_{xr}, b_{z\theta_1}, b_{r\theta_1}$ but we can set them to zero and
study the theory with only two components consistently.}.
This implies that the metric in IIA will have a
different fibration structure compared to the one that gave rise to the heterotic bckground.

$\bullet$ Case (3): $\Omega$ and ${\cal I}_{\theta_1\theta_2}$ also allow
$b_{x\theta_1}, b_{x\theta_2}, b_{y\theta_1}, b_{y\theta_2}$. However
this choice of orientifolding is not useful enough to get a global
IIA metric, as we discussed above.

The type IIA {\it local} metric will change accordingly to accomodate the
actions of $O6$ planes. The metric still remains a non--K\"ahler deformation
of a Calabi-Yau deformed conifold, but is now given as
\eqn\fiitaametnowb{\eqalign{& ds_{IIA}^2 = ~~ g_1~\left[dz
 + \Delta_1~{\rm cot}~\hat\theta_1~(dx -
b_{x\theta_2}~d\theta_2) + \Delta_2~{\rm cot}~ \hat\theta_2~(dy -
b_{y\theta_1}~d\theta_1)+ ...\right]^2~+ \cr & + g_2~ {[} d\theta_1^2
+ (dx - b_{x\theta_2}~d\theta_2)^2] + g_3~[ d\theta_2^2 + (dy -
b_{y\theta_1}~d\theta_1)^2{]} +  g_4~{\rm sin}~\psi~{[}(dx -
b_{x\theta_2}~d\theta_2)~d \theta_2 \cr & ~~~~~~~~~ + (dy -
b_{y\theta_1}~d\theta_1)~d\theta_1 {]} + ~g_4~{\rm
cos}~\psi~{[}d\theta_1 ~d\theta_2 - (dx - b_{x\theta_2}~d\theta_2)
(dy - b_{y\theta_1}~d\theta_1)].}}
One can also check that the metric remains invariant under
the orientifold operation $\Omega\cdot {\cal I}_{yz\theta_1}$ when
rotating by $\psi$ as in \gtone. This implies that we should set
$\psi = \langle\psi\rangle$ in \defofcoord. Allowing for non--constant $\psi$ (i.e.
extending $\psi$ to $z$) would result in a non--trivial constraint on the
warp factor $g_4$ in \fiitaametnowb:
\eqn\ntgfour{g_4~ = ~ g_4 (r, \theta_1, \theta_2)\vert_{r = r_0} ~=~
\sum_n a_n ~\theta_1^{2n+1}~ \tilde g_4(r_0, \theta_2)}
where $a_n$ are some non--zero numbers and $\tilde g_4$ is another
function of ($r= r_0, \theta_2$).

One might also contemplate a more generic relation between
$\psi$ and $z$ as $\psi ~=~ f(z)$. For the metric \fiitaametnowb\ to
remain invariant under orientifold action, the function $f(z)$ has
to satisfy
\eqn\fzen{f(z)~ + ~f(-z)~ = ~\pi.}
 Although
such a choice would put no constraints
on the warp factors $g_i$ in \fiitaametnowb, the functional
dependence would not be consistent with the simple choice of the
small coordinate shifts in type IIB \defofcoord. Therefore,
\ntgfour\ seems to be the only reasonable constraint we can
impose on the warp factor $g_4$.

The next subtleties are related to the choice of $B$ fields in type IIB.
The new choice of $B_{NS}$ fields ($b_{x\theta_2}$ and $b_{y\theta_1}$) will induce the following
$B_{NS}$ fields in the mirror type IIA background:
\eqn\bnfe{\eqalign{B_{\rm IIA} ~= ~ &\sqrt{\langle\alpha\rangle_1}~(dx -
b_{x\theta_2} d\theta_2)\wedge d\theta_1 -
\sqrt{\langle\alpha\rangle_2}~(dy -
b_{y\theta_1} d\theta_1)\wedge d\theta_2 +
A\sqrt{\langle\alpha\rangle_1}~ d\theta_1 \wedge dz ~ + \cr
& ~~~~~~~~~~~ + B\sqrt{\langle\alpha\rangle_2}~(dy~{\rm sin}~\psi
- {\rm cos}~\psi~ d\theta_2 - b_{y\theta_1}~{\rm sin}~\psi~d\theta_1)
\wedge dz}}
where $\langle\alpha\rangle_i$ and ($A,B$) were defined in \gtone, \realm. We wrote the $B_{NS}$ field using $\langle\alpha\rangle_i$
to emphasize the fact that some of the terms are pure gauge (see \realm\ for more details on this). Does
that means that we can gauge them away? It turns out that because of the presence of D6 branes, we
{\it cannot} gauge away all components of $B_{NS}$. The ungauged part of the $B_{NS}$ field will
appear as gauge fluxes on the D6 branes.

Since this issue is a little subtle, we will tread carefully in
the following. We can perform a $\psi$ rotation on ($y, \theta_2$)
as in \setoftrans\ by identifying $\langle\psi\rangle = \psi$ (see
also \gtone). This will change the $B_{NS}$ field into
\eqn\brot{\eqalign{B_{\rm IIA} ~ &= ~ (dx - b_{x\theta_2}
d\theta_2)\wedge d\tilde\theta_1 - (dy - b_{y\theta_1}
d\theta_1)\wedge d\tilde\theta_2 + (d\hat\theta_1 + d\hat\theta_2)
\wedge dz \cr & = ~ b_{x\theta_2} ~d\tilde\theta_1 \wedge
d\theta_2 + b_{y\theta_1} ~d\theta_1 \wedge d\tilde\theta_2 +
dx\wedge d\tilde\theta_1 - dy \wedge d\tilde\theta_2 +
(d\hat\theta_1 + d\hat\theta_2) \wedge dz}} where we have isolated
the B--field part that depends on the type IIB B--fields. The
quantities appearing here are defined as:
\eqn\thealp{\eqalign{&\tilde\theta_1 ~ = ~ -{1\o k\sqrt{b_1}}~
\left[{\rm ln}(k~{\rm cos}~\theta_1 + \hat\Delta_1)\right] \cr &
\hat\theta_1 ~ = ~ {1\o k\sqrt{b_1}}~ \left[ \sqrt{b_1}~{\rm
arctan}~{k~{\rm sin}~\theta_1\o \hat\Delta_1}\right] \cr & k ~=~
\sqrt{b_1-a_1\o a_1}, ~~~~~~ \hat\Delta_1 ~=~ \sqrt{1-k^2~{\rm
sin}^2~\theta_1} \cr & \langle\alpha\rangle_1 ~ = ~{1\o 1 +
(\Delta^0_1)^2~{\rm cot}^2~\theta_1 + (\Delta_2^0)^2~{\rm
cot}^2~\langle\theta_2\rangle} \equiv {1\o a_1 + b_1~{\rm
cot}^2~\theta_1}}} with an equivalent description for
$\tilde\theta_2$ and $\hat\theta_2$. Using this one can show that
the $b$--independent parts of \brot\ are pure gauge.

How many components of
the $b$--independent part of \brot\ can be gauged away? In the absence of $D6$ branes, all the components can be
gauged to zero. Ignoring the spacetime orientations of the six--branes and planes, we see that the $b$--independent
part of \brot\ that would appear as gauge flux $F$ on the wrapped $D6$ branes would be
\eqn\fflux{F_{z\theta_2} ~ = ~{}^{\rm lim}_{\epsilon \to 0}~~\Big[\epsilon^{-1}~
B \sqrt{\langle\alpha\rangle_2}\Big]~~d\theta_2 \wedge dz}
on a non--trivial two--cycle inside the three--cycle of the non--K\"ahler deformed conifold. This gauge
flux is very large and will provide a non--commutativity parameter to the
 gauge theory on wrapped $D6$ branes\foot{Observe that the $3+1$ dimensional ${\cal N} =1$ $SU(N)$ gauge theory is
still commutative.}.
All other
$b$--independent components of \brot\ are gauge equivalent to zero\foot{Observe that there are no possibilities
of any {\it pinning} effect \orione\ or any {\it dipole} behavior \difuli\ here. For topologically trivial cycles, we
would have encountered more involved scenarios.}.

Let us now consider the $b$--dependent part of \brot.
The type IIA B--field depends on the type IIB components ($b_{x\theta_2},
b_{y\theta_1}$) which could have non--zero field strength.
Because of the underlying mirror transformation the components of
\brot\ should be independent of ($x, y, z$). Furthermore, the $b$--dependent terms in \brot\
have wedge structures of the form
$d\theta_1 \wedge d\tilde\theta_2$ and $d\tilde\theta_1 \wedge d\theta_2$.
If the components ($b_{x\theta_2}, b_{y\theta_1}$) were functions of ($\theta_1, \theta_2$)
{\it only} and not of
$r$, the radial coordinate, we would find $dB_{\rm IIB} \ne 0, dB_{\rm IIA} = 0$, i.e.
the type IIA B--field \brot\ will have no field
strengths, whereas the type IIB B--field will have a non--zero field strengths\foot{Type IIB NS threeform field
$H_{\rm IIB} \equiv {\cal H}$ components will be
${\cal H}_{x\theta_1\theta_2} = \del_{[\theta_1} b_{\vert x\vert\theta_2]}$ and
${\cal H}_{y\theta_1\theta_2} = \del_{[\theta_2} b_{\vert y\vert\theta_1]}$.}.
Such B--field components are projected out {\it at} the orientifold
point, but away from the orientifold point\foot{In fact away from the orientifold point, all components of
the B--field can be allowed.}
we can have $b_{x\theta_2} = b_{x\theta_2}(\theta_1,
\theta_2)$ and $b_{y\theta_1} = b_{y\theta_1}(\theta_1, \theta_2)$.
Therefore, if we we want to gauge away as many components of $H_{\rm IIA}$ as possible, our first ans\"atze for the type IIB B--field components would be
\eqn\bbcoml{b_{x\theta_2} ~ = ~ \sum_{m, n}~b_{mn}~{\rm cot}^{m}~\theta_1~ {\rm cot}^{2n}~\theta_2,
~~~~~~ b_{y\theta_1} ~ = ~
\sum_{k,l}~c_{kl} ~{\rm cot}^{k}~\theta_1~{\rm cot}^{2l}~\theta_2}
where $b_{mn}$ and $c_{kl}$ are integers independent of ($x, y, z, \theta_i, r$). The precise numbers will be
determined by the background equations of motion. Plugging this ansatz \bbcoml\ into \brot, one of the
$b$--dependent component takes the following form:
\eqn\fofobde{b_{y\theta_1}~d\theta_1 \wedge d\tilde\theta_2 ~ = ~ \sum_{k,l}~ {c_{kl}}~d\Theta_1 \wedge d\Theta_2}
which is again a pure gauge. Similarly, the other $b$--dependent component can also be expressed as a pure gauge.
The variables $\Theta_i$ are defined in terms of $\theta_i$ as
\eqn\ththas{\Theta_1 ~ = ~ \int {\rm cot}^k~\theta_1, ~~~~~~~ \Theta_2 ~ = ~
\int {{\rm cot}^{2l}~\theta_2 \o a_2 + b_2~{\rm cot}^2~\theta_2}~.}
Expressing \brot\ in terms of $\Theta_i$, none of these components are parallel to the six--branes. Thus,
they would not contribute to fluxes in type IIA theory and could be completely gauged to zero. Then the only
non--trivial effect from type IIA B--field will be the appearence of the gauge flux \fflux\ on an internal two--cycle of the
non--K\"ahler deformed conifold\foot{The reader might wonder about switching on a type IIB component $b_{z\theta_2}$
which in type IIA would be mirror to a component parallel to the six--branes. However, such a component would be projected out in IIB by the orientifold operation. Away from the orientifold point, such a component will be allowed.}.

Above analysis assumed the seven branes of type IIB being far away so that we are no longer at
the orientifold point. This is consistent with the picture of integrating out all the flavors
to study ${\cal N} =1$ pure $SU(N)$ Yang--Mills theory. The choice of type IIB $B_{NS}$--field determines
the allowed components of the type IIB $B_{RR}$--field from the linearized equation of motion
\eqn\lineom{H_{RR} ~ = ~ \ast_6~H_{NS}}
where the Hodge--$\ast_6$ is w.r.t. the six dimensional
internal space. \lineom\ allows for the
following components of $H_{RR} \equiv h$: $h_{yzr}$ and $h_{xzr}$. This is equivalent to choosing
$\del B_{RR} \equiv \del B$ as $\del_rB_{xz}$ and $\del_rB_{yz}$. This means that one particular consistent choice of B--fields in type IIB {\it away} from the orientifold point is
\eqn\iibbchoice{B_{NS}:~\left[b_{x\theta_2}(\theta_1, \theta_2),  b_{y\theta_1}(\theta_1, \theta_2)\right]~~~~~~~
B_{RR}:~\left[B_{xz}(r), B_{yz}(r)\right]\,.}
So now the natural question would be:
which B--field components in type IIB and type IIA are allowed {\it at} the orientifold point?
Observe that the choice of B--fields away from the orientifold point \iibbchoice\ is not very
symmetric. It would seem more natural to switch on {\it similar} components for both $B_{NS}$ and $B_{RR}$
(see the solutions in \ks\ and \pandoz). A
possible choice at the orientifold point, that also satisfies \lineom, would be to define the $B_{NS}$--field as
$B_{NS}:~\left[b_{x\theta_2}(r),  b_{y\theta_1}(r)\right]$ and the field strengths as:
\eqn\koroth{\eqalign{&
H_{NS}:[{\cal H}_{x\theta_2 r}= \del_r b_{x\theta_2}, ~~~~~~~~{\cal H}_{y\theta_1 r}= \del_r b_{y\theta_1}] \cr
& H_{RR}:[h_{x\theta_2 z} = \ast_6 {\cal H}_{y\theta_1 r}, ~~~~~~~ {h}_{y\theta_1 z}= \ast_6 {\cal H}_{x\theta_2 r}]\,.}}
Are these choices of B--fields away from the orientifold point \iibbchoice\ and at the orientifold point \koroth\
consistent? We propose the following scenario:

\noindent $\bullet$ Locally, the radial dependence in \koroth\ will not be visible because of \defofcoord. Thus, $H_{NS}$ and $H_{RR}$ are very small in our region of interest.

\noindent $\bullet$ Once we go away from the orientifold point, then non--perturbative effects will change the
background \senbanks. In particular, $b_{x\theta_2}, b_{y\theta_1}$ will now also depend on $\theta_i$.

Thus, given a background \koroth\ we can deform it into \iibbchoice\ with non--perturbative
corrections. The deformation happens when we want to move the seven branes away from the wrapped D5 branes. This
is the case when one of the orientifold points coincides with the resolution two--cycle of the F--theory
base discussed in sec. 2.1. So we can still remain {\it at} the orientifold point if the resolution two--cycle
{\it does not} coincide with {\it any} of the orientifold points. This is again similar to the situation encountered
in the F--theory construction with Klebanov--Strassler base \gkp. The seven branes and orientifold points
do not coincide with the conifold point (where we have wrapped D5 branes).

Before moving ahead let us clarify one more point. The $B_{RR}$ fields for the two cases that we discussed above
i.e \iibbchoice\ and \koroth, would seem to imply that we have changed the orientation of the wrapped D5 branes.
Recall that our initial wrapped D5 branes were on a two-torus along ($y, \theta_2$) directions. However
the choice \koroth\ would seem {\it not} to allow such wrapped D5 branes. This is an illusion, because the
choice \koroth\ will change completely as soon as we go away from the orientifold point. 
Once we are away from the orientifold point, the sources for the 
wrapped D5 can be easily shown to be present there. However, there is another much deeper reason why we can allow 
all kind of configurations of wrapped D5 branes in our scenario. This has to do with the fact that we are defining the
background only locally. In the local version, the type IIB metric can be rewritten as
\eqn\bothmet{ds^2 = dr^2 + (dz + A~dx + B~dy)^2 + (d\theta_1^2 + dx^2) + (d\theta_2^2 + dy^2) + ....} 
where we can extract the values of $A$ and $B$ from \metformj\ as:
\eqn\anab{A ~= ~ \Delta_1^0~{\rm cot}~\langle\theta_1\rangle + {\cal O}(\theta_1), ~~~~~~~~ 
B ~= ~ \Delta_2^0~{\rm cot}~\langle\theta_2\rangle +  {\cal O}(\theta_2)}
which tells us that they are basically constants. This would imply that the fibration structure can be 
written as $d\tilde z ~ \approx ~dz + A~dx + B~dy$. Therefore from the above local metric, D5 branes wrapping
directions ($y, \theta_2$), can be easily traded with ($r, y$) directions and vice-versa. So the precise 
way by which we can go from the global type IIB model to the local metric \iibmet\ is to 
put non-trivial complex structure on the ($x, \theta_1$) and ($y, r$) tori, 
trade the local $r$ coordinate with local $\theta_2$ and then redefine the $d\tilde z$ 
fibration structure accordingly. This way
configuration {\it at} the orientifold point, and configuration {\it away} from the orientifold point 
will be identical.

Thus at the orientifold 
point, we can stick with our configuration \koroth. Now  
assuming such a scenario simplifies the ensuing analysis,
because we do not have to worry
about non--perturbative corrections from the splitting of orientifold--seven planes in
type IIB, the resulting O6 planes in type IIA will not split either. We then employ the following 
generic choice of type IIB RR field:
\eqn\roro{\eqalign{B_{RR} ~~& = ~c_1~ dx \wedge dz + c_2 ~dx \wedge dy + c_3 ~dx \wedge d\theta_2 ~ + \cr
       & ~~~~~~+ c_4 ~dy ~\wedge d\theta_1 + c_5 ~dz \wedge d\theta_1
       + c_6~ d\theta_1 \wedge d\theta_2}}
where we have allowed for two new components $c_2$ and $c_6$ that are naively allowed under
orientifold action. The coefficients $c_i$ are in general functions or $(r,~\theta_1,~\theta_2)$. The $B_{NS}$ components are still
($b_{x\theta_2}, b_{y\theta_1}$). With this choice of
$B_{RR}$, we obtain the following one--form field ${\cal A}$ in the mirror type IIA:
\eqn\oneform{{\cal A} = {2AB \o 1+A^2}(c_1 -\alpha B c_2)~ (dx-b_{x\theta_2} d\theta_2)
  + (c_1-2\alpha Bc_2) ~(dy-b_{y\theta_1} d\theta_1) - c_2 ~dz}
where we used the freedom of choosing the orientation
of $B_{NS}$ similar to \gtone. The fibration structure
of the metric \fiitaametnowb\ remains as before. The one--form \oneform\ will be
sourced by the wrapped D6 branes.

Apart from the one form, there will also be three--forms in the mirror IIA. They arise precisely
because of our choices of $B_{NS}$ components in IIB. They are given by:
\eqn\threenow{\eqalign{{\cal C}~=~& ~~~ C_{xy1}~ (dx-b_{x\theta_2}d\theta_2)\wedge (dy - b_{y\theta_1}d\theta_1)         \wedge d\theta_1 \cr
   &  + C_{xy2}~ (dx-b_{x\theta_2}d\theta_2) \wedge (dy - b_{y\theta_1}d\theta_1) \wedge d\theta_2  \cr
   &  + C_{xz1}~ (dx-b_{x\theta_2}d\theta_2)\wedge dz \wedge d\theta_1  +
    C_{yz2}~ (dy - b_{y\theta_1}d\theta_1)\wedge dz \wedge d\theta_2  \cr
   & + C_{xz2}~ (dx - b_{x\theta_2} d\theta_2)\wedge dz\wedge d\theta_2
    + C_{yz1}~ (dy - b_{y\theta_1}d\theta_1)\wedge dz\wedge d\theta_1}}
where the coefficients are defined in the following way:
\eqn\coffdeff{\eqalign{& C_{xy1}~= ~-c_5+f_1c_1+{2\alpha f_1 c_1 A^2B^2 \o 1+A^2} +2\alpha B(c_4-f_1c_2)\cr
& C_{xy2}~ = ~ {2\alpha AB^2 \o 1+A^2}\Big[-c_3+f_2(Bc_1-c_2)\Big] \cr
& C_{xz1}~ = ~ c_4-f_1c_2+{2f_1 c_1 A^2B \o 1+A^2}, ~~~~
C_{yz2}=-c_3+f_2(Bc_1-c_2) \cr
& C_{xz2}~ = ~{2 AB \o 1+A^2}\Big[-c_3+f_2(Bc_1-c_2)\Big],~~~~ C_{yz1} ~ = ~ A~f_1c_1}}
We can make the following interesting observations:

\noindent $\bullet$ All individual coefficients can be separated into $f_i$--dependent and $f_i$--independent parts.

\noindent $\bullet$ The terms containing $f_1$ are always accompanied by $d\theta_1$. The same statement is true
for the $f_2$--dependent terms. These terms (like $f_1 d\theta_1$ and $f_2 d\theta_2$) are exact functions, and can therefore be written in terms of $d\tilde\theta_i$ of \thealp.

\noindent $\bullet$ The $f_i$--dependent terms are also accompanied
by either $c_1$ or $c_2$ or both.

This raises the suspicion that maybe all RR--fluxes are pure gauge artifacts, too. We will show in the following that this is not the case, but that we obtain G--fluxes which will contribute to ${\cal M}$--theory fluxes. These, in turn, allow for a smooth flop transition, avoiding the singular point in the conifold transition.

It suffices first
to consider a simple case where some of the coefficients $c_i$ in \roro\ are allowed to vanish. What is the minimum
number of $c_i$ that have to be non--zero? Obviously,
the one--form \oneform\ depends only on $c_1$ and $c_2$. Since this has to be non--zero because of the wrapped
D6--branes, we will make the simplifying assumption that all other $c_i$ vanish. This doesn't quite work because
if $c_2$ is non-zero, then from \roro\ we see that this switches on a component $dx \wedge dy$. But from \koroth\ we
know that such a component doesn't exist, at least for the simplfied case that we wanted to analyse. Thus we take:
\eqn\firstc{{\rm{\bf Case~ 1:}}~~~~ c_1 ~=~c_1(\theta_2), ~~~~ c_2 ~=~ c_3 ~ = ~c_4 ~ = ~c_5 ~ = ~c_6 ~ = ~0}
where the $\theta_2$ dependence of $c_1$ is motivated from the choice of background $H_{RR}$ in 
\koroth\foot{There is a little more to it. The choice of $H_{RR}$ as in \firstc\ also means that we have 
chosen the {\it orientation} of the wrapped D5 branes in our type IIB model. As we saw earlier, it 
doesn't really matter how the D5 branes are wrapped as long as we are taking consistent background NS and RR fields
satisfying equations of motion and correct orientifold projections. Clearly \roro\ satisfies the second
constraint, and we can make sure that \firstc\ will satisfy both the constraints. Furthermore as we will see below,
many more cases can be entertained in our global picture.}. 
This implies for the
type IIB $H_{RR}$ fields:
\eqn\hroronow{h_{xz\theta_2} ~ = ~ {\del c_1\o \del\theta_2}, ~~~~~ h_{yz\theta_1} ~ = ~ 0}
which is consistent with the expected components of $H_{RR}$ in type IIB from \koroth\ if $b_{x\theta_2}$ also
vanishes, or is independent of the radial coordinate. Thus the choice \firstc\ implies that we choose in type IIB
the local metric \iibmet\ and the $B$-fields:
\eqn\bfafc{B_{NS} ~=~ b_{y\theta_1}(r)~dy \wedge d\theta_1, ~~~B_{RR} ~ = ~ c_1(\theta_2) ~dx \wedge dz}
and in mirror type IIA the following background
\eqn\iiaafter{\eqalign{& ds_{IIA}^2 = ~~ g_1~\left[dz
 + \Delta_1~{\rm cot}~\hat\theta_1~dx
 + \Delta_2~{\rm cot}~ \hat\theta_2~(dy -
b_{y\theta_1}~d\theta_1)+ ...\right]^2~+ \cr &~~~~~~~~~ + g_2~ {[} d\theta_1^2
+ dx^2] + g_3~[ d\theta_2^2 + (dy -
b_{y\theta_1}~d\theta_1)^2{]} +  g_4~{\rm sin}~\psi~{[}dx
~d \theta_2~ +  \cr & ~~~~~~~~~ + (dy -
b_{y\theta_1}~d\theta_1)~d\theta_1 {]} + ~g_4~{\rm
cos}~\psi~{[}d\theta_1 ~d\theta_2 - dx~
(dy - b_{y\theta_1}~d\theta_1)] \cr
& {\cal A} ~=~ {2ABc_1 \o 1+A^2}~ dx
  + c_1 ~(dy-b_{y\theta_1} d\theta_1)\cr
& B_{\rm IIA} ~ = ~ b_{y\theta_1}~ d\theta_1 \wedge d \tilde\theta_2}}
We now see that the choice \firstc\ simplifies the coefficients \threenow\ quite a bit. Infact, combining
\firstc, \thealp\ in \threenow\ we can re-express \threenow\ as ${\cal C} ~ = ~ \sum_{1=1}^6 ~C_i$ with
\eqn\threethen{\eqalign{& C_1 ~ \equiv ~ C_{[xy1]} ~ = ~ \Big(c_1+{2\alpha c_1 A^2B^2 \o 1+A^2}
\Big)~dx \wedge dy \wedge d\tilde\theta_1 \cr
& C_2 ~ \equiv ~ C_{[xy2]} ~ = ~ {2\alpha A B^2 c_1 \o 1 + A^2}~
d\hat\theta_2 \wedge dx \wedge (dy - b_{y\theta_1}d\theta_1) \cr
& C_3 ~ \equiv ~ C_{[xz1]} ~ = ~
{2 A B c_1 \o 1+A^2}~ d\hat\theta_1  \wedge dx \wedge dz \cr
&C_4 ~ \equiv ~ C_{[yz2]} ~ = ~ c_1~ d\hat\theta_2
\wedge (dy - b_{y\theta_1}d\theta_1)\wedge dz \cr
& C_5 ~ \equiv ~ C_{[xz2]} ~ = ~ {2AB c_1 \o 1 + A^2}~ d\hat\theta_2  \wedge
dx \wedge dz, ~~~ C_6 ~ \equiv ~ C_{[yz1]} ~ = ~ c_1~ dy \wedge dz \wedge d\hat\theta_1}}
where we have separated the terms with $d\tilde\theta_i$ and $d\hat\theta_i$ as defined in \thealp.

The existence of the ${\cal C}$ field component $C_{xz2}$ implies that we can define
a combination
\eqn\deco{{\cal V} ~ = ~ \sqrt{{\rm det}~G} + i \vert C_5 \vert}
where $G$ is the metric of the three--cycle along ($x, z , \theta_2$) directions \twoacomp\ and $\vert C_5\vert$
from \threethen\ is the coefficient along $d\zeta \equiv dx \wedge dz \wedge d\theta_2$ but now evaluated 
for $c_1 = c_1(\theta_1, \theta_2)$ by deforming away from the orientifold point. 
Therefore, the effective volume of the three cycle along ($x, z , \theta_2$) is ${\cal V}$ and not just
$\sqrt{{\rm det}~G}$ from \twoacomp. This implies that in the limit
\eqn\flop{ \sqrt{{\rm det}~G}~~\to~~0, ~~~~~ \vert {\cal V} \vert ~ \equiv ~ \sqrt{{\rm det}~G + \vert C_5 \vert^2}
~~\to ~~ \vert C_5 \vert}
which is non-zero. Thus, when we lift the type IIA background to ${\cal M}$--theory as in \gtone\ to
a seven dimensional manifold with $G_2$ structure, a flop operation is a completely {\it non-singular}
operation. This is perfectly consistent with the predictions of \amv\ and \edelstein\ and justifies the
analysis of \gtone.

Recall that the type IIA $B_{NS}$ fields were almost all pure gauge\foot{Of course this
does not  mean that they can all be gauged away. As discussed above, gauge flux along the brane world volume cannot be gauged away.}.
Let us now show that this is not the case for  $B_{RR}$. We can simply evaluate the field strenths of the fluxes.
The $G$--fluxes are found to be
\eqn\gfluxone{G_{xy\theta_1\theta_2} ~ = ~
\Bigg[{A^2\o 1+A^2} {\del \o \del\theta_2}(c_1 + 2\alpha c_1 B^2)~d\tilde\theta_1 \wedge d\theta_2
- 2 B^2 c_1 {\del \o \del\theta_1} \left({\alpha A^2\o 1+A^2}\right)~
d\theta_1 \wedge d\hat\theta_2 \Bigg]~ dx \wedge dy}
which is still non--zero even in the local limit where $A,~B$ and $\alpha$ are just constants.
The most important $G$--flux component is
\eqn\gfluxtwo{G_{xz\theta_1\theta_2} ~ = ~\Bigg[ {2 A^2\o 1+A^2}{\del \o \del\theta_2}(B c_1)~
 d\tilde\theta_1 \wedge d\theta_2
- 2Bc_1 {\del \o \del\theta_1} \left({A\o 1+A^2}\right)~ d\theta_1 \wedge d\hat\theta_2 \Bigg] ~
dx \wedge dz}
which shows that $\vert C_5 \vert$ cannot be gauged away. Finally, there is one more component that is
similar to the type IIB RR three form \hroronow. It is given by
\eqn\gfluxthree{G_{yz\theta_1\theta_2} ~ = ~ {\del c_2\o \del\theta_1}~d\theta_1 \wedge d\tilde\theta_2 \wedge
dy \wedge dz}
with all other components vanishing. The existence of these $G$--fluxes is precisely the reason why our ${\cal M}$--theory manifold does not have a $G_2$ holonomy, but only a $G_2$ structure.
In the next section we discuss the implications for $SU(3)$ and $G_2$ torsion classes in IIA and ${\cal M}$--theory, respectively.

For completeness let us also mention that our background may also
have non--zero five--form
\eqn\fiveform{C_{xyz\theta_1\theta_2} ~ = ~ c_6 + f_1 c_3 - f_2 (1+f_1) c_4 + B(c_5 -f_1c_1)f_2\,.}
For the simple background \firstc\ this is a pure gauge, and will therefore not affect
type IIA dynamics.

Our next simple case would be when we allow all the components in \koroth. This would imply that both the
$B_{NS}$ components ($b_{x\theta_2}, b_{y\theta_1}$) are non-zero. The $c_i$ in \roro\ will be
\eqn\phani{{\rm{\bf Case~ 2:}} ~~~~ c_1 ~=~c_1(\theta_2), ~~~c_4 ~ = ~c_4(z), ~~~~~~c_2 ~=~c_3 ~=~c_5 ~=~c_6 ~=~0}
and the type IIA metric will become \fiitaametnowb\ instead of \iiaafter. The type IIB threeform
will have both the components, with $h_{yz\theta_1}$ now given by $h_{yz\theta_1} ~=~ {\del c_4 \o \del z}$
and $h_{xz\theta_2}$ as before \hroronow.
It is also easy to see that the
${\cal C}$ components $C_{xy2}, C_{yz2}, C_{yz1}$ and $C_{xz2}$ will also remain as before \threethen. The only
changes will be to the following components: $C_{xy1}$ and $C_{xz1}$, and they will be given by
\eqn\changeflux{\delta C_{[xy1]}~ = ~ 2 \alpha B c_4~dx \wedge dy \wedge d\theta_1, ~~~ \delta C_{[xz1]} ~ = ~
c_4~dx \wedge dz \wedge d\theta_1}
These changes will be reflected on the $G$-fluxes also. As one would have expected, the changes in the
$G$-fluxes will be precisely from the additional $c_4$ terms. In fact we will have one new component
of $G$-flux in addition to the ones that we already evaluated in \gfluxone, \gfluxtwo, \gfluxthree\foot{These
components of $G$-fluxes will not change although the corresponding $C$-fields have changed.}. This is given by
\eqn\gfluxchange{ G_{[xyz\theta_1]} ~ = ~ - 2 \alpha B {\del c_4 \o \del z}~ dx\wedge dy \wedge dz \wedge
d\theta_1}
The five form, which is now given by
\eqn\ffnow{C_{[xyz\theta_1\theta_2]} ~ = ~ -\Big[c_4~ d\bar\theta_1
 \wedge d\tilde\theta_2 - c_1 ~ d\tilde\theta_1 \wedge d\hat\theta_2\Big] \wedge dx \wedge dy \wedge dz }
still remains a pure gauge. We have also defined $\bar\theta_1 ~=~ \theta_1 + \tilde\theta_1$. From above we
see that the case of flop in the M-theory lift of our background still remains non-singular. The flux
configurations are now much more involved, but the basic physics have not changed. One little concern here
would be the $c_4(z)$ part. This is explicitly a function of $z$ and therefore might create problems
for our T-duality transformations. One could overcome this problem by making a T-duality transformation on the
$H_{RR}$ directly following the rules given in \fawad. All we now require is to make $h_{yz\theta_1}$
independent of $z$ direction. For our case, this doesn't seem to be of any concern.

The above discussion then leads us to describe yet another case, that could in principle occur if we make the
field strengths in type IIB independent of the duality directions. This would be when in \roro\ we allow
\eqn\cinow{{\rm {\bf Case~3:}}~~~~ c_1 ~=~ c_1(\theta_2), ~~c_3 ~=~ c_3(z), ~~
c_4 ~=~ c_4(z), ~~ c_5 ~=~ c_5(y), ~~ c_2 ~= ~ c_6 ~=~ 0}
For this case, the $H_{RR}$ would be more complicated than the ones that we had earlier. The components of
$H_{RR}$ would become
\eqn\bolo{h_{xz\theta_2} ~ = ~ {\del c_1 \o \del \theta_2} - {\del c_3 \o \del z}, ~~~~~~~
h_{yz\theta_1} ~ = ~ {\del c_5 \o \del y} - {\del c_4 \o \del z}}
Both these components are to be arranged so that they satisfy the type IIB equation of motion. Since the
$B_{NS}$ components have remained unchanged we can easily satisfy the imaginary self-duality condition of
the type IIB three form fluxes. Rest of the analysis follows a straightforward route, using the mirror rules.
The type IIA metric and the one-forms remain same as before, but the three forms and their corresponding
$G$-fluxes change.

Before we go into the analysis of torsion classes, consider two other hypothetical scenarios. 
First one is a charge zero configuration. Due to the F-theory
construction behind our type II models, the global charges all cancel when we consider compact fourfold in
sec.\ 2.1. Imagine we continue the charge zero configuration even after we allow non-compact fourfolds.
Of course such a charge zero state will allow only a finite number of fractional D3 branes. Next, let us also
allow local charge cancellation. This way in \roro\ we can allow:
\eqn\chypo{{\rm {\bf Case~4:}}~~~~ c_3 ~=~ c_3(z), ~~
c_4 ~=~ c_4(z), ~~ c_1 ~\to~~0, ~~ c_5 ~=~ c_2 ~= ~ c_6 ~=~ 0}
For this case the $H_{RR}$ can be easily extracted from \bolo\ now. The three-forms are very interesting now.
They will be completely independent of $f_1$ or $f_2$. The non-zero components will be given by
\eqn\cnonzero{C_{xy2} ~=~ - {2\alpha c_3 A B^2 \o 1+A^2}, ~~~ C_{xz1}~=~ c_4, ~~~ C_{xz2}~=~ -{2c_3 AB\o 1+A^2},~~~
C_{yz2}~=~ -c_3}
with all other components vanishing. The $G$-fluxes for such a configuration are simpler than the earlier
examples. This is because of the coordinate structure of $c_3$ and $c_4$. One can easily show that the $G$-fluxes
will be independent of $c_4$, and will be given entirely by $c_3$ as
\eqn\ghypo{\eqalign{& G_{xyz\theta_2} ~ = ~ {2\alpha A B^2 \o 1+A^2} {\del c_3 \o \del z}, 
~~~~~~~~ G_{xy\theta_1 \theta_2} ~ =~
2c_3 B^2 {\del \o \del \theta_1}\left({\alpha A \o 1+A^2}\right)\cr 
& G_{xz\theta_1\theta_2} ~ = ~ 2B c_3
{\del \o \del \theta_1}\left({A \o 1+A^2}\right)}}
where the relevant component of $G$-flux
$G_{xz\theta_1\theta_2}$ vanishes in the local region implying that the corresponding
potential $C_{xz\theta_2}$ can be gauged away. The operation of flop in M-theory 
may however still remain non-singular 
because globally the $G$-flux component $G_{xz\theta_1 \theta_2}$ is not expected to vanish. 

The second one, although not very important for our main line of thought followed 
in this paper, is another generalisation that could be
done to \roro. What we mean is to add two more terms to \roro\ that are 
invariant under our global orientifold operation in type IIB theory. The 
only allowed ones are
\eqn\roroadd{{\rm {\bf Case~ 5:}}~~~~c_1 ~=~c_1(\theta_2), ~ c_i ~ = ~0, ~~~~
\delta B_{RR} ~ = ~  c_7~dr ~\wedge ~dx + c_8~dr ~\wedge ~ d\theta_1}
where the total type IIB background RR field will be $B_{RR} + \delta B_{RR}$ and $i ~=~2, ..., 6$.
We are also assuming that $c_7, c_8$ would in general be functions of 
($\theta_i, z$).

This change in type IIB $B_{RR}$ field will effect the mirror picture. It turns
out that the type IIA one form will not change at all from the value 
derived earlier in \oneform. However, 
the mirror type IIA three form will now change from the one derived earlier in 
\threenow\ to 
\eqn\bheg{\eqalign{\delta {\cal C}~ = ~& 
~~C_{ryz}~ dr \wedge(dy-b_{y\theta_1} d\theta_1)\wedge dz + 
 C_{rxz}~dr\wedge(dx-b_{x\theta_2} d\theta_2)\wedge dz \cr
& + C_{rxy}~dr\wedge(dx-b_{x\theta_2} d\theta_2)\wedge (dy-b_{y\theta_1}
  d\theta_1)}}
where as before, the total three form field will be ${\cal C} + 
\delta {\cal C}$, and the coefficients in \bheg\ are now given by
\eqn\bhegcoeff{C_{ryz} ~ = ~ c_7, ~~~~ C_{rxz}~=~ {2c_7 AB \o 1+A^2},
~~~~ C_{rxy}~=~ {2\alpha c_7 AB^2 \o 1+A^2}}  
Looking carefully at the additional components of the three form 
we observe that the three form component $C_{rxz}$ is exactly equal to the 
part of $C_{xz\theta_2}$ if we replace $c_7~ \leftrightarrow ~ -c_3$. In fact
all the above components are exactly the same under the substitution. What
does this mean? We already gave an answer to this question when we discussed 
the local equivalence between the $r$ and $\theta_2$ directions. Under this 
equivalence we clearly see why $c_7$ and $c_8$ terms of \roroadd\ do not
contribute to the one form \oneform. Furthermore it is also clear why $c_8$
term do not contribute to the three form above. 

The equivalences between ($r, \theta_2$) and ($c_7, - c_3$) are also 
consistent with the wrapped D5 brane scenario. Recall that because of the 
orientifold action the D5 branes can be thought to be along the ($r, y$)
directions in the internal space. This would imply that we should 
look for the combination 
\eqn\combnow{{\cal V}_2 ~ = ~ \sqrt{g_1 g_2 g_r} ~+~ {2i~c_7 AB \o 1+A^2}}
as our complexified volume, where $g_1, g_2, g_r$ are the warp factors in the 
type IIA metric \fiitaametnowb. This combination goes over to the 
original combination \deco\ once we go away from the orientifold limit
and consider $c_1$ in \roro\ as $c_1(\theta_1, \theta_2)$. In the same scenario we 
can define another effective volume along ($x,z, \theta_2$) direction as
\eqn\effxzy{{\cal V}_3 ~ = ~ \sqrt{g_1 g_2 g_3 - {g_1 g_4^2 {\rm sin}^2~\langle\psi\rangle \o 4}} 
~-~ {2i~c_3 AB \o 1+A^2}}
which has somewhat similar behavior as \combnow\ and would match precisely with \combnow\ if
$g_3 ~=~ g_r$ and $\langle\psi\rangle$ small. The connection to \deco\ will now be the following. 
Away from the orientifold point $b_{x\theta_2}$ and $b_{y\theta_1}$ will be functions of 
($\theta_1, \theta_2$) also. This would mean that we have the other two components 
$b_{x\theta_1}$ and $b_{y\theta_2}$. The D5 branes in type IIB can be assumed to warp directions
($y, \theta_2$) and then the volume element will be \deco. Our picture at the orientifold point with
volume element \combnow\ should be considered at this stage as a toy model that is helpful in clarifying all
the expected details that one would expect from a model away from the orientifold point. Thus 
\effxzy\ will serve as an {\it intermediate} scenario between \combnow\ and \deco.

Now from the earlier discussion we should expect that the $c_8$ term in \roroadd\
only contributes to the five form, exactly the way $c_6$ term 
contributes to the five form. This is easily verified by computing the 
mirror five form
\eqn\mifive{\delta {\cal C}_5 ~ = ~ -(c_8+c_7 f_1)~ dr\wedge(dx-
b_{x\theta_2} d\theta_2)\wedge (dy-b_{y\theta_1} d\theta_1)\wedge dz
  \wedge d\theta_1}  
where one can show that the other expected components of the five form 
$C_{rxyz\theta_2}$ and $C_{rxz\theta_1 \theta_2}$
are proportional to $\epsilon f_1$ and since $f_1$ goes like
$f_1 ~\sim~{1 \o \sqrt{\epsilon}}$ these 
components go to zero when $\epsilon ~\to ~0$. Finally, if we also take
\eqn\alsotake{ c_8~=~0, ~~~~~~~~~ h_{xzr} ~\equiv ~ {\del c_7 \o \del z}}
where $h_{xzr}$ is the type IIB three form, then one can show that the 
additional contribution to the five form \mifive\ is also a pure gauge. With 
this the analysis of type IIA will be complete.


\newsec{Torsion classes}

After all the detailed discussion about the global type IIA manifolds, we will 
now attempt to classify the IIA non--K\"ahler
manifolds that we constructed. As already pointed out in \gtone\
we do not find a half--flat manifold after performing three
T--dualities with fluxes. This might appear to contradict current
literature on this subject, but we will show the contrary. First, it has been
shown that lifting a 10dim manifold on a twisted circle (i.e. with
gauge field as in our case) can still give a supersymmetric M--theory
background, even if the 10dim manifold was not half--flat. But we
can only obtain torsion classes for our local background, which
does not show supersymmetry. We therefore neither expect a half--flat
manifold nor the one discussed in \minakas\ on the grounds of supersymmetry.
Furthermore, we actually do not expect a 11dim manifold with $G_2$ holonomy,
since our M--theory background has flux turned on.

\subsec{$SU(3)$ and $G_2$ structure manifolds}

It is by now widely known that string theory backgrounds with fluxes do not require
a Calabi--Yau manifold (with $SU(3)$ holonomy). The 4dim ${\cal N}=1$ supersymmetry
condition of a covariantly constant spinor is relaxed to the existence of a globally
defined spinor that is constant with respect to a {\sl torsional} connection. This reduces
the structure group of the 6dim manifold from $SO(6)$ to $SU(3)$ and the intrinsic torsion
decomposes under representations of $SU(3)$. In particular, the torsion lies in 5 classes \lust,
\salamon: $\tau_1\in{\cal W}_1\oplus{\cal W}_2\oplus{\cal W}_3\oplus{\cal W}_4\oplus{\cal W}_5$.

The failure of the torsional connection to be the Levi--Civita connection is measured in the
failure of fundamental 2--form and holomorphic 3--form to be closed. Defining a set of real
vielbeins $\{e_i\}$ one can define an almost complex structure as
\eqn\almcompl{\eqalign{ & E_1=e_1+i~e_2\cr
  & E_2=e_3+i~e_4\cr
  & E_3=e_5+i~e_6} }
which gives rise to a (1,1)--form w.r.t. this almost complex structure
\eqn\defJ{J=e_1\wedge e_2+e_3\wedge e_4+e_5\wedge e_6\,.}
Similarly, one defines a holomorphic 3--form w.r.t. this almost complex structure
\eqn\defOmega{\Omega=\Omega_+ + i~\Omega_-
  =(e_1+i~e_2)\wedge(e_3+i~e_4)\wedge(e_5+i~e_5)\,,}
where $\Omega_\pm$ are the real adn imaginary part of $\Omega$, repectively.
$J$ and $\Omega$ fulfill the compatibility relations
\eqn\compatible{\eqalign{& J\wedge \Omega_+=J\wedge \Omega_-=0\cr
  & \Omega_+ \wedge \Omega_-={2\o 3}J\wedge J\wedge J\,.}}
The torsion classes are then determined by the following forms:
\eqn\deftorsion{\eqalign{& {\cal W}_1 ~~\leftrightarrow~~ dJ^{(3,0)} \cr
  & {\cal W}_2 ~~\leftrightarrow~~ (d\Omega)^{(2,2)}_0 \cr
& {\cal W}_3 ~~\leftrightarrow~~ (dJ)^{(2,1)}_0 \cr
& {\cal W}_4 ~~\leftrightarrow~~ J\wedge dJ \cr
& {\cal W}_5 ~~\leftrightarrow~~ d\Omega^{(3,1)}\,,}}
where the subscript $0$ refers to the primitive part, i.e. in the cases in question $\beta
\in \bigwedge^{p,q}_0$ if $\beta\wedge J=0$.
It is immediately obvious that complex manifolds have to have vanishing
${\cal W}_1$ and ${\cal W}_2$ and K\"ahler manifolds are determined by $\tau_1\in {\cal W}_5$.
Decomposing $\Omega=\Omega_+ +i~\Omega_-$ we can write
more precisely
\eqn\deftorsform{\eqalign{& d\Omega_\pm \wedge J=\Omega_\pm \wedge dJ = {\cal W}_1^\pm
    ~ J\wedge J\wedge J\cr
  & d\Omega_\pm^{(2,2)} = {\cal W}_1^\pm~J\wedge J+{\cal W}_2^\pm\wedge J \cr
  & dJ^{(2,1)}~=~(J\wedge{\cal W}_4)^{(2,1)}+{\cal W}_3\,,}}
so ${\cal W}_1$ is given by two real numbers, ${\cal W}_1={\cal W}_1^+ +{\cal W}_1^-$,
${\cal W}_2$ is a (1,1) form and ${\cal W}_3$ is a (2,1) form.
With the definition of the contraction
\eqn\contraction{\rightharpoonup~:~\bigwedge{}^k ~T^*\otimes \bigwedge{}^n ~T^*
  \longrightarrow \bigwedge{}^{n-k} ~T^*}
and the convention $e_1\wedge e_2~\rightharpoonup~e_1\wedge e_2\wedge e_3\wedge e_4
  ~=~e_3\wedge e_4$ we define
\eqn\defwfour{\eqalign{& {\cal W}_4={1 \o 2}~ J \rightharpoonup dJ \cr
   &  {\cal W}_5={1\o 2}~ \Omega_+ \rightharpoonup d\Omega_+\,.}}
A half--flat manifold is specified by $\tau_1\in{\cal W}_1^+\oplus{\cal W}_2^+
\oplus{\cal W}_3$, which follows from $J\wedge dJ=0$ and $d\Omega_-=0$, but $d\Omega_+\ne 0$
(this lead to the terminology ``half--flat''). Note that the assignment of $\Omega_-$ and
$\Omega_+$ may be switched.

Similar statements hold true for M--theory on 7--manifolds, which would require $G_2$
holonomy to preserve 4dim ${\cal N}=1$ supersymmetry in the absence of flux. The fundamental
object now is a nowhere vanishing 3--form $\Phi$ and its failure to be closed and/or
co--closed determines the torsion. The relevant structure
group is $G_2$ and the intrinsic torsion  decomposes under this group. This results in 4
torsion classes for the 7--manifold:
$\tau_2\in{\cal X}_1\oplus{\cal X}_2\oplus{\cal X}_3\oplus{\cal X}_4$. There are given by
\eqn\gtwotorsion{\eqalign{& d\Phi~=~{\cal X}_1~(*\Phi)+{\cal X}_4\wedge\Phi+{\cal X}_3 \cr
  & d(*\Phi)~=~{4\o 3}~{\cal X}_4\wedge(*\Phi)+{\cal X}_2\wedge\Phi \,.}}
In \hitchin,\lust\ it was demonstrated how a manifold with $SU(3)$ structure can be lifted
on a trivial circle or interval to a
$G_2$ holonomy. One defines the $G_2$ invariant 3--form as
\eqn\defPhi{\Phi~=~\Omega_+ + J\wedge e_7}
where $e_7$ parametrizes the 7th direction, such that the resulting 7--manifold is
$M\times I$ with $I\subset \IR$. This produces Hitchin's flow equations if the
6--manifold is half--flat.

In contrast, we are more interested in the case discussed in \minakas. Starting with an
$SU(3)$--structure manifold $X$ they constructed a $G_2$--structure manifold $Y$ as a lift
over a twisted circle with dilaton and gauge field:
\eqn\Ymetric{ds_Y^2~=~e^{-2\alpha\phi}~ds_X^2~+~e^{2\beta\phi}~(dz+A)^2\,.}
We will adopt the string frame in which $\alpha=1/3$ and $\beta=2/3$. In this case one should
define the 3--form on the 7--manifold rather as
\eqn\defPhialt{\Phi~=~e^{-\phi}~\Omega_+~+~e^{-{2\o 3}\phi}~J\wedge e_7\,.}
This gives straightforward relations between the torsion classes ${\cal W}_i$ and
${\cal X}_j$ that generally involve the field strength $F=dA$ and the dilaton $\phi$.
It was shown in \minakas\ that requiring $G_2$ holonomy (i.e. $d\Phi=d(*\Phi)=0$ or
equivalently ${\cal X}_i=0$) leads to the following constraints on the $SU(3)$ torsion classes:
\eqn\minators{\eqalign{& {\cal W}_1^\pm ~=~ {\cal W}_2^-~=~ {\cal W}_3~=~ {\cal W}_4~=0~\cr
  &  {\cal W}_2^+~=~-e^\phi~F_0^{(1,1)}\,,~~~~~~~ {\cal W}_5~=~{1\o 3}~d\phi\,.}}
Note, that only in the string frame ${\cal W}_4=0$, otherwise it is also proportional to $d\phi$.
This shows that the 6--manifold does not need to be K\"ahler (if $F_0^{(1,1)}\ne 0$),
but it does not need to be half-flat either (it still could be if $d\phi=0$).

This short discussion was intended to clarify that half--flat manifolds are not the
only manifolds that can be lifted to a $G_2$ holonomy. One has to be specific about which type
of lift is chosen. It is immediately clear that our scenario requires the 7-th direction to
be a twisted circle, since the IIA background has a gauge field $A$. But since we have also
other background fluxes turned on, we obtain a torsional M--theory background after the lift.
Therefore, the manifold we propose in IIA is neither half--flat nor has it torsion restricted to
${\cal W}_2^+ \oplus {\cal W}_5$.

We now turn to the question what type of manifold we have constructed in IIA. Unfortunately, we
had to take a local limit of the metric to be able to perform T--dualities from IIB to IIA.
Therefore, the global information about our manifold is lost. It does strictly speaking not make
sense to discuss the metric \fiitaamet\ when one is looking for topological properties, such as torsion
classes. In the local background supersymmetry is not visible, but for the global supersymmetric
background we do not know the metric. At this point there is no way out of this dilemma and we
have to postpone the topological analysis of the supersymmetric background to future work.

However, we find it instructive to discuss the torsion of the local metric found in \gtone\
as an example. This is not the background obtained from the F--theory construction in IIB, but shows a different fibration structure. This does not have major consequences for the torsion classes, we therefore restrict ourselves to this case.

\subsec{Torsion classes before geometric transition}

In this section we will only discuss the IIA case. It is shown that with a quite generic
ansatz for the almost complex structure we can find a symplectic structure on the local metric,
but no half--flat structure\foot{This statement holds true if one considers the IIA background abtained from IIB with orientifold action, but the precise torsion classes differ.}.
One possible choice for vielbeins in IIA is
\eqn\IIAvielb{\eqalign{& e_1~=~dr\,,~~~~~~~~~~~~~~~~~~e_2~=~\sqrt{g_1}~\big(dz
    +A(dx-b_{x\theta_1} d\theta_1)+B(dy-b_{y\theta_2}d\theta_2)\big) \cr
  & e_3~=~{1\o 2}\sqrt{{4g_2g_3-g_4^2 \o g_2}}~d\theta_2\,, ~~~~~~~~
    e_4~=~{1\o 2}\sqrt{{4g_2g_3-g_4^2 \o g_2}}~(dy-b_{y\theta_2}d\theta_2) \cr
  & e_5~=~\sqrt{g_2}\left(\sin{\psi_0} (dx-b_{x\theta_1} d\theta_1) + \cos{\psi_0}~
      d\theta_1 + {g_4 \o 2g_2}~d\theta_2\right) \cr
  & e_6~=~\sqrt{g_2}\left(\cos{\psi_0} (dx-b_{x\theta_1} d\theta_1) - \sin{\psi_0}~
      d\theta_1 - {g_4 \o 2g_2}~(dy-b_{y\theta_2}d\theta_2)\right)}}
Following the discussion in \tp\ we write down the most generic candidate for a fundamental
2--form
\eqn\ansatzJ{J~=~\sum_{i<j}~a_{ij}~ e^i\wedge e^j\,,}
where the coefficients $a_{ij}$ could in principle depend on all coordinates.
This has to be compatible with an almost complex structure
${\cal J}^A_B=\delta^{AC} J_{CB}$, i.e. we require ${\cal J}^2=-1$. If we furthermore make
the assumption that $a_{13}=a_{14}=0$,\foot{This seems sensible because it holds for all known
conifold geometries \tp, and basically only implies that the complex vielbein containing $dr$
does not contain $e_3$ and $e_4$. It can still contain $dy$ and $d\theta_2$, since those also
appear in $e_5$ and $e_6$.} the almost complex structure takes a particularly simple form \tp.
The complex vielbeins can be written as
\eqn\defcomplvielb{\eqalign{& E_1~=~e_1+i~e_2 \cr
    & E_2~=~e_3+i~(X~e_4-P~e_6)\cr
    & E_3~=~e_5+i~(X~e_6+P~e_4)\,.}}
with the restriction $P^2+X^2=1$. In the following we will make the simplifying assumption
that $P$ (and therefore $X$) is a function of $r$ only.

With this setup, $J$ and $\Omega$ are defined and one can calculate $dJ$ and $d\Omega$.
One immediately notices that ${\cal W}_4=0$, because
\eqn\jwedgedj{J\wedge dJ~=~-{4g_2g_3-g_4^2 \o 2}~\Big(P(r)P'(r)+X(r)X'(r)\Big)
  ~dr\wedge dx \wedge dy\wedge d\theta_1 \wedge d\theta_2\,,}
which is identically zero because of $P(r)^2+X(r)^2=1$. It is nevertheless not a half-flat
manifold, because there is no choice of $P(r)$ that makes either $d\Omega_+=0$ or
$d\Omega_-=0$.

We can, however, choose $P(r)$ to give a symplectic structure. Consider ${\cal W}_1^\pm$
given by $d\Omega_\pm\wedge J={\cal W}_1^\pm~J\wedge J\wedge J $:
\eqn\wone{\eqalign{& {\cal W}_1^+~=~{\sqrt{g_1}~(4g_2g_3-g_4^2) \o 2\sqrt{1-P(r)^2}}~P'(r)~
  dr\wedge dx\wedge dy\wedge dz\wedge d\theta_1\wedge d\theta_2 \cr
  & {\cal W}_1^-~=~-{1\o 4}~\sqrt{g_1}~\left(P(r)(g_4^2-4g_2g_3)+g_4\sqrt{4g_2g_3-g_4^2}
    ~\sqrt{1-P(r)^2}\right)\times\cr
  & ~~~~~~~~~~~~~\big(b_{x\theta_1}'(r)+b_{y\theta_2}'(r)\big)~
    dr\wedge dx\wedge dy\wedge dz\wedge d\theta_1\wedge d\theta_2\,.}}
Note that in our local background the function $g_i$ are simply constants and we have assumed
the IIB $B_{\rm NS}$--field components to have $r$--dependence only. Obviously, ${\cal W}_1^+$
vanises if $P(r)$ is constant (the prime denotes derivative w.r.t. $r$). ${\cal W}_1^-$ vanishes
if
\eqn\solP{P~=~{g_4 \o 2 \sqrt{g_2g_3}}~=~{\rm constant}.}
It turns out, that this is also the only value for which ${\cal W}_3$ vanishes, and in this
case $dJ=0$. Let us stress again, that there is no choice for $P(r)$ that would give
${\cal W}_5=0$ or ${\cal W}_2^\pm=0$.
The remaining torsion classes could only vanish if the IIB $B_{\rm NS}$ field was constant.
In that case we would trivially recover a CY, since then all metric components would be constant.
For completeness, let us also give ${\cal W}_5$ and ${\cal W}_2^\pm$ with the choice \solP\ for
$P$:
\eqn\beforetorsion{\eqalign{& {\cal W}_2^+~=~-{i\o 8\sqrt{g_2g_3g_5}}\bigg[
    2A\sqrt{g_1g_3g_5}~ b_{x\theta_1}'~e_1\wedge (\cos\psi_0~e_3+\sin\psi_0~e_4)\cr
  & ~~~~~~~+2A\sqrt{g_1g_3}~g_4~ b_{x\theta_1}'e_1\wedge(\cos\psi_0~e_5-\sin\psi_0~e_6) \cr
  & ~~~~~~~+4A\sqrt{g_1g_2}~g_3~ b_{x\theta_1}'e_2\wedge(\cos\psi_0~e_4-\sin\psi_0~e_3) \cr
  & ~~~~~~~+g_4\sqrt{g_5}\sin2\psi_0~b_{x\theta_1}'~(e_3\wedge e_4-e_5\wedge e_6) +
    g_5\sin2\psi_0~b_{x\theta_1}'~(e_3\wedge e_6-e_4\wedge e_5) \cr
  & ~~~~~~~-4g_2g_3~(\cos2\psi_0~b_{x\theta_1}'+B~b_{y\theta_2}')~(e_3\wedge e_5+e_4
    \wedge e_6)\cr
  & ~~~~~~~-2B\sqrt{g_1}~b_{y\theta_2}'~\big(\sqrt{g_2g_5}~e_2\wedge e_6+g_2\sqrt{g_3}~e_1
    \wedge e_5
    +g_4\sqrt{g_2}~e_2\wedge e_4\big)\bigg]}}
 \eqn\beftor{\eqalign{&{\cal W}_2^-~=~{i\o 16}\bigg[4A\sqrt{{g_1\o g_2}}~b_{x\theta_1}'~
  (\cos\psi_0~e_2\wedge e_3+\sin\psi_0~e_2\wedge e_4) + 8A\sqrt{{g_1g_3\o g_5}}~\sin\psi_0~
    b_{x\theta_1}'~e_1\wedge e_3\cr
  & ~~~~+4A{\sqrt{g_1}~g_4\o \sqrt{g_2g_5}}~\sin\psi_0~ b_{x\theta_1}'~e_2\wedge e_6
    -4\sin2\psi_0~ b_{x\theta_1}'~(e_3\wedge e_5+e_4\wedge e_6) \cr
  & ~~~~+ 4B\sqrt{{g_1\o g_3}}~ b_{y\theta_2}'~e_1\wedge e_6 + 4 \sqrt{{g_1\o g_2g_5}}
    \big(Ag_4\cos\psi_0~ b_{x\theta_1}'-2Bg_2~ b_{y\theta_2}'\big)~e_2\wedge e_5 \cr
  & ~~~~+{(-g_5\cos\psi_0+g_4^2)~ b_{x\theta_1}'+4g_2g_3~ b_{y\theta_2}' \o g_2g_3}~
    \left(e_3\wedge e_6-e_4\wedge e_5+{g_4\o \sqrt{g_5}}~\big(e_3\wedge e_4-e_5\wedge e_6\big)
    \right) \cr
  & ~~~~+4\sqrt{{g_1\o g_3}}~(-2Ag_3\cos\psi_0~b_{x\theta_1}'+Bg_4~b_{y\theta_2}')~
    e_1\wedge e_4\bigg]}}
  \eqn\betoreb{\eqalign{&{\cal W}_5~=~\sqrt{{g_2g_3 \o g_5}}~(b_{x\theta_1}'+b_{y\theta_2}')~e_2+
    2A\sqrt{{g_1g_3\o g_5}}\sin\psi_0~b_{x\theta_1}'~e_5\cr
  & ~~~~~~~+{\sqrt{g_1}\o 4\sqrt{g_2g_5}}~(2Ag_4\cos\psi_0~b_{x\theta_1}'-4Bg_2~b_{y\theta_2}')~
    e_3+A\sqrt{{g_1\o g_2}}\cos\psi_0~b_{x\theta_1}'~e_6}}
where we have defined $g_5=4g_2g_3-g_4^2$. The vielbeins $\{e_i\}$ are defined in \IIAvielb.

\subsec{Torsion classes after geometric transition}

Very similar remarks hold true for the local metric after transition.
We can find a symplectic structure, but ${\cal W}_2^\pm$ and ${\cal W}_5$ are nonzero.
The metric after transition was obtained in \gtone\ to be:
\eqn\IIBmetric{\eqalign{& ds^2~=~dr^2+e^{2\phi}~\big(dz+A(dx-b_{x\theta_1}d\theta_1)
    +B(dy-b_{y\theta_2} d\theta_2)\big)^2 \cr
  & ~~~~~~~~~~+\left({g_2\o 2}-\sqrt{{g_2\o g_3}}~{g_4\o 4}\right)\Big(d\theta_1^2 +
    \big(dx-b_{x\theta_1}d\theta_1\big)^2\Big) \cr
  & ~~~~~~~~~~+ \left({g_2\o 2}+\sqrt{{g_2\o g_3}}~{g_4\o 4}\right)\Big(d\theta_2^2 +
    \big(dy-b_{y\theta_2}d\theta_2\big)^2\Big)\,.}}
Taking the ansatz \defcomplvielb\ but now with real vielbeins
\eqn\IIAvielb{\eqalign{& e_1~=~dr\,,~~~~~~~~~~~~~~~~~~e_2~=~e^\phi~\big(dz
    +A(dx-b_{x\theta_1} d\theta_1)+B(dy-b_{y\theta_2}d\theta_2)\big) \cr
  & e_3~=~{1\o 2}\sqrt{g_+}~d\theta_2\,, ~~~~~~~~
    e_4~=~{1\o 2}\sqrt{g_+}~(dy-b_{y\theta_2}d\theta_2) \cr
  & e_5~=~{1\o 2}\sqrt{g_-}\left(\sin{\psi_0} (dx-b_{x\theta_1} d\theta_1) + \cos{\psi_0}~
      d\theta_1\right) \cr
  & e_6~=~{1\o 2}\sqrt{g_-}\left(-\cos{\psi_0} (dx-b_{x\theta_1} d\theta_1) + \sin{\psi_0}~
      d\theta_1 \right)}}
also gives ${\cal W}_4=0$ automatically. Again, ${\cal W}_1^+$ can only vanish if $P(r)$
is constant and solving ${\cal W}_1^-=0$ gives $P(r)=0$. There is no choice of $P(r)$ that
would allow for ${\cal W}_5=0$ or ${\cal W}_2^\pm=0$. With the choice $P=0$ the remaining
torsion classes are
\eqn\aftertorsion{\eqalign{& {\cal W}_2^+~=~-{i\o 4}~\left[{2Ae^\phi\o \sqrt{g_-}}~
    b_{x\theta_1}'~\big(\sin\psi_0~(e_1\wedge e_3+e_2\wedge e_4)+\cos\psi_0~(e_1\wedge e_4
    -e_2\wedge e_3)\big)\right. \cr
  &~~~~~~~~~~ +\sin2\psi_0~b_{x\theta_1}'~(e_3\wedge e_5+e_4\wedge e_6)
     +\cos2\psi_0~b_{x\theta_1}'~(e_4\wedge e_5-e_3\wedge e_6)\cr
  &~~~~~~~~~~\left. -{2Be^\phi\o\sqrt{g_+}}~b_{y\theta_2}'~(e_1\wedge e_6-e_2\wedge e_5)
     + b_{y\theta_2}~(e_4\wedge e_5-e_3\wedge e_6)\right] \cr
  & {\cal W}_2^-~=~-{i\o 4}~\left[{2Ae^\phi\o \sqrt{g_-}}~
    b_{x\theta_1}'~\big(\cos\psi_0~(e_1\wedge e_3+e_2\wedge e_4)-\sin\psi_0~(e_1\wedge e_4
    -e_2\wedge e_3)\big)\right. \cr
  &~~~~~~~~~~ +\cos2\psi_0~b_{x\theta_1}'~(e_3\wedge e_5+e_4\wedge e_6)
     -\sin2\psi_0~b_{x\theta_1}'~(e_4\wedge e_5-e_3\wedge e_6)\cr
  &~~~~~~~~~~\left. -{2Be^\phi\o\sqrt{g_+}}~b_{y\theta_2}'~(e_1\wedge e_5+e_2\wedge e_6)
     - b_{y\theta_2}'~(e_3\wedge e_5+e_4\wedge e_6)\right] \cr
  & {\cal W}_5~=~{2Ae^\phi\o \sqrt{g_-}}~b_{x\theta_1}'~\big(\cos\psi_0~e_6-
    \sin\psi_0~e_5\big)-{Be^\phi\o\sqrt{g_+}}~ b_{y\theta_2}'~e_3
    -{1\o 2}~(b_{x\theta_1}'- b_{y\theta_2}')~e_2}}
where we have defined
\eqn\gpm{g_\pm~=~2g_2\pm\sqrt{{g_2\o g_3}}~g_4}
and $\phi$ is the IIA dilaton from before transition. In the limit that we took when
performing the T--dualities to go from IIB to IIA, this is exactly the same as the
IIB dilaton and constant (at least in the local limit).
We see that the geometric transition maps the torsion classes $W_2^{\pm}$ and
$W_5$ into themselves. By using the Hitchin flow equations, this means that the
corresponding $G_2$ torsion classes are mapped into themselves. But we know that
the flop just replaces the usual $x^{11}$ lifting direction with the fiber
direction inside the Hopf fibration $S^3/\IZ_N$ over $S^2$. These two circles are
used to lift $SU(3)$ torsion classes to $G_2$ torsion classes and this tells us
that the $G_2$ torsion classes are not changed during the flop.

\newsec{Conclusions and Future Directions}

Geometric transitions are a powerful method to connect string
theory and field theory. They have been used extensively to check
and predict effective results in field theory from string theory.
In this project, which is a continuation of our previous work
\gtone, \realm, we addressed several questions related to the
geometric transition framework and reached conclusions which point
out some subtle modification of the original framework of \vafai.

First, we showed that, in terms of SUGRA solutions, an acceptable
way to to construct a SUSY solution is to consider a global
picture with additional branes and planes. Away from the wrapped
D5 branes on resolution two cycles of our manifold, there are
additional seven branes and orientifold seven planes during the
geometric transition. Such a configuration has two immediate
advantages: on the one hand their presence allows a consistent
lift to a SUSY F--theory picture and on the other hand they do not
directly influence the effective theory as they can be viewed as
massive flavors that are integrated out. If we consider them as
part of the effective theory (by raising the energy scale of the
theory), the field theory would be an $SU(N)$ gauge theory with
fundamental flavors. This is a somewhat different configuration
from the original framework of \vafai\ (see also \radu) and would
change the geometric transition identifications. These further
complications will translate into a currently unknown metric for
D5--D7 system.

Second, the local picture that we proposed in this paper captures
the dynamics of ${\cal N} =1~ SU(N)$ gauge theory without any
flavors. The local picture also explains the fact that in the
mirror IIA picture our non--K\"ahler manifold is not expected to
resemble any of the examples proposed in the literature \minu.
Globally, in type IIA there are extra D6 branes which correspond
to massive flavors which are integrated out to obtain the local
model. In both type IIA and type IIB, the global framework
involves subtle orientifold actions that have roots in our
proposed F--theory setup\foot{It would be interesting to 
see whether the global M-theory
lift of our type IIA background is connected to any of the solutions presented in 
\dall. A way to search for such a solution is to consider the solutions of \dall\ and
look for four dimensional slices of the seven dimensional manifolds that would resemble 
Atiyah-Hitchin spaces. In addition to that, if the local form of the metric  
resembles the local M-theory metric presented in \gtone, then this solution can be a candidate for 
our global M-theory picture. More details on this will be presented in the sequel.}.

Another important topic in our work was to deepen the evidence of
a heterotic transition, even though the usual topological string
methods are still missing for the (0,2) theories. We managed to
show that there is a surprising analogy between two hitherto
different solutions, namely, the Maldacena--Nunez solution \mn\ in
type IIB and the heterotic dual to the type IIB solution proposed
in \realm. In fact, an interesting generalization of the
Maldacena--Nunez solution with several warp factors was mapped to
a global heterotic solution which then takes us away from the
orientifold point. The existence of such a mapping indicates the
possibility that our global heterotic model may indeed be dual to
the theory on wrapped heterotic NS5 branes. However, to make this
more concrete we need to provide more evidence, and the simple
identification between two metrics is probably not too rigorous.

Our discussion is just the tip of an iceberg. Clearly, much work
remains in order to clarify above points. It would be interesting
to find information about the global picture by integrating in the
massive flavors. This would imply changes in $H_{NS}$ and $H_{RR}$
due to the presence of D7 branes which consequently change the
dual gauge theory\foot{Similar results were obtained for the case
of the solution of \ks\ in \ouyang.}. This method should remove
the non--SUSY problems of \pandoz\ and offer us consistent global
geometry.

Our work should also be viewed as connecting gauge theories with
the dynamics of non--K\"ahler manifolds. This would open up an
extensive direction of research. An immediate next step would be
to further connect our results with the generalized complex
geometries of Hitchin \hitchin\ and Gualtieri \gualtieri, which
unify metric and NS field. It should also be possible to consider
the heterotic solution in the context of Chiral de Rham complex
and chiral differential operators. This would provide a method for
performing genuine heterotic topological computations with
important application in describing moduli spaces of ${\cal N} =
1$ theories. 

\centerline{\bf Acknowledgements}

\noindent We would like to thank Allan Adams, Katrin Becker,
Robert Brandenberger, Atish Dabholkar, Ji-Xiang Fu, Rhiannon Gwyn,
Shamit Kachru, Jan Louis, Dave Morrison, Eric Sharpe, Nemani
Suryanarayana, Andy Strominger, Wati Taylor, Li-Sheng Tseng,
Cumrun Vafa and Shing-Tung Yau for useful discussions and
correspondences. M. B. and A. K. would like to thank the
Department of Physics at Harvard University and the Radcliffe
Institute for Advanced Studies at Harvard University for
hospitality during the final stages of this work. The work of M.
B. was supported by NSF grant NSF-0354401, the University of Texas
A\&M and the Radcliffe Institute for Advanced Studies at Harvard.
The work of K. D. is supported by in part by the McGill university
start-up grant and NSERC. The work of S. K. is supported by NSF grants
NSF-DMS-02-96154 and NSF-DMS-02-44412. The work of A. K. is
supported by DFG--the German Science Foundation, DAAD--the German
Academic Exchange Service, the European RTN Program
MRTN-CT-2004-503369 and the University of Maryland. And the work
of R. T is supported in part by PPARC.

\noindent {\bf Preprint Numbers:}~ILL-(TH)-05-5, LTH-682, MIFP-05-29,
ZMP-HH/05-31

\listrefs


\bye

\centerline{\bf Contents}\nobreak\medskip{\baselineskip=12pt
\parskip=0pt\catcode`\@=11

\noindent {1.} {Introduction} \leaderfill{1} \par
\noindent {2.} {Geometric Transitions, Fluxes and Gauge Theories} \leaderfill{6} \par
\noindent {3.} {The Type IIB Background From M-Theory Dual} \leaderfill{9} \par
\noindent {4.} {Mirror Formulas using Three T-Dualities} \leaderfill{12} \par
\noindent \quad{4.1.} {Metric Components} \leaderfill{12} \par
\noindent \quad{4.2.} {$B_{NS}$ Components} \leaderfill{14} \par
\noindent \quad{4.3.} {Background Simplications} \leaderfill{16} \par
\noindent \quad{4.4.} {Rewriting the Deformed Conifold Background} \leaderfill{17} \par
\noindent {5.} {Chain 1: The Type IIA Mirror Background} \leaderfill{21} \par
\noindent \quad{5.1.} {Searching for the $d\theta_1 d\theta_2$ term} \leaderfill{26} \par
\noindent \quad{5.2.} {Physical meaning of $f_1$ and $f_2$} \leaderfill{36} \par
\noindent \quad{5.3.} {$B$-fields in the Mirror set-up} \leaderfill{37} \par
\noindent \quad{5.4.} {The Mirror Manifold} \leaderfill{39} \par
\noindent {6.} {Chain 2: The M-Theory Description of the Mirror} \leaderfill{42} \par
\noindent \quad{6.1.} {One-Forms in M-Theory} \leaderfill{42} \par
\noindent \quad{6.2.} {M-Theory lift of the Mirror IIA Background} \leaderfill{44} \par
\noindent {7.} {Chain 3: M-Theory Flop and Type IIA Reduction} \leaderfill{52} \par
\noindent \quad{7.1.} {The Type IIA Background} \leaderfill{57} \par
\noindent \quad{7.2.} {Analysis of Type IIA Background and Superpotential} \leaderfill{59} \par
\noindent {8.} {Discussion and Future Directions} \leaderfill{63} \par
\noindent \quad{8.1.} {Future Directions} \leaderfill{63} \par
\noindent Appendix {1.} {Algebra of $\alpha$} \leaderfill{66} \par
\noindent Appendix {2.} {Details on $G_2$ Structures} \leaderfill{68} \par
\catcode`\@=12 \bigbreak\bigskip}

\newsec{Introduction}

During the last years there has been tremendous progress toward
constructing the string theory dual descriptions of large N gauge
theories. The first steps in this direction were made by
considering the conformal ${\cal N} = 4$, $D=4$ super Yang-Mills
theory \mal\ and later on more realistic field theories with
${\cal N} = 1$ supersymmetry and confinement were described in
\ks,\ps,\mn\ and \vafai. In a slightly different context, the
connection between gauge theories and topological string theory
was discussed in \gv\ for type IIA strings and in \dvu\ for type
IIB strings, the latter leading to the powerful Dijkgraaf-Vafa
conjecture by which non-perturbative computations in field
theories can be performed using perturbative expansions in matrix
models.

The seminal work of Vafa \vafai\ very nicely showed how to embed
the topological duality of \gv\ into the framework of the AdS/CFT
correspondence. The basic idea of \vafai\ was to consider the
${\cal N} = 1$ theory resulting from type IIA superstring theory
in the deformed conifold background $T^{*}S^3$ in the presence of
$N$ D6 branes wrapped around the Lagrangian $S^3$ cycle and
filling the external space and compute the corresponding
superpotential of this theory. On the other hand it was known that
the superpotential for the field theory living in the four
noncompact directions of the D6 branes is described by a
Chern-Simons gauge theory on $S^3$ \wittenchern, whose
superpotential can be computed in terms of topological field
theory amplitudes. Therefore, a connection between large $N$
Chern-Simons theory/topological string duality to ordinary
superstring theory and the AdS/CFT correspondence was
established. The idea was soon extended to many more rather
interesting models in \civ,\eot,\civu,\civd,\radu,\radd,\radt\
and \radp.

In all the above mentioned models the superpotentials computed by
topological strings are mapped by the AdS/CFT like correspondences to
geometries generated by fluxes on the dual closed string side.
Since we start with D6 branes, we expect to have  RR two-form
fluxes in the dual closed string solution. These fluxes thread
through a holomorphic $P^1$ cycle inside a blow up of a conifold.
The superpotential is a product between the RR two-form fluxes and
the K\"ahler form associated with the
four-cycle that is Hodge dual to $P^1$. It turns out \vafai\
that open/closed string duality requires another term in the
superpotential. This term originates from the field theory gluino
condensate, as topological open string computations imply the
existence of a term linear in the gluino condensate \ber, which
gets mapped into the size of the holomorphic two-cycle. Therefore
one has to have a term linear in the size of the holomorphic
two-cycle which can be obtained if this is multiplied by a
four-form. Furthermore, this four-form should be of NS type.

The quest for this four-form has been the subject of intense
scrutiny in the last years. As there is no D-brane which can
create it, the flux originates from changes in the geometry. The
best way to understand its appearance is to go to the mirror type
IIB picture and consider the superpotential \tv  \eqn\tve{W =
\int \Omega \wedge (H_{RR} + \varphi H_{NS}),~~\varphi = \chi +
i~e^{-\Phi},} where $\chi$ is the axion, $\Phi$ is the dilaton,
$H_{RR}$ is the RR three-form flux and $H_{NS}$ is the NS
three-form flux. From here we can go to the mirror type IIA
picture
 and the fluxes map into the RR two-form flux and an NS four-form
 flux respectively. Of course, the main question is why would the
 NS fluxes appear when one
starts with brane configurations involving only D branes? A
partial answer to this question was given in \louis, where the
origin of the NS four-form $F_4^{NS}$ was related to the fact that
the (3,0) form $\Omega=\Omega^++i\Omega^-$ is not closed for the
type IIA compactification, and therefore $d\Omega^+\sim
F_4^{NS}$. The fact that $d \Omega \ne 0$ allows an extra term in
the superpotential $\int d \Omega \wedge J$ , $J$ being the
fundamental two-form\foot{Notice that this term also plays a crucial role
in the S-duality conjecture of \vafai.}.
Manifolds with the property that the real
part of the (3,0) form is not closed, while the imaginary part
satisfies $d\Omega^-=0=d(J\wedge J)$ are called half flat
manifolds. Half flat manifolds are examples of non-K\"ahler
manifolds\foot{The manifold that we will eventually get in type
IIA side later in this paper, will however be more general than
the half-flat manifold in the sense that both $d\Omega^\pm \ne 0$.}.

At this point one might wonder, {{why would a non-K\"ahler
manifold appear as a result of a geometric transition from a
brane configuration with D6 branes wrapped on a cycle of a Calabi
Yau manifold?}} One of the goals of this paper is to give an
answer to this question. In short the answer is this: {\it{the
fluxes live on a non-K\"ahler manifold because the D6 branes
actually are themselves wrapped on a cycle inside a non-K\"ahler
manifold and the geometric transition is a flop inside a $G_2$
manifold with torsion!}}

To arrive to the non-K\"ahler geometry where the D branes are
wrapped we start with the type IIB solution corresponding to D5
branes wrapped on the resolution cycle $P^1$ of the resolved
conifold \pandoz. The supergravity solution involves, besides the
RR three-form, a NS three-form and a RR five-form. These fields
are required by the string equations of motion. Similarly as in
\civ\ the presence of the NS three-form before the transition is a
signal that a NS three-form should also exist after the geometric
transition has taken place.

As the resolved conifold is a toric manifold we can easily
identify three $S^1$ coordinates and take a T-duality in these
directions.\foot{The outcome of these T-dualities is different
from the ones of \dotu,\dotd\ and  \dott\ where one T-duality takes
a type IIB picture to a type IIA brane configuration.}
 \foot{There
have been previous attempts to relate the resolved conifold and
the deformed conifold by starting with the deformed conifold
\minasianone. As the deformed conifold does not admit a $T^3$
fibration using this manifold as a starting point may seem more
problematic, although, we have been informed that there are
some papers that overcome this problem \hori.} By applying Buscher's T-duality formulas we observe
that the NS three-form transforms into the type IIA metric in such
a way that the resulting manifold is not K\"ahler. \foot{This is a
generalization of the notion of the mirror symmetry, where the B field
in the type IIB picture is traded for a non  K\"ahlerity in type IIA.
A similar observation has been made in \mind\ where the generalized mirror
symmetry exchange was between a non closed $J + i B$ and a nonclosed
$\Omega$.} The
D5 branes get mapped into D6 branes which are wrapped on a three-cycle
inside a {\it{ non-K\"ahler deformation of a deformed conifold}}.

The non-K\"ahler deformation of a deformed conifold is then
locally lifted to M theory. The lift leads to a $G_2$
manifold\foot{By this we mean a manifold endowed with an almost $G_2$
structure. The holonomy however is contained in $G_2$. For more details on the
almost $G_2$ structure the reader may want to refer to the work of \salamon\ and the references given
in Appendix 2.}
which is
a deformation of the manifold constructed in \brand, \cveticone,
and as such is
described in terms of some left invariant one forms\foot{For an earlier
discussion of special holonomy spaces like $spin(7)$ and the corresponding
one-forms, the reader may want to see \cvetictwo.}.
These
one forms reduce to the ones of \brand, \cveticone\ in the absence
of B fields and they can be exchanged giving rise to a flop
transformation. The resulting type IIA geometry describes a
{\it{non-K\"ahler deformation of a resolved conifold}}. The
non-K\"ahler deformation has a non-closed (3,0) form whose
derivative can be used to construct the additional contribution
to the superpotential $\int d \Omega \wedge J$. These non-K\"ahler
deformations of resolved and deformed conifolds, and the $G_2$ manifolds
resulting from their lift to M-theory are, to the best of our knowledge,
first concrete examples. In earlier literature these manifolds were
anticipated as solutions of type II and M-theories although no concrete
examples were presented.

To summarize, we propose {\it{a new geometric transition in the
type IIA theory}} which relates D-branes wrapped on cycles of
non-K\"ahler manifolds and fluxes on other non-K\"ahler manifolds.
When lifted to M theory this transition describes a flop inside a
$G_2$ manifold with torsion. \foot{This is related to the recent
results of \nekra. This paper discusses a topological string model
concluding that the non-integrability of the complex structure is
related to the existence of Lagrangian NS branes called ``NS
two-branes'' in \vafai. In our language, the non-K\"ahlerity
condition will be related the existence of the ``NS two-brane''.
It would be extremely interesting to relate this non-K\"ahlerity
appearing in the supergravity description to a corresponding
effect in the Chern-Simons theory.} This would lead to a deeper
understanding of non-K\"ahler geometries. These geometries have
only recently been discussed in some detail in the context of
heterotic strings in \beckerD, \lust, \bbdg, \lustu, \bbdgs,
\lustd, \bd\ and \lisheng\ and also of type II and M theories in \micu, \minu, \dal,
\mind, \beru, \dalu.

Our new geometric transition would enrich the ``landscape
picture'' advocated in \susskind,\douo, \banks\ in the sense of extra
identifications between branes on cycles of non-K\"ahler
geometries and fluxes on cycles of related non-K\"ahler
geometries.

This paper is organized as follows: In section 2 we give a very
brief review on the subject of geometric transitions and outline
of the calculation that we will perform in this paper. Our
starting point is the type IIB metric describing $D5$ branes
wrapping a $P^1$ of a resolved conifold and through a series of
T-duality transformations and a flop we shall be able to describe
the geometric transition taking place in the type IIA mirror in
great detail. In section 3 we give an alternative way to derive
the metric using fourfold compactifications in M-theory in the
presence of fluxes. Section 4 discusses the mirror formulas that
we will use to get the full background in the type IIA theory. In
the absence of fluxes, it is known that the mirror type IIA
picture involves $D6$ branes wrapping an $S^3$ of a deformed
conifold \vafai.
In section 4.4 we write the metric of the deformed
conifold in a simpler way by making a coordinate transformation.
We will discuss the reason why a deformed conifold may not
have a $T^3$ fibration. Section 5 begins the study of the mirror
manifold. We will present an explicit way to get the mirror
manifold in the type IIA theory. We will show that the naive
$T^3$ direction of the resolved conifold does not lead to the
right mirror metric, which can nevertheless be determined by a
set of restricted coordinate transformations. These aspects will
be discussed in sections 5.1 and 5.2.
In section 5.3 we will determine the
$B$ field background and the final metric for the mirror manifold
will appear as eqn. (5.63). In section 6 we begin
our ascent to M-theory. In the absence of any fluxes in the type
IIB picture, we expect a manifold with $G_2$ holonomy after
lifting to M-theory. In the presence of fluxes, we will also get
a seven dimensional manifold which now has a $G_2$ structure and
torsion\foot{In this paper we will interchangeably use $G_2$ structure
and $G_2$ holonomy. The seven dimensional manifold that we construct
will have an almost $G_2$ structure, although we will not check the
holonomy here. More detailed discussions will be relegated to part
II of this paper. We thank K. Behrndt and G. Dall'Agata for correspondences
on this issue. See also \behr, \monar.}.
The generic study of $G_2$ holonomy manifolds has been
done earlier using left invariant one-forms \amv.
For our case we will
also have one forms that are appropriately shifted by the
background $B$ fields of the type IIA theory. Locally these one
forms look exactly like the ones without fluxes. However,
globally the system is much more involved, as we do not have any
underlying $SU(2)$ symmetry. The M-theory lift of the mirror type
IIA manifold is given as eqn (6.21), (6.22).

After lifting to M-theory we shall discuss the flop taking place
in the resulting $G_2$ manifold. This will be studied in
section 7 using the one forms that we devised earlier. We shall
show that for torsional $G_2$ manifolds the flop is a little
subtle. We discuss this in detail and compute the form of the
metric after the flop. The result for the metric is given as eqn.
(7.17) of section 7.1. Knowing the M-theory metric after the flop,
helps us to get the corresponding type IIA metric easily by
dimensional reduction. The resulting manifold in the type IIA
theory is non-K\"ahler and the metric is given in eqn. (7.18). We
show that the metric is basically a non-K\"ahler deformation of
the resolved conifold. Interestingly, this is a rather similar
situation as in the type IIA manifold {\it before} the geometric
transition has taken place, whose
metric is a non-K\"ahler deformation of the deformed
conifold. In this section we further study some properties of
these manifolds like non-K\"ahlerity and the underlying
superpotential. We leave a detailed discussion on various aspects
of the whole duality chain for part II of this paper. We end with
a discussion in section 8.

\newsec{Geometric Transitions, Fluxes and Gauge Theories}

We will summarize here some useful facts about geometric
transitions, fluxes and field theory results.

Geometric transitions are examples of generalized AdS/CFT
correspondence which relate D-branes in the open string picture
and fluxes in the closed string picture. There are several types
of geometric transitions depending on the framework in which we
formulate them. The type IIB geometric transition, which starts
with $D5$ branes wrapping a $P^1$ of a resolved conifold, has a
parallel counterpart in which $D5$ branes wrap a vanishing two
cycle of a conifold. This is the Klebanov-Strassler model \ks. In
fact, these two models have identical behaviors in the IR of the
corrresponding four-dimensional ${\cal N} =1$ gauge theories.
However the UV behaviors are different. In the UV the geometric
transition models give rise to six-dimensional gauge theories,
whereas the Klebanov-Strassler model remains four dimensional.
Both these models show cascading behavior. The cascading behavior
in the geometric transition models manifest as an infinite
sequence of flop transitions \civd. The corresponding brane
constructions for these models have been developed earlier in
\dotu. The precise equivalences between the two models were shown
in \dotd\ using (a) T-dual brane constructions, and (b) M-theory
four-fold compactifications. We will not go into the details of
this in the present paper but instead delve directly to the
supergravity aspects of geometric transitions in both type II and
M-theories, with the starting point being $D5$ wrapped on
resolution $P^1$ cycle of a resolved conifold. (For the
Klebanov-Strassler model, the duality chain is not obvious).
Readers interested in the details of the equivalence should look
up the above mentioned references. Some related work has also been
done in \giveon.

The figure below gives an overview of the geometric transitions
and flops that will be discussed in this paper:

\vskip.3in

\centerline{\epsfbox{geometry.eps}}\nobreak

\vskip.3in

\noindent Let us elaborate this figure. The type IIB theory,
depicted at the bottom left level, is the most studied one in
different approaches \mn, \ks\ and \vafai. Here one starts with
the field theory living on D5 branes wrapped on the resolution
$P^1$ cycle of a resolved conifold. In the strong coupling limit
of the field theory the $P^1$ cycle shrinks but the theory avoids
the singularity by opening up an $S^3$ cycle inside a deformed
conifold. In the figure, this is given by a dotted line pointing
to the box on the lower right side of the picture, where a
geometric transition has taken place. The deformed geometry
encodes the information about the strongly coupled field theory
as the size of the $S^3$ cycle is identified with the gluino
condensate in the field theory \vafai.

There is an analog version of this process for the type IIA string
which is depicted in the middle line of the figure. This time one
starts with the field theory living on D6 branes wrapped on the
$S^3$ cycle inside a deformed conifold. In the strong coupling
limit of the field theory the $S^3$ cycle shrinks but the theory
avoids the singularity by opening up a $P^1$ cycle inside a
resolved conifold (given by the next dotted line). As before, the
complexified volume of the $P^1$ cycle is identified with the
gluino condensate appearing in the field theory.

The transition looks mysterious if seen from ten dimensions but
its understanding can be simplified by going up to eleven
dimensions where it appears as a flop transition inside a $G_2$
manifold \amv\ as seen in the upper line of the figure. For the
above mentioned case of D6 branes wrapped on an $S^3$ cycle, the
$G_2$ manifold appears as a cone over a $Z_N$ quotient of $S^3
\times \tilde{S}^3$ and the flop switches the two $S^3$ cycles.
The type IIB transition has also been lifted to an M theory
picture involving a warped fourfold compactification \dotd.
Alternatively, by using one T-duality, the geometric transition
has also been discussed in the brane configuration language in
\dotu\ and \dott.

Even though there is a compelling evidence in favor of using
geometric transitions to describe strongly coupled field
theories, there are still some unanswered questions. One of them
is related to the form of the superpotential (1.1). Intuitively
one would think that if we start with D branes, the supergravity
solution should involve only RR fluxes;  but, as it turns out,
 we need NS fluxes, too. In the language of the Klebanov-Strassler model \ks,
both RR and NS fluxes appear naturally as we have fractional D branes but this
is not obvious in
the language of \vafai.

One goal of this paper is to fill this gap and clarify the
presence of the NS fluxes. We shall start with a known solution
for the type IIB  configuration with wrapped D5 branes on the
resolved conifold by including both NS and RR fluxes. This is
depicted in the lower left box of the figure. We first go up one
step by using three T-dualities and as a result we get a
non-K\"ahler type IIA geometry located in the middle left box of
the figure. Then we go up to M theory, obtaining a $G_2$ manifold
with torsion. We follow the upper arrow by performing a flop
inside the  $G_2$ manifold and then descend to obtain another
non-K\"ahler type IIA geometry. We shall leave the last step and
a detailed discussion on various aspects of the duality chain,
 for a future publication as we are
mostly concerned here with the type IIA geometric transition.

\noindent In the figure the dark arrows represent the directions
that we will be following in this paper. The dotted arrows
represent the connection between two geometries that are related
by a geometric transition. The duality cycle will therefore be
powerful enough to give the precise supergravity background for
{\it all} the examples studied in the literature so far. Notice
also the fact that the key difference between the work presented
here and some of the related work \minasianone\
 is that our starting point involves $D5$ branes
wrapping a resolved conifold and {\it not} a deformed conifold
with fluxes. We believe that a deformed conifold with fluxes {\it
does not} have any obvious $T^3$ fibration and so a simple
Strominger, Yau and Zaslow (SYZ) \syz\ analysis may not be easy to
perform (see however \hori). We will elaborate more on this as we go along.

\noindent We will begin by describing the first box in this
picture (located on the lower left part of the figure): the type
IIB background.

\newsec{The Type IIB Background From M-Theory Dual}

The type IIB background with D5 wrapping an $S^2$ of a resolved
conifold has been discussed in detail in \pandoz. The metric of
the system is shown to follow from the standard D3 brane metric
at a point on the non-compact manifold. In this section we will
give an {\it alternative} derivation of this result from M-theory
with fluxes. One of the major advantages of using M-theory as
opposed to the type IIB theory is the drastic reduction of the
field content. The bosonic field content of M-theory consists of
only the metric and four-form fluxes. Furthermore preserving
supersymmetry in lower dimensions puts some constraints on the
fluxes. The constraint equations are generically {\it linear} and
therefore one can avoid the complicated second order equations
that we would get by solving equations of motion.

To be more specific, we shall consider a {\it non-compact}
fourfold in  M-theory with G-fluxes. The non-compact fourfold is a
$T^2$ fibration over a resolved conifold base. The $T^2$
fibration will be trivial (for the time being) and therefore the
manifold is a product manifold. At this point one might get a
little worried by the fact that the trivial fibration may force
the Euler characteristic to vanish and therefore may disallow
fluxes. The Euler characteristic is indeed zero but by having a
non-compact manifold we still have the possibility to have
non-zero fluxes. There is one immediate advantage of having a
trivial fibration. The $T^2$ torus doesn't degenerate at any
point and therefore when we shrink the $T^2$ torus to zero size
to go to the type IIB framework, there will be no seven branes
and orientifold seven planes. This will simplify the
subsequent analysis. Also, having a non-compact manifold allows
us to put as many branes in the setup as we like. For the compact
case, the number of branes (and fluxes) is constrained by an
anomaly cancellation rule \rBB, \svw.

Let us consider a fourfold that is of the form ${\cal M}_6 \times
T^2$, where ${\cal M}_6$ denotes the resolved conifold which is
oriented along $x^{4,5,6,7,8,9}$, with one of the $S^2$ along
$x^{4,5}$ and the other $S^2$ along $x^{8,9}$. The second $S^2$
degenerates at the radial distance $x^7 = 0$, whereas the other
$S^2$ has a finite size. The coordinate $x^6$ is the usual $U(1)$
fibration. Using angular coordinates in terms of which the metric
is generically written, the two $S^2$'s have coordinates
$\theta_1, \phi_1$ and $\theta_2, \phi_2$. The radial coordinate
is $r \equiv x^7$ and the $U(1)$ coordinate is $\psi \equiv x^6$,
the latter being non-trivially fibered over the two $S^2$'s. The
product torus $T^2$ is oriented along $x^3$ and $x^{11}$, with
$x^{11} \equiv x^a$ being the M-theory direction.

In the presence of G-fluxes with components $G_{457a}$ and
$G_{3689}$, the  backreaction on the metric has been worked out in
\rBB. The metric picks up a warp factor $\Delta$ which is a
function of the internal (radial) coordinate only. The generic
form of the metric is \eqn\mthmetric{ds^2 = \Delta^{-1}
ds^2_{012} + \Delta^{1/2} ds^2_{{\cal M}_6 \times T^2},} where
$ds^2_{012}$ denotes the Minkowski directions. In case that two
covariantly constant spinors of definite chirality on the
internal space can be found, supersymmetry requires  the internal
G-fluxes to be primitive and hence self-dual in the eight
dimensional sense\foot{This is the case we shall be interested
in. The generalization to non-chiral spinors on the internal
space was worked out in \ms, \fg\ and \bbs. In this case the
primitivity condition is replaced by a more general equation.}.
Finally, there is also a spacetime component $G_{012m}$, where
$x^m$ is one of the internal space directions. This component of
the flux is given in terms of warp factor as $G_{012m} = \del_m
\Delta^{-3/2}$. The warp factor, $\Delta$, in turn satisfies the
equation $\quabla \Delta^{3/2} =$ sources.

Let us {\it replace} all the fluxes with M2 branes. This
situation has been considered earlier in section 4 of \bbdgs. The
metric of the system is the metric of $N$ M2 branes at a point on
the fourfold ${\cal M}_6 \times T^2$  \eqn\metofmtwo{ds^2 =
H_2^{-2/3} ~ds^2_{012} + H_2^{1/3}~ ds^2_{{\cal M}_6 \times T^2},}
where $H_2$ is the harmonic function of the M2 branes. As
discussed in \bbdgs, there is a one-to-one connection between the
two pictures: the harmonic function in the M2 brane framework is
related to the warp factor describing the flux via the relation
$H_2 = \Delta^{3/2}$. One can easily show that the source
equations work out correctly using the above identification
\bbdgs.

Imagine now that instead of M2 branes we have $M$ M5 branes in
our framework. These M5 branes wrap three-cycles inside the
fourfold. The three-cycles are an $S^1$ product over an $S^2$
base. We will assume that the M5 branes wrap the directions
$x^{a,4,5}$ inside the fourfold. The metric ansatz for the
wrapped M5 brane is \eqn\mfivemet{ds^2 = H_5^\alpha~ds^2_{012} +
H_5^\beta~ds^2_{S^2 \times S^1} + H_5^\gamma~ds^2_{36789},} where
$H_5$ is the harmonic function of the wrapped M5 branes, and
$\alpha,\beta,\gamma$ are constants. Observe that generically
$\beta \ne \gamma$ and therefore the metric of wrapped M5 branes
is warped differently along the $S^2 \times S^1$ and the
remaining $x^{3,6,7,8,9}$ directions.

The above discussion was in the absence of any fluxes. Let us
switch on three-form potentials $C_{a45}$ and $C_{378}$. In the
presence of these potentials the M5 branes will contain the
following world-volume term \townsend \eqn\sourceterm{S = -{1\o 4}
\int \Gamma^{il}\Gamma^{jm} \Gamma^{kn} (F_{ijk} -
C_{ijk})(F_{lmn} - C_{lmn}),} in addition to the usual source
term that contributes a {\it five} dimensional delta function in
the supergravity equations of motion. We have also denoted the
field strength of the self dual two form propagating on the
$M5$ branes as $F_{ijk}$.
In the
absence of three-form sources this term would be absent and the
equations of motion will only contain a five dimensional delta
function source related to the M5 branes. In the presence of $C$
fields, the above action will induce an M2 brane charge and
therefore the delta function contribution to the action will
become eight dimensional. Therefore the background configuration
will be the usual M2 brane background, implying that $\beta =
\gamma$. When reduced to the type IIB theory by shrinking the
fiber torus to zero size, the metric will be precisely the D3
brane metric obtained in \pandoz. Here we have provided a
derivation of this metric from M-theory. For the M2 brane
background the warp factor will satisfy the usual equation
\rBB,\svw\ \eqn\wareqsa{\ast \quabla H_5 - 4\pi^2 X_8 + {1\o 2} G
\wedge G = -4\pi^2 \sum_{i=1}^n \delta^8(y-y_i),} where the Hodge
duality is over the unwarped metric, $n$ is the number of
fractional M2 branes situated at points $y_i$ in the fourfold and
$X_8$ is a polynomial in powers of the curvature.

{}From the M-theory point of view, our choice of G-fluxes
immediately reduces  to the $H_{NS}$ and $H_{RR}$ fluxes in the
type IIB theory. The fact that supersymmetry requires the G-fluxes
to be primitive, implies that the NS and the RR fluxes should be
dual to each other. This duality gives rise to {\it linear}
equations in the type IIB theory. These linear equations are
given in \pandoz\ (see for example, eq 4.10, 4.11 and 4.12 of
\pandoz). As expected, notice that the primitivity of G-fluxes in
M-theory implies that the form should be of type (2,2) \rBB. This
means that the NS and the RR fluxes when combined to form a
three-form ${\cal G} \equiv H_{NS} + \varphi H_{RR}$ with
$\varphi$ being the usual axion-dilaton scalar, will be a (2,1) form.
This is in agreement with the analysis of \ks\ and \pandoz.

To summarize, starting with M-theory we have rederived the form of
the metric describing wrapped M5 metric in the type IIB theory.
The choice of fluxes fixes the complex structure to some
particular value $\tau$. Using this we define a one form $dz =
dx^3 + \tau dx^a$. The final background can be presented in a
compact form using the notation of \pandoz\ (with $\alpha = -2/3,
\beta = \gamma = 1/3$) \eqn\wrapmfivebg{\eqalign{& ds^2 =
H_5^{-2/3}~ds^2_{012} + H_5^{1/3}~ ds^2_{{\cal M}_6 \times T^2,}
\cr & G = e_{\theta_1} \wedge e_{\phi_1} \wedge G_1 +
e_{\theta_2} \wedge e_{\phi_2} \wedge G_2,}} where $e_{\theta_i}$
and $e_{\phi_i}$ with $i = 1,2$ are defined in \pandoz.
The forms
$G_1$ and $G_2$ also have an explicit representation. They can be
defined in terms of the remaining forms $dz, dr$ and $e_\psi$
(defined in \pandoz) as \eqn\defg{\eqalign{& G_1 = {\bar \tau} f'
~dz \wedge dr + {\bar \tau} ~dz \wedge e_\psi - {\rm c.c,} \cr &
G_2 = {\bar \tau} g' ~dz \wedge dr - {\bar \tau} ~dz \wedge
e_\psi - {\rm c.c,}}} where $f$ and $g$ are solutions to the
linear equations following from supersymmetry. Observe also the
fact that we haven't yet fixed the value of $\tau$, the complex
structure of the $x^{3,a}$ torus\foot{The complex structure of the
base will be discussed later.}.
It is not too difficult to see
that the complex structure can be fixed to $\tau = i$ so that we
end up with a square torus. The analysis follows closely to the
one discussed in \beckerD, so we will not repeat it here. The fact
that fluxes are not constant here (as opposed to the constant
fluxes in \beckerD) does not alter the result for the complex
structure.



\newsec{Mirror Formulas using Three T-dualities}

In this section we will determine the formulas for the metric and
fluxes $B_{NS}$ and $B_{RR}$ of a general type IIA manifold that
is the mirror of a six dimensional type IIB manifold that is a
$T^3$-fibration over a three dimensional base. According to
Strominger, Yau and Zaslow \syz\ the mirror manifold can be
determined by performing three T-dualities on the fiber. We shall
be using the T-duality formulas of \tduality. Later on we will
use our general result for the particular case of the resolved
conifold.

\subsec{Metric Components}

We will start by determining the metric components of the mirror
manifold. Let us call the $T^3$ directions of the lagrangian $T^3$ fibered
manifold with which we start as $x, y$ and $z$. We will be
performing three T-dualities \tduality\ along these directions in
the order $x, y, z$. The starting metric in the type IIB theory
has the following components \eqn\meetcok{\eqalign{ ds^2 = &~
j_{\mu \nu}dx^\mu ~dx^\nu + j_{x\mu} dx ~dx^\mu +  j_{y \mu} dy~
dx^\mu +  j_{z\mu} dz ~ dx^\mu +  j_{xy} dx ~dy \cr
 & ~ + j_{xz} dx ~dz +  j_{zy} dz ~ dy +  j_{xx} dx^2 +  j_{yy}dy^2 +  j_{zz}~dz^2}}
where $\mu, \nu \ne x, y, z$, and the $j's$ are for now arbitrary.
After a straightforward calculation we obtain the form of the
metric of the mirror manifold \eqn\finmetccc{\eqalign{ds^2 = &
\left( G_{\mu\nu} - {G_{z\mu}G_{z\nu} - {\cal B}_{z\mu} {\cal
B}_{z\nu} \over G_{zz}} \right) dx^\mu~dx^\nu +2 \left( G_{x\nu} -
{G_{zx}G_{z\nu} - {\cal B}_{zx} {\cal B}_{z\nu}
 \over G_{zz}} \right) dx~dx^\mu  \cr
& ~ + 2\left( G_{y\nu} - {G_{zy}G_{z\nu} - {\cal B}_{zy} {\cal B}_{z\nu}
 \over G_{zz}}\right) dy~dx^\nu +
2\left( G_{xy} - {G_{zx}G_{zy} - {\cal B}_{zx} {\cal B}_{zy} \over
G_{zz}}\right) dx~dy  \cr & ~ + {dz^2\over G_{zz}} + 2{{\cal
B}_{\mu z} \over G_{zz}} dx^\mu~dz + 2{{\cal B}_{xz} \over G_{zz}}
dx~dz + 2{{\cal B}_{yz} \over G_{zz}} dy~dz \cr &~+  \left( G_{xx}
- {G^2_{zx} - {\cal B}^2_{zx} \over G_{zz}} \right) dx^2 + \left(
G_{yy} - {G^2_{zy} - {\cal B}^2_{zy} \over G_{zz}} \right)
dy^2.}} The various components of the metric can be written as
\eqn\gmunu{\eqalign{G_{\mu\nu} = &~~ {j_{\mu\nu}j_{xx} -
j_{x\mu}j_{x\nu} + b_{x\mu}b_{x\nu} \over j_{xx}} -
{(j_{y\mu}j_{xx} - j_{xy} j_{x \mu} + b_{xy} b_{x\mu})
(j_{y\nu}j_{xx}
 - j_{xy} j_{x \nu} + b_{xy} b_{x\nu}) \over
j_{xx}(j_{yy}j_{xx}- j_{xy}^2 + b_{xy}^2)} \cr & ~~ +
{(b_{y\mu}j_{xx} - j_{xy} b_{x \mu} + b_{xy}
j_{x\mu})(b_{y\nu}j_{xx}
 - j_{xy} b_{x \nu} + b_{xy} j_{x\nu})\over
j_{xx}(j_{yy}j_{xx}- j_{xy}^2 + b_{xy}^2)},}}

\eqn\gmuz{\eqalign{G_{\mu z} = &~~ {j_{\mu z}j_{xx} -
j_{x\mu}j_{xz} + b_{x \mu}b_{xz} \over j_{xx}} - {(j_{y\mu}j_{xx}
- j_{xy} j_{x \mu} + b_{xy} b_{x\mu}) (j_{yz}j_{xx} - j_{xy} j_{x
z} + b_{xy} b_{xz}) \over j_{xx}(j_{yy}j_{xx}- j_{xy}^2 +
b_{xy}^2)} \cr & ~~ + {(b_{y\mu}j_{xx} - j_{xy} b_{x \mu} +
b_{xy} j_{x\mu})(b_{yz}j_{xx} - j_{xy} b_{x z} + b_{xy}
j_{xz})\over j_{xx}(j_{yy}j_{xx}- j_{xy}^2 + b_{xy}^2)},}}

\eqn\gzz{\eqalign{G_{zz} = &~~ {j_{zz}j_{xx} - j^2_{xz} +
b^2_{xz}\over j_{xx}} - {(j_{yz}j_{xx} - j_{xy} j_{xz} + b_{xy}
b_{xz})^2 \over j_{xx}(j_{yy}j_{xx}- j_{xy}^2 + b_{xy}^2)} \cr &
~~ + {(b_{yz}j_{xx} - j_{xy} b_{x z} + b_{xy} j_{xz})^2 \over
j_{xx}(j_{yy}j_{xx}- j_{xy}^2 + b_{xy}^2)},}}

\eqn\gymu{ G_{y \mu} = -{b_{y \mu} j_{xx} - b_{x \mu} j_{xy} + b_{xy}
 j_{\mu x} \over j_{yy}j_{xx}- j_{xy}^2 + b_{xy}^2},
~ G_{y z} = -{b_{y z} j_{xx} - b_{x z} j_{xy} + b_{xy} j_{z x}
\over j_{yy}j_{xx}- j_{xy}^2 + b_{xy}^2},}

\eqn\gyy{G_{yy} = {j_{xx} \over j_{yy}j_{xx}- j_{xy}^2 +
b_{xy}^2},~ G_{xx} = {j_{yy} \over j_{yy}j_{xx}- j_{xy}^2 +
b_{xy}^2}, ~G_{xy} = {-j_{xy} \over j_{yy}j_{xx}- j_{xy}^2 +
b_{xy}^2},}

\eqn\gmux{G_{\mu x} = {b_{\mu x} \over j_{xx}} + {(j_{\mu y} j_{xx} -
 j_{xy} j_{x \mu} + b_{xy} b_{x \mu}) b_{xy} \over
j_{xx}(j_{yy}j_{xx}- j_{xy}^2 + b_{xy}^2)}
+ {(b_{y \mu} j_{xx} - j_{xy} b_{x \mu} + b_{xy} j_{x \mu}) j_{xy}
 \over j_{xx}(j_{yy}j_{xx}- j_{xy}^2 + b_{xy}^2)},}

\eqn\gzx{ G_{z x} = {b_{z x} \over j_{xx}} + {(j_{z y} j_{xx} -
j_{xy} j_{x z} + b_{xy} b_{x z}) b_{xy} \over
j_{xx}(j_{yy}j_{xx}- j_{xy}^2 + b_{xy}^2)}
 + {(b_{y z} j_{xx} - j_{xy} b_{x z} + b_{xy} j_{xz}) j_{xy}
  \over j_{xx}(j_{yy}j_{xx}- j_{xy}^2 + b_{xy}^2)}.}
In the above formulae we have denoted the type IIB
$B$ fields as $b_{mn}$, whose explicit
form will be computed in the next section.
We will use this more general formula for the particular case of
the resolved conifold \pandoz\ a little later. Our next goal is
to determine the NS fluxes on the mirror manifold for the most
general case.

\subsec{$B_{NS}$ Components}

\noindent  For the generic case we will switch on all the
components of the $B$ field \eqn\bcompolk{\eqalign{ b = & ~~
b_{\mu\nu} ~ dx^\mu \wedge dx^\nu + b_{x \mu} dx \wedge dx^\mu +  b_{y \mu}
~ dy~\wedge dx^\mu + b_{z \mu} ~ dz \wedge dx^\mu \cr & ~~ + ~ b_{xy}
~ dx \wedge dy +
 b_{xz} ~ dx  \wedge dz +  b_{zy}~ dz  \wedge dy.}}
In the later sections we will concentrate on the special
components that describe a resolved conifold with branes.
\noindent After applying again the T-dualities, the NS component
of the $B$ field in the mirror set-up will take the form
\eqn\bbfielc{\eqalign{ {\tilde B} = & ~~ \left( {\cal B}_{\mu\nu}
+ {2 {\cal B}_{z[\mu} G_{\nu]z} \over G_{zz}} \right) dx^\mu
\wedge dx^\nu + \left( {\cal B}_{\mu x} + {2 {\cal B}_{z[\mu}
G_{x]z} \over G_{zz}}\right)
 dx^\mu \wedge dx  \cr
& ~~ \left( {\cal B}_{\mu y} + {2 {\cal B}_{z[\mu} G_{y]z} \over G_{zz}}
 \right) dx^\mu \wedge dy
+ \left( {\cal B}_{xy} + {2 {\cal B}_{z[x} G_{y]z} \over G_{zz}}
\right) dx \wedge dy \cr & ~~ + {G_{z \mu} \over G_{zz}} dx^\mu
\wedge dz + {G_{z x} \over G_{zz}} dx \wedge dz + {G_{z y} \over
G_{zz}} dy \wedge dz.}} Here the $G_{mn}$ components have been
given above, and the various ${\cal B}$  components can now be
written as \eqn\bmunu{\eqalign{ {\cal B}_{\mu\nu} = & ~~
{b_{\mu\nu} j_{xx} + b_{x \mu} j_{\nu x} - b_{x \nu} j_{\mu x}
\over j_{xx}} \cr & + ~~ {2 (j_{y[\mu}j_{xx} - j_{xy}j_{x[\mu} +
b_{xy} b_{x[\mu}) (b_{\nu]y}j_{xx} - b_{\nu]x}j_{xy} - b_{xy}
j_{\nu]x}) \over j_{xx}(j_{yy}j_{xx}- j_{xy}^2 + b_{xy}^2)},}}

\eqn\bmuz{\eqalign{ {\cal B}_{\mu z} = & ~~ {b_{\mu z} j_{xx} +
b_{x \mu} j_{z x} - b_{x z} j_{\mu x} \over j_{xx}} \cr & + ~~ {2
(j_{y[\mu}j_{xx} - j_{xy}j_{x[\mu} + b_{xy} b_{x[\mu})
(b_{z]y}j_{xx} - b_{z]x}j_{xy} - b_{xy} j_{z]x}) \over
j_{xx}(j_{yy}j_{xx}- j_{xy}^2 + b_{xy}^2)},}}

\eqn\bmuy{{\cal B}_{\mu y} = {j_{\mu y} j_{xx} - j_{xy} j_{x \mu} + b_{xy} b_{x \mu}
 \over j_{yy}j_{xx}- j_{xy}^2 + b_{xy}^2}, ~~~
{\cal B}_{z y} = {j_{z y} j_{xx} - j_{xy} j_{x z} + b_{xy} b_{x z} \over j_{yy}j_{xx}-
 j_{xy}^2 + b_{xy}^2},}

\eqn\bmux{ {\cal B}_{\mu x} = {j_{\mu x} \over j_{xx}} - {j_{xy} (j_{\mu y} j_{xx} -
j_{xy} j_{x \mu} + b_{xy} b_{x \mu}) \over
j_{xx}(j_{yy}j_{xx}- j_{xy}^2 + b_{xy}^2)} + {b_{xy} (b_{x\mu}j_{xy} - b_{y\mu}j_{xx} -
b_{xy}j_{xz})
 \over j_{xx}(j_{yy}j_{xx}- j_{xy}^2 + b_{xy}^2)},}

\eqn\bzx{ {\cal B}_{z x} = {j_{z x} \over j_{xx}} - {j_{xy} (j_{z y} j_{xx} -
j_{xy} j_{xz} + b_{xy} b_{x z}) \over
j_{xx}(j_{yy}j_{xx}- j_{xy}^2 + b_{xy}^2)} + {b_{xy} (b_{xz}j_{xy} - b_{yz}j_{xx} -
 b_{xy}j_{xz})
\over j_{xx}(j_{yy}j_{xx}- j_{xy}^2 + b_{xy}^2)},}

\eqn\bxy{{\cal B}_{xy} = {-b_{xy} \over j_{yy}j_{xx}- j_{xy}^2 + b_{xy}^2}.}

In the above analysis, there is one subtlety related to the
compactness of the $x,y,z$ directions. The type IIB $B$ fields
defined wholly along these directions, i.e. ${\cal B}_{yz}, {\cal
B}_{zx}$ and ${\cal B}_{xy}$, should be {\it periodic}. This would
mean, for example, if we specify a value of ${\cal B}_{yz}$  as
(say) $\alpha_{yz}$ then this should also be equal to
$-\alpha_{yz}$ because of periodicity. This implies that the
values of ${\cal B}_{yz}, {\cal B}_{zx}$ and ${\cal B}_{xy}$
found in \bmuy, \bzx\ and \bxy\ are {\it ambiguous} up to a
possible sign. Later on we shall use consistency conditions to fix
the sign.

Furthermore, observe that we haven't yet discussed how the RR $B$
fields look like in the mirror set-up.  We will eventually
compute the form of these fields when we perform the M-theory lift
of the type IIA mirror. We also need to see how the
fermions transform under mirror symmetry. This will be important
in order to understand the complex structure of the mirror
manifold and to check whether it is integrable or not. This will
be discussed in the sequel to this paper. The string coupling
constant in the type IIB theory and the one in the type IIA mirror
are related in the following way \eqn\ccconss{g_A = {g_B \o
\sqrt{(j_{xx}j_{yy} - j_{xy}^2 + b_{xy}^2)~G_{zz}}},} where we
have defined $G_{zz}$ in \gzz. This coupling constant is in
general a function of the internal coordinates.

\subsec{Background Simplifications}

The background given in the above set of formulas can be written
in a {\it compact} form which will be helpful to see the fibration
structure more clearly
\eqn\metcomp{\eqalign{ds^2 = & {1\over G_{zz}} (dz + {\cal B}_{\mu z}
dx^\mu + {\cal B}_{xz} dx
+ {\cal B}_{yz} dy)^2 -{1\over G_{zz}}
(G_{z\mu} dx^\mu + G_{zx} dx + G_{zy} dy )^2 \cr
& + G_{\mu\nu} dx^\mu dx^\nu + 2 G_{x \nu} dx dx^\nu + 2 G_{y \nu} dy dx^\nu +
2 G_{xy} dx dy + G_{xx} dx^2 + G_{yy} dy^2.}}
The above compact form can be simplified even further for the particular
example we are interested in. More concretely the $G_{mn}$
components ($m,n = \mu, x, y, z$) become rather simple if one
assumes the following choices of $j_{mn}, b_{mn}$
\eqn\chofjb{j_{\mu x} = j_{\mu y} = j_{\mu z} = 0; ~~~ b_{xy} =
b_{zx} = b_{zy} = 0; ~~~ b_{\mu\nu} = 0.}
In this case the negative components in the metric vanish. The
above assumption implies that the type IIB metric of a D5
wrapping an $S^2$ of the resolved conifold has no off-diagonal
components. This can be easily checked and we shall elaborate this
further later on. The type IIB $b$ field choice tells us that
off-diagonal components are allowed but the cross terms vanish. This
can also be verified easily. With this choice of type IIB metric
the metric components of the mirror take the following form

\eqn\nowgmunu{G_{\mu\nu} = j_{\mu\nu} + {b_{x\mu} b_{x \nu} \over j_{xx}} +
 {(b_{y\mu}j_{xx} - b_{x\mu} j_{xy})(b_{y\nu} j_{xx} - b_{x\nu} j_{xy}) \over
j_{xx}(j_{yy} j_{xx} - j_{xy}^2)},}

\eqn\nowgzxy{G_{\mu z} ~=~ G_{zx} ~= ~G_{zy} ~= ~0,}

\eqn\nowgyyzzxx{G_{xx} = {j_{yy} \over j_{yy} j_{xx} - j^2_{xy}},
~~ G_{yy} = {j_{xx}\over j_{yy} j_{xx} - j_{xy}^2}, ~~G_{zz} =
j_{zz}- {j_{xz}^2 \over j_{xx}} - {(j_{yz}j_{xx} - j_{xy}
j_{xz})^2 \over j_{xx}(j_{yy} j_{xx} - j_{xy}^2)},}

\eqn\nowgxy{G_{xy} = {-j_{xy} \over j_{yy}j_{xx} - j_{xy}^2},}

\eqn\nowgmuxz{G_{\mu x} = {b_{\mu x} \over j_{xx}} + {(b_{y
\mu}j_{xx} - b_{x \mu} j_{xy})j_{xy} \over j_{xx}(j_{yy}j_{xx} -
j_{xy}^2)}, ~~ G_{y \mu} = -{b_{y \mu}j_{xx} - b_{x \mu} j_{xy}
\over j_{yy}j_{xx} - j_{xy}^2}.}
On the other hand, the ${\cal B}$ fields appearing in the metric
and fluxes (not to be confused with the ${\tilde B}$ fields in
the type IIA picture) take the form

\eqn\nowbmunu{{\cal B}_{\mu \nu}~ = ~ {\cal B}_{y \mu} ~ = ~
{\cal B}_{x \mu} ~ = ~ {\cal B}_{xy} ~ = 0,}

\eqn\nowbmuzxy{ {\cal B}_{\mu z} = b_{\mu z} + {b_{x\mu} j_{xz}
\over j_{xx}} - {(j_{yz}j_{xx} - j_{xy}j_{xz})(b_{\mu y} j_{xx} -
b_{\mu x} j_{xy}) \o j_{xx}(j_{yy}j_{xx} - j_{xy}^2)},}

\eqn\nowbzx{{\cal B}_{zx} = {j_{zx} \over j_{xx}} -
{j_{xy}(j_{zy}j_{xx} - j_{xy}j_{xz}) \over j_{xx}(j_{yy}j_{xx} -
j_{xy}^2)}, ~~~~ {\cal B}_{y z} = -{j_{yz}j_{xx} - j_{xy} j_{zx}
\over j_{yy} j_{xx} - j_{xy}^2}.}
Using the above choices of $G_{mn}$ and ${\cal B}_{mn}$ one can
easily show that all components of the mirror NS flux vanish,
${\tilde B}_{mn} = 0$.

Naively one would expect that the background of the mirror
manifold that we just derived corresponds to a deformed conifold
of \ks\ in the presence of fluxes. In order to see the relation to
the deformed conifold  our mirror background can be simplified
further. But before doing so, let us rewrite the deformed conifold
background of \ks\ in a suggestive way so that a comparison can be
made.

\subsec{Rewriting the Deformed Conifold Background}

The metric of a D6 brane wrapping a three cycle of a deformed
conifold has been discussed earlier in \edelstein. Let us
recapitulate the result. To obtain the metric, one defines a
K\"ahler potential ${\cal F}$ as a function of $\rho^2 \equiv
{\rm tr} (W^\dagger W)$ with $W$ being a complex $2 \times 2$
matrix satisfying ${\rm det}~W = -\epsilon^2/2$. The quantity
$\rho$ is basically the radial parameter and $\epsilon$ is a real
number. The generic form of the Ricci flat K\"ahler background is
determined from (see \candelas\ for details) \eqn\conimet{ds^2 =
{\cal F}'~{\rm tr} (dW^\dagger dW) + {\cal F}''~\vert {\rm
tr}(W^\dagger dW)\vert^2,} where the primes are defined as ${\cal
F}' = d{\cal F}/d\rho^2$ and the determinant of the deformed
conifold metric is given by $\epsilon^{-8}(\rho^4 -
\epsilon^4)^2$. For a Ricci flat metric ${\cal F}'$ becomes equal
to \eqn\feqto{ {(\sqrt{2}\epsilon)^{-{2\over 3}} (2\epsilon^2
\rho^2 \sqrt{\rho^4 - \epsilon^4} - 2 \epsilon^6~ {\rm
ch}^{-1}(\rho^2/\epsilon^2))^{1/3} \over
\sqrt{\rho^4-\epsilon^4}}.} In this form it is not too difficult
to write the metric of a bunch of D6 branes wrapping the three
cycle of a deformed conifold. If we take the limit $\rho \to
\epsilon$ the metric becomes the metric of an $S^3$ space. The
generic form of the wrapped D6 metric is \eqn\gendsixmet{ds^2 =
{\cal A}_0~ ds^2_{0123} + {\cal A}_1~ d\rho^2 + {\cal A}_2
~ds_1^2 + {\cal A}_3 ~ds_2^2 + {\cal A}_4~ ds_4^2 + {\cal A}_5~
ds^2_3,} where ${\cal A}_i$ are some specific functions of the
radial coordinate $\rho^2$ and the metric components $ds_i$ are
given by \eqn\metcomptwo{\eqalign{ & ds_1^2 = (d\psi + {\rm cos}
\theta_1~d\phi_1 + {\rm cos} \theta_2~d\phi_2)^2, \cr & ds_2^2 =
d\theta_1^2 + {\rm sin}^2\theta_1~d\phi_1^2, ~~~ ds_4^2 =
d\theta_2^2 + {\rm sin}^2\theta_2~d\phi_2^2, \cr & ds_3^2 = 2~{\rm
sin} \psi~ (d\phi_1 d \theta_2~{\rm sin} \theta_1 + d\phi_2 d
\theta_1~{\rm sin} \theta_2) + 2~{\rm cos} \psi ~( d\theta_1
d\theta_2 - d\phi_1 d\phi_2 ~{\rm sin} \theta_1 {\rm sin}
\theta_2).}} The appearance of ${\rm sin} \psi$ and ${\rm cos}
\psi$ in the above metric is a little disconcerting as the expected
$U(1)$ symmetry acting on $\psi$ as $\psi \to \psi + c$ is not
present \minasianone. This means that the deformed conifold {\it
cannot} be written as a simple $T^3$ fibration over a three
dimensional base. An immediate consequence of this is that the
usual SYZ technique cannot be applied. On the other hand, D5
branes wrapped on a resolved conifold do have the required $U(1)$
isometries related to constant shifts in $\psi, \phi_1, \phi_2$,
so that we can perform three T-dualities and obtain the mirror
manifold, as we are doing in this paper. The $T^3$ fibration
corresponds to the $\psi, \phi_1, \phi_2$ torus. In the notations
of the previous subsection, $z = \psi, x = \phi_1, y = \phi_2$
for the type IIB resolved conifold (we will give a more precise
mapping soon).

In order to compare the deformed conifold metric with the metric
obtained for our mirror manifold it is convenient to perform a
change of coordinates
 $\theta_2, \phi_2$: \eqn\tranthe{
\pmatrix{{\rm sin}~\theta_2~ d\phi_2 \cr d\theta_2} \to
\pmatrix{{\rm cos}~ \psi & {\rm sin}~ \psi \cr -{\rm sin}~ \psi &
{\rm cos}~ \psi} \pmatrix{{\rm sin}~\theta_2 ~d\phi_2 \cr
d\theta_2},} with the other coordinates $\theta_1$ and $\phi_1$
remaining unchanged. Although identical in spirit, the above
transformation is {\it different} from equation (2.2) of
\minasianone. Under the transformation \tranthe, the metric
component $ds_3^2$ changes to \eqn\dsthree{ds_3^2 ~ \to~ 2
d\theta_1~d\theta_2 - 2{\rm sin}~\theta_1~{\rm
sin}~\theta_2~d\phi_1 ~d\phi_2,} so that the $\psi$ dependence is
completely removed. However this doesn't mean that we regain the
$U(1)$ isometry. The $\psi$ dependence enters into $ds_1^2$ in
such a way that a shift in the other coordinates fails to remove
it. Observe that the metric components $ds_2^2$ and $ds_4^2$
remain unaltered.

To summarize: the above observation tells us that a simple $T^3$
fibration of a deformed  conifold does not exist. In a new
coordinate system (which is discussed in \minasianone) it might
be possible to regain some of the $U(1)$ isometries, although a
SYZ description of the corresponding mirror appears futile (see \hori\
for some proposals to reconcile this were given).
Nevertheless, the above change of coordinates will be useful to
understand the type IIA mirror background and compare the
metric with the expected deformed conifold metric.

In the following we will rewrite the metric \gendsixmet\ in such a
way that it can be mapped to the mirror manifold obtained by using
three T-dualities. The readers interested in the type IIA mirror
background may want to skip this part and go directly to the next
section. If we define a warp factor $h \equiv h(\rho)$ then the
generic metric of D6 branes wrapping a three cycle of a deformed
conifold can be written as \eqn\nordi{ds^2 = h^\alpha~
ds^2_{0123} + h^\beta~ dr^2 + h^\gamma~ ds_1^2 + h^\delta~(ds_2^2
+ ds_4^2) + h^\rho~ds_3^2,} where ${\alpha,\beta,\gamma,\rho}$
are the various numerical powers for the wrapped D6 branes
(compare with the previous formula \gendsixmet)  and $ds_i$ ($i =
1, 2, 3$) are defined earlier. We have also taken a slightly
simplified case where corresponding to ${\cal A}_3 = {\cal A}_4$
in our earlier notation, as this is the case we are interested in.
The above metric can now be written in a more suggestive way \ohta
\eqn\metsugges{\eqalign{ds^2 = & h^\alpha~ds^2_{0123} +
(h^{\beta/2}dr)^2 + h^\gamma ~(d\psi + {\rm cos}~\theta_1~d\phi_1
+ {\rm cos}~\theta_2~d\phi_2)^2 + \cr & ~~~~ + (h^\delta -
h^\rho)~ ({\rm sin}~\psi~{\rm sin}~\theta_1~d\phi_1 + {\rm
cos}~\psi~d\theta_1 - d\theta_2)^2 ~ + \cr & ~~~~ + (h^\delta -
h^\rho)~ ({\rm cos}~\psi~{\rm sin}~\theta_1~d\phi_1 - {\rm
sin}~\psi~d\theta_1 + {\rm sin}~\theta_2~d\phi_2)^2 ~ + \cr &
~~~~ +  (h^\delta + h^\rho)~ ({\rm sin}~\psi~{\rm
sin}~\theta_1~d\phi_1 + {\rm cos}~\psi~d\theta_1 + d\theta_2)^2 ~
+ \cr & ~~~~ + (h^\delta + h^\rho)~ ({\rm cos}~\psi~{\rm
sin}~\theta_1~d\phi_1 - {\rm sin}~\psi~d\theta_1 - {\rm
sin}~\theta_2~d\phi_2)^2.}} Our next goal is to rewrite this
metric in terms of the T-dual coordinates $x, y, z$ that we used
earlier to get the fields of the mirror manifold, by using the
identification $z = \psi, x = \phi_1, y = \phi_2, \mu,\nu =
\theta_1, \theta_2$. Omitting the $r$ and the $ds^2_{0123}$ term
the metric \metsugges\ can be written as
\eqn\metwritxyz{\eqalign{ds^2 = & ~~~h^\gamma ~(dz + f^1_{zx}~dx
+ f^2_{zy}~dy)^2 + (h^\delta - h^\rho)~(f^3_{xx}~dx +
f^4_{x\mu}~dx^\mu)^2 + \cr &+ (h^\delta - h^\rho)~(f^5_{yy}~dy +
f^6_{xy}~dx + f^7_{x\mu}~dx^\mu)^2 + (h^\delta +
h^\rho)~(f^8_{xx}~dx + f^9_{x\mu}~dx^\mu)^2 + \cr &~~~~~ ~~~~ +
(h^\delta + h^\rho)~(f^{10}_{yy}~dy + f^{11}_{xy}~dx +
f^{12}_{x\mu}~dx^\mu)^2},} where $f^i_{mn}, ~i= 1, 2,...,12, ~~m,n
= x, y, z$ can be easily related to the coefficients in
\metsugges. Having written the metric in the form of \metsugges\
or \metwritxyz\ still does not tell us that D6 branes wrapped on
$S^3$ of a deformed conifold should have a $T^3$ fibration,
because of the appearance of ${\rm sin}~\psi$ and ${\rm
cos}~\psi$ in the product. On the other hand writing the metric in
the form \metwritxyz\ will help us to relate it to the mirror
metric that we derived in the previous section. We will do this
in the next section.

Let us comment a little more on the transformation \tranthe. As
mentioned earlier, this transformation would remove the $\psi$
dependence in $ds_3$ and bring it to the form \dsthree. However
the $d\psi$ fibration structure will now change because ${\rm
cos}~\theta_2~d\phi_2$ changes under \tranthe. The change will
generically introduce some terms proportional to $d\theta_2$ in
the $d\psi$ fibration structure. The precise change will be
\eqn\psichange{{\rm cot}~\theta_2~dy ~\to ~ {\rm
cot}~{\bar\theta_2}~({\rm cos}~\psi~dy + {\rm
sin}~\psi~d\theta_2),} where $\bar\theta$ is the change in
$\theta$ under the transformation \tranthe. Now the change
\psichange\ explicitly introduces the $\psi$ dependence in the
fibration structure but removes it from the other parts of the
metric. In the {\it delocalized} limit, the $\psi$ values are
basically constant (of order $2\pi$) and therefore can be
approximated by constants. {\it This is the only assumption that
we will consider at this stage}. Under this assumption the
$d\theta$ dependent term appearing in the fibration structure
\psichange\ can be absorbed by a shift in $d\psi$ as $d\psi ~\to
~d(\psi - a~{\rm ln~ sin}~{\theta_2})$, where we have approximated
$\bar\theta$ by $\theta$ and $a$ is a constant. Under this
transformation and in the delocalization limit the $d\psi$
fibration structure does not change too much from its original
value. In this way we can recover a simplified form of the
deformed conifold metric. Observe that this doesn't mean that we
generate a $U(1)$ isometry in a theory that didn't have an
isometry before transformation. We can only get the metric with
$\psi$ isometry in the delocalization limit.

In an alternative scenario, we can get rid of the $\psi$
dependences in $ds_3$ by restricting to a specific value of
$\psi$, i.e $\psi = 2\pi$. This choice of $\psi$ can be easily
obtained in the mirror dual picture (that we are going to discuss
in the next section). In the mirror we expect to see a resolved
conifold. The mirror of this can be determined from performing
three T-dualities along the $\psi, \phi_1$ and $\phi_2$
directions. T-dualities require that we delocalize the  $\psi,
\phi_1$ and $\phi_2$ directions. Since the resolved conifold
metric is already independent of these directions, delocalizing
simply amounts to setting $\psi = \phi_1 = \phi_2 = 2\pi$. A
somewhat related discussion on the transformation of $ds_3^2$ has
been given in \ohta. Therefore we will consider the specific
delocalized limit of the deformed conifold where the $\psi$
dependences will be removed by resorting to the specific value
$\psi = 2\pi$.

For completeness and since we will need these expressions for
later comparison we will list the expressions for the vielbeins
describing $D6$ wrapped on an $S^3$ of a deformed conifold
\eqn\vieiib{\eqalign{& e^1_{\phi_1} = \sqrt{h_+}~{\rm cos}~\psi ~
{\rm sin}~\theta_1, ~~~ e^1_{\theta_1} = -\sqrt{h_+}~{\rm
sin}~\psi, ~~~~e^1_{\phi_2} = \sqrt{h_+}~{\rm sin}~\theta_2 \cr &
e^2_{\phi_1} = \sqrt{h_+}~{\rm sin}~\psi ~ {\rm sin}~\theta_1,
~~~e^2_{\theta_1} = \sqrt{h_+}~{\rm cos}~\psi,
~~~~~~e^2_{\theta_2} = \sqrt{h_+}\cr & e^3_{\phi_1} =
\sqrt{h_-}~{\rm cos}~\psi ~ {\rm sin}~\theta_1, ~~~
e^3_{\theta_1} = -\sqrt{h_-}~{\rm sin}~\psi, ~~~~e^3_{\phi_2} =
\sqrt{h_-}~{\rm sin}~\theta_2 \cr & e^4_{\phi_1} =
\sqrt{h_-}~{\rm sin}~\psi ~ {\rm sin}~\theta_1, ~~~
e^4_{\theta_1} = \sqrt{h_-}~{\rm cos}~\psi, ~~~~~~e^4_{\theta_2} =
-\sqrt{h_-}\cr & e^5_\psi = \sqrt{h^\gamma}, ~~ e^5_{\phi_1} =
\sqrt{h^\gamma}~ {\rm cos}~\theta_1, ~~ e^5_{\phi_2} =
\sqrt{h^\gamma}~{\rm cos}~\theta_2, ~~e^6_r = \sqrt{h^\beta}}}
where $h_+ = h^\delta + h^\rho$ and $h_- = h^\delta - h^\rho$.
{}From here properties of the background such as the fundamental
form, holomorphic three form etc., can be easily extracted.

\newsec{Chain 1: The Type IIA Mirror Background}

In this section we will determine the exact form of the mirror
manifold and apply our generic formulas to the special case $D5$
branes wrapping a resolved conifold in the type IIB theory. We
will find that the manifold is not quite a deformed conifold in
the presence of fluxes as one would have naively expected, rather
it will turn out to be a non-K\"ahler manifold that could even be
non-complex.

To determine the precise form of the manifold let us first present
the metric for D5 branes wrapped on an $S^2$ of a resolved
conifold. It is given in \pandoz\ in the following form
\eqn\metresconi{\eqalign{ds^2 = & ~h^{-1/2} ds^2_{0123} + h^{1/2}
\Big[\gamma' dr^2 + {1\o 4} \gamma' r^2 (d\psi + {\rm
cos}~\theta_1 d\phi_1 + {\rm cos}~\theta_2 d\phi_2)^2 + \cr &
~~~+ {1\o 4}\gamma(d\theta_1^2 + {\rm sin}^2~\theta_1 d\phi_1^2)
+ {1\o 4}(\gamma + 4a^2) (d\theta_2^2 + {\rm sin}^2~\theta_2
d\phi_2^2)\Big],}} where we have used the notations of \pandoz\
and $\gamma$ is defined as a function of $r^2$ only\foot{We have
also defined the radial coordinate $r$ and $\gamma'$ in the
following way: $\gamma' \equiv {d\gamma \o dr^2} = {2\o 3}
{\sqrt{\gamma + 6a^2}\o \gamma + 4a^2}$ and the Ricci flatness
condition gives rise to the equality $r = \sqrt{\gamma
\sqrt{\gamma + 6a^2}}$.}. The presence of wrapped
$D5$ branes in the metric is
signalled by the harmonic function $h$ whose functional form can be
extracted from \pandoz.
Observe that the parameter $a$ creates
an asymmetry between the two spheres denoted by $\theta_1,
\phi_1$ and $\theta_2, \phi_2$. As discussed in \pandoz, for
small $r$ (the radial coordinate) the $S^3$ denoted by $\psi,
\theta_1, \phi_1$ shrinks to zero size whereas the other sphere
remains finite with radius $a$. This is the resolving parameter.
As can be easily seen, when the resolving parameter goes to zero
size, the manifold becomes a conifold, and the metric works out
correctly. Furthermore, the curvature remains regular all
through. Notice also that the metric \metresconi\ has three
isometries related to constant shifts in $\psi, \phi_1$ and
$\phi_2$ as $\psi \to \psi + c_1, \phi_1 \to \phi_1 + c_2, \phi_2
\to \phi_2 + c_3$. But there are no isometries along $r,
\theta_1$ and $\theta_2$ directions because of the warp factors
and the $d\psi$ fibration structure. Therefore there is a natural
$T^3$ structure associated with $\psi, \phi_1, \phi_2$
directions. This $T^3$ is actually a special lagrangian submanifold \karch. A direct way to
see this would be to evaluate the condition required for a cycle to be lagrangian.
Alternatively, one can see that the $\phi_1, \phi_2$ directions lead to a brane-box
configuration after two T-dualities \karch. This configuration preserves susy. Furthermore, a
T-duality along $\psi$ direction has been shown earlier to lead to a susy
preserving configuration \dmconi. Thus
$\psi, \phi_1, \phi_2$ lead to a lagrangian submanifold that preserves susy after three T-dualities.
However, as we will soon see, the T-duality
directions are not the naively expected isometry directions. The
T-dualities in this scenario are a little subtle as we will
elaborate soon. For the time being we will continue to consider
the naive $T^3$ as our SYZ directions.

Let us now write the metric in terms of the T-duality coordinates
that we used in the previous section. We shall use the following
mapping to relate the above metric to the one
presented earlier
\eqn\miapmet{\eqalign{& (x, y, z) ~\to ~(\phi_1, \phi_2, \psi),
\cr & (dx, dy, dz) ~ = ~ \left({1\o 2}{\sqrt{h^{1/2}\gamma}}~{\rm
sin}~\theta_1~d\phi_1, {1\o 2} {\sqrt{h^{1/2}(\gamma+
4a^2)}}~{\rm sin}~\theta_2~d\phi_2, {1\o 2} {r\sqrt{\gamma'
h^{1/2}}}~ d\psi \right).}}
The physical meaning of
$x,y,z$ can be given as follows: under a single T-duality along $\psi$,
the system maps to an intersecting brane configuration \dotu, \dotd,
\ohta; $x, y$ and $z$ form the coordinates of the branes. More precisely,
we are in fact converting the two spheres with coordinates
($\phi_1, \theta_1$) and ($\phi_2, \theta_2$) to tori with coordinates
($x, \theta_1$) and ($y, \theta_2$) respectively. Recall that a sphere
is topologically the same as a tori with a {\it degenerating} cycle
(i.e. if we shrink one of the cycles of the $T^2$ to zero size then this would
be topologically the same as a sphere) and
therefore this mapping would be locally indistinguishable.
Furthermore, this mapping will
be particularly useful to perform many simplifying manipulations later in the
paper which are otherwise messy in the original coordinate system.

\noindent With this map,
we can now write the
various components of the wrapped D5 metric:
\eqn\comedfi{\eqalign{& j_{zz} = 1 , ~~j_{xx} =  1 +  {\gamma' \o
\gamma}~ r^2 ~{\rm cot}^2~\theta_1, \cr & j_{yy} =  1 +  {\gamma'
\o \gamma + 4a^2}~ r^2~ {\rm cot}^2~\theta_2, \cr & j_{zx} =
{\sqrt{\gamma'\o \gamma}}~ r~ {\rm cot}~\theta_1, ~~ j_{zy} =
\sqrt{\gamma' \o \gamma + 4a^2} ~r ~{\rm cot}~\theta_2, \cr &
j_{xy} =
 {\gamma' \o \sqrt{\gamma(\gamma + 4a^2)}}~ r^2~
 {\rm cot}~\theta_1 ~{\rm cot}~\theta_2, ~~j_{rr} = \gamma' h^{1/2}, \cr
& j_{\theta_1\theta_1} =  {1\o 4}\gamma h^{1/2}, ~~
j_{\theta_2\theta_2} = {1\o 4} h^{1/2} (\gamma + 4a^2)}} with the
rest of the components zero. The $B_{NS}$ fields on the other
hand have the following components (see also sec. 4 of \pandoz):
\eqn\bfico{b = {\cal J}_1 ~d\theta_1 \wedge dx + {\cal J}_2~
d\theta_2 \wedge dy} with the rest of the components zero and
${\cal J}_i$ are now functions of the radial and the angular coordinates,
i.e ($r, \theta_1, \theta_2$). In \pandoz\ the $B$ field was only function
of the radial coordinate. Here since we converted all the spheres in the
metric to tori, we will keep $B$ as a generic function of ($r, \theta_1,
\theta_2$) to preserve supersymmetry.
Notice also that the choice of the $B$ field and the metric is
consistent with the assumptions that we made in
the previous section, namely $b_{xy} = b_{yz} = b_{zx} = 0$ and
the cross term $j_{(x,y,z)\mu} = 0$. The small
and the large $r$ behavior of $\gamma$ is \pandoz:
\eqn\smlarbe{\gamma_{r\to 0} = {1\o \sqrt{6} a} r^2 - {1\o 72
a^4} r^4 + {\cal O}(r^6), ~~~~ \gamma_{r \to \infty} = r^{4/3} -
2 a^2 + {\cal O}(r^{-4/3}).} In the above set of components
\comedfi, if we ignore the overall $h^{1/2}$ dependences, observe
that for small $r$ (which we will concentrate on mostly) $\gamma$
is a small quantity, and thus terms like $j^{-1}$ (which we will
encounter soon) could be expanded in powers of $\gamma$ (or $r$),
because the $\theta_i$ dependences can be made generically small.
We will however try to avoid making approximations and
concentrate on the exact values as far as possible.

Before moving ahead one comment is in order. The metric of $D5$
wrapping an $S^2$ of a resolved conifold has  no $j_{\theta_1
\theta_2}$ component, i.e. no $d\theta_1 d\theta_2$ cross term.
However, our anticipation will be to have such a cross term in
the mirror, see e.g. \metcomptwo. We know that T-dualities {\it
cannot} generate such terms (in the absence of $B$ fields). In
the presence of $B$ fields, as we show below, cross term of the
form $d\theta_1 d\theta_2$ do get generated. However these cross
terms combine together with $dx$ and $dy$ terms (as will become
obvious soon) and therefore do not generate the {\it single}
$d\theta_1 d\theta_2$ term. We will discuss a way to generate
this later.

The expected mirror manifold will have the following form of the
metric \metcomp: \eqn\metcomthree{\eqalign{ds^2 = & {1\over
G_{zz}} (dz + {\cal B}_{\mu z} dx^\mu + {\cal B}_{xz} dx + {\cal
B}_{yz} dy)^2 + \cr & + G_{\mu\nu} dx^\mu dx^\nu + 2G_{x \nu} dx
dx^\nu + 2G_{y \nu} dy dx^\nu + 2G_{xy} dx dy + G_{xx} dx^2 +
G_{yy} dy^2.}} The $d\psi$ fibration structure is more or less
consistent in form, so lets check whether the components work out
fine. By denoting \eqn\defalpha{\alpha^{-1} = j_{xx} j_{yy} -
j^2_{xy} + b^2_{xy} = j_{xx} j_{yy} - j^2_{xy},} we write:
\eqn\bfibg{\eqalign{& {\cal B}_{xz} = -\alpha ~j_{xz} =
-\sqrt{\gamma'\o \gamma}~ \alpha~r~ {\rm cot}~\theta_1 \cr &
{\cal B}_{yz} = -\alpha ~j_{yz}  = -\sqrt{\gamma' \o \gamma +
4a^2} ~\alpha~r ~{\rm cot}~\theta_2 \cr & {\cal B}_{\mu z} =
b_{\mu z} + \alpha(b_{x\mu} j_{xz} + b_{y \mu} j_{yz}).}} This
can combine together with \bfibg\ to give the following fibration
structure: \eqn\actfibstr{(dz - b_{z\mu}~dx^\mu) - \alpha~j_{xz}
(dx - b_{x\theta_1}~d\theta_1) - \alpha~j_{yz} (dy -
b_{y\theta_2}~d\theta_2)} where we have kept the $B$ field
component $b_{\mu z}$. The above form of the fibration is highly
encouraging because it looks similar to (4.36). And since $\alpha
= 1 +$ higher orders, up to those terms we seem to be getting the
fibration structure in somewhat expected form.

\noindent Let us now look at other terms. \eqn\othtern{\eqalign{&
G_{xx} =  \alpha~j_{yy}, ~~~~ G_{yy} =  \alpha~j_{xx} \cr &
G_{\mu\nu} = j_{\mu\nu} + \alpha(j_{yy}~b_{x\mu}~b_{x\nu} +
j_{xx}~b_{y\mu} ~b_{y\nu} - j_{xy}(b_{y\mu} ~b_{x\nu} + b_{x\mu}
~b_{y\nu}))}} The existence of cross terms in the above formula
is very important. This tells us that we can have components like
$G_{\theta_1 \theta_2}$. Such terms do exist in the usual
deformed conifold metric and are {\it absent} in the  resolved
conifold metric. The other terms will be:
\eqn\ottretop{\eqalign{& G_{xy} = {-j_{xy} \o j_{yy}j_{xx} -
j^2_{xy} + b^2_{xy}} = -\alpha~j_{xy} \cr & G_{x\mu} =
\alpha(j_{yy}~b_{\mu x} + j_{xy} b_{y\mu}), ~~~~ G_{y \mu} =
\alpha(j_{xx}~b_{\mu y} + j_{xy} b_{x \mu})}} Again, the
existence of cross term is important, they will give rise to
components like $G_{\phi_1 \theta_2}$ and $G_{\phi_2 \theta_1}$.
Such terms are {\it not} present in the resolved conifold
setting, but do exist in the deformed conifold metric! Finally
there is the $zz$ component \eqn\gzzc{G_{zz} = \alpha} The above
term is again of order one. Let us furthermore introduce the
shorthand notation \eqn\defAandB{j_{xz} = A = \Delta_1~{\rm
cot}~\theta_1, ~~~~ j_{yz} = B = \Delta_2~{\rm cot}~\theta_2,}
with $\Delta_1$ and $\Delta_2$ being warp factors. Now combining
everything together we get the following mirror manifold:
\eqn\mirman{\eqalign{ds^2 = &~~ g_1~\left[(dz - b_{z\mu}~dx^\mu)
- \Delta_1~{\rm cot}~\theta_1~ (dx - b_{x\theta_1}~d\theta_1) -
\Delta_2~{\rm cot}~\theta_2~(dy - b_{y\theta_2}~d\theta_2)+
..\right]^2 + \cr &~~~~~ +  g_2~ d\theta_1^2 + g_3~d\theta_2^2   +
g_4~(dx - b_{x\theta_1}~d\theta_1)^2 +\cr & ~~~~~ + g_5~(dy -
b_{y\theta_2}~d\theta_2)^2 -  g_7~(dx -
b_{x\theta_1}~d\theta_1)(dy - b_{y\theta_2}~d\theta_2)}} where
$g_i \equiv g_i(r, \theta_1, \theta_2)$ are some functions of $r,
\theta_1, \theta_2$ coordinates and can be easily determined from
our analysis above. They are given as \eqn\gis{\eqalign{&
g_1=\alpha^{-1}, ~~~~~g_2={\gamma \sqrt{h} \o 4}, ~~~~~~ g_3=
{(\gamma+4a^2) \sqrt{h} \o 4}, \cr & g_4=\alpha~j_{yy}, ~~~~
g_5=\alpha j_{xx}, ~~~~~~ g_7=2\alpha~j_{xy}.}}

At this point let us compare our metric \mirman\ to the metric
of  the wrapped D6-branes on $S^3$ of a deformed conifold. The
generic form of that metric is given by (we take the {\it
delocalized} metric in \metcomptwo) \eqn\dsixdco{\eqalign{ds^2 =
& ~~ {\tilde g}_1~(dz + \tilde\Delta_1~{\rm cot}~\theta_1~dx +
\tilde\Delta_2~{\rm cot}~\theta_2~dy + ..)^2 + \cr & ~~ {\tilde
g}_2~ [d\theta_1^2 + dx^2] + {\tilde g}_3~[d\theta_2^2 + dy^2] +
{\tilde g}_4~[d\theta_1 ~d\theta_2 - dx~dy]}} where ${\tilde
g}_i$ are again some functions of $r, \theta_1, \theta_2$ that
could be easily evaluated. Let us now compare the two metrics:

\vskip.1in

\noindent $\bullet$ As a general rule, everywhere where we would expect
$dx$ or $dy$, they are replaced in \mirman\ by the appropriate
${\cal B}$--dependent fibration stucture
$(dx-b_{x\theta_1}d\theta_1)$ or $(dy-b_{y\theta_2}d\theta_2)$,
respectively. In fact, this non--trivial
fibration will be responsible for making the manifold \mirman\ a
non-K\"ahler space, as we will show later. If we define $d\hat
x=dx-b_{x\theta_1}d\theta_1$ and $d\hat y=dy-b_{y\theta_2}d\theta_2$
we find agreement between \mirman\ and \dsixdco\ in all terms involving
$dx$ and $dy$, up to differing warp factors.

\noindent $\bullet$ We also see that the $d\theta_1~d\theta_2$ cross
term is now entirely absorbed in the fibration structure and there is
{\it no single} $d\theta_1~d\theta_2$ term in \mirman.
Whatever $d\theta_1 d\theta_2$
terms are generated actually combine with $dx$ and $dy$ to give
us the ${\cal B}$-dependent fibration, and therefore no {\it extra}
$d\theta_1 d\theta_2$ term appears.

\noindent $\bullet$ Apart from the $dx$-- and $dy$--fibration mentioned above
the $d\psi$ (or $dz$ in our notation)
fibration structure of both the metrics have similar form modulo
some warp factors and relative signs.
The relative
signs between the $dx, dy$ terms in \mirman\ and \dsixdco\
can be fixed if we fix ${\cal B}_{yz},
{\cal B}_{zx}$ and ${\cal B}_{xy}$ in \bmuy, \bzx, and \bxy\ as
{\it minus} of themselves. In the following we will assume
that we have fixed the signs of the $B$ fields. This way the
fibration structures of \mirman\ and \dsixdco\ would tally.

\noindent $\bullet$ {}From the transformation \tranthe\ we expect
the  coefficients $g_3$ and $g_5$ to be the same. But a careful
analysis \gis\ shows that they are in fact different. The
coefficients $g_2$ and $g_4$ are also different, which could in
principle be because \tranthe\ do not act on them. But $g_{3,5}$
should be the same if we hope to recover the deformed conifold
scenario.

\noindent In the following we will try to argue a possible way
to  generate the $\theta$ cross terms. We will see that the
T-duality directions are slightly different from the naively
expected directions (which we called $x,y,z$). In the process we
will also fix the coefficients $g_i$. We begin with the search
for the $d\theta_1 ~d\theta_2$ term in the metric.

\subsec{Searching for the $d\theta_1 d\theta_2$ term}

The absence of the $d\theta_1 d\theta_2$ term in \mirman\ is a
near miss. As one can see that the metric that we get in \mirman\
is almost the metric of a deformed conifold (in the delocalized
limit) when we switch off the $B$ fields except that we are
missing the $d\theta_1 d\theta_2$ term. Furthermore this term
should come in the metric with the precise coefficient $\alpha
j_{xy}$.

The absence of this term however raises some doubts about the
directions that we made our T-dualities. But since we almost
reproduced the correct form of the metric, we cannot be too far
from the right choice of the isometry directions. Now whatever
new isometry directions we choose in the resolved
conifold\foot{We have been a little sloppy here. By resolved
conifold we will always mean $D5$ wrapped on $S^2$ of the
resolved conifold unless mentioned otherwise.}
 side should keep the present form of the mirror metric intact. This is a
 strong restriction because we cannot change the
$d\psi, dx$ and $dy$ fibration structure any more as they are
already in the expected format. The only things that we could
fiddle with are the $d\theta_i$ terms in the mirror side.
Therefore the question is: what changes in the resolved side are
we allowed to perform that would {\it only} affect the $d\theta_i$
parts of the mirror manifold?

An immediate guess would be to change the $\theta_i$ terms so as
to generate a $j_{\theta_1 \theta_2}$ directly in the resolved
conifold setup in type IIB theory \metresconi. A way to get this
would be to go to a new coordinate system given by:
\eqn\necosy{\theta_1 ~\to ~ \theta_1 + \gamma~\theta_2, ~~~~~
\theta_2 ~\to ~ \theta_2 + \beta~\theta_2,} where $\gamma, \beta$
are small integers. This will give us the necessary cross term
and will change the other terms to \eqn\otterto{{\rm
cot}~\theta_1~dx ~\to ~ (\gamma~\theta_2 + {\rm
cot}~\theta_1 + ...)~dx, ~~~ dx ~\to ~ (1 + \gamma~\theta_2~{\rm
cot}~\theta_1 + ...)~dx,} and similar changes to the $dy$ terms. In
other words the warp factors in front of the $dx,dy$ terms will
change and the metric will have a $j_{\theta_1 \theta_2}$ term.

Making a mirror transformation now to the resolved conifold
metric will generate the requisite $d\theta_1 d\theta_2$ term
term, but the coefficient of this is an arbitrary number.
Therefore will not explain the $\alpha j_{xy}$ coefficient that
we require. Instead of this, we can perform the following
infinitesimal rotation on the sphere coordinates of the type IIB
metric \metresconi: \eqn\rotonsp{ \pmatrix{dx \cr d\theta_1}~~\to
~~ \pmatrix{1 & \epsilon_1 \cr -\epsilon_1 & 1}\pmatrix{dx \cr
d\theta_1}, ~~~~~ \pmatrix{dy \cr d\theta_2}~~\to ~~ \pmatrix{1 &
\epsilon_2 \cr -\epsilon_2 & 1}\pmatrix{dy \cr d\theta_2}.} This
will keep the metric of the two spheres ($\theta_1, \phi_1$) and
($\theta_2, \phi_2$) invariant, but will generate $j_{z\mu},
j_{x\mu}, j_{y\mu}$ and $j_{\mu\nu}$ components. Interestingly,
one can easily verify from the T-duality rules that this changes
only the  $G_{\mu\nu}$ and $G_{z\mu}$ components with all other
metric components $G_{mn}$ invariant. The change in $G_{\mu\nu}$
can be written as \eqn\metchgmunu{\eqalign{G_{\mu\nu} =
j_{\mu\nu} & ~+ \alpha~[j_{yy}~b_{x\mu}b_{x\nu} +
j_{xx}~b_{y\mu}b_{y\nu} - j_{xy}(b_{y\mu}b_{x\nu} +
b_{x\mu}b_{y\nu})] ~+ \cr & ~ -\alpha~[j_{yy}~j_{x\mu}j_{x\nu} +
j_{xx}~j_{y\mu}j_{y\nu} - j_{xy}(j_{y\mu}j_{x\nu} +
j_{x\mu}j_{y\nu})].}} The second line is by replacing $b
\leftrightarrow j$. Observe that only the second line in
\metchgmunu, which we shall call $G_{\mu\nu}^{\rm new}$,
introduces new components in the metric. Similarly, $G_{z\mu}$
can be written as: \eqn\gmuz{G_{\mu z} = j_{\mu z} -
\alpha~[j_{yy}~j_{x\mu}j_{xz} + j_{xx}~j_{y\mu}j_{yz} -
j_{xy}(j_{y\mu}j_{xz} + j_{x\mu}j_{yz})].} Both these components
will contribute to $d\theta_1^2, d\theta_2^2$ and $d\theta_1
d\theta_2$ (by keeping only the $j$ components) in addition to
the terms that we already have in \mirman, as: \eqn\condstwo{ds^2
\to ds^2 + G_{\mu\nu}^{\rm new}~dx^\mu dx^\nu - {G_{z\mu}
G_{z\nu} \o G_{zz}}~ dx^\mu dx^\nu.}

This is almost what we might require, but again the coefficient
is arbitrary. And there seems no compelling reason for a
particular coefficient to show up in the metric. Both the above
analysis have failed to provide a specific reason for the $\alpha
j_{xy}$ coefficient in the metric. Therefore it is time now to
look at other possibilities as we have exploited the
transformations on $x, y$ and $\theta_i$, but haven't yet
considered the possibilities of a $\psi$ (or $z$) transformation.
Can we change the $dz$ terms without spoiling the consistency of
the mirror manifold?

A little thought will tell us that the allowed changes should be
done directly to the line element, so that we can define
distances properly in both type IIB as well as the mirror type
IIA. We can start defining some transformation on $x, y, z$ and
$\theta_i$ using some (as yet) unknown functions. However, this
procedure is rather involved, because eventually we have to
determine $dx, dy, dz$ and $d\theta_i$ thereby giving rise to set
of PDE's. Therefore it will be easier if we make transformations
{\it directly} on $dx, dy, dz$ and $d\theta_i$. Transformations
on the infinitesimal shifts are, on the other hand, not always
integrable. Therefore to avoid this problem, let us start by
making transformations on {\it finite} shifts $\delta x, \delta
y, \delta z, \delta \theta_i$. We will later integrate these
expressions to get the transformations on the coordinates $x, y,
z$ and $\theta_i$. As we will see below, this will still be a
rather involved procedure, so to simplify things a little bit, we
will consider shifts only for a fixed $r$. In other words, we
will ignore $\delta r$ variations (i.e {\it delocalise} along the
$r$ direction).
We like to remind the reader
that this is just to simplify the ensuing calculations. A more
detailed analysis will be presented elsewhere.

For {finite} shifts $\delta x, \delta y, \delta z, \delta
\theta_i$ of the coordinates of the resolved conifold a typical
distance $d$ on the resolved conifold can be written in terms of
distances $d_1, d_2$ on the two spheres (or more appropriately, tori,
although we will continue calling them spheres)
with coordinates ($x,
\theta_1$) and ($y, \theta_2$) as: \eqn\lineel{\|~ d ~\| =
\sqrt{d_1^2 + d_2^2 + (\delta z + \Delta_1~{\rm
cot}~\theta_1~\delta x + \Delta_2~{\rm cot}~\theta_2~\delta y)^2}}
where we have ignored the variations along the radial direction
$r$ just for simplicity. We have also defined the distances along
the two spheres as: \eqn\linesp{d_1 = {1\o 2} \sqrt{\gamma
\sqrt{h} ~(\delta\theta_1)^2 + 4(\delta x)^2}, ~~~~ d_2 = {1\o 2}
\sqrt{(\gamma + 4 a^2) \sqrt{h} ~(\delta\theta_2)^2 + 4(\delta y)^2}.} Now
the  allowed change in the resolved side that would affect only
the $\delta\theta_i$ parts will be to change $\delta z$
to\foot{This can be motivated as follows.
The generic shift between two
nearby points can be written as: $z_{(1)} - z_{(2)} = (\tilde
z_{(1)}-\tilde z_{(2)}) +
(\rho_{1(1)}~\theta_{1(1)}-\rho_{1(2)}~\theta_{1(2)}) +
 (\rho_{2(1)}~\rho_{2(1)}-\alpha_{2(2)}~\theta_{2(2)})$.
Now defining
$\rho_{j(1)}~\theta_{j(1)}-\rho_{j(2)}~\theta_{j(2)} =
\rho_j (\theta_{j(1)} - \theta_{j(2)})$, we get the
corresponding equation. Also since $\delta\theta_i = \theta_{i(1)} - \theta_{i(2)}$,
there is no problem with the periodicity here. This will be clear later when we integrate these
shifts and write the final result in terms of ${\rm cos}~\theta_i$ and ${\rm sin}~\theta_i$, which are
periodic variables.}
\eqn\psipsi{\delta z ~\to~ \delta z +
\rho_1~\delta\theta_1 + \rho_2~\delta\theta_2.} This typically
means that we are slanting the $z$ direction along the $\theta_i$
directions. This would convert the line element  in \lineel\ to
\eqn\conichange{\|~ d ~\|^2 = d_1^2 + d_2^2 + \left[\delta\tilde
z + \sum_{i=1}^2 (\Delta_i~{\rm cot}~\theta_i~\delta x_i +
\rho_i~\delta\theta_i)\right]^2} with $\tilde z$ being the new
$z$ direction and $\rho_i$ are generic functions of ($r,
\theta_i$). We have also defined $x_1= x$ and $x_2 = y$. Now
without a loss of generality we can write the $\rho_i$'s as:
$\rho_1 = f_1~\Delta_1~{\rm cot}~\theta_1$ and $\rho_2 =
f_2~\Delta_2~{\rm cot}~\theta_2$ where $f_i$ are now generic
functions of ($r,\theta_i$) that we have to determine. Using
this, the line element will take the final form \eqn\fibcha{\|~ d
~\|^2 = d_1^2 + d_2^2  + \left[\delta\tilde z + \sum_{i=1}^2
(\Delta_i~{\rm cot}~\theta_1~\delta x_i + f_i~\Delta_i~{\rm
cot}~\theta_i~\delta\theta_i)\right]^2} The above changes could
also be viewed as having  generated the following new components
of the metric in the resolved side: $j_{\tilde\psi \theta_1},
j_{\tilde\psi \theta_2}, j_{x \theta_1}, j_{y\theta_2}$ and
$j_{\theta_1 \theta_2}$. As we discussed before these components
will only change the $\delta\theta_i$ part of the metric, i.e.
the $\delta\theta_i^2 + \delta\theta_2^2$ part and will keep all
the fibrations in the mirror picture intact. The changes in the
mirror metric can therefore be calculated from \condstwo.

Until now the arguments have been more or less parallel to the
arguments that we provided earlier for the changes in the $\delta
x, \delta y$ or $\delta\theta_i$ terms. However, as we show
below, the changes in the $\delta z$ part actually allows us to
fix the form of the functions $f_i$. To see how this is possible,
make the following changes in $\delta x, \delta y$ coordinates:
\eqn\dxdychange{\delta x ~\to~ \delta x - f_1~\delta\theta_1,
~~~~~ \delta y ~\to ~ \delta y - f_2~\delta \theta_2.} The effect
of this change is rather immediately obvious: it removes the
effect of the changes made earlier by the $\delta z$
transformation. However this change in $\delta x, \delta y$ can
also be assumed  as though $j_{x\theta_1}$ and $j_{y\theta_2}$
cross terms have been added in the metric. Taking into account
all the above changes, and also allowing a finite shift along the
radial direction $\delta r$, the line element on the resolved
conifold will take  the following form:
\eqn\metresconichange{\eqalign{\|~d~&\|^2  ~= ~ h^{1/2} ~\gamma'
(\delta r)^2 + (\delta \tilde z + \Delta_1~{\rm cot}~\theta_1~
\delta \tilde x + \Delta_2~{\rm cot}~\theta_2 ~\delta\tilde y)^2
+ (\delta\tilde x)^2 +  (\delta\tilde y)^2 +\cr & + {1\o
4}~(\gamma \sqrt{h}+4f_1^2)~(\delta\theta_1)^2  - 2f_1
\delta\tilde x~ \delta\theta_1 + {1\o 4}~(\gamma \sqrt{h} +
4a^2\sqrt{h} + 4f_2^2)~(\delta\theta_2)^2 - 2 f_2~\delta\tilde y~
\delta\theta_2.}} Observe that in the above line element cross
components have developed and the distances along the $\theta_i$
directions have changed by warp factors. Both these changes are
given in terms of the unknown functions $f_i$ (to be determined
soon)\foot{The physical meaning of $f_i$ will be discussed in the
next subsection. For the time being we will view $f_i$ as being a consequence
of generic coordinate transformations, whose integral form will be presented
later in this section.}.

It is now important to consider some special limits of the
functions  $f_i$, as one can easily show that for small and
finite values of $f_1$ and $f_2$, a $\delta\theta_1
\delta\theta_2$ term does {\it not} get generated in the mirror
picture. Assuming that the $f_i$ are large, one could use a
special regularization scheme that would generate this term. The
coordinates $\tilde z, \tilde x_i, \theta_i$ are now the
coordinates in which the line element of the mirror manifold is
in a known format. Therefore the above transformation from ($z,
x, y$) ~$\to$~($\tilde z, \tilde x, \tilde y$) for the resolved
conifold means that we are writing the line element in the
coordinates of the mirror. To simplify the ensuing calculations,
let us use the notation introduced in \defAandB. With this
definition, we can re-express the line element \metresconichange\
as new components for the resolved conifold metric. The various
components can now be written as: \eqn\varcompo{\eqalign{&
j_{\tilde x \tilde x} = 1 + A^2, ~~ j_{\tilde y \tilde y} = 1 +
B^2, ~~ j_{\tilde x \theta_1} = - f_1, ~~ j_{\tilde y \theta_2} =
f_2 \cr & j_{\theta_1 \theta_1} = {1\o 4}(4 f_1^2 + {\gamma
\sqrt{h}}), ~~ j_{\theta_2 \theta_2} = {1\o 4}(4f_2^2 + {\gamma
\sqrt{h}} + 4a^2\sqrt{h})  \cr &  j_{\tilde x \tilde z} = A, ~~
j_{\tilde y \tilde z} = B, ~~ j_{\tilde z \tilde z} = 1-
\epsilon, ~~ j_{\tilde x \tilde y} = AB}} with the radial and the
spacetime components remaining the same as earlier. Observe that
we have shifted the $j_{\tilde z \tilde z}$ component by a small
amount $\epsilon$ in \varcompo. Letting $\epsilon \to 0$ and
$f_i\to\infty$ will result in a finite $d\theta_1 d\theta_2$--term
in the mirror metric. This is our regularization scheme, so to
speak. Ergo, \varcompo\ is the correct IIB starting metric, which
we T--dualize along $\tilde x, \tilde y$ and $\tilde z$ to obtain
the mirror manifold. It can also be easily verified that both the
$B_{NS}$ and the $H_{RR}$ (given in the next section) remain
completely unchanged in forms because of their wedge structures and
antisymmetrisations. The only change there is that now
everything is written by tilde-coordinates.

Let us now see the possible additional metric components that we
can  get after we make a mirror transformation. From the
T-duality rules we see that the new components are:
\eqn\newcom{\eqalign{& G_{\tilde z\theta_1} = -\alpha j_{\tilde
x\theta_1}[j_{\tilde x\tilde z} j_{\tilde y\tilde y} - j_{\tilde
x\tilde y} j_{\tilde y\tilde z}] = \alpha~ f_1~A \cr & G_{\tilde
z\theta_2} = -\alpha j_{\tilde y\theta_2}[j_{\tilde y\tilde z}
j_{\tilde x\tilde x} - j_{\tilde x\tilde y} j_{\tilde x\tilde z}]
= \alpha ~f_2~B \cr & G_{\theta_1 \theta_1} = j_{\theta_1
\theta_1} + \alpha~j_{\tilde y\tilde y}[b_{\tilde x\theta_1}^2 -
j_{\tilde x\theta_1}^2] = {\gamma\sqrt{h}\o 4} +
\alpha~(1+B^2)~b_{\tilde x\theta_1}^2 + \alpha~f_1^2 A^2 \cr &
G_{\theta_2 \theta_2} = j_{\theta_2 \theta_2} + \alpha~j_{\tilde
x\tilde x}[b_{\tilde y\theta_2}^2 - j_{\tilde y\theta_2}^2] =
{(\gamma + 4a^2)\sqrt{h} \o 4} + \alpha~(1+A^2)~b_{\tilde
y\theta_2}^2 + \alpha~f_2^2 B^2 \cr & G_{\theta_1 \theta_2} =
-\alpha j_{\tilde x\tilde y} [ b_{\tilde x\theta_1} b_{\tilde
y\theta_2} - j_{\tilde x\theta_1} j_{\tilde y\theta_2}] =
-\alpha~AB~b_{\tilde x\theta_1}~ b_{\tilde y\theta_2} +
\alpha~f_1~f_2~AB \cr & G_{\tilde z\tilde z} = \alpha(j_{\tilde
x\tilde x}j_{\tilde y\tilde y}j_{\tilde z\tilde z} -j_{\tilde
x\tilde x} j^2_{\tilde y\tilde z} - j_{\tilde y\tilde
y}j^2_{\tilde x\tilde z} - j_{\tilde z\tilde z} j^2_{\tilde
x\tilde y} + 2 j_{\tilde x\tilde y}j_{\tilde y\tilde z}j_{\tilde
z\tilde x}) = {\alpha - \epsilon}.}} The $B$ field dependent
terms in \newcom\ would reorganize themselves according to the
fibration structure that we discussed earlier. What remains now
is to see whether the additional terms (which depend on $f_i$)
can be used effectively. The distance along the
 new $\theta_1 \theta_2$ directions will be \condstwo:
\eqn\metonetwo{\eqalign{ds^2_{\theta_1 \theta_2} & = 2
\left(G_{\theta_1\theta_2}^{\rm new} - {G_{\tilde z\theta_1}
G_{\tilde z\theta_2} \o G_{\tilde z\tilde z}}
 \right)~\delta\theta_1~\delta\theta_2 \cr
& = - 2\alpha ~ f_1 f_2~j_{\tilde x\tilde y}\left[{\epsilon \o
\alpha - \epsilon}\right] ~\delta\theta_1~\delta\theta_2=
 - 2f_1f_2 ~j_{\tilde x\tilde y}
  ~\epsilon~\delta\theta_1~\delta\theta_2.}}
At this point we can use our freedom to define   the functions
$f_1$ and $f_2$. As discussed earlier, we see that for finite
values of the functions $f_i$ the above metric component is
identically zero in the limit $\epsilon\to 0$. However if we
define $f_i \equiv \epsilon^{-1/2} \beta_i$ such that $\beta_1
\beta_2 = -\alpha$, then \metonetwo\ implies
\eqn\metfin{ds^2_{\theta_1 \theta_2} = 2\alpha~j_{\tilde x\tilde
y}~\delta\theta_1 \delta\theta_2} which is what we require. In
this way we recover the elusive $\theta_1\theta_2$ component of
the metric.

At this point one might ask whether different choices for $f_1
f_2$ could be entertained. From the mirror metric \mirman\ we
observed that $-$ in the absence of $B$-fields $-$ the metric
resembles the deformed conifold in the delocalized limit (of
course with the $d\theta_1 d\theta_2$ absent). When we restore
back this term (via the above analysis) we should recover the
exact deformed conifold setup. This is possible, if in the
absence of $B$ fields, the product $f_1 f_2$ is proportional to
$\alpha$. Therefore, in the presence of fluxes, we believe that
this will continue to hold.

To show that the choice of $\beta_1 \beta_2 = -\alpha$ is
consistent, we have to determine $f_i$ individually. To see this
let us first bring the metric for the $\theta^2$ terms in
\newcom\ into a more canonical form:
\eqn\scaletheta{\delta\theta_1 ~\to~ {2 \o
\sqrt{h^{1/2}\gamma}}~\delta\tilde\theta_1, ~~~~~~~
\delta\theta_2 ~\to~ {2 \o
\sqrt{h^{1/2}(\gamma+4a^2)}}~\delta\tilde\theta_2.} Assuming now
that the above equation can be integrated to give a relation
between $\theta_i$ and $\tilde\theta_i$, the change in all terms
with a $b$--fibration can easily be absorbed in \bfico\ as:
\eqn\tildeb{\eqalign{b  & = ~~{2 {\cal J}_1 \o
\sqrt{h^{1/2}\gamma}}~d\tilde\theta_1 \wedge d\tilde x + {2 {\cal
J}_2 \o \sqrt{h^{1/2}(\gamma+4a^2)}}~d\tilde\theta_2 \wedge
d\tilde y \cr &= ~~ \tilde{b}_{\theta_1 x}~ d\tilde\theta_1
\wedge d\tilde x + \tilde{b}_{\theta_2 y}~ d\tilde\theta_2 \wedge
d\tilde y,}} where we have taken the infinitesimal limit to write
$b$ with one forms $d\theta_i$. The relation \tildeb\  implies
for the $\theta$ dependent metric components:
\eqn\gththco{\eqalign{G_{\tilde\theta_1 \tilde\theta_2} & =
-\alpha ~AB~\tilde{b}_{x\theta_1} \tilde{b}_{y\theta_2} +
\alpha~{4AB~f_1 f_2\o h^{1/2}\sqrt{\gamma(\gamma+4a^2)}} \cr
G_{\tilde\theta_1 \tilde\theta_1} & = 1 + \alpha (1+B^2)~
\tilde{b}_{x\theta_1}^2 + \alpha~{4\o h^{1/2}\gamma}~f_1^2 A^2,
\cr G_{\tilde\theta_2 \tilde\theta_2} & = 1 + \alpha (1+A^2)
~\tilde{b}_{y\theta_2}^2+ \alpha~{4\o h^{1/2}
(\gamma+4a^2)}~f_2^2 B^2.}} Note, that this changes \metonetwo\
to \eqn\metonetilde{\eqalign{ds^2_{\theta_1 \theta_2} & = -
2~{4f_1f_2 ~AB\o h^{1/2}\sqrt{\gamma (\gamma+4a^2)}}
~\epsilon~\delta\tilde\theta_1~\delta\tilde\theta_2,}} so we now
want to require \eqn\betaonetwo{\beta_1 \beta_2 = - {\alpha\o
4}~h^{1/2}\sqrt{\gamma(\gamma+4a^2)}.} To find out if this is
consistent with the other metric components, take a look at
\eqn\thetatwotwo{\eqalign{ds^2_{\tilde\theta_2 \tilde\theta_2}~
&~ = ~\left(G_{\tilde\theta_2 \tilde\theta_2}~
 - {G_{\tilde z\tilde \theta_2}
 G_{\tilde z\tilde\theta_2} \o G_{\tilde z\tilde z}}\right)
~(\delta\tilde\theta_2)^2 \cr &~ = ~\left(1 + \alpha (1+ A^2)
\tilde{b}_{y\theta_2}^2 - {4B^2~\beta_2^2 \o
h^{1/2}(\gamma+4a^2)}\right)~(\delta\tilde\theta_2)^2.}} {}From
the above analysis we see that the $\tilde b_{y\theta_2}$ term
will join $\delta\tilde y$ in the fibration as shown in \mirman,
and the rest of the term (which depends on $\beta_2$) will act as
a warp factor. The $\delta\tilde y$ term has warp factor $g_4$.
Can we use this to determine $\beta_2$?

At first this may seem impossible as we can have any coefficients
in front of $(\delta\tilde\theta_2)^2$ and $(\delta\tilde y -
\tilde{b}_{y\theta_2}~\delta\tilde\theta_2)^2$ or
$(\delta\tilde\theta_1)^2$ and the corresponding $(\delta\tilde x
- \tilde{b}_{x\theta_1}~\delta\tilde\theta_1)^2$. A little
thought will tell us that this is not quite true. Of course we
are allowed to have {\it any} coefficients in front of
$(\delta\tilde\theta_1)^2$ and $(\delta\tilde x - \tilde
b_{x\theta_1}~ \delta\tilde\theta_1)^2$, but the case for
$(\delta\tilde\theta_2)^2$ and
 $(\delta\tilde y - \tilde b_{y\theta_2}
~\delta\theta_2)^2$ is different. This is because of the
transformation \tranthe.  Under this transformation (in the
absence of fluxes) the coefficients of $\delta\theta_2$ and
$\delta y$ should be same so that under \tranthe\ the line
element does not change. Now we expect similar thing when we
switch on fluxes if we denote $\delta\hat y = \delta\tilde y -
\tilde b_{y\theta_2}~\delta\tilde\theta_2$, and define an
equivalent transformation between ($\delta\hat y,
\delta\tilde\theta_2$). Notice that there is no such constraint
on the $\delta\tilde x, \delta\tilde\theta_1$ term. Therefore
from the above argument and \gis\ we have the following equality:
\eqn\ftwos{ 1 - {4\o h^{1/2}(\gamma+4a^2)}~\beta_2^2 B^2 = \alpha
(1+A^2) ~~~\Rightarrow ~~ \beta_2 = \pm {1\o
2}\sqrt{\alpha~h^{1/2}(\gamma+4a^2)}.} Thus we fix $\beta_2$ or
equivalently $f_2$. As expected the line element for the
$\theta_1^2$ component can now be written in terms of $\beta_1$
as \eqn\metoneone{ds^2_{\theta_1 \theta_1} = \left(1 + \alpha~(1
+ B^2) \tilde{b}_{x\theta_1}^2 - {4A^2~\beta_1^2\o
h^{1/2}\gamma}\right)(\delta\tilde\theta_1)^2.} The $b$ dependent
term can be equivalently absorbed in the $\delta\tilde x$
fibration structure as $\delta{\hat x} \equiv \delta\tilde x +
\tilde b_{x\theta_1}~\delta\tilde\theta_1$. The rest of the
remaining term serve as warp factor for the
$(\delta\tilde\theta_1)^2$ term. What happens now if we argue an equality
between the coefficients of the
$\delta\tilde\theta_1$ and $\delta{\hat x}$ terms?
This would imply:
\eqn\foneequa{1 - \beta_1^2~{4\o h^{1/2}\gamma}~ A^2 = \alpha
(1+B^2) ~~ \Rightarrow ~~ \beta_1 = \pm {1\o
2}\sqrt{\alpha~h^{1/2}\gamma}.} {}From \ftwos\ and \foneequa\ we
see that we can indeed choose the signs in a way that fulfills
\betaonetwo ! This is consistent with our assumption, implying
that in this setup we will see the sphere metrics appear with one
unique warp factor. This is again expected in the case without
fluxes. What we see here is that, this remains true for the case
with fluxes also.

To summarize, we see that the T-duality directions  are $\tilde
z, \tilde x, \tilde y$ on the resolved side, and the line element
is more or less the same as \lineel\ with additional cross
components \metresconichange. To go from \lineel\ to
\metresconichange\ we have performed some transformations on the
finite shifts $\delta z, \delta x, \delta y$ and
$\delta\theta_i$. Defining three vectors $V_i$ as
\eqn\vectors{V_1 = \pmatrix{\delta z\cr \delta\theta_1\cr
\delta\theta_2}, ~~~ V_2 = \pmatrix{\delta x\cr \delta\theta_1\cr
\delta\theta_2}, ~~~ V_3 = \pmatrix{\delta y\cr \delta\theta_1\cr
\delta\theta_2}} and three matrices as: \eqn\matdef{\eqalign{ &
M_1 = \pmatrix{1& \Delta_1~f_1~{\rm cot}~\theta_1&
\Delta_2~f_2~{\rm cot}~\theta_2\cr 0& 1& 0\cr 0& 0& 1}, ~~~ M_2 =
\pmatrix{1& -f_1& 0\cr 0& 1& 0\cr 0& 0& 1},\cr & M_3 =
\pmatrix{{\rm cos}~\psi & 0 & {\rm sin}~\psi - f_2~{\rm cos}~\psi
\cr 0 & 1& 0\cr -{\rm sin}~\psi & 0 & {\rm cos}~\psi + f_2~{\rm
sin}~\psi}}} we can convert the line element \lineel\ to
\metresconichange\ via the transformations: \eqn\transf{V_1~\to
~M_1~V_1, ~~~~ V_2~\to ~M_2~V_2, ~~~~ V_3~\to~M_3~V_3.} This will
generate the T-duality directions $\tilde\psi, \tilde\phi_1,
\tilde\phi_2$, or $\tilde z, \tilde x, \tilde y$, as mentioned
above\foot{One has to further delocalize the $\delta z$ fibration
structure by putting $\psi = 2\pi$ in the final result so that
T-dualities could be performed. As mentioned earlier, this is the
only assumption that we will consider.}.

We can now try to go to the infinitesimal shifts given by the
forms $dx_i, d\theta_i$. At this point we therefore assume  that
we can replace \eqn\relshif{(\delta\tilde
x_i,~\delta\tilde\theta_i,~ \delta\tilde z) ~ \to ~ (d\tilde
x_i,~ d\tilde\theta_i,~ d\tilde z)} in the final line element.
This would imply that the line element with finite shifts can
serve as the metric for the mirror manifold. This is somewhat
strong assumption as the transformations that we made on the
finite shifts $\delta x_i, \delta z, \delta \theta_i$ may not
always be extrapolated to transformations on infinitesimal
shifts. It turns out that the transformation that we made can be
extrapolated if we assume that these transformations are
restricted on the plane ($x, y, z, \theta_i$), with the
constraint $\delta r = 0$.

To see this first view the $z$ coordinate (or the $\psi$
coordinate) to be determined in terms of $\psi_1$ and $\psi_2$ as
$z \equiv \psi_1 - \psi_2$. In fact, as we will soon encounter in
the M-theory section, the coordinate $z$ which will eventually be
the $z$ coordinate of the type IIA mirror manifold, and the
M-theory eleventh direction $x_{11}$ can be written in terms of
$\psi_1$ and $\psi_2$ as: \eqn\psionetwom{dz \equiv d\psi_1 -
d\psi_2, ~~~~~~ dx_{11} = d\psi_1 + d\psi_2.} This means that in
the type IIB picture we can distribute the coordinates on two
different $S^3$'s parametrized by: ($\psi_1, x, \theta_1$) and
($\psi_2, y, \theta_2$). Existence of two $S^3$ here is just for
book keeping, and will eventually be related to the real $S^3$'s
of the mirror manifold.

Let us now consider one $S^3$ with coordinates ($\psi_1, x,
\theta_1$). This $S^3$ is at a fixed $r$ and fixed ($\psi_2, y,
\theta_2$). From the analysis done above, we see that the
transformations generically lead to an integral of the form
(compare e.g. to \fibcha): \eqn\intg{\int (b^2 + a^2~{\rm
cot}^2~\theta_1)^{{m\o 2}}~{\rm cot}^n~\theta_1 ~d\theta_1} where $a$ and
$b$ are some constants on the sphere $S^3$, $n = 0, 1$ and
$m = 1, -1$.\foot{In the notations of Appendix 1, the constants $a,b$ can be
extracted from $\langle \alpha \rangle_1^{\pm{1\o 2}}$
measured at a constant radius.
There is also an
overall constant related to the expectation value $\langle \alpha \rangle$ that
we ignore here. Replacing $\theta_1$ by $\theta_2$ the constants $a,b$
should now be extracted from $\langle \alpha \rangle_2^{\pm{1\o 2}}$.}
It turns out that the shifts $\delta\psi_1, \delta x$ etc. can be
integrated to yield the following transformations on the
coordinates $\psi_1, x, \theta_1$: \eqn\coordinate{\eqalign{&
\psi_1 ~\to~ \psi_1 - {\left[1 - \left({a^2 - b^2 \o a^2}
\right)~{\rm sin}^2~\theta_1\right]^{m+1\o 4}
 \o \sqrt{\epsilon}~(a^{-1} {\rm sin}~\theta_1)^{m+1\o 2}} -
{\left(a^2 - b^2 \right)^{m\o 2} \o \sqrt{\epsilon}}
~{\rm sin}^{-1}~\left[\left({a^2 -
b^2 \o a^2} \right)^{1\o 2}~ {\rm sin}~\theta_1 \right] \cr & x ~\to ~ x -
{\left(a^2 - b^2 \right)^{m\o 2} \o \sqrt{\epsilon}}
{\rm ln}~{\left[ \left({a^2 - b^2 \o a^2}
\right)^{1\o 2} ~{\rm cos}~\theta_1 + \sqrt{1 - \left({a^2 - b^2 \o
a^2} \right)~{\rm sin}^2~\theta_1}
 \right]^{m+1 \o 2} \o \left[{\sqrt{1 - \left({a^2 - b^2 \o a^2}
\right)~
{\rm sin}^2~\theta_1} + \left({a^2 - b^2 \o a^2} \right)^{1-m \o 2}~
{\rm cos}~\theta_1 \o \sqrt{1 - \left({a^2 - b^2 \o a^2}
\right)~
{\rm sin}^2~\theta_1} - \left({a^2 - b^2 \o a^2} \right)^{1-m \o 2}
~{\rm cos}~\theta_1}\right]^{1\o (m+1)\sqrt{a^2-b^2}}}}}
with $\theta_1$ transforming as
$\theta_1 ~\to~ c\cdot \theta_1$ where $c$ is another constant. (Observe that these
expressions have the required periodicity).
In the above transformations we have inherently assumed that $a >
b$. What happens for the case $a < b$?

This case turns out to be rather involved, but nevertheless
do-able. The transformation can again be integrated for both $m = \pm 1$
to give us
the following results (here we present the result only for $m = 1$):
\eqn\aleb{\eqalign{& \psi_1 ~\to~ \psi_1 +
\left(b^2 - a^2 \o a~\sqrt{\epsilon} \right)~{\rm ln}~{\left[\left(b^2 - a^2 \o
a^2 \right) ~{\rm sin}~\theta_1 + \sqrt{1 + \left(b^2 - a^2 \o
a^2 \right)~{\rm sin}^2~\theta_1}\right] \o {\rm exp}~\left[
{a^2~\sqrt{1 + \left(b^2 - a^2 \o a^2 \right)~{\rm
sin}^2~\theta_1} \o (b^2 - a^2)~ {\rm sin}~\theta_1}\right]}  \cr
& x ~\to ~ x - {a\o 2\sqrt{\epsilon}}~ {\rm ln}~ \left[{\sqrt{1 + \left(b^2 -
a^2 \o a^2 \right)~{\rm sin}^2~\theta_1}
 + {\rm cos}~\theta_1 \o \sqrt{1 + \left(b^2 -
a^2 \o a^2 \right)~{\rm sin}^2~\theta_1}
- {\rm cos}~\theta_1}\right] + \left(b^2 -
a^2 \o a~\sqrt{\epsilon} \right)~{\rm sin}^{-1}~ {\left(b^2 - a^2 \o a^2
\right)~{\rm cos}~\theta_1 \o \sqrt{1 + \left(b^2 - a^2 \o a^2
\right)^2}}}} with $\theta_1$ transformation remaining the same.
For the other $S^3$ the $y$ and $\theta_2$ transformation would
look similar to the above transformations on $x$ and $\theta_2$
transformations respectively. However $\psi_2$ transformation
will differ by relative signs. More details on the effect of these
transformations on the mirror metric is given in Appendix 1.

Taking the above transformations
 into account, and then performing the three T-duality
transformations we get the final mirror manifold (in the
delocalised limit) with the following form of the metric written
with $dx, dy, dz$ and $d\theta_i$:
\eqn\mirmanchange{\eqalign{ds^2 = &~~ g_1~\left[(dz -
b_{z\mu}~dx^\mu) + \Delta_1~{\rm cot}~\hat\theta_1~ (dx -
b_{x\theta_1}~d\theta_1) + \Delta_2~{\rm cot}~\hat\theta_2~(dy -
b_{y\theta_2}~d\theta_2)+ ..\right]^2 + \cr &~~~~~~~ + g_2~
[d\theta_1^2  + (dx - b_{x\theta_1}~d\theta_1)^2] +
g_3~[d\theta_2^2 + (dy - b_{y\theta_2}~d\theta_2)^2] +  \cr
& ~~~~~~~ + g_4~ [ d\theta_1~d\theta_2 - (dx - b_{x\theta_1}~d\theta_1)(dy -
b_{y\theta_2}~d\theta_2)]}} where we have used un-tilded
coordinates to avoid clutter (we will continue using this
coordinates in the rest of the paper unless mentioned otherwise).
The dotted part in the $dz$ fibration are the corrections to
$\theta_i$  terms from the scaling etc. We have written ${\rm
cot}~\hat\theta_i$ instead of ${\rm cot}~\theta_i$ to emphasize
the change in $\theta_i$. Its is interesting to note that (as we
saw earlier) this is the only change in $\theta$ because of
\scaletheta. All other changes due to scalings etc. have been
completely incorporated! Observe also that we now require only
four warp factors $g_1, g_2, g_3$ and $g_4$ instead of six that
we had earlier in \gis. The precise warp factors  can now be
written explicitly as: \eqn\gisnow{\eqalign{& g_1 = \alpha^{-1},
~~~g_2 = \alpha~j_{yy}, ~~~ g_3 = \alpha~j_{xx}, ~~~ g_4 =
2\alpha~j_{xy} \cr &~~~~~~~~~~~ \Delta_1 = \sqrt{\gamma'\o
\gamma} ~r~\alpha, ~~~ \Delta_2 = \sqrt{\gamma'\o \gamma + 4a^2}~r~\alpha}}
where $\alpha$ and $j$ are defined in \defalpha\ and \comedfi\ (with
${\rm cot}~\theta_i$ changed to ${\rm cot}~\hat\theta_i$, and we use the
same symbol $\Delta_i$ to denote the new $\Delta_i$),
respectively, and the $b$ fields have been rescaled according to
\tildeb. With these values the metric \mirmanchange\ can be
compared to \dsixdco.

\subsec{Physical meaning of $f_1$ and $f_2$}

The transformations that we performed in the previous subsection
to bring the metric in the form \metresconichange\ using the
functions $f_1$ and $f_2$ can be given some physical
meaning\foot{The discussion in this section is motivated from the
conversations that one of us (R.T) had with the UPenn group,
especially V. Braun and M. Cvetic.}. As discussed earlier, the
conversion from ($\phi_1, \theta_1$) and ($\phi_2, \theta_2$) to
($x, \theta_1$) and ($y, \theta_2$) coordinates respectively is to
write the metric as the metric of tori. Now the metric of the tori
can be generically written in terms of complex structures $\tau_i$
where $i = 1,2$ represent the two tori. We can define
\eqn\cstr{dz_1 = dx - \tau_1~d\theta_1, ~~~~~~~ dz_2 = dy -
\tau_2~d\theta_2} as the two coordinates of the two tori. The
metric \lineel\ can therefore be written in terms of $dz_i$ as:
\eqn\linaga{ds^2 = (dz + \Delta_1~{\rm cot}~\theta_1 ~dx +
\Delta_2~{\rm cot}~\theta_2 ~dy)^2 + \vert dz_1 \vert^2 + \vert
dz_2 \vert^2,} with the complex structure not yet specified. The
transformation that we performed in the previous section, would
therefore correspond to the following choice of the complex
structure of the base tori: \eqn\csbase{\tau_1 = f_1 + {i\o 2}
\sqrt{\gamma \sqrt{h}}, ~~~~~~~~ \tau_2 = f_2 + {i\o 2}
\sqrt{(\gamma + 4 a^2) \sqrt{h}}} where we have already defined
$\gamma, a$ in earlier sections. Observe that when $f_1 = 0 = f_2$
then the base is a torus with complex structure \eqn\basecs{\tau_1
= {i\o 2} \sqrt{\gamma \sqrt{h}},~~~~~~ \tau_2 = {i\o 2}
\sqrt{(\gamma + 4a^2) \sqrt{h}}} In this limit the metric has no
cross terms. This is basically the metric that we started off
with. The transformations in the previous subsection are therefore
to convert $\tau_i \to \tau_i + f_i$ via $SL(2, R)$
transformations on the two tori\foot{For example using local
$SL(2, R)$ matrices $\pmatrix{1&f_i\cr 0&1}$.}. In the limit when
$f_i$ are very large, the base of the six dimensional manifold
($\theta_1, \theta_2, r$) is very large compared to the $T^3$
fiber ($x, y, z$). This situation is consistent with the fact that
the generalized SYZ transformations require similar condition
\syz\ for mirror rules to work properly (see also \louis). On the
other hand, this limit is precisely the opposite to the one where
the geometric transition takes place. This is one reason why we
have to go through non-trivial manipulations to get to the final
metric of the mirror manifold\foot{We thank the referee for
pointing this out.}.

\subsec{$B$ Fields in the Mirror Setup}

There is another possibility that we haven't entertained yet.
This is to allow new $B$ field components in the resolved side.
Observe that the analysis presented above was done from the
resolved conifold setup when we only had $B_{NS}$ fields with
components $b_{x\theta_1}, b_{y\theta_2}$ and $b_{z\mu}$. What
happens if we switch on a cross component $b_{xy}$ that has legs
on both the spheres in the type IIB resolved conifold setup? Can
this generate a $d\theta_1 d\theta_2$ term? First, of course this
will not convert to a component of the metric under a mirror
transformation and will appear in the type IIA framework as a
$B$ field. This $B$ field will become a threeform field in
M-theory. We will discuss this later. Second, the mirror metric
will change. This change can be easily evaluated $-$ and we shall
do this below $-$ but before that lets see whether it is indeed
possible to switch on such component in the type IIB setup.

To analyze this, we go back to our fourfold scenario that we had
in section 3.  In the fourfold setup, no matter what choice we
make for the components of the $G$ fluxes, the metric will retain
its warped form and the only thing that could change will be the
exact value of the warp factor. Therefore we can choose an
additional component, say $G_{587a}$, and get the corresponding
$b_{xy}$ flux in type IIB.

Now under this choice of $B$ field, the metric will have the
additional  term $-{1\o G_{zz}}(G_{z\mu} dx^\mu + G_{zx} dx +
G_{zy} dy )^2$. One can show that the form of the $d\psi$
fibration structure remains the same, although the values will
differ by additional $b_{xy}$ terms. In the notations of the
earlier sections, let us assume the form of the metric to be:
\eqn\metfassu{\eqalign{ ds^2 = &~~(dz + {\rm fibration})^2 +
{\cal G}_{xx}~dx^2+ {\cal G}_{yy}~dy^2
 + {\cal G}_{\theta_1 \theta_1}~d\theta_1^2 +
 {\cal G}_{\theta_2 \theta_2}~d\theta_2^2 + 2{\cal G}_{xy}~dx dy + \cr
& ~~~~~~~ 2{\cal G}_{x\theta_1}~dx d\theta_1 + 2{\cal G}_{y
\theta_2}~dy d\theta_2 + 2{\cal G}_{x \theta_2}~dx d\theta_2 +
2{\cal G}_{y \theta_1} ~dy d\theta_1 + 2{\cal G}_{\theta_1
\theta_2}~d\theta_1 d\theta_2.}} Using the mirror rules given
earlier, one can work out all these     components. Since
$b_{xy}$ is non-zero, the analysis gets a little involved. If we
define again $\alpha^{-1} =j_{xx}j_{yy}-j_{xy}^2+b_{xy}^2$, this
time $b_{xy}$ being different from zero, we arrive exactly at the
form \mirman: \eqn\metnowinty{\eqalign{&{\cal G}_{xx}~dx^2+
2{\cal G}_{x\theta_1}~dx d\theta_1 + {\cal G}_{\theta_1
\theta_1}~d\theta_1^2 = \alpha ~j_{yy}~[d\theta_1^2 + ~(dx -
b_{x\theta_1}~d\theta_1)^2] \cr & {\cal G}_{yy}~dy^2 + 2{\cal
G}_{y \theta_2}~dy d\theta_2 +{\cal G}_{\theta_2 \theta_2}~
d\theta_2^2 = \alpha~j_{xx}~[d\theta_2^2 + (dy - b_{y\theta_2}~
d\theta_2)^2] \cr & {\cal G}_{xy}~dx~dy+ {\cal G}_{\theta_1
\theta_2}~d\theta_1 d\theta_2 + {\cal G}_{x \theta_2} ~dx
d\theta_2+ {\cal G}_{y \theta_1}~dy d\theta_1 = \cr &
~~~~~~~~~~~~~~-\alpha~j_{xy} (dx - b_{x\theta_1}~d\theta_1) (dy -
b_{y\theta_2}~d\theta_2)].}} Where we have given the precise warp
factors of every terms. Therefore, introducing a new component of
the $B$ field has not changed the form of the metric and has
failed to generate the $d\theta_1 d\theta_2$ term. What changes,
in the final mirror picture, is that we will now have a non zero
$B_{NS}$ flux in type IIA setup. We will comment on this later.

The above choice of $B$ fields in the resolved conifold side is
highly  unnatural (although it may be allowed from the
supergravity analysis). A more natural way to generate a $B$
field in the mirror side has already been taken into account when
we switched to the tilde-coordinates. In fact the cross terms in
the resolved conifold metric \metresconichange\ i.e. the $j_{x
\theta_1}$ and the $j_{y \theta_2}$ terms will be responsible to
give a non-zero $B$ field in the mirror picture. This way we can
give another physical meaning to the shifts that we performed in
\dxdychange. The background $B$ field in the type IIA background
can now be written in terms of the deformed conifold coordinates
as: \eqn\bintwoa{\eqalign{{\tilde B} = &~{2f_1\o
\sqrt{h^{1/2}\gamma}}~d\tilde x \wedge d{\tilde\theta_1} + {2f_2
\o \sqrt{h^{1/2}(\gamma + 4a^2)}}~d\tilde y \wedge
d{\tilde\theta_2}  \cr &~ + \left({2Af_1 \o
\sqrt{h^{1/2}\gamma}}~d\tilde\theta_1 + {2Bf_2 \o
\sqrt{h^{1/2}(\gamma + 4a^2)}}~d\tilde\theta_2\right) \wedge
d{\tilde z}}} with all other components vanishing in the limit
$\epsilon \to 0$. These $B$ fields are in general large, because
the transformations \dxdychange\ that we performed in the
resolved conifold setup is large. In the limit where we define
$\tilde B \equiv \epsilon^{-1/2} \hat B$, the finite part $\hat
B$ will be given by: \eqn\hatb{{\hat B \o \sqrt{\alpha}} = d{x}
\wedge d\theta_1 - d{y} \wedge d\theta_2 + (A~d\theta_1 -
B~d\theta_2) \wedge d{z}} modulo an overall sign if we employ the
opposite choice in \ftwos\ and \foneequa. Again, we have omitted
the tildes in the final expression. Notice also the following interesting
facts:

\noindent $\bullet$ We can replace $dx$ and $dy$ by the corresponding one forms
$d\hat x$ and $d\hat y$ because of the wedge structure (at constant $b$).
We will soon use this property (in the next sub-section) to get another
form of the $B$ fields that is more adapted to our mirror set-up.

\noindent $\bullet$ When we use an integrable complex structure for the two tori
in \csbase, i.e when we use $\langle\alpha\rangle_1, \langle\alpha\rangle_2$
instead of $\alpha$ (see Appendix 1), the $B_{NS}$ field takes the following form:
\eqn\bnow{{\hat B} = \sqrt{\langle\alpha\rangle_1} ~d{x}
\wedge d\theta_1 - \sqrt{\langle\alpha\rangle_2}~ d{y} \wedge d\theta_2 +
(A~\sqrt{\langle\alpha\rangle_1}~d\theta_1 -
B~\sqrt{\langle\alpha\rangle_2}~d\theta_2) \wedge d{z}.}
As one can easily see, this is a pure gauge! Thus even though we
have large $B$ field in this scenario, the effect of this is nothing as it is
a gauge artifact. More on this will appear in a forthcoming paper \toappear.

\subsec{The Mirror Manifold}

{}From the detailed analysis in the above two subsections,  we
can summarize the following: our metric of the mirror manifold
has strong resemblance to the metric of D6 wrapped on deformed
conifold, but they differ because of non-trivial B-dependent
fibration of some of the terms. As we see, this is the key
difference between the two metrics (apart from the non-trivial
warp factors). The manifold \mirmanchange\ is generically non-K\"ahler
whereas the metric \dsixdco\ is Ricci-flat and K\"ahler. The
metric evaluated in \mirmanchange\ is actually after we perform a
coordinate transformation (or have $\psi = 2\pi$ fixed) and
therefore we see no $\psi$ dependence in the final picture. In
the {\it usual} coordinate system, our ansatz therefore, for the
exact metric in type IIA will be to take the ``usual'' $D6$ brane
wrapped on the deformed conifold (i.e. eq. \metcomptwo) and
replace the ($d\psi, d \phi_1, d\phi_2, {\tilde g}(r)$), or in
the notation that we have been using: ($ dz, dx, dy, {\tilde g}$)
by \eqn\redpsietc{\eqalign{& dz ~\to ~ dz - {b}_{z\mu}~dx^\mu \cr
& dx ~\to ~ dx - b_{x\theta_1}~d\theta_1 \cr & dy ~\to ~ dy -
b_{y\theta_2}~d\theta_2 \cr & {\tilde g}_i(r, \theta_1, \theta_2)
~ \to ~ g_i(r, \theta_1, \theta_2)}} with the remaining terms
unchanged. Observe that before this replacement (i.e in the absence of fluxes)
\dsixdco\ is exactly
\metcomptwo\ up to ${\rm cos}~\psi$ and ${\rm sin}~\psi$
dependences. We believe, as discussed above, this has to do with
the delocalization of the $\psi$ coordinate. In the presence of fluxes,
the final answer for
the type IIA metric therefore will be to convert
\mirmanchange\ into\foot{\noindent This ansatz
can actually be given a little more rigorous derivation. To see
this from \mirmanchange, perform the following local
transformation $$\pmatrix{d{\hat y}\cr d\theta_2} ~\to
~\pmatrix{{\rm cos}~\psi & -{\rm sin}~\psi \cr {\rm sin}~\psi &
{\rm cos}~\psi}~\pmatrix{d{\hat y} \cr d\theta_2}$$ \noindent
where $d{\hat y} \equiv dy + b_{y\theta_2}~d\theta_2$. The change
in the $dz$-fibration structure will be in such a way as to
restore back ${\rm cot}~\theta_2~dy$ via the {\it reverse}
transformation a-la \psichange.}: \eqn\fiiamet{\eqalign{ds_{IIA}^2
= &~~ g_1~\left[(dz - {b}_{z\mu}~dx^\mu) + \Delta_1~{\rm
cot}~\hat\theta_1~(dx - b_{x\theta_1}~d\theta_1) + \Delta_2~{\rm
cot}~ \hat\theta_2~(dy - b_{y\theta_2}~d\theta_2)+ ..\right]^2
\cr & ~~~~~~~~~~~~~~ +~ g_2~ {[} d\theta_1^2 + (dx -
b_{x\theta_1}~d\theta_1)^2] + g_3~[ d\theta_2^2 + (dy -
b_{y\theta_2}~d\theta_2)^2{]}  \cr & ~~~~~~~~~~~~~~ + ~ g_4~{\rm
sin}~\psi~{[}(dx - b_{x\theta_1}~d\theta_1)~d \theta_2 + (dy -
b_{y\theta_2}~d\theta_2)~d\theta_1 {]}\cr & ~~~~~~~~~~~~~~ +
~g_4~{\rm cos}~\psi~{[}d\theta_1 ~d\theta_2 - (dx -
b_{x\theta_1}~d\theta_1) (dy - b_{y\theta_2}~d\theta_2)],}} where
$g_i$ are again given by \gisnow. Similarly the finite part of
the background $B$ field can be transformed from \hatb\ to the
following form involving the ${\rm sin}~\psi$ and ${\rm cos}~\psi$
dependences as: \eqn\hatbnow{{\hat B \o \sqrt{\alpha}} = dx
\wedge d\theta_1 - dy \wedge d\theta_2 + A~d\theta_1\wedge dz -
B~({\rm sin}~\psi ~dy - {\rm cos}~\psi~d\theta_2) \wedge dz.} The
type IIA coupling on the other hand can no longer be constant
even though in the type IIB side we start with a constant
coupling. The constant coupling on the type IIB side is
generically fixed by RR and NS fluxes via a superpotential
(though not always). If we start with a type IIB coupling $g_B$
(constant or non-constant), the type IIA theory is given by a
non-constant coupling $g_A$, that depend on the coordinates of
the internal space as \eqn\ccons{g_A = {g_B \o \sqrt{1-
{\epsilon\o \alpha}}}.} Observe that a small coupling in the type
IIB side implies a small coupling on the mirror manifold.
Therefore any perturbative calculation in type IIB side will have
a corresponding perturbative dual in the mirror side. This is
another advantage that we get from the mirror manifolds.

The above analysis more or less gives the complete background for
the mirror case (the RR background will be dealt with shortly).
For later comparison we would however need the three form NSNS
field strength defined as $H = d\hat B$. This is basically the
finite part of the three form, and is given by \foot{Written in terms of
$\sqrt{\langle\alpha\rangle_1}$ and $\sqrt{\langle\alpha\rangle_2}$ this is exactly zero,
and therefore serves as a gauge artifact.}:
\eqn\hfine{\eqalign{H & = -\sqrt{\alpha^3} A~(A~dA + B~dB) \wedge
d\theta_1 \wedge dz + \sqrt{\alpha^3} (A~dA + B~dB) \wedge dy
\wedge d\theta_2 \cr &  + \sqrt{\alpha}~dA \wedge d\theta_1
\wedge dz + \sqrt{\alpha^3}~B~(A~dA + B~dB) \wedge ({\rm
sin}~\psi ~dy - {\rm cos}~\psi~d\theta_2) \wedge dz \cr &  -
\sqrt{\alpha^3}~(A~dA + B~dB)\wedge dx \wedge d\theta_1 ~
 -\sqrt{\alpha} ~dB \wedge ({\rm sin}~\psi ~dy - {\rm cos}~\psi~d\theta_2) \wedge dz,}}

\noindent where we have used the following simplifying
definitions  in the above form of $H$: \eqn\sifoo{dA =
\del_{\theta_i}A ~d\theta_i + \del_r A~dr, ~~~~ d\sqrt{\alpha} =
-\sqrt{\alpha^3}(A~dA + B~dB)} with similar definition for $dB$.
Notice that $H$ involves all components of the mirror manifold
and therefore will be spread over the whole space.

\noindent We now need to show the following things:

\noindent (1) The manifold is explicitly non-K\"ahler i.e. $dJ \ne
0$, where $J$ is the fundamental two form. The manifold should
also be non-Ricci flat but have an $SU(3)$ holonomy. Recall that $SU(3)$
holonomy {\it doesn't} imply Ricci-flatness.

\noindent (2) The complex structure should in general be
non-integrable. Therefore the manifold should be non-complex and
non-K\"ahler. The properties of such manifolds have been
discussed earlier in \louis, \dal.

\noindent (3) We have to calculate the superpotential and show
that the holomorphic three form $\Omega$ is in general {\it not}
closed.

\noindent Some of these details will be addressed in later
sections of this paper. We will leave a more elaborate discussion
for part II. We now go to the M-Theory analysis.

\newsec{Chain 2: The M-theory Description of the Mirror}

Now that we have obtained the mirror metric, it is time to go to
the second chain of fig. 1 and lift the type IIA configuration to
M-theory. Initial studies have been done in \amv. We will use
their ideas to go to another $G_2$ holonomy manifold which is
related to the previous one by a flop. But before moving ahead,
we will require some geometric details of the background. These
geometric details will help us to formulate the background in a
way so that the procedure of flop will be simple to see.

\subsec{One Forms in M-theory}

We first need to define one forms in M-theory. These one forms are
different from the ones presented in say \amv, \brand, \cveticone\
as they
have contributions from the $B$ fields in the resolved conifold
side. These $B$ fields are in general periodic variables and
therefore let us denote them by angular coordinates $\lambda_1,
\lambda_2$ as ${\rm tan}~\lambda_1 \equiv a_1 b_{x\theta_1}, {\rm
tan}~\lambda_2 = a_2 b_{y\theta_2}$, with $a_1, a_2$ constants.
These one-forms are however only defined {\it locally} because
the $B$ fields that we will use in the definition are not
globally defined variables. The existence of these one forms can
be argued from the consistency of the metric\foot{Observe that
locally the metric \fiiamet\ is exactly the metric of $D6$ branes
wrapping an $S^3$ of a deformed conifold, and therefore will have
similar one-forms as in \amv, \brand, \cveticone\
by which we can express
the metric. Globally the issue is much more involved.}. They are
given by: \eqn\oneformsM{\eqalign{& \sigma_1 = {\rm
sin}~\psi_1~dX + {\rm sec}~\lambda_1~{\rm cos}~(\psi_1 +
\lambda_1)~d\Theta_1 \cr & \sigma_2 = {\rm cos}~\psi_1~dX - {\rm
sec}~\lambda_1~{\rm sin}~(\psi_1 + \lambda_1)~d\Theta_1 \cr &
\sigma_3 = d\psi_1 + n_1~{\rm cot} ~\hat\Theta_1~dX - n_2~{\rm
tan}~\lambda_1~{\rm cot} ~\hat\Theta_1~d\Theta_1}} where we have
defined new coordinates $\psi_1, \psi_2, X, \Theta_1,
\hat\Theta_1$.
 Their relation to $\psi, x, \theta_1, \hat\theta_1$ will be determined
as we proceed with our calculation.
 We have
also included two functions $n_1$ and $n_2$ in the definition of
$\sigma_3$. These are functions of all the coordinates, and will
be analyzed later. Using these, we can define another set of one
forms with $X, \Theta_1$ etc. replaced by $Y, \Theta_2$ etc. as:
\eqn\seconefor{\eqalign{& \Sigma_1 = -{\rm sin}~\psi_2~dY +  {\rm
sec}~\lambda_2~{\rm cos}~(\psi_2 -
 \lambda_2)~d\Theta_2 \cr
& \Sigma_2 = -{\rm cos}~\psi_2~dY - {\rm sec}~\lambda_2~{\rm
sin}~(\psi_2 - \lambda_2)~d\Theta_2 \cr & \Sigma_3 = d\psi_2 -
n_3~{\rm cot} ~\hat\Theta_2~dY +
 n_4~{\rm tan}~\lambda_2~{\rm cot} ~\hat\Theta_2~d\Theta_2.}}
We have chosen the respective signs with the foresight of making
a simple identification with our original variables possible. The
above set of one forms will suffice to define the corresponding
seven dimensional manifolds in M-theory. Although, not important
for our work here, we can make some interesting simplifications.
The quantities $b_{x\theta_1}$ and $b_{y\theta_2}$, as discussed
above, are basically periodic variables, and for small
$\lambda_i$ we can define another angular coordinates $\beta_1$
and $\beta_2$ that modify the original $\psi_1, \psi_2$ as
\eqn\lambpsi{\beta_1 = \psi_1 - b_{x\theta_1}, ~~~~~~ \beta_2 =
\psi_2 - b_{y\theta_2}.} With these choices of angles, we can
define another set of one forms in M-theory in the following way:
\eqn\anosetone{\eqalign{& \tilde\sigma_1 = {\rm sin}~\psi_1~dX +
{\rm cos}~\beta_1~d\Theta_1  \qquad \tilde\Sigma_1 = {\rm
sin}~\psi_2~dY + {\rm cos}~\beta_2~d\Theta_2 \cr & \tilde\sigma_2
= {\rm cos}~\psi_1~dX + {\rm sin}~\beta_1~d\Theta_1 \qquad
\tilde\Sigma_2 = {\rm cos}~\psi_2~dY + {\rm
sin}~\beta_2~d\Theta_2,}} with $\tilde\sigma_3$ and
$\tilde\Sigma_3$ being identical to $\sigma_3$ and $\Sigma_3$,
respectively. This way of writing the one-forms helps us to
compare them to the one-forms given in \amv, \brand, and \cveticone.
Observe
that for small background values of $b_{x\theta_1}$ and
$b_{y\theta_2}$ the above set \anosetone\ is the same as
\oneformsM\ and \seconefor. Furthermore, the field strength
vanishes locally, and these one forms satisfy the $SU(2)$
algebra. This is true because over a small patch the
$b_{x\theta_1}$ and $b_{y\theta_2}$ values are constants.
Therefore we can approximate \eqn\dxdxpatch{d\hat x \equiv dx -
b_{x\theta_1}~d\theta_1 = d(x - b_{x\theta_1}~\theta_1), ~~~~~
d\hat y \equiv dy - b_{y\theta_2}~d\theta_2 = d(y -
b_{y\theta_2}~\theta_2)} as exact one forms. In this way
\oneformsM\ and \seconefor\ appear like the usual one forms for
the $G_2$ manifold and satisfy an $SU(2) \times SU(2)$ symmetry.
Globally, there is no $SU(2) \times SU(2)$ symmetry because our
manifold is no longer a K\"ahler manifold. This is also clear
from the one-forms  \oneformsM\ and \seconefor. We will use this
local identification many times to compare our results to the
ones from literature. Also, having an exact form for $d\hat x$
and $d\hat y$ locally means vanishing type IIB $B$ fields. In
this way we will be able to extend our results to the case with
torsion.

\subsec{M-theory Lift of the Mirror IIA Background}

To perform the M-theory lift, we need the field strength
$F_{mn}$ which comes from the mirror dual of the three-form
$H_{RR} \equiv {\cal H}$ and five-form $F_5$
 in type IIB theory with $D5$ on a resolved conifold. The five
form appears because a $D5$ wrapped on an $S^2$ with $B_{NS}$
fluxes gives rise to a $D3$ brane source as we discussed in the
beginning of this paper. The background value of the RR potential
is given in \pandoz\ as: \eqn\hrrbg{\eqalign{& {\cal H} = c_1~(dz
\wedge d\theta_2 \wedge dy - dz \wedge d\theta_1 \wedge dx) +
c_2~{\rm cot}~\hat\theta_1~dx \wedge d\theta_2 \wedge dy -
c_3~{\rm cot}~\hat\theta_2 ~dy \wedge d\theta_1 \wedge dx \cr &
F_5 = K(r)~(1 + \ast) ~dx \wedge dy \wedge dz \wedge d\theta_1
\wedge d\theta_2,}} with $c_i$ being constant coefficients and
$K(r)$ being a function of the transverse coordinate. This means
that we have the following components of the RR three form:
${\cal H}_{\mu x z}, {\cal H}_{\mu y z}$ and ${\cal H}_{\mu x
y}$. The T-duality rules for the RR field strengths with
components along the T-dual directions are given in \fawad:
\eqn\fadhu{\eqalign{& {\tilde F}^{(n)}_{ijk....} = F^{(n+1)}_{x
ijk....} - n B_{x[i} F^{(n-1)}_{jk....]} + n(n-1)j_{xx}^{-1}
B_{x[i}j_{\vert x\vert j} F^{(n-1)}_{x\vert k....]} \cr & {\tilde
F}^{(n)}_{xij....} = F^{(n-1)}_{ij....} - (n-1)j_{xx}^{-1}
j_{x[i}F^{(n-1)}_{x\vert jk....]}}} where $n$ denote the rank of
the form, $x$ is the T-duality direction and $B$ is the NS field.
Notice also that in the above relation, the duality direction $x$
is inert under anti-symmetrization. Under a mirror
transformation, the RR three-form will give rise to the following
gauge potentials in type IIA theory: \eqn\gaugepot{\eqalign{&
F_{z\theta_1} = {\cal H}_{xy\theta_1}, ~~~~ F_{z\theta_1} = {\cal
H}_{xy\theta_2} \cr & F_{y\theta_1} = -{\cal H}_{xz\theta_1} +
{\cal H}_{xy\theta_1}\left[{j_{yz}j_{xx} - j_{xy}j_{xz} \o j_{yy}
j_{xx} - j_{xy}^2} - {\cal B}_{zy}\right]\cr & F_{x\theta_2}=
{\cal H}_{yz\theta_2} - {\cal H}_{xy\theta_2} \left[
{j_{xy}(j_{yz}j_{xx} - j_{xy}j_{xz}) \o j_{xx}(j_{yy} j_{xx} -
j_{xy}^2)}
 +{\cal B}_{zx} -{j_{xz}\o j_{xx}}\right]}}
which simplifies, after choosing the signs of ${\cal B}_{mn}$ in
a way that makes the fibration structure in the mirror
consistent, as: \eqn\simsol{\eqalign{& F_{z\theta_1} = -c_3~{\rm
cot}~\hat\theta_2, ~~~~~~~~~~~~~~ F_{z\theta_2} = -c_2{\rm
cot}~\hat\theta_1 \cr & F_{y\theta_1} = c_1 - 2c_3~\alpha~B~{\rm
cot}~{\hat\theta_2}, ~~ F_{x\theta_2} = c_1 - 2c_2~\alpha~A~{\rm
cot}~{\hat\theta_1}.}} {}From above we will eventually extract
gauge potentials $A_x$ and $A_y$ (the $A_z$ potential can be
absorbed in the definition of $dx_{11}$)\foot{There is a simple
reason for this. We had earlier defined $dz \equiv d\psi_1 -
d\psi_2$ and $dx_{11} \equiv d\psi_1 + d\psi_2$. Thus $dx_{11} =
d\psi_1 - d\psi_2 + 2 d\psi_2 = dz + 2d\psi_2$. Therefore, any
additional $dz$ dependent terms should be absorbed in $dx_{11}$
by changing the coefficient in front of $dz$ in $dx_{11}$. For
example $dx_{11} + A_z~dz = (1+A_z)~dz + 2 d\psi_2 = d\tilde
x_{11}$ when $A_z$ is a pure gauge.}.

The analysis done above only gave us some of the gauge fields in
type IIA theory. To get the other components, we need to get the
mirror dual of the five form $F_5$. One can easily show that the
$F_5$ part contributes to $F_{\theta_1 \theta_2}$. The precise
value turns out to be \eqn\fiveformv{\eqalign{F_{\theta_1
\theta_2} & =~ K(r) - b_{x\theta_1} {\cal H}_{yz\theta_2} -
b_{y\theta_2} {\cal H}_{xz\theta_1} - {\cal B}_{z\theta_2}~{\cal
H}_{xy\theta_1} - {\cal B}_{z\theta_1}~{\cal H}_{xy\theta_2} \cr
& = ~ K(r) - c_1~(b_{x\theta_1} - b_{y\theta_2}) -
2c_3~\alpha~B~b_{y\theta_2}~{\rm cot}~\hat \theta_2 +
2c_2~\alpha~A~b_{x\theta_1}~{\rm cot}~\hat\theta_1}} where again
the sign in ${\cal B}_{z\theta_1}$ and ${\cal B}_{z\theta_2}$ have
been chosen opposite to the definitions employed in sec. 4.2. The
above mentioned gauge potentials will eventually appear as metric
components in M-theory. As is well known, the M-theory metric
components will typically look like \eqn\mthmetlook{G^M_{\mu\nu}
= e^{-{2\phi \o 3}} g^{IIA}_{\mu\nu} - e^{4\phi\o 3} A_\mu A_\nu,
~~~~~ G^M_{\mu~11} = - e^{4\phi \o 3} A_\mu} where $A_\mu$ are
the gauge fields and $\phi$ is the type IIA dilaton. In fact, we
can use the above definitions to absorb some terms in $dx^{11}$
when we define the gauge fields. Up to some warp factors our
ans\"atze for the gauge fields will therefore be:
\eqn\gaugefields{\eqalign{& A_x = \Delta_3 ~{\rm
cot}~\hat\theta_1, ~~~~~~~~~A_{\theta_1} =
-\Delta_3~b_{x\theta_1}~{\rm cot}~\hat\theta_1 \cr & A_y = -
\Delta_4~{\rm cot}~\hat\theta_2, ~~~~~ A_{\theta_2} =
\Delta_4~b_{y\theta_2}~{\rm cot}~\hat\theta_2}} where
$\Delta_{3,4}$ are some specific functions of $\theta_i, x, y$
and $z$.   Now combining everything together we can write the
part of the M-theory metric originating from the gauge fields as:
\eqn\gaupot{A\cdot dX \equiv \Delta_3~{\rm cot}~\hat\theta_1~(dx
- b_{x\theta_1}~d\theta_1) - \Delta_4~{\rm cot}~\hat\theta_2~(dy
- b_{y\theta_2}~d\theta_2)} where we have delocalized the
direction $\psi$. The above potentials \gaupot\ are basically the
wrapped $D6$ brane sources that have been converted to geometry
giving rise to the M-theory metric \eqn\mteorymet{\eqalign{ds^2
&= e^{-{2\phi \o 3}}
(h^{-1/2}~ds_{0123}^2+h^{1/2}\gamma'~dr^2)\cr & + e^{-{2\phi \o
3}}~ds^2_{IIA} + e^{4\phi \o 3}~(dx_{11} + \Delta_3~{\rm
cot}~\hat\theta_1~d\hat x - \Delta_4~{\rm cot}~\hat\theta_2~d\hat
y)^2}} with $x_{11}$ being the eleventh direction, and
$ds^2_{IIA}$ is the metric given in \fiiamet. We have also used
the definition of $\hat x$ and $\hat y$ (introduced in section 5)
to write the fibration structure in a compact form. Recall also
that we are using the un-tilded coordinates henceforth.

The metric \mteorymet\ is basically the M-theory metric that we
are  looking for. To see how the $G_2$ structure appears from
this, we need to write the metric using the one forms that we
gave in the previous section. This will also help us to perform a
flop in the metric.

\noindent {}From the M-theory metric \mteorymet\ we see that the
total fibration structure is \eqn\fibstr{ g_1~e^{-{2\phi \o 3}}
(dz + \Delta_1~{\rm cot}~\hat\theta_1~d\hat x + \Delta_2~{\rm
cot}~\hat\theta_2~d\hat y)^2 + e^{4\phi \o 3}~(dx_{11} +
\Delta_3~{\rm cot}~\hat\theta_1~d\hat x - \Delta_4~{\rm
cot}~\hat\theta_2~d\hat y)^2.} In writing this, the careful
reader might notice that we have used ${\rm cot}~\hat\theta_i$ in
\hrrbg. However, we have two options here: we can either absorb
the scaling etc. in the definition of warp factors or keep the
warp factors as they are and change $\theta$ in ${\rm cot}\theta$
to accommodate this. Furthermore, the scaling of $\theta$, which
only affects this term, is actually of ${\cal O}(1)$ and could be
ignored in the subsequent calculations. We will however not
assume any approximations and continue using $\hat\theta_i$ in
the fibration. Observe that all other one forms are defined wrt
$\theta_i$.

If we now identify $dz \equiv  d\psi_1 - d\psi_2$ and $dx_{11}
\equiv d\psi_1 + d\psi_2$, then one can easily see that the above
fibration can be written in terms of the one forms that we
devised earlier in \oneformsM\ and \seconefor, as
\eqn\ondev{\alpha_3^2~ (\sigma_3 + \Sigma_3)^2  + \alpha_4^2~
(\sigma_3 - \Sigma_3)^2} with $\alpha_i$ being the relevant warp
factors; and we also identify ($X, Y, \Theta_1, \Theta_2$) to
($x, y, \theta_1, \theta_2$). The coefficients $a_i$ are simply
the identity, so $\tan\lambda_1=b_{x\theta_1}$ and
$\tan\lambda_2=b_{y\theta_2}$. This way of writing the $z$-- and
$x_{11}$-- fibration also forces us to set $n_1=n_2=\Delta_1$ and
$n_3=n_4=\Delta_2$. \ondev\ is consistent with the expected form
for the M-theory lift of the $D6$ configuration. Let us now look
at the other possible combinations of the one forms.

The first combination is to consider the sum of the squares of
the difference between the one forms. In other words, we will
consider: \eqn\sumofdif{(\sigma_1 - \Sigma_1)^2 + (\sigma_2 -
\Sigma_2)^2.} This give rise to the following algebra:
\eqn\algeb{\eqalign{ & ({\rm sin}~\psi_1~dX + {\rm
sec}~\lambda_1~{\rm cos}~(\psi_1 + \lambda_1)~d\Theta_1 + {\rm
sin}~\psi_2~dY -
 {\rm sec}~\lambda_2~{\rm cos}~(\psi_2 - \lambda_2)~d\Theta_2)^2 + \cr
& ({\rm cos}~\psi_1~dX - {\rm sec}~\lambda_1~{\rm sin}~(\psi_1 +
\lambda_1)~d\Theta_1 + {\rm cos}~\psi_2~dY + {\rm
sec}~\lambda_2~{\rm sin}~(\psi_2 - \lambda_2)~d\Theta_2)^2  = \cr
&~~~ d\Theta_1^2 + d\Theta_2^2 + (dX - {\rm
tan}~\lambda_1~d\Theta_1)^2 + (dY - {\rm
tan}~\lambda_2~d\Theta_2)^2 + \cr & ~~-2~{\rm cos}~(\psi_1 -
\psi_2)~[d\Theta_1~d\Theta_2 - (dX - {\rm
tan}~\lambda_1~d\Theta_1)~(dY - {\rm tan}~\lambda_2~d\Theta_2)] +
\cr & ~~-2~ {\rm sin}~(\psi_1 - \psi_2)~[d\Theta_1~(dY - {\rm
tan}~\lambda_2~d\Theta_2) + d\Theta_2 ~ (dX - {\rm
tan}~\lambda_1~d\Theta_1)].}} This is more or less the expected
form, but differs from \fiiamet\ by some
 relative signs. To fix the signs we need to evaluate the other possible combination of
one forms, i.e. $(\sigma_1 + \Sigma_1)^2 + (\sigma_2 +
\Sigma_2)^2$.
This time the algebra will yield:
\eqn\algebyiel{\eqalign{& ({\rm sin}~\psi_1~dX + {\rm sec}~\lambda_1~{\rm cos}~(\psi_1 +
 \lambda_1)~d\Theta_1 + {\rm sin}~\psi_2~dY +
{\rm sec}~\lambda_2~{\rm cos}~(\psi_2 + \lambda_2)~d\Theta_2)^2 + \cr
& ({\rm cos}~\psi_1~dX - {\rm sec}~\lambda_1~{\rm sin}~(\psi_1 + \lambda_1)~d\Theta_1 +
{\rm cos}~\psi_2~dY - {\rm sec}~\lambda_2~{\rm sin}~(\psi_2 + \lambda_2)~d\Theta_2)^2  = \cr
& ~~~ d\Theta_1^2 + d\Theta_2^2 + (dX - {\rm tan}~\lambda_1~d\Theta_1)^2 +
(dY - {\rm tan}~\lambda_2~d\Theta_2)^2 + \cr
& ~~ + 2~{\rm cos}~(\psi_1 - \psi_2)~[d\Theta_1~d\Theta_2 +
(dX - {\rm tan}~\lambda_1~d\Theta_1)~(dY - {\rm tan}~\lambda_2~d\Theta_2)] + \cr
& ~~ + 2~{\rm sin}~(\psi_1 - \psi_2)~[- d\Theta_1~(dY - {\rm tan}~\lambda_2~d\Theta_2) +
 d\Theta_2 ~ (dX - {\rm tan}~\lambda_1~d\Theta_1)].}}
It differs from \algeb\ only by overall minus signs in the
$\cos\psi$ and $\sin\psi$ terms. To get the exact form of the
metric that we have in \mteorymet, we write
\eqn\mmetgeneric{ds^2 = {\alpha}_1^2 ~\sum_{a = 1}^2 (\sigma_a +
\xi\Sigma_a)^2 + {\alpha}_2^2 ~ \sum_{a = 1}^2 (\sigma_a - \xi
\Sigma_a)^2 + {\alpha}_3^2 ~(\sigma_3 + \Sigma_3)^2 +
{\alpha}_4^2 ~ (\sigma_3 - \Sigma_3)^2 + {\alpha}_5^2 ~dr^2,}
where we have introduced the factor $\xi$ for $\Sigma_1$ and
$\Sigma_2$ to account for the different warp factors that the
spheres $(x,\theta_1)$ and $(y,\theta_2)$ have\foot{In the notations that
we used here, they are tori of course. But it is easy to get to the sphere case.
Observe also that when we switch off the $\lambda_i$ the metric reduces to the
well known $G_2$ form.}.
By comparison
with \fiiamet\ and \mteorymet\ we determine the warp factors
${\alpha}_i$ with $\xi=\sqrt{g_3/g_2}$ to be:
\eqn\warpyiden{\eqalign{{\alpha}_1 &= {1\o 2} e^{-{\phi\o
3}}\sqrt{\xi^{-1}g_4 + 2g_2}, ~~~ {\alpha}_2 = {1\o 2} e^{-{\phi
\o 3}}\sqrt{2g_2-\xi^{-1}g_4},\cr {\alpha}_3 &= e^{2\phi\o 3}, ~~~
{\alpha}_4 = e^{-{\phi\o 3}}\sqrt{g_1}, ~~~ {\alpha_5} =
e^{-{\phi\o 3}}\sqrt{\gamma'\sqrt{h}}.}} Observe also that,
writing the metric as \mmetgeneric, one can easily identify it to
the metric presented in \amv, \brand, \cveticone\
where the vielbeins in
terms of $\sigma_a$ and $\Sigma_a$ are written (see for example
eq. 2.7 of \brand). This will be useful in the following.

To extend our analysis further, we will now require all the
components  of the seven dimensional metric. We denote the
M-theory metric as $ds^2 = {\cal G}_{mn}~dx^m~dx^n$, where $m,n =
x, y, z, \theta_1, \theta_2, r, a$ and $x^a \equiv x^{11}$. The
various components are now given by: \eqn\mecoone{\eqalign{&
(1)~~{\cal G}_{xx} = g_1~e^{-{2\phi\o 3}}~ \Delta_1^2~ {\rm
cot}^2~\hat\theta_1 + g_2~e^{-{2\phi\o 3}} + e^{4\phi \o 3}~
\Delta_3^2 ~ {\rm cot}^2~\hat\theta_1 \cr & (2)~~ {\cal G}_{yy} =
g_1~e^{-{2\phi\o 3}}~ \Delta_2^2~ {\rm cot}^2~\hat\theta_2 +
g_3~e^{-{2\phi\o 3}} + e^{4\phi \o 3}~ \Delta_4^2 ~{\rm
cot}^2~\hat\theta_2 \cr & (3)~~ {\cal G}_{\theta_1 \theta_1} =
e^{-{2\phi \o 3}}~(g_1~ \Delta_1^2~ {\rm cot}^2~\hat\theta_1 +
g_2)~b^2_{x\theta_1} + g_2~e^{-{2\phi \o 3}} +
b_{x\theta_1}^2~e^{4\phi \o 3}~\Delta_3^2~{\rm
cot}^2~\hat\theta_1 \cr & (4)~~ {\cal G}_{\theta_2 \theta_2} =
e^{-{2\phi \o 3}}~(g_1~ \Delta_2^2 ~{\rm cot}^2~\hat \theta_2 +
g_3)~b^2_{y\theta_2} + g_3~e^{-{2\phi \o 3}} +
b_{y\theta_2}^2~e^{4\phi \o 3}~\Delta_4^2~{\rm
cot}^2~\hat\theta_2 \cr & (5) ~~ {\cal G}_{x \theta_1} =-
e^{-{2\phi \o 3}}~(g_1~ \Delta_1^2~ {\rm cot}^2~\hat\theta_1 +
g_2)~b_{x\theta_1} - b_{x\theta_1}~e^{4\phi \o 3}~\Delta_3^2 ~
{\rm cot}^2~\hat\theta_1 \cr & (6)~~ {\cal G}_{y \theta_2} =-
e^{-{2\phi \o 3}}~(g_1~ \Delta_2^2~ {\rm cot}^2~\hat\theta_2 +
g_3)~b_{y\theta_2} -  b_{y\theta_2}~e^{4\phi \o 3}~\Delta_4^2 ~
{\rm cot}^2~\hat\theta_2 \cr & (7)~~ {\cal G}_{x \theta_2} =-
e^{-{2\phi \o 3}}~\left(g_1~\Delta_1~\Delta_2~{\rm cot}~\hat
\theta_1~{\rm cot}~\hat\theta_2 - {g_4 \o 2}~{\rm
cos}~\psi\right) ~b_{y\theta_2}  + {g_4 \o 2}~e^{-{2\phi \o
3}}~{\rm sin}~\psi + \cr & ~~~~~~~~~~~~~~~~~~ + e^{4\phi \o
3}~\Delta_3~\Delta_4~{\rm cot}~\hat\theta_1~{\rm
cot}~\hat\theta_2~b_{y\theta_2} \cr & (8)~~ {\cal G}_{y \theta_1}
= - e^{-{2\phi \o 3}}~\left(g_1~\Delta_1~\Delta_2~{\rm cot}~
\hat\theta_1~{\rm cot}~\hat\theta_2 - {g_4 \o 2}~{\rm
cos}~\psi\right) ~b_{x\theta_1}  + {g_4 \o 2}~e^{-{2\phi \o
3}}~{\rm sin}~\psi + \cr & ~~~~~~~~~~~~~~~~~~  + e^{4\phi \o
3}~\Delta_3~\Delta_4~{\rm cot}~\hat\theta_1~{\rm
cot}~\hat\theta_2~b_{x\theta_1} \cr & (9) ~~ {\cal
G}_{\theta_1\theta_2} = e^{-{2\phi \o
3}}~\left(g_1~\Delta_1~\Delta_2~{\rm cot}~\hat\theta_1~{\rm
cot}~\hat\theta_2 - {g_4 \o 2}~{\rm cos}~\psi\right)
~b_{x\theta_1}~b_{y\theta_2}  + {g_4 \o 2}~e^{-{2\phi \o 3}}~{\rm
cos}~\psi +\cr & ~~~~~~~~~~~~~~~~~~ - {g_4 \o 2}~e^{-{2\phi \o
3}}~{\rm sin}~\psi~(b_{x\theta_1} + b_{y\theta_2}) - e^{4\phi \o
3}~\Delta_3~\Delta_4~{\rm cot}~\hat\theta_1~{\rm
cot}~\hat\theta_2~b_{x\theta_1}~b_{y\theta_2} \cr & (10)~~ {\cal
G}_{xy} = e^{-{2\phi \o 3}}~\left(g_1~\Delta_1~\Delta_2~{\rm cot}~
\hat\theta_1~{\rm cot}~\hat\theta_2 - {g_4 \o 2}~{\rm
cos}~\psi\right) - e^{4\phi \o 3}~\Delta_3~\Delta_4~{\rm
cot}~\hat\theta_1~{\rm cot}~\hat\theta_2 \cr & (11) ~~ {\cal
G}_{zx} = g_1~e^{-{2\phi \o 3}}~\Delta_1~{\rm cot}~\hat\theta_1
~~(12) ~~{\cal G}_{zy} = g_1~e^{-{2\phi \o 3}}~\Delta_2~{\rm
cot}~\hat\theta_2 ~~ (13)~~{\cal G}_{zz} = g_1~e^{-{2\phi \o
3}}\cr & (14) ~~ {\cal G}_{z \theta_1} = -g_1~e^{-{2\phi \o
3}}~\Delta_1~{\rm cot}~\hat\theta_1~b_{x\theta_1} ~~ (15)~~ {\cal
G}_{z \theta_2} = -g_1~e^{-{2\phi \o 3}}~\Delta_2~{\rm
cot}~\hat\theta_2~b_{y\theta_2} \cr & (16)~~ {\cal G}_{rr} =
e^{-{2\phi \o 3}}~\gamma'~\sqrt{h} ~~~ (17) ~~{\cal G}_{ax} =
\Delta_3~e^{4\phi \o 3}~ {\rm cot}~\hat\theta_1 ~~~ (18) ~~{\cal
G}_{ay} = - \Delta_4~e^{4\phi \o 3}~ {\rm cot}~\hat\theta_2 \cr &
(19) ~~{\cal G}_{a\theta_1} = -\Delta_3~e^{4\phi \o 3}~ {\rm
cot}~\hat\theta_1~b_{x\theta_1} ~~~~~ (20)~~{\cal G}_{a\theta_2}
= \Delta_4~e^{4\phi \o 3}~ {\rm cot}~\hat\theta_2~b_{y\theta_2}
~~(21)~~{\cal G}_{aa} = e^{4\phi \o 3}}}
The remaining seven components are all vanishing for this
specific case: \eqn\vancom{ {\cal G}_{rx} = {\cal G}_{ry} = {\cal
G}_{rz} = {\cal G}_{r\theta_1} = {\cal G}_{r\theta_2} = {\cal
G}_{ra} = {\cal G}_{az} = 0} with the spacetime metric
$e^{-{2\phi \o 3}}~h^{-{1\o 2}}$. Therefore \mecoone\ and
\vancom\ are basically the 28 components of our metric.

Having gotten the precise components and the one forms that
describe our  background, we can now get back to some of the
questions that we raised earlier in the type IIA section. Our
first question was to verify the non-K\"ahler nature of the type
IIA picture \fiiamet. To do this we need the vielbeins. They can
be calculated from the one forms \oneformsM\ and \seconefor.
With the assumption \eqn\ourassum{n_1 = n_2 = \Delta_1 =
\Delta_3, ~~~~~~~ n_3 = n_4 = \Delta_2 = \Delta_4} the vielbeins
are now easy to determine. They can be extracted from the metric
components \mecoone, \vancom\ and the one forms \oneformsM,
\seconefor\ if we put $\phi = 0$ in \mecoone, and are defined as
\eqn\viel{e^a \equiv e^a_\mu ~dx^\mu = e^a_x~dx + e^a_y~dy +
e^a_z~dz + e^a_{\theta_i}~d\theta_i + e^a_r~dr,} where $a = 1,
..., 6$ and $e^a_\mu$ are given by (recall that
$\xi=\sqrt{g_3/g_2}$): \eqn\vielcomp{\eqalign{& {e^1_x \o \sqrt{2
g_2 - g_4/\xi}} = {1\o 2} {\rm sin}~\psi_1 = {e^3_x \o
\sqrt{g_4/\xi + 2 g_2}}, ~~~ {e^1_{\theta_1} \o \sqrt{2
g_2-g_4/\xi}} = {{\rm cos}~(\psi_1 + \lambda_1) \o 2~{\rm
cos}~\lambda_1} = {e^3_{\theta_1} \o \sqrt{g_4/\xi + 2 g_2}} \cr
& {e^1_y \o \sqrt{2 g_2- g_4/\xi}} = {1\o 2} {\rm sin}~\psi_2 =
{- e^3_y \o \sqrt{g_4/\xi + 2 g_2}}, ~~~
 {- e^1_{\theta_2} \o \sqrt{2 g_2 - g_4/\xi}} = {{\rm cos}~(\psi_2 -
 \lambda_2) \o 2~{\rm cos}~\lambda_2} =  {e^3_{\theta_2} \o \sqrt{g_4/\xi + 2 g_2}} \cr
& {e^2_x \o \sqrt{2 g_2 - g_4/\xi}} = {1\o 2} {\rm cos}~\psi_1 =
 {e^4_x \o \sqrt{g_4/\xi + 2 g_2}}, ~~~
{-e^2_{\theta_1} \o \sqrt{2 g_2 - g_4/\xi}} =
 {{\rm sin}~(\psi_1 + \lambda_1) \o 2~{\rm cos}~\lambda_1} =
  {-e^4_{\theta_1} \o \sqrt{g_4/\xi + 2 g_2}} \cr
& {e^2_y \o \sqrt{2 g_2 - g_4/\xi}} = {1\o 2} {\rm cos}~\psi_2 =
 {- e^4_y \o \sqrt{g_4/\xi + 2 g_2}}, ~~~
 {e^2_{\theta_2} \o \sqrt{2 g_2 - g_4/\xi}} =
 {{\rm sin}~(\psi_2 - \lambda_2) \o 2~{\rm cos}~\lambda_2} =
  {-e^4_{\theta_2} \o \sqrt{g_4/\xi + 2 g_2}}\cr
& e^5_x = \sqrt{g_1}~\Delta_1 ~{\rm cot}~\hat\theta_1, ~~e^5_y =
\sqrt{g_1}~\Delta_2 ~{\rm cot}~ \hat\theta_2, ~~ e^5_{\theta_i} =
 -\sqrt{g_1}~\Delta_i~{\rm tan}~\lambda_i ~{\rm cot}~\hat\theta_i,
~~e^5_z =\sqrt{g_1}}} with the remaining components all vanishing
except $e^6_r$ given by $e^6_r = \sqrt{\gamma'\sqrt{h}}$, where
we have defined $\gamma'$ in the beginning of section 5. These
are the precise vielbeins for our case, and the metric \fiiamet\
can also be written in terms of \vielcomp. In fact, as is well
known, both the metric and the fundamental two form $J$ for our
case can be written using the vielbeins \vielcomp\ as:
\eqn\metj{ds^2 = \sum_{a = 1}^6~ e^a \otimes e^a, ~~~~~~~~~ J =
\sum_{a,b} ~e^a \wedge e^b.} One may also define complex
vielbeins as $(e^1+ie^2), ~(e^3+ie^4)$ and $(e^5+ie^6)$, then we
get $J=(e^1\wedge e^2)+(e^3\wedge e^4)+(e^5\wedge e^6)$. This
reads in components as: \eqn\jcomponents{\eqalign{J &
=(\alpha_2^2-\alpha_1^2)~\sin\psi dx\wedge dy -
(\alpha_1^2+\alpha_2^2)~dx \wedge d\theta_1 +
(\alpha_1^2+\alpha_2^2)~dy \wedge d\theta_2 \cr & +
(\alpha_2^2-\alpha_1^2)~[\cos\psi -
\sin\psi~\tan\lambda_2]~dx\wedge d\theta_2 -
(\alpha_2^2-\alpha_1^2)~[\cos\psi -
\sin\psi~\tan\lambda_1]~dy\wedge d\theta_1 \cr & -
(\alpha_2^2-\alpha_1^2)~[\cos\psi~(\tan\lambda_1+\tan\lambda_2) +
\sin\psi~(1-\tan\lambda_1~\tan\lambda_2)]~d\theta_1\wedge
d\theta_2 \cr & +
\alpha_4\alpha_5\Delta_1\cot\hat\theta_1~dx\wedge dr +
\alpha_4\alpha_5\Delta_2\cot\hat\theta_2~dy\wedge dr +
\alpha_4\alpha_5~dz\wedge dr \cr &
-\alpha_4\alpha_5\Delta_1\cot\hat\theta_1\tan\lambda_1~d\theta_1\wedge
dr - \alpha_4\alpha_5\Delta_2\cot\hat\theta_2\tan\lambda_2
~d\theta_2\wedge dr.}} From here one can easily compute $dJ$ and
show that in general $dJ \ne 0$ because of terms like
$d\lambda_i$. This implies that the metric \fiiamet\ is in
general not K\"ahler.

Having reached the explicit form of the background, it is now time
to pause a little to discuss some of the expected mathematical
properties of these backgrounds. The six dimensional manifold
that we gave in \fiiamet\ has an $SU(3)$ holonomy, is
non-K\"ahler and in general could be non-complex. Mathematical
properties of such manifolds have been discussed in some details
in \salamon, \gauntlett, \lust, \louis\ though an explicit
example have never been given before. To our knowledge, the
examples that we gave in this paper, are probably the first ones.

The holonomy of these manifolds are measured not wrt to the
usual  Riemannian connection, but wrt so-called torsional
connection. Earlier concrete examples of this were given in \sav,
\beckerD, \GP, \bbdg, \bbdgs. These examples dealt mostly with
heterotic theory and were non-K\"ahler but compact with integrable
complex structures. The examples that we gave here are mostly
non-compact, though compact examples could also be constructed
with some effort. Both the heterotic cases and the type II cases
are examples of torsional manifolds. As discussed in \salamon,
\louis\ and \lust, the torsional manifolds are classified by {\it
torsion-classes} ${\cal W}_i$, with $i = 1, 2, ...., 5$. In fact,
the torsion ${\cal T}$ belongs to the five classes:
\eqn\torso{{\cal T} \in {\cal W}_1 \oplus {\cal W}_2 \oplus {\cal
W}_3 \oplus  {\cal W}_4 \oplus {\cal W}_5} which in turn is
related to the $SU(3)$ irreducible representations. To determine
the torsion classes for our case, we need the ($3,0$) form
$\Omega$: \eqn\Omegad{\Omega \equiv \Omega_+ + i \Omega_- = (e^1
+ i e^2) \wedge (e^3 + i e^4) \wedge (e^5 + i e^6)} where $e^i$
are computed in \vielcomp. From the definition of the fundamental
two form $J$ in  \metj\ it can be observed that both
$\Omega_{\pm}$ are annihilated by $J$ and $\Omega_+ \wedge
\Omega_- = {2\o 3} J \wedge J \wedge J$. For our case,
\eqn\omepm{\eqalign{ & \Omega_+ = e^1 \wedge e^3 \wedge e^5 - e^2
\wedge e^4 \wedge e^5 -  e^1 \wedge e^4 \wedge e^6 - e^2 \wedge
e^3 \wedge e^6 \cr & \Omega_- = e^1 \wedge e^4 \wedge e^5 +  e^2
\wedge e^3 \wedge e^5 +  e^1 \wedge e^3 \wedge e^6 -  e^2 \wedge
e^4 \wedge e^6}} Using the three forms, and the fundamental two
form $J$ all the torsion classes can be constructed. The
fundamental two form $J$ and the holomorphic ($3,0$) form
$\Omega$ are not covariantly constant, as we had observed earlier
in \rstrom, \beckerD, \bbdg, \bbdgs, and they obey
\eqn\jandomega{\eqalign{& {\cal D}_m J_{np} = \nabla_m J_{np} -
{\cal T}_{mn}^{~~~r}J_{rp} - {\cal T}_{mp}^{~~~r}J_{nr} = 0 \cr &
{\cal D}_m \Omega_{nmp} = \nabla_m \Omega_{npq} - {\cal
T}_{mn}^{~~~r}\Omega_{rpq} - {\cal T}_{mp}^{~~~r}\Omega_{nrq} -
{\cal T}_{mq}^{~~~r}\Omega_{npr} = 0.}} The fact that the complex
structure is also not integrable might be argued from the
definition of the complex vielbeins: $e^i + i e^{i+1}$, that we
gave earlier. A more detailed analysis of the mathematical
discussion that we gave here will be relegated to part II. The
fact that the fundamental two form is not covariantly constant
implies that $dJ$ involves (3,0) and a (2,1) $+ {\rm c.c}$
pieces. The definition of $d$ becomes:
\eqn\defofd{d\omega^{(p,q)} = d\omega^{(p-1,q+2)} +
d\omega^{(p,q+1)} +  d\omega^{(p+1,q)} +  d\omega^{(p+2,q-1)}}
which, for an integrable complex structures will not have the
$d\omega^{(p-1,q+2)} +  d\omega^{(p+2,q-1)}$ pieces.

The lift of these six dimensional manifolds in M-theory gives rise
to $G_2$ manifolds in M-theory equipped with a $G_2$ structure.
The $G_2$ structure is specified by a three form $\tilde\Omega$
that could be easily evaluated from the vielbeins. We have
already defined six vielbeins earlier. Let us define $e^7 =
\sigma_3 + \Sigma_3$. Using this we can use the $SU(3)$ structure
to determine a $G_2$ structure on the seven manifold as:
\eqn\thrM{\eqalign{\tilde\Omega & = J \wedge e^7 + \Omega_+ \cr &
= e^1 \wedge e^3 \wedge e^5 - e^2 \wedge e^4 \wedge e^5 -  e^1
\wedge e^4 \wedge e^6 - e^2 \wedge e^3 \wedge e^6 \cr & ~~~~ + e^1
\wedge e^2 \wedge e^7 + e^3 \wedge e^4 \wedge e^7 +  e^5 \wedge
e^6 \wedge e^7.}} This determines the $G_2$ structure induced
from the $SU(3)$ structure of \fiiamet. The $G$ fluxes will give
rise to a three form $G_3 = \ast_7 G$ on the seven manifold. This
$G_3$ is {\it not} related to the torsion, the torsion three form
being given by \eqn\torthrf{{\cal\tau} = -\ast d\tilde\Omega -
\ast\left[{1\o 3} \ast(\ast d\tilde\Omega \wedge \tilde\Omega)
\wedge \tilde\Omega \right].} To see how the connections in both
the theories behave, the reader may want to look into \ivan. The
equivalent to the torsion classes are now the four modules \gray:
\eqn\torMM{{\cal T} \in \chi_1 \oplus \chi_2 \oplus \chi_3 \oplus
\chi_4} where $\chi_i$ are the ${\bf 1, 14, 27, 7}$ of $SO(7)$.
For our case we do not have a closed three form and therefore the
manifold will have a $G_2$ structure. A $G_2$ structure of the
type $\chi_1 \oplus \chi_3 \oplus \chi_4$ is in general
integrable with a Dolbeault cohomology given in \graytwo. For
further details see Appendix 2.

There are many questions that arises now regarding the seven
dimensional manifold that we presented. They will be tackled in the
sequel to this paper. To continue,
we will assume that the manifold has an
explicit $G_2$ structure. It goes without saying, of course that,
since we followed strict mirror rules the background that we get
should always preserve the set of conditions required.

\newsec{Chain 3: M-theory Flop and Type IIA Reduction}

Having the $G_2$ manifold, we can perform
a  flop transition and reduce to the corresponding type IIA picture.
Before we perform the flop, we revisit the one-forms
of the previous section. We will continue following
the steps laid out in \brand, \cveticone,
since the local behavior of our metric is
almost that of \brand, \cveticone.

To proceed further, first define a set of $2 \times 2$ matrices
$N_1$ and $N_2$ in the following way:
\eqn\nonemat{N^{\lambda_1}_1 \equiv \pmatrix{ \sigma_3 &
e^{i\psi_1} ~[e^{i\lambda_1}~{\rm sec}~\lambda_1~d\theta_1 - i
~dx] \cr e^{-i\psi_1}~[e^{-i\lambda_1}~{\rm
sec}~\lambda_1~d\theta_1 + i~ dx] & -\sigma_3}}
\eqn\ntwomat{N^{\lambda_2}_2 \equiv \pmatrix{ \Sigma_3 & \xi
e^{i\psi_2}~[e^{-i\lambda_2} ~{\rm sec}~\lambda_2~d\theta_2 + i~
dy] \cr \xi e^{-i \psi_2}~[e^{i\lambda_2}~ {\rm
sec}~\lambda_2~d\theta_2 - i~ dy] & -\Sigma_3}} where $\sigma_3$
and $\Sigma_3$ are the one form appearing in \oneformsM\ and
\seconefor, and we have again introduced the factor $\xi$ to
account for the asymmetry in the $\hat x$ and $\hat y$ terms. The
off--diagonal terms in the matrices are contributions from the
torsional part of the metric in the type IIA theory. Observe
that, in the absence of $B$ fields (in the original type IIB
theory), these matrices will take the following known form:
\eqn\matchange{\eqalign{N^0_1 & \equiv\pmatrix{ d\psi_1 +
\Delta_1~{\rm cot}~\hat\theta_1~dx & e^{i\psi_1} (d\theta_1 - i~
dx) \cr e^{-i \psi_1}(d\theta_1 + i~ dx) & -d\psi_1 -
\Delta_1~{\rm cot}~\hat\theta_1~dx} \cr N^0_2 & \equiv\pmatrix{
d\psi_2 - \Delta_2~{\rm cot}~\hat\theta_2~dy & \xi e^{i\psi_2}
(d\theta_2 + i~ dy) \cr \xi e^{-i \psi_2}(d\theta_2 - i~ dy) &
-d\psi_2 + \Delta_2~{\rm cot}~\hat\theta_2~dy}}}
which can be easily derived from the one forms given in \brand, \cveticone.

The matrices \nonemat\ and \ntwomat\ can be combined in various
ways to create new $2 \times 2$ matrices for our space. A generic
combination will be \eqn\gencov{N^{\lambda_1 \lambda_2}_{[a, b]}
= a~N^{\lambda_1}_1 - b~N^{\lambda_2}_2} with integral $a,b$.
Using various choices of $a,b$ we can express our eleven
dimensional metric. Locally however, one can show, that there are
two choices given by $N^{\lambda_1 \lambda_2}_{[1, 1]}$ and
$N^{\lambda_1 \lambda_2}_{[0, 1]}$ that are specifically useful
to write the M-theory metric. In the absence of $B$ fields (in
the original type IIB picture) these matrices are related to the
left invariant one forms $\omega$ and $\tilde\omega$ for the
$G_2$ spaces, i.e. \eqn\leftone{N^{00}_{[1, 1]}~ \to~
\tilde\omega, ~~~~~~~ N^{00}_{[0, 1]} ~ \to ~ \omega} as an
equality up to $SU(2)$ group elements. Of course, since for our
case there is no underlying $SU(2)$ symmetry globally (our
manifold being non-K\"ahler from the start i.e. directly in the
type IIA case) these relations are only in local sense. In the
presence of $B$ fields $-$ or non trivial fibrations in the
mirror/M-theory set-up $-$ the metric can still be written in
terms of $N^{\lambda_1 \lambda_2}_{[a, b]}$  even though they are
no longer related to $\omega$ and $\tilde\omega$. In fact, if we
remove the restriction on $a,b$ in \gencov\ as simple constants
and allow more generic values for them, we can express our
M-theory metric completely in terms of \gencov\ in the following
way: \eqn\mmetgec{ds^2 = - \left[{\rm det}~N^{\lambda_1
\lambda_2}_{[\beta, \beta]} - {\rm Tr}^2~(N^{\lambda_1
\lambda_2}_{[\gamma, \gamma]}\cdot \Gamma_3)\right] - \left[{\rm
det}~N^{\lambda_1 \lambda_2}_{[\delta, -\delta]} - {\rm
Tr}^2~(N^{\lambda_1 \lambda_2}_{[\epsilon, -\epsilon]}\cdot
\Gamma_3)\right]} where $\Gamma_3$ is the third Pauli matrix, and
$\beta, \gamma, \delta$ and $\epsilon$ can be extracted from
\warpyiden\ as: \eqn\bacvalve{\eqalign{& \beta = {e^{-{\phi \o
3}}\o {2}}~\sqrt{2g_2 - g_4/\xi}, ~~~~~\gamma = {e^{-{\phi \o
3}}\o {4}}~\sqrt{4g_1 - 2g_2 + g_4/\xi}\cr & \delta = {e^{-{\phi
\o 3}}\o {2}}~\sqrt{2g_2+g_4/\xi}, ~~~~~ \epsilon = {e^{-{\phi \o
3}}\o {4}}~\sqrt{4 e^{2\phi} - 2g_2 -g_4/\xi}.}}
In the absence of fluxes
i.e. $\lambda_i = 0$, on the other hand, it is conjectured that
the M-theory metric can
be written in terms of $N^{\lambda_1 \lambda_2}_{[a, b]}$ as \amv
\eqn\mmefg{ds^2 = - {\rm det}~N^{00}_{[A, A]} - {\rm
det}~N^{00}_{[B, -B]}} where the values of $A$ and $B$ are
related to the radial distances $r$ only. We see that
this type of metric is not realized in out set-up
because \mmetgec\ do not reduce to \mmefg\ by making
$\lambda_i = 0$.
To compare our result \mmetgec\ to the one without fluxes, we
need the metric of the $G_2$ manifold with terms of the form
${\rm det}~N^{00} - {\rm Tr}^2~(N^{00}\cdot \Gamma_3)$. Taking
this limit is of course possible and an example of this has been
given in \brand, \cveticone. Our manifold would therefore resemble this
scenario, although one has to be careful here. The identification
is only local where the $\lambda_i$ values are approximately
constant. The matrices $N_i$ in the presence and in the absence
of fluxes differ by $\lambda_i$ terms, giving rise to one forms
that do not have any underlying $SU(2)$ symmetry. To summarize
the situation, our metric \mmetgec\ in the presence of fluxes
gives rise to new $G_2$ holonomy manifolds that have not been
studied before. In the absence of fluxes, \mmetgec\ gives rise to
one of the examples studied in \brand\ (see e.g. equation (3.5),
and discussion) which in our notation would look like:
\eqn\gukbran{ds^2 = - \left[{\rm det}~N^{00}_{[a, a]} - {\rm
Tr}^2~(N^{00}_{[b, b]}\cdot \Gamma_3)\right] - \left[{\rm
det}~N^{00}_{[c, -c]} - {\rm Tr}^2~(N^{00}_{[d, -d]}\cdot
\Gamma_3)\right]} with $a, b, c$ and $d$ are given by the radial
coordinate $r$ as: \eqn\abcdvalue{\eqalign{& a = \sqrt{{4 r^2 +
12 r - 27 \o 48}}, ~~~ b = {\sqrt{{36 r - 4 r^2 - 81}}
 \o 24}, ~~~ c = \sqrt{4 r^2 - 12 r - 27  \o
48} \cr
& ~~~~~~~~~~~~~~~~~~~~ d = \sqrt{{16 r^4 -
48  r^3 - 148 r^2 + 108 r + 324 \o 48(4r^2 - 9)}}}}
As mentioned in \brand, \cveticone, there are alternative expressions for the
values of $a, b, c$ and $d$ obtained by scaling the metric. Then
the scale factor will appear in the set of relations \abcdvalue.
This will be useful for us to get rid of $\epsilon^{-1}$ factors
in the type IIA three form $d\tilde B$ with $\tilde B$ given earlier
in \bintwoa.
We will discuss this soon. For more
details the readers can see sec. 3 of \brand. This is the point
where our conclusions would differ from the results of \amv\  where
\mmefg\ is presented as the M-theory lift of the type IIA configuration.

To proceed further, we need to perform the flop transition and
go  to the corresponding type IIA picture. On the other hand, if
we followed \amv, we
would see from the one forms \oneformsM\ and \seconefor\ that we
now have two possible directions along which we can compactify
and come down to type IIA: $dx$ and $d\theta_1$ (or
correspondingly $dy$ and $d\theta_2$). But the $\theta_1$
direction is not globally defined as it comes with the
corresponding type IIA $b_{x\theta_1}$ field. So the
compactification to type IIA should rather be performed along
$dx$. To
see this, let us first write the M-theory metric in the following
suggestive way: \eqn\metsugges{\eqalign{ds^2 = & ~ e^{-{2\phi \o
3}} g_1 (dz + \gamma_1~d\theta_1 + \gamma_2~d\hat y)^2 + e^{4\phi
\o 3} (dx_{11} +\gamma_3~d\theta_1 + \gamma_4~d\hat y)^2 + \cr &
~+~e^{-{2\phi \o 3}}~g_3~(d\theta_2^2 + d\hat y^2) + e^{-{2\phi
\o 3}}~g_2~{\rm cot}^2~\lambda_1~d\theta_1^2 + g_5~(dx + {\cal
A})^2 - g_5~{\cal A}^2 + \cr & ~+~e^{-{2\phi \o 3}}~ g_4~{\rm
cos}~(\psi + \lambda_1)~{\rm sec}~\lambda_1~d\theta_1~d\theta_2 +
e^{-{2\phi \o 3}}~g_4~{\rm sin}~(\psi + \lambda_1)~{\rm
sec}~\lambda_1~d\hat y ~d\theta_1}} where we have already defined
$d\hat y \equiv dy - {\rm tan}~\lambda_2~d\theta_2$ earlier, and
the $dx$ fibration structure is represented as $dx + {\cal A}$,
${\cal A}$ being the corresponding one form that will appear in
type IIA as gauge fields. The other variables appearing in
\metsugges\ can be defined as follows:
\eqn\defmet{\eqalign{\gamma_1 & = -\Delta_1 ~{\rm
cot}~\hat\theta_1~{\rm tan}~\lambda_1, ~~~~~~~ \gamma_2 =\Delta_2
~{\rm cot}~\hat\theta_2, ~~~~~~~
 \gamma_3 = -\Delta_3 ~{\rm cot}~\hat\theta_1~{\rm tan}~\lambda_1 \cr
\gamma_4 & =-\Delta_4 ~{\rm cot}~\hat\theta_2, ~~~~ g_5 \equiv
e^{4\Phi \o 3} = (g_1~\Delta_1^2 ~{\rm cot}^2~\hat\theta_1 +
g_2)~e^{-{2\phi \o 3}} + e^{4\phi \o 3} \Delta_3^2~{\rm
cot}^2~\hat\theta_1 \cr {\cal A}& = A_1 ~dz + A_2 ~d\theta_1 +
A_3 ~d\theta_2 + A_4 ~d\hat y + A_5 ~dx_{11}}} where $\Phi$ is the
type IIA dilaton and $A_i$ are the components of the gauge fields
defined in the following way: \eqn\comgaud{\eqalign{& A_1 =
{e^{-{2\phi \o 3}}~g_1~\Delta_1~{\rm cot}~\hat\theta_1 \o (g_2 +
g_1~\Delta_1^2 ~{\rm cot}^2~\hat\theta_1)~e^{-{2\phi \o 3}} +
e^{4\phi \o 3} \Delta_3^2~{\rm cot}^2~\hat\theta_1}, ~~~ A_2 = -
{\rm tan}~\lambda_1 \cr & A_3 = {{1\o 2} e^{-{2\phi \o 3}}~g_4~
{\rm sin}~\psi\o (g_2 + g_1~\Delta_1^2 ~{\rm
cot}^2~\hat\theta_1)~e^{-{2\phi \o 3}}
 + e^{4\phi \o 3} \Delta_3^2~{\rm cot}^2~\hat\theta_1} \cr
& A_5 = { e^{4\phi \o 3} ~\Delta_3~ {\rm cot}~ \hat\theta_1 \o
(g_2 + g_1~\Delta_1^2 ~{\rm cot}^2~\hat\theta_1)~e^{-{2\phi \o
3}} + e^{4\phi \o 3} \Delta_3^2~{\rm cot}^2~\hat\theta_1}\cr &
A_4 = {{\rm cot}~\hat\theta_1{\rm cot}~\hat\theta_2~(e^{-{2\phi
\o 3}}~g_1~ \Delta_1~\Delta_2 -  e^{4\phi \o
3}~\Delta_3~\Delta_4) - {1\o 2} e^{-{2\phi \o 3}}~g_4~{\rm
cos}~\psi \o (g_2 + g_1~\Delta_1^2 ~ {\rm
cot}^2~\hat\theta_1)~e^{-{2\phi \o 3}} + e^{4\phi \o 3}
\Delta_3^2~ {\rm cot}^2~\hat\theta_1}}}
The metric \defmet\ would
basically be our answer, but if we look closely  we see that
\defmet\ do not resemble the expected resolved conifold metric in
the absence of $b_{x\theta_1}, b_{y\theta_2}$.
Therefore instead of reducing along $dx$, which would not give us the
expected resolved conifold metric in the absence of
$b_{x\theta_1}, b_{y\theta_2}$ without further coordinate
transformation, we will consider performing a flop in M-Theory
and then reduce along $dx_{11}$.
To consider the flop and the subsequent change in the metric we
have to do a transformation to our M-theory metric \mmetgeneric.
Before moving ahead, let us clarify one minor thing regarding the
scalings of the metric \mmetgeneric. As discussed in \brand, \cveticone,
scaling the metric from ${\cal G} ~\to ~ a^2 ~{\cal G}$ keeps the
$G_2$ structure intact. We can use this freedom of rescaling the
metric to remove the $\epsilon^{-{1\o 2}}$ factor in \bintwoa. The
coordinates of the $G_2$ manifold are given in terms of $r, x, y,
z, \theta_1, \theta_2$ and $x_{11}$. Let us scale the radial
coordinate $r$ as $r ~\to ~ \epsilon^{1\o 6}~r$. From \miapmet,
we see that $dz$ immediately scales as $\epsilon^{1\o 6}~dz$,
even though $d\psi$ may not scale. This is important, since now
the ${\rm sin}~\psi$ and ${\rm cos}~\psi$ in the metric \fiiamet\
will remain unchanged. If we now rescale $$x, y, z, \theta_i,
x_{11}  ~\to ~ \epsilon^{1\o 6}~x, \epsilon^{1\o 6}~y,
\epsilon^{1\o 6}~z, \epsilon^{1\o 6}~\theta_i, \epsilon^{1\o
6}~x_{11}$$ \noindent the metric \fiiamet\ or the complete
M-theory metric \mmetgeneric\ will have an overall scale of
$\epsilon^{1\o 3}$ (provided, of course, that we also scale the
$B$ field ${\tilde b}_{mn}$ in the fibration accordingly).
This metric preserves $G_2$ holonomy as
before, but now the three form flux $C_3$ coming from the two
form $B_{NS}$  in \hatbnow\ as $C_3 = \hat B \wedge dx_{11}$ will
be finite. Thus, we can make the background finite using the
rescaling freedom, and therefore this gives us confidence in
considering only the finite part of the background \hatbnow\ and
the metric \fiiamet\ without any $\epsilon$ dependences anywhere.

After this detour, it is now time to consider the issue of flop
on  the M-theory metric \mmetgeneric. This way we will be able to
connect the final answer, after dimensional reduction, to the
type IIA metric implied above in \metsugges. A simple way to
guess the answer would be to restore the case without any
torsion. In the absence of torsion the type IIA reduction should
be a resolved conifold with fluxes and no $D6$ branes. This would
imply that the metric looks like two $S^2$ spheres (one of them
with a finite size as $r$, the radial coordinate, approaches
zero). If we now use our one forms \oneformsM\ and \seconefor,
one way to generate this would be if we consider the
transformation on $N^{\lambda_1 \lambda_2}_{[a,b]}$ as:
\eqn\floptra{N^{\lambda_1 \lambda_2}_{[1,0]} ~~\to ~~N^{\lambda_1
\lambda_2}_{[{1+f \o 2}, {1\o 2}]}, ~~~~~~~ N^{\lambda_1 \lambda_2}_{[0,
-1]} ~~\to ~~ N^{\lambda_1 \lambda_2}_{[{1-f \o 2}, {1\o 2}]}} upto
possible conjugations. Locally, the above relation will imply a
similar relation in the absence of type IIB fluxes. We have also
kept a parameter $f$ in \floptra. Therefore,
the transformation \floptra\ will convert our case \mmetgec\ to:
\eqn\afterflop{ds^2_{\rm Flop} = -\left[{\rm det}~N^{\lambda_1
\lambda_2}_{[\beta, 0]} - {\rm Tr}^{2}~(N^{\lambda_1
\lambda_2}_{[\gamma, 0]}\cdot \Gamma_3)\right] - \left[{\rm
det}~N^{\lambda_1 \lambda_2}_{[f\delta, \delta]} - {\rm
Tr}^{2}~(N^{\lambda_1 \lambda_2}_{[f \epsilon, \epsilon]}\cdot
\Gamma_3)\right]} upto possible rescaling, and $\beta, \gamma,
\delta$ and $\epsilon$ have already been given in \bacvalve.

In the limit $f \to 0$ the above metric gives the right sphere
part but fails to give the fibration structure correctly. This
implies that a global definition {\it a-la} \floptra\ may not be
possible here. What went wrong? A careful study of \afterflop\
reveals that in the summation of the one forms $\Sigma_a -
f~\sigma_a, ~a = 1, 2,3$ we had used the same $f$ for all the
three terms. In the limit where $f \to 0$ this gives the right
sphere metric but wrong fibration. A way out of this can be
immediately guessed by having a different factor in the third
term, i.e. having $\Sigma_3 - g~\sigma_3$, and the rest with $f$.
Locally, this is exactly the one predicted by \brandtwo, and
therefore, using out patch argument, we can extend this to all
other patches. This implies the following metric after we make a
flop in M-theory: \eqn\afterflopagain{\eqalign{ds^2_{\rm Flop} = &
-\left[{\rm det}~N^{\lambda_1 \lambda_2}_{[\beta, 0]} - {\rm
Tr}^{2}~(N^{\lambda_1 \lambda_2}_{[\gamma, 0]}\cdot
\Gamma_3)\right] \cr & - \left[{\rm det}~N^{\lambda_1
\lambda_2}_{[f\delta, \delta]} + {\rm Tr}^{2}~(N^{\lambda_1
\lambda_2}_{[{f \delta \o 2}, {\delta \o 2}]}\cdot \Gamma_3) -
{\rm Tr}^{2}~(N^{\lambda_1 \lambda_2}_{[{g \alpha_3 \o 2},
{\alpha_3 \o 2}]}\cdot \Gamma_3) \right]^{g \to 1}_{f \to 0}}}
where the variables have been defined in \bacvalve\ and
\warpyiden. Thus, before flop the metric is given by \mmetgec\
and after flop it is given by \afterflopagain.

\subsec{The Type IIA Background}

To obtain the type IIA theory we can reduce either  via $dz$ or
via $dx_{11}$. This will not lead us back to the type IIA theory
we started with because of the change induced in the metric by
the flop. To have a one to one correspondence with the type IIA
picture before flop, let us reduce along direction $dx_{11}$. The
new $G_2$ metric can now be written in the following suggestive
way: \eqn\gtwosuggest{\eqalign{ds^2 & =  e^{4\phi \o 3}\left[dz +
\Delta_1~{\rm cot}~\hat\theta_1~(dx - b_{x\theta_1}~d\theta_1) +
\Delta_2 ~{\rm cot}~\hat\theta_2~(dy -
b_{y\theta_2}~d\theta_2)\right]^2  \cr +& e^{-{2\phi \o
3}}\left({g_2\o 2} - {g_4 \o 4 \xi}\right)\left[d\theta_1^2 + (dx
- b_{x\theta_1}~d\theta_1)^2\right]  + e^{-{2\phi \o
3}}\left({g_2\o 2} + {g_4 \o 4\xi}\right)\left[d\theta_2^2 + (dy -
b_{y\theta_2}~d\theta_2)^2\right] \cr & ~~~~~ + {1\o 4}
e^{-{2\phi \o 3}}~g_1~\left[dx_{11} + 2\Delta_1~ {\rm
cot}~\hat\theta_1~(dx - b_{x\theta_1}~d\theta_1)\right]^2}} which
clearly shows that the base is locally a resolved conifold.
Note, that we have again used the freedom to absorb $A_z$ into $dx_{11}$.
The
$dx_{11}$ term in \gtwosuggest\ is basically the fibration over
which we have to reduce to get to type IIA theory. As mentioned
above, we can also reduce along $dz$, as the $dx_{11}$ and $dz$
directions can be easily exchanged among each other. The metric
\gtwosuggest\ is thus the right $G_2$ metric after flop and could
be compared to \metsugges. A redefinition of the coordinates of
\metsugges\ and some coordinate transformation would relate
\metsugges\ to \gtwosuggest\ and would also simplify the form of
the gauge potential given earlier in \comgaud. The final type IIA
metric after a dimensional reduction turns out to be:
\eqn\iiafinal{\eqalign{ds^2 = & {1\o 4}\left(2g_2 - {g_4\o
\xi}\right)\left[d\theta_1^2 + (dx - b_{x\theta_1}~d\theta_1)^2
\right]  + {1\o 4} \left(2g_2 + {g_4\o
\xi}\right)\left[d\theta_2^2 + (dy -
b_{y\theta_2}~d\theta_2)^2\right]  \cr & + e^{2\phi}\left[dz +
\Delta_1~{\rm cot}~\hat\theta_1~(dx - b_{x\theta_1}~d\theta_1) +
\Delta_2 ~{\rm cot}~\hat\theta_2~(dy -
b_{y\theta_2}~d\theta_2)\right]^2}} which is precisely the metric
of a resolved conifold when we switch off $b_{x\theta_1}$ and
$b_{y\theta_2}$ (or consider it locally over a patch where
$b_{x\theta_1}$ and $b_{y\theta_2}$ are constants). In the
presence of $b_{x\theta_1}$ and $b_{y\theta_2}$ we get the
``usual'' metric but shifted by the generic ansatz that we
proposed in \redpsietc. Therefore, we can now make a precise
statement: {\it the metric before geometric transition is given by
\fiiamet, and after the transition is given by \iiafinal}. The
type IIA metric has two spheres whose radii are proportional to
\eqn\radoft{r_1 = {1\o 2} \sqrt{2g_2 - g_4\sqrt{g_3 g^{-1}_2}}, ~~~~
r_2 = {1\o 2}
\sqrt{2g_2 +g_4\sqrt{g_3 g^{-1}_2}}.}
One of them would shrink to zero
size while the other doesn't when we approach the origin. The
type IIA coupling is now given by \eqn\iicofinal{ g_A = 2^{-{3\o
2}} e^{-{\phi \o 2}} \alpha^{-{3\o 4}}} where $\alpha$ is defined
in \defalpha. Observe that the coupling is not a constant but is
a function of the internal coordinates, and it doesn't blow up
anywhere in the internal space. This background is the expected
background after we perform a geometric transition on \fiiamet.
This means that the $D6$ branes in \fiiamet\ should completely
disappear and should be replaced by fluxes in the type IIA
picture. From the $G_2$ manifold that we had in \gtwosuggest, we
see that this is indeed the case, and the gauge fluxes are given
by: \eqn\gaufinalii{{\cal A}\cdot dX =  2\Delta_1~{\rm
cot}~\hat\theta_1~(dx - b_{x\theta_1}~d\theta_1)} which, as one
can easily check, looks like the remnant of $D6$ brane sources
modified appropriately by our ans\"atze \redpsietc. There are also
$B_{NS}$ fields that originate from the dimensional reduction of
the three form fields in M-theory. Since we are reducing along the
direction $dx_{11}$ they would be the same $\hat B$ field that we
had in \hatbnow. The only difference will be that the finite part
(which is of course $\hat B$ itself) is now the exact solution as
we had removed the $\epsilon^{-1/2}$ dependence by scaling our
$G_2$ manifold before flop. The $B_{NS}$ can be written down
directly from \hatbnow\ as:
 \eqn\hatbnowfinal{{B \o \sqrt{\alpha}} =
 dx \wedge d\theta_1 - dy \wedge d\theta_2 +
 A~d\theta_1\wedge dz - B~({\rm sin}~\psi ~dy -
{\rm cos}~\psi~d\theta_2) \wedge dz,} which will again be a pure gauge artifact.
Combining \iiafinal,
\iicofinal, \gaufinalii\ and \hatbnowfinal, we recover the
precise background after geometric transition in type IIA picture.

\subsec{Analysis of Type IIA Background and Superpotential}

In this section we will try to verify the non-K\"ahler nature of
our background and the corresponding superpotential. Other detail
aspects, for example non integrability of complex structure,
torsion classes etc., will be left for part II of this paper. To
check the non-K\"ahlerity of this background we will have to
determine the corresponding vielbeins. They can be easily
extracted from \iiafinal, and are given by:
\eqn\vielsfinal{\eqalign{e & = \pmatrix{e^1_x & e^1_y & e^1_z &
e^1_{\theta_1}& e^1_{\theta_2}& e^1_r \cr \noalign{\vskip -0.20
cm}  \cr e^2_x & e^2_y & e^2_z & e^2_{\theta_1}& e^2_{\theta_2}&
e^2_r \cr \noalign{\vskip -0.20 cm}  \cr e^3_x & e^3_y & e^3_z &
e^3_{\theta_1}& e^3_{\theta_2}& e^3_r \cr \noalign{\vskip -0.20
cm}  \cr e^4_x & e^4_y & e^4_z & e^4_{\theta_1}& e^4_{\theta_2}&
e^4_r \cr \noalign{\vskip -0.20 cm}  \cr e^5_x & e^5_y & e^5_z &
e^5_{\theta_1}& e^5_{\theta_2}& e^5_r \cr \noalign{\vskip -0.20
cm}  \cr e^6_x & e^6_y & e^6_z & e^6_{\theta_1}& e^6_{\theta_2}&
e^6_r} \cr \noalign{\vskip -0.20 cm}  \cr & = \pmatrix{0 & 0 & 0
& r_1 & 0 & 0 \cr \noalign{\vskip -0.20 cm}  \cr
 r_1 & 0 & 0 & -r_1~b_{x\theta_1} & 0 & 0 \cr
\noalign{\vskip -0.20 cm}  \cr 0 & 0 & 0 & 0 & r_2 & 0 \cr
\noalign{\vskip -0.20 cm}  \cr 0 &  r_2 & 0 & 0 &
-r_2~b_{y\theta_2} & 0 \cr \noalign{\vskip -0.20 cm}  \cr e^\phi
\Delta_1~ {\rm cot}~\hat\theta_1 & e^\phi \Delta_2~ {\rm
cot}~\hat\theta_2  & e^\phi & - e^\phi \Delta_1~ {\rm
cot}~\hat\theta_1 b_{x\theta_1}  & - e^\phi \Delta_2~{\rm
cot}~\hat\theta_2 b_{y\theta_2} & 0 \cr \noalign{\vskip -0.20
cm}  \cr
 0 & 0 & 0 & 0 & 0 & e^6_r}}}
where $r_1$ and $r_2$ are the radii of the two spheres as defined
earlier. We have kept $e^6_r$ undefined here. But this can also
be easily seen to be the usual vielbein for the resolved conifold
case in the type IIB picture. Now to check the non-K\"ahlerity we
have to construct the fundamental two form ${\cal J}$ using these
vielbeins. Before evaluating this, observe that in the absence of
$b_{x\theta_1}$ and $b_{y\theta_2}$ the manifold should be
K\"ahler with a K\"ahler form $J$. In the presence of
$b_{x\theta_1}$ and $b_{y\theta_2}$ the fundamental form ${\cal
J}$ can be written as a linear combination of the usual K\"ahler
form $J$ and additional  $b_{x\theta_1}$ and $b_{y\theta_2}$
dependent terms, as \eqn\twoform{{\cal J} = J +
e^{\phi}~(\Delta_1~{\rm cot}~\hat\theta_1~b_{x\theta_1}~e^6_r
\wedge d\theta_1 +
 \Delta_2~{\rm cot}~\hat\theta_2~b_{y\theta_2}~e^6_r \wedge d\theta_2)}
where $dJ = 0$. From above it is easy to see that $d{\cal J} \ne
0$ in general  because of non- zero $db_{x\theta_1}$ and
$db_{y\theta_2}$. Therefore the manifold \iiafinal\ is a
non-K\"ahler manifold. For completeness, let us also write down
all the components of the type IIA metric:
\eqn\twoacomp{\eqalign{g & = \pmatrix{g_{xx} & g_{xy} & g_{xz} &
g_{x\theta_1} & g_{x\theta_2} \cr \noalign{\vskip -0.20 cm}  \cr
g_{xy} & g_{yy} & g_{yz} & g_{y\theta_1} & g_{y\theta_2} \cr
\noalign{\vskip -0.20 cm}  \cr g_{xz} & g_{yz} & g_{zz} &
g_{z\theta_1} & g_{z\theta_2} \cr \noalign{\vskip -0.20 cm}  \cr
g_{x\theta_1} & g_{y\theta_1} & g_{z\theta_1} &
g_{\theta_1\theta_1}& g_{\theta_1\theta_2} \cr \noalign{\vskip
-0.20 cm}  \cr g_{x\theta_2} & g_{y\theta_2} & g_{z\theta_2} &
g_{\theta_1\theta_2} & g_{\theta_2\theta_2}} \cr \noalign{\vskip
-0.25 cm}  \cr & = \pmatrix{C_1 & e^{2\phi} AB & e^{2\phi} A &
-b_{x\theta_1} C_1 & -e^{2\phi} b_{y\theta_2} AB \cr
\noalign{\vskip -0.20 cm}  \cr
 e^{2\phi} AB & D_1 & e^{2\phi} B & -e^{2\phi} b_{x\theta_1} AB
 & -b_{y\theta_2} D_1 \cr
\noalign{\vskip -0.20 cm}  \cr e^{2\phi} A & e^{2\phi} B  &
e^{2\phi} &  -e^{2\phi} b_{x\theta_1} A & -e^{2\phi}
b_{y\theta_2} B \cr \noalign{\vskip -0.20 cm}  \cr -b_{x\theta_1}
C_1 &  -e^{2\phi} b_{y\theta_2} AB & -e^{2\phi} b_{x\theta_1} A &
C + b^2_{x\theta_1}C_1 & e^{2\phi} AB b_{x\theta_1} b_{y\theta_2}
\cr \noalign{\vskip -0.20 cm}  \cr
 -e^{2\phi} b_{y\theta_2} AB & -b_{y\theta_2} D_1  & -e^{2\phi}
 b_{y\theta_2} B  & e^{2\phi} AB b_{x\theta_1} b_{y\theta_2}  &
D + b^2_{y\theta_2}D_1}}} where $A$ and $B$ have been defined
earlier in \defAandB. The other
 variables appearing in \twoacomp\ can be defined as follows:
\eqn\defiiac{C_1 = C + A^2~e^{2\phi}, ~~ D_1 = D + B^2~e^{2\phi},
~~ C = {g_2\o 2} - {g_4 \o 4\xi}, ~~ D = {g_2\o 2} + {g_4 \o
4\xi}} Thus, \twoacomp\ is the final answer for the type IIA
background without any $D6$ branes and with two-- and three--form
field strengths. There are many questions that arise from the
explicit background that we have in \iiafinal\ and \twoacomp. Let
us elaborate them:

\noindent $\bullet$ The first issue is related to the choice
of complex structure for our manifold. The complex structure is
written in terms of the fermions, and therefore we have to see
how the fermions transform under three T-dualities. From the
generic analysis of \kachruone, we see that the T-dual fermions
give rise to a complex structures that is in general not integrable
(in other words, the Nijenhaus tensor does not vanish). Therefore we
will get a non-complex manifold.

\noindent $\bullet$ The next issue is related to the
non-K\"ahlerity of our manifold. The naive expectation
(also from the results of \louis) would be that the manifold
we get in type IIA will be
{\it half-flat}. This comes from the fact that \iiafinal\ is
non-K\"ahler and also non-complex. Half-flat manifolds are
classified by torsion classes. For our case all these can be
explicitly derived from the metric. Below we will show that the
naive expectation is {\it not} realized in string theory and
our manifold will be more general than a half-flat manifold.

\noindent $\bullet$ The third issue is the asymptotic behavior of
our metric. We haven't yet checked whether the metric that we derived above is
non-degenerate and non-singular. Although unrelated, a similar
metric with identical $B$ dependent fibration structure found in \sav,\bbdg,\bbdgs,
showed a good asymptotic behavior and was non-degenerate and non-singular. We expect
similar thing to happen here too, though a full study will be presented in the
sequel to this paper.

\noindent $\bullet$ Last but not the least, we need to determine
the superpotential that governs our type IIA background. Before
doing so, let us start by recalling which the fields are present in
M--theory. The NS field \hatbnow\ is lifted to a three-form
$C$ by adding a leg in the $x^{11}$ direction. Its derivative is
$G = d C$. For the compactification on a $G_2$ manifold X with
the invariant 3-form $\tilde\Omega$ defined earlier in \thrM,
the form of the superpotential was
first proposed in \gukov\  as: \eqn\potu{W = \int_{X} \tilde\Omega \wedge
G.} This form of the superpotential has been corrected in \ach,
\bw\ in order to make the right hand side a complex quantity. The
superpotential becomes\foot{Of course, the superpotential is
for a {\it generic} $G_2$ structure. If the structure group is a
subgroup of $G_2$ then the superpotential will be a truncation of the one
that we mention here. These details have been addressed recently in \bertwo.
We thank K. Behrndt for correspondence on this issue.}
\eqn\potc{W = \int_{X} ( \tilde\Omega + i C) \wedge
G.} Then, when reducing from 11 dimensions to 10 dimensions as in
\amv, \brandtwo, instead of just obtaining the volume of the
resolved conifold $J$, we get the complexified volume $J + i B$,
and this is the quantity that enters in the 10 dimensional potential
to give
\eqn\potzu{\tilde{W}_{1} = \int_{X_6}  (J + i B) \wedge d H_{3}.}
This is true because $\tilde\Omega$ descends to $J$ as it loses one leg
but $G$ descends as an RR 4-form. The RR 4-form should originate from
D4 branes. Since we know that our brane configurations did not
contain any D4 branes, there is no contribution from
$dH_{3}$ in the superpotential.

But this is not the full story because of the properties of the
compactification manifold. By considering the dimensional
reduction on the manifold \iiafinal, we have to use the fact that
the manifold does not have a closed (3,0) form. Let us first
recall the results for the case without torsion \tp, \pandoz. In
that case the condition that the (3,0) form is closed\foot{This
statement is equivalent to saying that the manifold is Ricci
flat.} is related to a  differential equation for a function
depending on the radial coordinate. The result was a one
parameter family of Calabi-Yau metrics on the resolved conifold.

In our case, the situation is different. We consider equation
\iiafinal\ and we read off the vielbeins from \vielsfinal.
The holomorphic 3-form is built as before as: \eqn\holof{\Omega =
(e_1 + i e_2) \wedge (e_3 + i e_4)  \wedge (e_5 + i e_6).} {}From
\holof\ we see that the condition for the (3,0) form to be closed
implies a differential equation which involves the functions
$b_{x \theta_{1}}, b_{y \theta_2}$, as they appear in
$\hat{x},~\hat{y},~\hat{z}$. As the functions  $b_{x \theta_{1}},
b_{y \theta_2}$ are arbitrary, the differential equation will not
have a solution for generic values of $b_{x \theta_{1}}, b_{y
\theta_2}$, so our situation is different from the one of \tp. It
also differs from the situation of \pandoz\ in the sense that our
manifold is not Ricci flat because the same differential equation
does not have a solution for generic values of $b_{x \theta_{1}},
b_{y \theta_2}$. For our case one can explicitly evaluate the
three-forms. Using the definitions of $d\hat x$ and $d\hat y$ we can
express our result as:
\eqn\ugapm{\eqalign{& {\Omega_+ \o \sqrt{4g_2^2 - {g_4^2 ~\xi^{-2}}}} =
e^6_r~d\theta_1 \wedge d\theta_2 \wedge dr - e^\phi~d \theta_1 \wedge
d\hat y \wedge (dz + \Delta_1~{\rm cot}~\hat\theta_1~d\hat x)~ + \cr
& ~~~~~~~~~~~~~~~~~ - e^6_r~d\hat x \wedge d\hat y\wedge dr + e^\phi~
d\theta_2 \wedge d\hat x \wedge (dz +
\Delta_2~{\rm cot}~\hat\theta_2~d\hat y)\cr
& {\Omega_- \o \sqrt{4g_2^2 - {g_4^2~\xi^{-2}}}} = e^\phi~d\theta_1 \wedge
d\theta_2 \wedge (dz + \Delta_1~{\rm cot}~\hat\theta_1~d\hat x +
\Delta_2~{\rm cot}~\hat\theta_2~d\hat y)~+ \cr
& ~~~~~~~~~~~~~~~~~ + e^6_r~(d\theta_1 \wedge d\hat y \wedge dr +
d\hat x \wedge d\theta_2 \wedge dr) - e^\phi~d\hat x \wedge d\hat y \wedge
dz}}
where $e^6_r$ is the associated vielbein for the $r$ direction.
{}From above we see that both $d\Omega_+$ and $d\Omega_-$ will not vanish
for the background that we have.
The existence of  $b_{x \theta_{1}}, b_{y \theta_2}$
implies the  non-closeness of the holomorphic 3-form $\Omega$.
Therefore our manifold is a specific non-complex, non-K\"ahler
manifold that is not half-flat.
Manifolds with an $\Omega$ which is not closed have been studied
in \louis\ where four forms $F^{2,2} \propto (d \Omega)^{2,2}$
correspond to harmonic forms measuring flux, and they can be
expanded in some basis\foot{For integrable complex structures one
could expand in $h^{1,1}$ basis, although if the manifold is
simultaneously non-K\"ahler this would be tricky. Recall also that we are
using $d\Omega \equiv d\left[\Omega^{(3,0)}\right] =
d\Omega^{(2,2)} + d\Omega^{(3,1)} +
d\Omega^{(4,0)}$ for the non-complex manifold.}.
The four
forms can then be combined with the independent holomorphic 2
forms to give contributions to the superpotential as
\eqn\extrasup{(J + i B) \wedge d \Omega.} This way we
encounter a first concrete example where the superpotential gets
an extra piece from the non  closed holomorphic 3-form. The case in
\louis\ involved an $\Omega$ with only the real part non--
closed and the manifold was a half flat manifold. Our case is
more general, as both $d \Omega_{+}$ and $d \Omega_{-}$ can be
non zero as functions of $b_{x \theta_{1}}, b_{y \theta_2}$.

\newsec{Discussion and future directions}

The subject of the present work was to clarify issues concerning
NS fluxes in geometric transitions and the non-K\"ahler geometries
arising in
the mirror pictures. Our starting point was the observation of
Vafa \vafai\ that the closed string dual to D6 branes wrapped
on a deformed conifold
is not a K\"ahler geometry. To obtain this mysterious departure from
K\"ahlerity, we started\foot{In terms of the dual ${\cal N} = 1$ gauge theory
this is the IR of the gauge theory. In terms of geometry this is the region
where $r$, the radial parameter, is small.}
from a IIB picture with D5 branes wrapped on a
$P^1$ cycle inside the resolved conifold and went to the mirror
picture by performing three T-dualities on the fiber $T^3$. The
result was a
non-K\"ahler geometry whose metric could be given precisely as a
non-K\"ahler deformation of a deformed conifold.
We then lifted this to M theory where the result was a new $G_2$ manifold with
torsion. A flop inside the $G_2$ manifold and a reduction to type IIA
brought us to the closed IIA picture with a non-K\"ahler deformation of the
resolved conifold. The latter non-K\"ahlerity can then be traced back
to the existence of the NS flux in the initial type IIB picture.
Our final result is not quite a half-flat manifold as anticipated
earlier \louis, but
is actually more general because the imaginary part of the
holomorphic 3-form is also not closed.

On the way we also solved some puzzles regarding the
T-duality between branes wrapped on the deformed and resolved conifold.
Previous attempts started from the deformed
conifold and the problems
encountered were related to the fact that this is not a toric variety.
We started with the resolved conifold which is a toric variety and
identified the $T^3$ fibration.

\subsec{Future directions}

There are many unanswered questions that we left for future work. A sample of
them are as follows:

\noindent $\bullet$ In the
figure we have drawn, there is an extra step which should be covered,
the mirror symmetry
that goes from the closed IIA to closed type IIB and it would be
interesting to see whether this would give rise to
 a K\"ahler geometry with
NS flux or to a non-K\"ahler geometry. Unfortunately, there is an
immediate problem that one would face while doing the mirror transformation.
The background that we have has lost the isometry along the $z$ direction.
Recall that the type IIB background that we started with in the beginning
of the duality chain had complete isometries along the $x, y$ and $z$
directions. In the final type IIA picture the metric does surprisingly have
all the isometries, but the $B$ field breaks it. Maybe a transformation
of the form \tranthe\ could be used here to get the mirror metric.

\noindent $\bullet$ To get the cross terms in the type IIA
mirror metric (on the
$D6$ brane side) we had used only a set of restricted coordinate
transformations which only lie on the two $S^3$ directions of the
corresponding type IIB picture. It will now be interesting to
see whether this could be generalized for the case where we could consider
$\delta r$ variations. In particular, for a particular $S^3$ parametrized
by ($\psi_1, \theta_1, x$), one should now consider both $\delta r$ and
$\delta \theta_2$ variations for a given variation of $x, \theta_1$ and
$\psi_1$. Similar discussion should be done for the other $S^3$ parametrized
by ($\psi_2, \theta_2, y$).

\noindent $\bullet$ In the type IIA mirror background we have $B$ fields
both before and after geometric transitions. On the $D6$ brane side, the
presence of a $B$ field amounts to having non-commutativity on the world
volume of $D6$ branes. However, this $B$ field is in general not a
constant, and therefore may not have such a simple interpretation. This is
somewhat related to a discussion on $C$-deformation in \ooguri. The
$C$-deformations in general violates Lorentz invariance. It will be
interesting to see if there is any connection to our result, or if the
precise background that we propose does indeed realize the Lorentz violation
and $C$-deformations.

\noindent $\bullet$  As discussed above, the manifold that we have
in 11 dimensions has a $G_2$
structure. However we haven't evaluated the holonomy of the
manifold and it should be
interesting to do so in order to check that the supersymmetry is preserved.
One immediate thing to check would be whether the manifold could
become complex. In other words whether the complex structure is
integrable or not.
This is an interesting question and can only be answered after we trace the
behavior of fermions when we do the mirror transformation.
Generic studies done
earlier have shown that in general these mirror manifolds {\it do not} have
an integrable complex structure. In addition to this, there is also the
question of the {\it choice} of the complex structure. Recall that in the
type IIB theory which we started out with,
the background fluxes generate a superpotential that {\it fixes} the complex
structure. This would
imply that the K\"ahler structures are all fixed in the type IIA picture.
On the other hand, if we start with a type IIB framework with, say,
$h^{1,1} =1$
one might also be able to fix the complex structure in the mirror  just
by fixing the K\"ahler structure in the type IIB side via (non-perturbative)
corrections to the type IIB superpotential. Thus the choice of
complex structure in type IIA will be uniquely fixed. It will be
important to see if the choice of complex structure that we made here is
consistent with the value fixed by the superpotential.

\noindent $\bullet$ There is another important aspect of geometric
transition that one needs to carefully verify. This has to do with
the disappearance of $D6$ branes when we perform the
geometric transition. Since $D6$ branes support gauge fluxes, the
disappearance of $D6$ branes would imply that after geometric transition
there cannot be any localized gauge fluxes. To show this aspect, the
M-theory lift will be very useful. Recall that the
world volume couplings (and interactions) of
$D6$ branes can be extracted from M-theory lagrangian using the normalizable
harmonic (1,1) form of the corresponding Taub-NUT space \imamura.
Now that we have fluxes and also a background non-K\"ahler geometry in
type IIA theory (or in other words a torsional $G_2$ manifold in M-theory)
the analysis of the normalizable harmonic form is much more complicated.
In the presence of fluxes this has been considered in \robbins. It was found
there that the harmonic forms themselves change by the backreaction
of the fluxes on geometry. It will now be important to evaluate this
harmonic form and show that it is normalizable.
This would allow a localized gauge
flux to appear in the type IIA scenario, proving that the $G_2$
metric is indeed the lift of the $D6$ brane on a non-K\"ahler geometry.
This harmonic
form should either vanish or become non normalizable {\it after}
we do a flop. This will show that we do not expect a localized
gauge flux in type IIA theory and therefore the $D6$ branes have
completely disappeared! This will confirm Vafa's scenario. However as
discussed in \brand, the issue of normalisable form is subtle here. In the
absence of torsion, the usual form is badly divergent even before the
transition. In the presence of torsion (or fluxes in the type IIB
theory) the back reaction of fluxes on geometry might make this
norm well behaved, as has been observed in \robbins\ in a different context.
This will be useful for comparison.

\noindent $\bullet$ The $G_2$ manifold is explicitly non-K\"ahler (because it
is odd dimensional), and appears from a fibration over another
six-dimensional non-K\"ahler manifold. As we saw earlier, this manifold
is also neither complex nor half-flat.
Thus we have a new $G_2$ manifold whose slice is a specific six dimensional
non-K\"ahler space. Therefore one should be able to use
Hitchin flow equations \hitchin\
to construct the $G_2$ manifold from the six dimensional non-K\"ahler
space. When the base is a half-flat manifold, this has already been done
in the work of Hitchin \hitchin. The flow equations take the following
form:
\eqn\flowe{ dJ = {\del \Omega_+ \o \del t}, ~~~~~ d\Omega_- =
- J \wedge {\del J \o \del t}}
where $t$ is a real parameter that determines the $SU(3)$ structure
of the base and is related to the vielben $e_7$.
For our case, since we know almost every detail of the
metric, it will be interesting to check whether the $G_2$ manifold
follows from the flow equations\foot{For some more details on Hitchin's flow
equations related to a generic lift of six-manifolds to seven-manifolds, the
readers may want to consult \dal. We thank G. Dall'Agata for correspondence
on this issue.}.

\noindent $\bullet$ The non-K\"ahler manifold that we get in type IIA theory
both before and after geometric transtions should fit in the
classification of torsion classes \salamon. Since they are more generic than
the half-flat manifolds, the torsion classes will be less constrained. It will
also be interesting to find the full $G_2$ structure for the M-theory
manifolds.

\noindent $\bullet$ The analysis that we performed in this paper starting with
$D5$ wrapped on resolution $P^1$ cycle of a resolved conifold  is only the
IR description of the corresponding gauge theory. The full analysis should
involve additional $D3$ branes in the type IIB picture, so that the cascading
behavior can be captured. However the cascade being an infinite sequence of
flop transitions, makes simple supergravity description of the full theory
a little more involved. It will be interesting to pursue this
direction to see how the type IIA mirror phenomena works. We hope to address
this issue in near future.

\vskip.2in

\centerline{\bf Acknowledgments}

Its our pleasure to thank Lilia Anguelova, Klaus Behrndt, Volker Braun,
Mirjam Cvetic,
Gianguido Dall' Agata,
Katrin Becker, Shamit Kachru, Amir-Kian Kashani-Poor, Dieter Luest,
Leopoldo Pando-Zayas and Cumrun Vafa
for many interesting discussions and
useful correspondences. The work of M.B. is supported by NSF grant
PHY-01-5-23911 and an Alfred Sloan Fellowship. The work of K.D.
is supported in part by a Lucile and Packard Foundation
Fellowship 2000-13856. A.K. would like to acknowledge support
from the German Academic Exchange Service (DAAD) and the
University of Maryland.~R.T. is supported by DOE Contract
DE-AC03-76SF0098 and NSF grant PHY-0098840.

\vskip.2in

\newsec{\bf Appendix 1: Algebra of $\alpha$}

In the earlier sections we have defined $\alpha$ as
$\alpha = {1\o 1 + A^2 + B^2}$ where $A,B$ are given in \defAandB. This
quantity $\alpha$ is a very crucial quantity as the finite transformation
depends on it. While integrating the finite shifts we saw that we need to
approximate the transformation on a particular sphere, parametrised by
($\psi_i, x_i, \theta_i$), as though the other sphere components are
constants. In other words, to study the transformation on, say, sphere 1
we take the $\theta_2$ terms as a constant (or fixed at an average
value). In other words, any generic $\theta_i$ will be denoted as
\eqn\genthtea{\theta_1 = \langle\theta_1\rangle
+ \vartheta_1, ~~~~ \theta_2 =
\langle\theta_2\rangle + \vartheta_2}
where $\langle\theta_{1,2}\rangle$
are the average values of the $\theta$ coordinates.
To see how the $\alpha$ factor responds to this, let us first
define three useful quantities:
\eqn\thrusefi{\eqalign{&\langle \alpha \rangle_1 =
{1 \o 1 + \Delta^2_1~{\rm cot}^2~\theta_1 +
\Delta^2_2~{\rm cot}^2~\langle\theta_2\rangle}, ~~
\langle \alpha \rangle_2 = {1 \o 1 + \Delta^2_1~{\rm cot}^2~
\langle\theta_1\rangle +
\Delta^2_2~{\rm cot}^2~\theta_2}\cr
& ~~~~~~~~~~~~~~~~~ \langle\alpha\rangle~ = {1 \o 1 +
\Delta^2_1~{\rm cot}^2~\langle\theta_1\rangle +
\Delta^2_2~{\rm cot}^2~\langle\theta_2\rangle}}}
Using the above definitions, one can easily show that $\sqrt{\alpha}$ has the
following expansion:
\eqn\alpexpa{\sqrt{\alpha} = \sqrt{\langle \alpha \rangle_1} \left( 1 +
{\rm x}~\Delta_2^2~\langle \alpha \rangle_1 ~
{\rm cot}^2~\langle\theta_2\rangle \right)^{-{1\o 2}}}
where we have kept the radial variations as constant as before, and the
quantity x appearing above being given by the following exact expression:
\eqn\valx{{\rm x} = -{4~ {\rm cosec}~2
\langle\theta_2\rangle~{\rm tan}~\vartheta_2~( 1
+ {\rm cot}~2\langle\theta_2\rangle~
{\rm tan}~\vartheta_2) \o 1 + {\rm cot}~\langle\theta_2\rangle
~{\rm tan}~\vartheta_2~ (2 + {\rm cot}~
\langle\theta_2\rangle~{\rm tan}~\vartheta_2)}.}
If the warp factor $\Delta_2$ is chosen in such a way that in \alpexpa\ the
quantity in the bracket is always small, then $\alpha$ will have the
following expansions at all points in the internal space:
\eqn\alepdg{\sqrt{\alpha} = \sqrt{\langle \alpha \rangle_1} - {1\o 2} ~{\rm x}~
\Delta_2^2~{\rm cot}^2~
\langle\theta_2\rangle~\langle \alpha \rangle_1^{3/ 2} + ....}
and therefore could be approximated
simply as $\sqrt{\langle \alpha \rangle_1}$. Similar
argument will go through for $\langle \alpha \rangle_2$ for the other sphere.
Another
alternative way to write $\alpha$ is
\eqn\altalp{\sqrt{\alpha} = {\langle\alpha\rangle \o \sqrt{\langle \alpha
\rangle_1}} + ... = {\langle\alpha\rangle \o \sqrt{\langle \alpha \rangle_2}}
+ ...}
where again the dotted terms can be easily determined for different spheres.
The above two pair of expressions:
$\sqrt{\alpha} = \sqrt{{\langle\alpha\rangle}_{1,2}}$ and
$\sqrt{\alpha} =
{\langle\alpha\rangle \o \sqrt{{\langle\alpha\rangle}_{1,2}}}$
are responsible for
the two different set of coordinate transformations in \coordinate\ and
\aleb\ with $m = \pm 1$.
Observe also that under the above approximations, some of the
components of the deformed conifold metric will now look like:
\eqn\defcnop{\eqalign{ds^2_{\theta_1 \theta_2} &= -2 \sqrt{\langle \alpha
\rangle_1 \langle \alpha \rangle_2}~j_{xy}~d\theta_1~d\theta_2,\cr
ds^2_{\theta_2 \theta_2} &= \langle \alpha \rangle_2 (1 + A_1^2)~d\theta_2^2,
\cr
ds^2_{\theta_1 \theta_1} &= \langle \alpha \rangle_1 (1 + B_1^2)~d\theta_1^2}}
where $A_1$ and $B_1$ are the values of $A$ and $B$ at the average values of
$\theta_1$ and $\theta_2$ respectively.

\vskip.2in

\newsec{\bf Appendix 2: Details on $G_2$ structures}

The threeform $\Omega$ that we described earlier in the context
of type IIA manifold can be fixed by a subgroup
$SU(3)$ of $SO(6)$. In fact both $J$,
the fundamental form and $\Omega$ can be fixed simultaneously by $SU(3)$. The
torsion classes that we mentioned earlier are basically the measure of the
non-closedness of $\nabla J$ and $\nabla \Omega$. Detailed discussions on this
are in \grayone, \salamon. As we saw earlier, the type IIA manifold is neither
K\"ahler nor complex. Therfore the torsion is generic. When the complex
structure becomes integrable the torsional connection is known as Bismut
connection \bismut.

Similarly the threeform $\tilde\Omega$ can be fixed by a
subgroup of $GL_7$. This is the exceptional Lie group $G_2$ which is a
compact simple Lie subgroup of $SO(7)$ of dimension 14. The existence of a
$G_2$ structure is equivalent to the existence of the fundamental three form
$\tilde\Omega$. The following interesting cases have been studied in the
literature for the torsion free $G_2$ case (for a more detailed review on this
the reader may look into the last reference of \ivan. A short selection
on $G_2$ manifolds are in \salamontwo, \monar, \kath, \joyce):

\noindent $\bullet$ When
$\nabla \tilde\Omega = 0$, then holonomy is contained in
$G_2$ with a Ricci flat $G_2$ metric \bonan.

\noindent $\bullet$  When $d\tilde\Omega = d \ast \tilde\Omega = 0$ then the
fundamental form is harmonic and the corresponding $G_2$ manifold is
called parallel. First examples were constructed in \salamontwo. Later
on, first compact examples were given in \joyce, \kovalev.

\noindent $\bullet$ When
$\varphi \equiv \ast(\tilde\Omega \wedge \ast d\tilde\Omega) =0$
then the $G_2$ structure is known to be balanced, where $\varphi$ is known as
the Lee form. If the Lee form is closed, then the $G_2$ structure is locally
conformally equivalent to a balanced one.
When $d\ast \tilde\Omega = \varphi \wedge \tilde\Omega$ then
the $G_2$ structure is called integrable.

In the presence of torsion we have already described the changes that one
would expect from the above mentioned conditions. In terms of the torsion
three form $\tau$ given in \torthrf,
the connection shifts from $\nabla$ mentioned above
to the usual expected form $\nabla + {1\o 2} \tau$. Now the manifold is no
longer Ricci flat and the curvature tensor takes the following form:
\eqn\curva{{\cal R}_{ijkl} = R_{ijkl} - {1\o 2} \tau_{ijm}\tau_{klm}
-{1\o 4} \tau_{jkm}\tau_{ilm} - {1\o 4} \tau_{kim} \tau_{jlm}}
where $R$ measures the curvature wrt the Riemannian connection. For the
case when the base is half-flat, a construction of $G_2$ manifolds satisfying
some of the features mentioned above is given in \ivan. For our case we
do not have a half-flat base, and therefore the manifold that we presented in
sections 6 and 7 are new examples of $G_2$ manifolds with torsion that
satisfy all the string equations of motion.

\listrefs

\bye